\documentclass[%
 reprint,
 superscriptaddress,
noshowpacs,nopreprintnumbers,
nobibnotes,
 amsmath,amssymb,
 aps,
pra,
floatfix,
a4paper,
]{revtex4-1}

\pdfoutput=1

\usepackage{graphicx}
\usepackage{dcolumn}
\usepackage{bm}
\usepackage{hyperref}
\usepackage{longtable}
\usepackage{multirow}

\def\aj{AJ}
\def\apj{ApJ}
\def\araa{ARA\&A}
\def\aap{A\&A}
\def\mnras{MNRAS}

\makeatletter
\def\p@subsection{}
\makeatother

\renewcommand{\vec}[1]{\mathbf{#1}}

\graphicspath{{fig/}}

\begin{document}
\noindent
\textsf{Astronomy Reports}\\

\title{Frequency-Dependent Core Shifts in Ultracompact Quasars}

\author{\firstname{P.~A.}~\surname{Voitsik}}
\email[E-mail: ]{voitsik@asc.rssi.ru}
\affiliation{Astro Space Center, P. N. Lebedev Physical Institute, Russian
Academy of Sciences, ul. Profsoyuznaya 84/32, Moscow, 117997, Russia}

\author{\firstname{A.~B.}~\surname{Pushkarev}}
\affiliation{Crimean Astrophysical Observatory, Russian Academy of
Sciences, Nauchnyi, Crimea, 298409, Russia}
\affiliation{Astro Space Center, P. N. Lebedev Physical Institute, Russian
Academy of Sciences, ul. Profsoyuznaya 84/32, Moscow, 117997, Russia}

\author{\firstname{Y.~Y.}~\surname{Kovalev}}
\affiliation{Astro Space Center, P. N. Lebedev Physical Institute, Russian
Academy of Sciences, ul. Profsoyuznaya 84/32, Moscow, 117997, Russia}
\affiliation{Moscow Physical Technical Institute, Institutskii proezd 9,
Dolgoprudnyi, Moscow oblast', 141700, Russia}
\affiliation{Max Planck Institut for Radio Astronomy, 69 Auf dem H\"ugel,
D-53121 Bonn, Germany}

\author{\firstname{A.~V.}~\surname{Plavin}}
\affiliation{Astro Space Center, P. N. Lebedev Physical Institute, Russian
Academy of Sciences, ul. Profsoyuznaya 84/32, Moscow, 117997, Russia}
\affiliation{Moscow Physical Technical Institute, Institutskii proezd 9,
Dolgoprudnyi, Moscow oblast', 141700, Russia}

\author{\firstname{A.~P.}~\surname{Lobanov}}
\affiliation{Max Planck Institut for Radio Astronomy, 69 Auf dem H\"ugel,
D-53121 Bonn (Endenich), Germany}
\affiliation{Institute for Experimental Physics, University of Hamburg,
Luruper Chaussee 149, D-22761 Hamburg, Germany}

\author{\firstname{A.~V.}~\surname{Ipatov}}
\affiliation{Institute of Applied Astronomy, Russian Academy of Sciences,
Kutuzov Embankment 10, St. Petersburg, 191187, Russia}


\begin{abstract}
Results of a pilot project with the participation of the ``Kvazar-KVO''
radio interferometry array in observations carried out with the European 
VLBI Network are presented.  The aim of the project was to conduct and analyze
multi-frequency (1.7, 2.3, 5.0, 8.4~GHz) observations of the parsec-scale 
jets of 24 active galactic nuclei. Three observing sessions were successfully
carried out in October 2008. Maps of the radio intensity distributions
have been constructed in all four frequencies using phase referencing.
A method for measuring the frequency-dependent shift of the position of
the VLBI core by applying relative astrometry to observations of close 
triplets of radio sources has been developed. The fundamental possibility
of detecting core shifts in ultra-compact sources for which traditional
methods based on the achromatic positions of optically thin regions of
the jet are not suitable is demonstrated. The conditions for successful
measurement of this shift are discussed; these are determined by the 
closeness of the calibrator used, the effective resolution of the system,
the quality of the filling of the $uv$ plane, the relative orientations of
the jets in the triplets, and the brightnesses of the sources.
\end{abstract}

\maketitle

\section{INTRODUCTION}
In images of extragalactic relativistic jets obtained using Very Long Baseline
Interferometry (VLBI), the ``core'' is the name given to the compact,
bright feature at the visible base of the jet. The position of the core
is determined by absorption in the radiating plasma (synchrotron
self-absorption) or in the ambient material \cite{Blandford_Konigl_1979,Konigl_1981,Lobanov_1998}.
At any given observing frequency
$\nu$, the core is located in the region of the jet with optical depth
$\tau(\nu) \approx 1$, leading to a shift in the absolute position of the
core $r_\text{core}\propto \nu^{-1/k_{\text{r}}}$ \cite{Lobanov_1998}. In the case of
synchrotron self-absorption with equipartition between the energy densities
of the relativistic particles and the magnetic field, $k_\text{r} = 1$ \cite{Blandford_Konigl_1979}.
However, in the presence of external absorption or gradients in the density
and pressure, $k_\text{r}$ can differ from unity \cite{Lobanov_1998}.

The apparent shifts of the cores of active galactic nuclei (AGNs) have
direct astrophysical and astrometric applications in relation to compact
radio sources. This effect can be used to estimate various physical
parameters of compact relativistic jets. At the same time, the shift of the
core with frequency can influence measurements and estimates based on
multi-frequency VLBI observations: (1)~the construction of spectral-index
maps~\cite{Lobanov_1998,Kovalev_2008}, (2)~measurements of Faraday
rotation~\cite{Hovatta_2012,Krav2016,Krav2017},
(3)~astrometric and geophysical measurements at 4 and 13~cm~\cite{Ma_1998,Petrov2009}, and
(4)~comparison of radio and optical coordinate systems~\cite{PK_letter2017,KPP2017,PK2017}.

Understanding and taking into account the effects of absorption in such
investigations requires systematic studies of core shifts for a
representative sample of compact radio sources, in particular, those used
for astrometric applications. To achieve this, we organized a pilot experiment
on the European VLBI network (EVN), with the following aims: measuring
the core shifts of ultracompact quasars via relative astrometry for selected
triplets of sources, comparison with the results obtained for traditional
methods based on comparison of optically thin parts of the jets observed at
different frequencies, gaining experience in the organization of larger-scale
astrometric measurements of core shifts in the future, testing the
participation of telescopes in the Russian ``Kvazar-KVO'' array in EVN
observations and estimating the gain provided by this participation.

\section{OBSERVATIONS AND DATA REDUCTION}

\subsection{Source Sample}

We selected eight ultracompact extragalactic radio sources for these
observations, meeting the following criteria: (1) that the source be
included in the main list of objects in the catalog of the International
Celestial Reference Frame (ICRF)~\cite{Ma_1998}; (2) that the structure index of
the source be equal to 1 or 2~\cite{Ma_1998}, i.e., that the source structure be
dominated by the core; (3)~that the maximum correlated spectral flux density
of the source at 8~GHz exceeds 1.2~Jy. Based on these criteria, we also
added the source 1749+096, which is the most compact and bright of the
candidate ICRF sources~\cite{Ma_1998}. We also selected two phase calibrators at angular
separations of no more than $4^{\circ}$ for each of these sources.

\subsection{EVN Observations}

Observations of our targets were carried out on the EVN in three 12-hour
sessions during October 2008: October 19--20 at S and X bands (central
frequencies 2.27 and 8.38~GHz), October 22--23 at C band (4.97~GHz), and
October 29--30 at L band (1.66~GHz). At each frequency, we recorded eight
frequency channels (so-called IFs), each with a bandwidth of 8~GHz. Both
right- and left-circular polarizations were recorded in L and C bands, while
only right-circular polarization was recorded in S and X bands. The aggregate
bit rate was 512~Mbits/s. The data were correlated at the Joint Institute for
VLBI in Europe (JIVE).

Three 32-m telescopes of the Russian ``Kvazar-KVO'' VLBI network took
part in these EVN observations, appreciably improving the $uv$ coverage
in the East--West direction (see Section~\ref{s:kvazar}). However, the failure
of the Hartebeesthoek telescope in South Africa significantly limited
the North--South resolution of the interferometric array. Unfortunately,
this had a very substantial effect on the accuracy of the astrometric
measurements, since the baselines to Hartebeesthoek would have provided
the highest angular resolution in the resulting data.

\subsection{Data Reduction}

The preliminary reduction of the data was carried out in the AIPS package~\cite{Greisen_2003},
and included the following steps: removal of bad data based on information
received from the telescopes and a visual inspection of the data; application
of phase corrections related to the passage of the radio signal through the
ionosphere, carried out using the task TECOR; calibration of the amplitudes
using system temperatures and gain curves measured at the telescopes
using the task APCAL; preliminary calibration of the phases using a global
fringe-fitting procedure (the task FRING). Independent solutions for the
group delays and fringe frequencies were found for each frequency channel (IF).
Corrections for the shape of the complex passband were applied using the task
BPASS. Further, the task SPLIT was used to apply all the derived corrections
to the data, average over frequency within each IF, and export the $uv$ data
in a format suitable for the subsequent analysis.

Maps for all the sources were constructed using the CLEAN algorithm
realized in the \emph{Difmap} package~\cite{Shepherd_1997}. Global amplitude corrections were
obtained for each frequency channel and each antenna by comparing the total
intensity CLEAN model with the initial calibrated data. The resulting
corrections were averaged over all the sources. Amplitude corrections
${>}10\%$ were then applied to the input data using the task CLCOR in AIPS.

The final step in the reduction of the data in AIPS was realizing a
phase-referencing regime relating the phases for the weaker sources in each
triplet to those of a calibrator. We chose the brightest, most compact source
in each triplet as the phase calibrator. Further, we found a phase solution
for each such calibrator using the task FRING, taking into account the source
structure, which was then applied to both the data for the calibrator itself
and the data for the two other sources in the triplet.

\subsection{Reconstruction of the VLBI Maps}

The final VLBI maps for the 24 target sources obtained using natural weighting
of the data are presented in Fig.~\ref{fig:maps}. Contour maps at 1.7, 2.3, 5.0, and
8.4~GHz are presented for each source. The dynamic ranges of these maps,
defined as the ratio of the intensity peak to the noise level in the map,
vary from $10^3$ for weak sources to $10^4$ for the brightest sources at
1.7, 5.0, and 8.4~GHz. The dynamic ranges of the maps at 2.3~GHz are lower,
especially for sources weaker than 1~Jy, since the phased Westerbork Synthesis
Radio Telescope~--- one of the most sensitive elements of the EVN~--- did not
take part in the observations at this frequency. The typical noise level
of the reconstructed images is 0.3~mJy/beam.

The parameters of the VLBI maps are summarized in Table~\ref{tab:maps}, namely, the source
name, central frequency of the synthesized image in GHz, peak intensity in
mJy/beam, residual noise level in mJy/beam, total flux in mJy, determined as
the sum of all CLEAN components, major and minor axes corresponding to the
full-width at half-maximum of the restoring beam in milliarcseconds (mas), and
the position angle of the beam in degrees.

\subsection{Modeling of the Sources}
\label{s:modeling}

In order to determine the position of the core in the map, we modeled the
brightness distribution for each source using a set of circular Gaussian
components. The model was fit to the data in the spatial-frequency domain
using the \emph{Difmap} program via $\chi^2$ minimization. The number of
components was chosen to be the minimum necessary to describe all significant
detected elements of the source structure at a given frequency (usually three
to six).

\subsection{Importance of the Kvazar-KVO Telescopes for the EVN Results}
\label{s:kvazar}

\begin{figure}
 \includegraphics[width=\linewidth,trim=0cm 0.8cm 0cm 0cm]{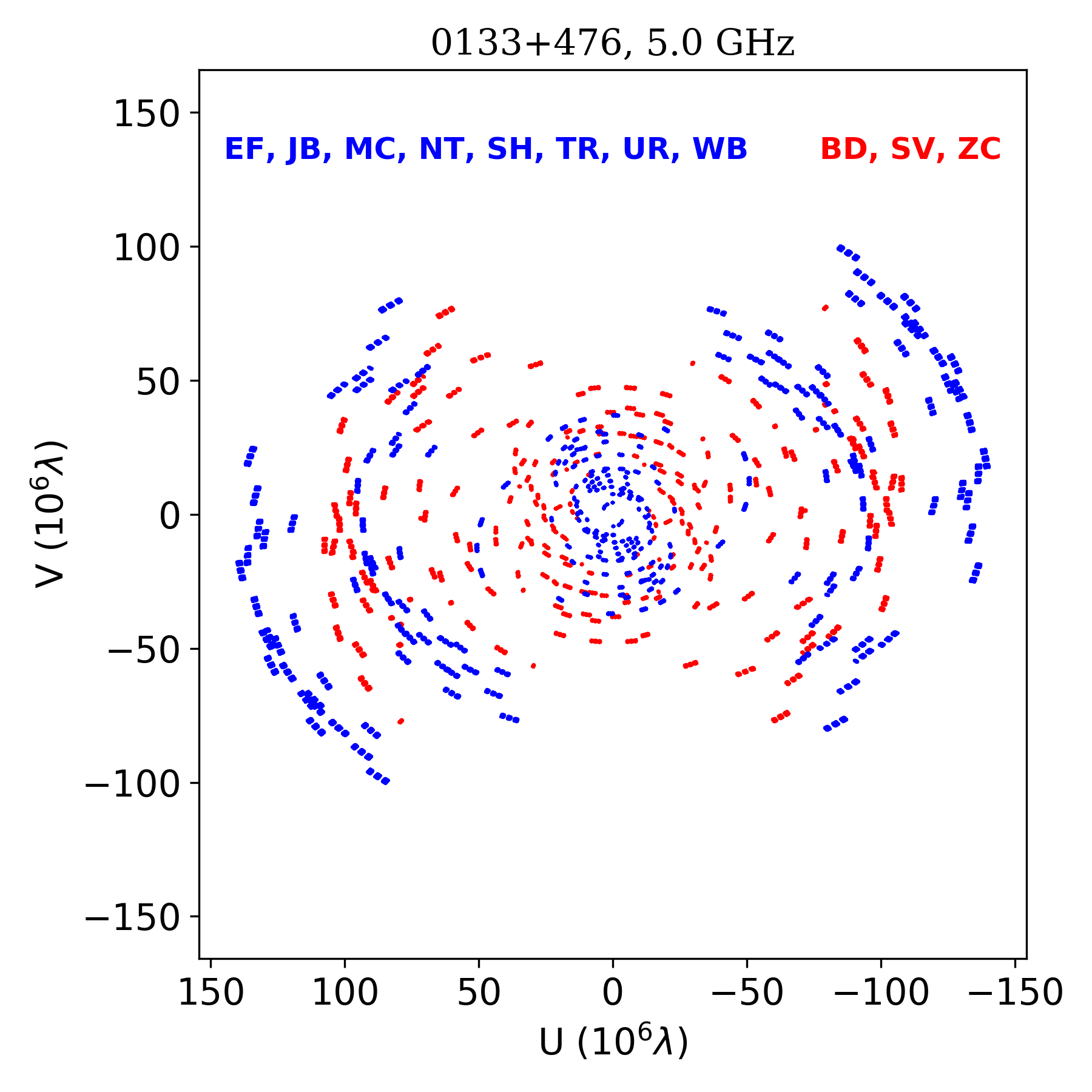}
 \caption{Example of $uv$-coverage for 0133+476 at 5~GHz. The red points denote
projected baselines involving Kvazar-KVO telescopes and the blue points show
the remaining projected baselines.}
 \label{fig:uv}
\end{figure}

\begin{figure}
 \includegraphics[width=0.95\linewidth,trim=0cm 0.8cm 0cm 0cm]{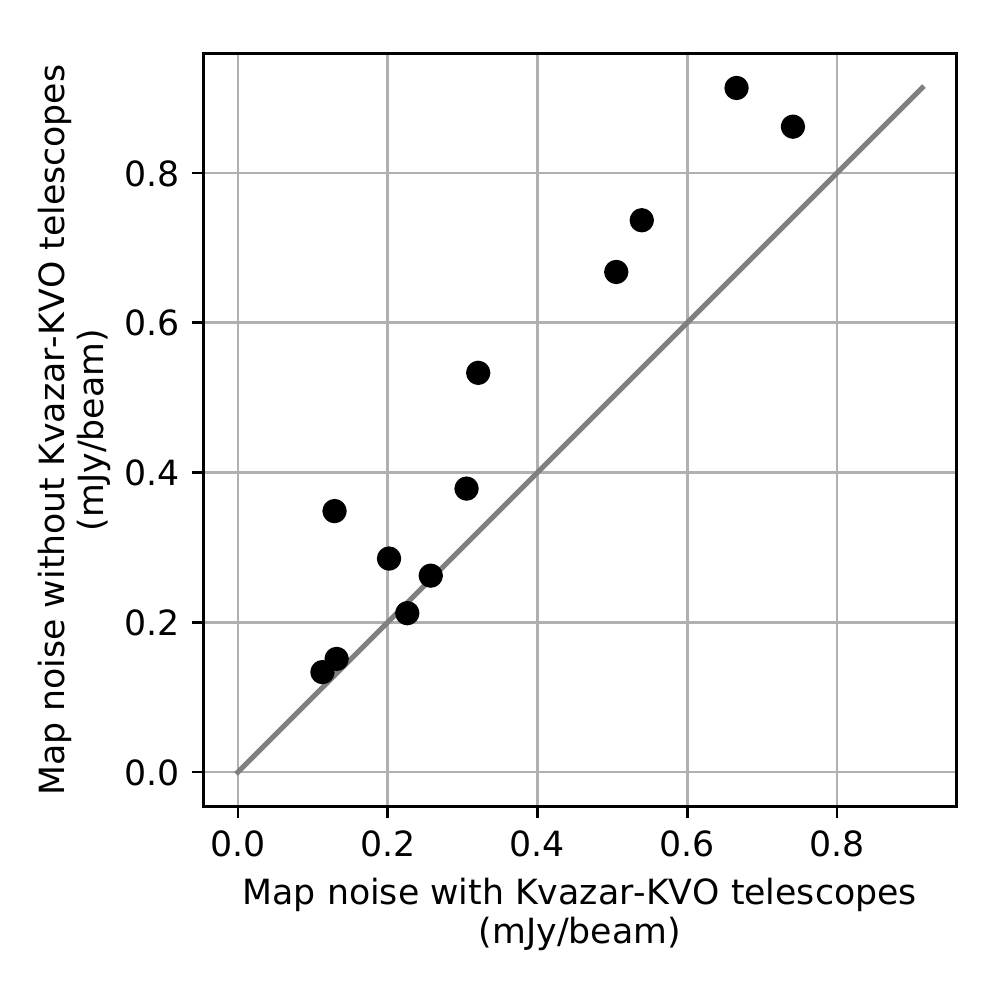}
 \caption{Mean residual noise level in the images constructed with and
without the Kvazar--KVO telescopes. The median noise levels are
0.28 and 0.36~mJy/beam, respectively.}
 \label{fig:kvazar_noise}
\end{figure}

Figure~\ref{fig:uv} shows that the telescopes of the Russian Kvazar--KVO
array appreciably improve the filling of the $uv$ plane for
the EVN observations, especially at medium-length baselines. The participation
of three Kvazar-KVO antennas together with the eight EVN
stations increases the number of baselines by about a factor of two,
enhancing the reliability of the image reconstruction and appreciably
lowering the noise level in the maps. To estimate the contribution made
to the data from the Kvazar telescopes, we carried out all the data
reduction a second time excluding the Badary, Zelenchuk, and Svetloe telescopes
and constructed images of the three sources 0125+487, 0133+476, and 0151+474
at all frequencies. We chose sources with relatively high declinations for
this analysis in order to reduce the influence of the poor $uv$-coverage in the
North--South direction. Figure~\ref{fig:kvazar_noise} compares the residual noise levels in the
images reconstructed with and without the Kvazar telescopes. In the latter
case, the residual noise level in the images is about 30$\%$ higher, on
average. The residual noise in the L, C, and X-band images of 0151+474
remained practically the same. The absence of such additional interferometric
baselines is especially important when constructing images of weak sources,
whose extended structure may not be detected due to insufficient sensitivity.

\subsection{Phase Calibration}

The phase of the visibility function at the correlator output is the
sum of contributions from various effects:
\[
 \varphi = \varphi_\text{struct} + \varphi_\text{pos} + \varphi_\text{atmo} +
\varphi_\text{inst} \,,
\]
where $\varphi_\text{struct}$ is the phase associated with the source
structure, $\varphi_\text{pos}$ the phase  associated with the position
of the source coordinates relative to the phase center, $\varphi_\text{atmo}$
the phase due to the atmosphere (ionosphere and troposphere), and
$\varphi_\text{inst}$ the phase due to delays in the apparatus and
imperfection of the frequency standards.

The terms $\varphi_\text{struct}$ and $\varphi_\text{pos}$ are determined
by the source and the terms $\varphi_\text{atmo}$ and $\varphi_\text{inst}$
by the telescopes and equipment used. The global fringe-fitting minimizes
the phase when it introduces global corrections for the antenna phases. Thus,
the sum $\varphi_\text{atmo} + \varphi_\text{inst} + \varphi_\text{pos}$
is determined. There remains only the term $\varphi_\text{struct}$, which
depends on the projected baselines ($uv$ points).

After calibration with the phase calibrator, the phase of the target source
will be
\begin{multline}
 \varphi^{t} =  \varphi_\text{struct}^\text{t} + (\varphi_\text{pos}^\text{t} -
\varphi_\text{pos}^\text{c}) +
(\varphi_\text{atmo}^\text{t} - \varphi_\text{atmo}^\text{c}) \\
 \nonumber{}+ (\varphi_\text{inst}^\text{t} - \varphi_\text{inst}^\text{c}) \,.
\end{multline}
Here, the subscript ``c'' denotes the calibrator and the subscript ``t'' the
target. The term $\varphi_\text{struct}^\text{c}$ takes into account the
model for the calibrator obtained during our imaging. The term
$\varphi_\text{inst}^\text{t} - \varphi_\text{inst}^\text{c} \approx
0$, since the instrumental delays are independent of the source and vary
weakly with time.  The phase difference $\varphi_\text{atmo}^\text{t} -
\varphi_\text{atmo}^\text{c} \approx 0$ for sources that are nearby each
other on the sky over sufficiently short time intervals. In this case,
\[
 \varphi^\text{t} \approx \varphi_\text{struct}^\text{t} + (\varphi_\text{pos}^\text{t} -
\varphi_\text{pos}^\text{c}) .
\]

Thus, after applying the solution obtained for the calibrator, the phase of
the target source contains only information about the source structure and
the relative positions of the calibrator and target.

\section{MEASURING THE FREQUENCY-DEPENDENT CORE SHIFT}

Let us introduce the following vector quantities: \linebreak
$\vec{S}_\text{ph.c.}$ represents the coordinates of the phase center
used in the correlation of the data, $\vec{S}_\text{core}$ the
coordinates of the VLBI core on the sky, $\vec{S}_\text{center}$ the
coordinates of the map center after fringe fitting, and
$\vec{X}_\text{core}$ the relative coordinates of the VLBI core in the
map. The coordinates of the core and the map center may be related in a
non-trivial way, and only for a point source can we write exactly
$\vec{S}_{\text{core}} = \vec{S}_{\text{center}}$.

Using the definitions of the above quantities, the coordinates of the core
of the calibrator can be written
\begin{equation}
 \label{eq:1}
 \vec{X}_{\text{core}}^{\text{cal}} = \vec{S}_{\text{core}}^{\text{cal}} -
\vec{S}_{\text{center}}^{\text{cal}} \,.
\end{equation}
The relative coordinates of the core of the target source are
\begin{equation}
 \vec{X}_{\text{core}}^{\text{tar}} = [\vec{S}_{\text{core}}^{\text{tar}} -
\vec{S}_{\text{ph.c.}}^{\text{tar}}] -
 [\vec{S}_{\text{center}}^{\text{cal}} - \vec{S}_{\text{ph.c.}}^{\text{cal}}] \,.
\end{equation}
The difference in the relative coordinates of the core of the target source
at frequencies $\nu_1$ and $\nu_2$ is
\begin{equation}
\label{eq:3}
\begin{split}
 \vec{X}_{\text{core}}^{\text{tar}}(\nu_2) &-
 \vec{X}_{\text{core}}^{\text{tar}}(\nu_1) \\
 =
 [\vec{S}_{\text{core}}^{\text{tar}}(\nu_2) -
  \vec{S}_{\text{core}}^{\text{tar}}(\nu_1)] &-
 [\vec{S}_{\text{center}}^{\text{cal}}(\nu_2) -
  \vec{S}_{\text{center}}^{\text{cal}}(\nu_1)] \,.
\end{split}
\end{equation}

The coordinates of the phase center ($\vec{S}_{\text{ph.c.}}$) have
canceled out, since they do not depend on the observing frequency. Using
\eqref{eq:1}, the last expression in square brackets in \eqref{eq:3} can be written
\begin{equation}
\begin{split}
 \vec{S}_{\text{center}}^{\text{cal}}(\nu_2) &-
 \vec{S}_{\text{center}}^{\text{cal}}(\nu_1) \\
=
  \vec{S}_{\text{core}}^{\text{cal}}(\nu_2) -
  \vec{S}_{\text{core}}^{\text{cal}}(\nu_1) &-
 [\vec{X}_{\text{core}}^{\text{cal}}(\nu_2) -
  \vec{X}_{\text{core}}^{\text{cal}}(\nu_1)] \,.
\end{split}
\end{equation}
Equation~(\ref{eq:3}) then acquires the form
\begin{equation}
 \label{eq:5}
\begin{split}
 [\vec{X}_{\text{core}}^{\text{tar}}(\nu_2) -
  \vec{X}_{\text{core}}^{\text{tar}}(\nu_1)] -
 [\vec{X}_{\text{core}}^{\text{cal}}(\nu_2) -
  \vec{X}_{\text{core}}^{\text{cal}}(\nu_1)] \\
=
 [\vec{S}_{\text{core}}^{\text{tar}}(\nu_2) -
  \vec{S}_{\text{core}}^{\text{tar}}(\nu_1)] -
 [\vec{S}_{\text{core}}^{\text{cal}}(\nu_2) -
  \vec{S}_{\text{core}}^{\text{cal}}(\nu_1)] \,.
\end{split}
\end{equation}

Thus, the left-hand side of \eqref{eq:5} contains directly measurable quantities,
and the right-hand side the desired variation of the coordinates of the
VLBI core with frequency. It follows from~\eqref{eq:5} that it is not possible to
directly separate the frequency-dependent core shifts of the calibrator
and target sources without using some additional \emph{a priori} information.

\subsection{Measurement of the Core Shift when the Jet Direction is Known}
\label{s:method_astrometry}

Studies have shown that the shift of the VLBI core position with frequency
is usually along the direction of the relativistic jet of the source~\cite{Pushkarev_2012}.
This makes it possible to measure the core shifts independently for each
source in a triplet. We used the following model in these calculations:
\[
\begin{split}
\vec{X}_{\text{core}, i}^{j} = & ((\vec{S}_\text{apex} - \vec{S}_\text{ph.c.})_i +
\Delta r_{\text{core},i}^{j} \vec{d}_i) \\
&- (\vec{S}_\text{center} - \vec{S}_\text{ph.c.})^j \,,
\end{split}
\]
where the subscripts $i \in \{1,2,3\}$ denote the number of a source in the
triplet, $j \in \{L,S,C,X\}$ denotes the frequency, $\vec{S}_\text{apex}$
is the true position of the base (apex) of the jet, $\vec{d}$ is the jet
direction, and $\Delta r_\text{core}$ is the core shift we are seeking.
Having for a given triplet measured values of the vectors
$\vec{X}_{\text{core}, i}^{j}$ and \emph{a priori} estimates of the
vectors $\vec{d}_i$, we can obtain the \emph{a posteriori} distribution
of the probability density of the vectors $(\vec{S}_\text{apex} -
\vec{S}_\text{ph.c.})_i$ and $(\vec{S}_\text{center} -
\vec{S}_\text{ph.c.})^j$, as well as the desired quantity
$\Delta r_{\text{core},i}^{j}$. We used the Markov Chain Monte Carlo
method for these computations, realized in the \emph{PyMC3} library~\cite{Salvatier_2016}.
The core shift $\Delta r_{\text{core}}$ was calculated relative to its
X-band position. This approach makes it possible to estimate the core shift
even if the jet direction for the source is unknown, assuming a uniform
\emph{a priori} distribution $\vec{d}$.

We also developed a method for measuring the core shift for a pair of
sources closely spaced on the sky. As was shown in formula (5) of the
previous section, for two sources with referenced phases, only the difference
vector for the frequency-dependent shift of the VLBI core can be measured
directly, $\vec{CS}_{\text{rel}} = \vec{CS}_1 - \vec{CS}_2$.
When the jet direction is known, this difference vector can unambiguously
be separated into components as $\vec{CS}_{\text{rel}} =
\Delta r_{\text{core},1} \vec{d}_1 - \Delta r_{\text{core},2}
\vec{d}_2$, where $\vec{d}_1$ and $\vec{d}_2$ are unit vectors
in the directions of the jets of the first and second source. The quantities
$\Delta r_{\text{core},1}$ and $\Delta r_{\text{core},2}$ represent
the desired core shifts for the two sources.

We calculated the jet direction as the mean position angle of the innermost
Gaussian components in our source model relative to the 8.4~GHz core.
We were not able to determine the jet direction for the most compact source
in our sample, the blazar 0235+164, from our data. For this source, we
adopted the jet position angle from~\cite{Kutkin_2018} based on numerous 43-GHz observations
on the Very Long Baseline Array (VLBA). For another compact source, 0440$-$003,
we used data from the MOJAVE project at 15~GHz~\cite{Lister_2016,Hovatta_2014} to identify the jet
direction, since the resolution provided by our 8~GHz EVN observations was
insufficient.

\begin{figure}
\includegraphics[width=\linewidth,trim=0.2cm 2cm 3cm 3.3cm]{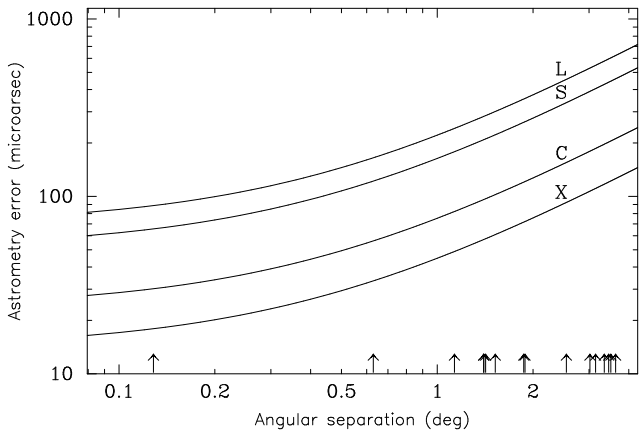}
\caption{Accuracy of the relative astrometry at 1.7~GHz (L), 2.3~GHz (S),
5.0~GHz (C), and 8.4~GHz (X) as a function of the angular separation of the
calibrator and target. The arrows show the angular separations between the
sources in the triplets.}
\label{fig:rel_astrometry_errors}
\end{figure}

\subsection{Accuracy of the Core Shift Measurements}
\label{s:errors}

The question of the accuracy of relative radio astrometry was first
considered in~\cite{Shapiro79} for the two bright, closely spaced quasars 3C~345 and
NRAO~512 observed quasi-simultaneously with VLBI on a single baseline.
It was shown in this pioneering work that the uncertainties in the relative
coordinates of the sources can comprise only a small fraction of a
milliarcsecond. The accuracy of astrometric measurements using modern
multi-antenna aperture synthesis systems such as the VLBA or EVN cannot
be calculated analytically, but it is known that this accuracy is limited
primarily by uncorrected effects due to the propagation of the radio signals
in the troposphere and ionosphere~\cite{Reid14}. Thus, among the numerous factors
influencing the accuracy of relative astrometry, the main ones are the
resolution of the interferometric array, $\lambda/D_\text{max}$, the number
of antennas in the array and the accuracies of each of their positions,
the angular separation of the target source and the calibrator $\Delta\Theta$,
the overall duration of the experiment, the stability of the troposphere and
ionosphere determining the source--calibrator cycle time, and the
signal-to-noise ratio for the target. The dominant source of systematic
errors is likely observations at large zenith angles $z$, since the uncorrected
tropospheric delay grows as $\text{sec}z$. Therefore, observations at
large zenith angles should be avoided, while also taking into consideration
the fact that limiting the observations to small zenith angles can appreciably
lower the quality of the filling of the $uv$ plane, and accordingly also the
reliability of the image reconstruction.

Computer simulations of artificial sets of 8.4~GHz VLBI observations with the
VLBA and EVN have also been used to estimate the astrometric accuracy of
phase-referencing observations~\cite{Pradel06}. It was shown that the typical astrometric
errors are lowest for bright, point-like sources at intermediate declinations,
and comprise about 50~microarcsecond ($\mu$as)  for $\Delta\Theta=1^{\circ}$,
growing to 300~$\mu$as for higher and lower declinations. The astrometric
errors are nearly constant at about $14~\mu$as for very closely spaced
sources at intermediate declinations.

We estimated the accuracy of our relative astrometry $\sigma_a$ by applying
the frequently used relation $2\Delta\Theta(\lambda/D_\text{max})$~\cite{Reid14},
where $\Delta\Theta$ is measured in radian. We modified this relation in
order to take into account the fact that the errors never go to zero,
even for very closely spaced sources~\cite{Pradel06}, namely,
\[
\sigma_a = 2\Delta\Theta(\lambda/D_\text{max}) + 14 \lambda/\lambda_{\text{3.6cm}} \,.
\]
Figure~\ref{fig:rel_astrometry_errors} shows estimates of the astrometric errors obtained using this
relation for various frequencies as a function of the angular separation of
the sources in the triplets of our sample and a typical value
$D_\text{max}=8300$~km, corresponding to the maximum baselines realized
between Shanghai and Jodrell Bank or Shanghai and Noto. Although some of
the objects in our sample have intermediate declinations, the real uncertainties
in the relative astrometry could be higher, since, although these sources are
very compact, they are nevertheless not point-like. Due to the nature of
the method we have used to measure the VLBI core shifts, we must add
uncertainty in the jet direction to the uncertainties in the relative
astrometry:
\[
\sigma_\varphi = \frac{\sqrt{\sigma_{r,\text{core}}^2 +
\sigma_{r,\text{jet}}^2}}{r} \,,
\]
\noindent
where $r$ is the distance from the core to the innermost jet component at
our highest frequency (8.4~GHz), and $\sigma_{r,\text{jet}}$ and
$\sigma_{r,\text{core}}$ are the uncertainties in the positions of the
innermost jet component and the core at 8.4~GHz. Note that the formal
uncertainties in the positions of Gaussian components are usually very small,
and do not reflect the real uncertainty in the jet direction. We calculated the
jet direction as the mean position angle of several innermost jet components
relative to the core, and adopted the uncertainty of this mean as
$\sigma_\varphi$. For sources with only one jet component, we used the
conservative estimate for the uncertainty in the jet direction $10^{\circ}$.
The median value of $\sigma_\varphi$ was $6^{\circ}$.

The uncertainty in the core shift obtained by expanding the difference
vector $\vec{CS}_{\text{rel}}$ into components can be estimated
analytically. This uncertainty depends appreciably on the angle between
the component vectors (jet directions) $\Delta \varphi$ and is comprised of
two parts:
\[
\sigma_1 \propto
\frac{\sigma_\varphi|\vec{CS}_{\text{rel}}|}{\sin^2 \Delta
\varphi}
\]
due to uncertainty in the jet directions and
\[
\sigma_2 =
\frac{\sqrt{\sigma_a^2+\sigma_{r,\text{core}}^2}}{|\sin\Delta
\varphi|} \approx \frac{\sigma_a}{|\sin \Delta \varphi|}
\]
due to uncertainty in the difference vector $\vec{CS}_{\text{rel}}$
itself, which depends on the astrometric error
\[
\sigma_a= \left(\frac{2\Delta\Theta}{D} +
\frac{14}{\lambda_{\text{3.6cm}}}\right) \left(\lambda_1^2 +
\lambda_2^2\right)^{1/2}
\]
and the total uncertainty in the positions of the cores of the sources
\begin{multline*}
\sigma_{r,\text{core}} = \\
= \sqrt{\sigma_{r,\text{core}_1,\nu_1}^2 +
\sigma_{r,\text{core}_2,\nu_1}^2 +
\sigma_{r,\text{core}_1,\nu_2}^2 +
\sigma_{r,\text{core}_2,\nu_2}^2} \,.
\end{multline*}

In our analysis, we neglected the uncertainties in the core coordinates
$\sigma_{r,\text{core}}$, since these are small compared to $\sigma_a$.
Thus, the uncertainty in the measured core shift grows rapidly in the case
of small angles between the jet directions $\Delta \varphi$, making the
triplet method fairly sensitive to the condition that the jet directions
for high-accuracy measurements be close to orthogonal.

\subsection{Comparison of Methods for Determining the Core Shift}

The main difficulty in measuring the shift in the position of the core is
accurately aligning maps of the radio brightness obtained at different
frequencies. This problem arises due to the loss of information about the
absolute coordinates of the source during the standard reduction of VLBI
data, including phase self-calibration during the mapping process.

One method that makes it possible to overcome this problem is based on
the method of self-referencing~\cite{Lobanov_1998,Kovalev_2008,Sokolovsky_2011}, where the
alignment of the images
at different frequencies is carried out using a bright jet component whose
emission is optically thin, so that its position is achromatic. There is also
a more universal approach to the realization of self-referencing, where the
images are aligned based on the results of a two-dimensional cross correlation
of optically thin regions of the jet~\cite{Walker_2000}. This method has been applied together
with modeling of the source structure using a number of Gaussian components
to determine the VLBI core shifts of four BL~Lac objects~\cite{Sullivan_2009}, a sample of
190 sources~\cite{Pushkarev_2012}, and other large source samples~\cite{Plavin2018}.
Inadequacies of
this method include the systematics of the measurements in the case of
strong spectral-index gradients along the jet, the presence of model
assumptions about the coordinates of the VLBI core, and limitations to its
applicability to sources with fairly rich structure suitable for the
cross correlation analysis.

Another method for measuring core shifts, which we have used in our current
study, is based on relative VLBI astrometry~\cite{MS1984}. Its main advantage is
that it applies fewer model assumptions and can be used for compact sources
with minimum structure, where the self-referencing method cannot be applied.
However, this method also has inadequacies: the phase calibrator relative
to which the position of a target is measured also has a core shift, which
must be taken into account, limiting the accuracy of such measurements.
Only pairs of sources whose inner jets have appreciably different directions
are free from this problem. This method also requires good filling of the
$uv$ plane and high angular resolution to work well.

\begin{figure*}
 \includegraphics[width=0.9\textwidth]{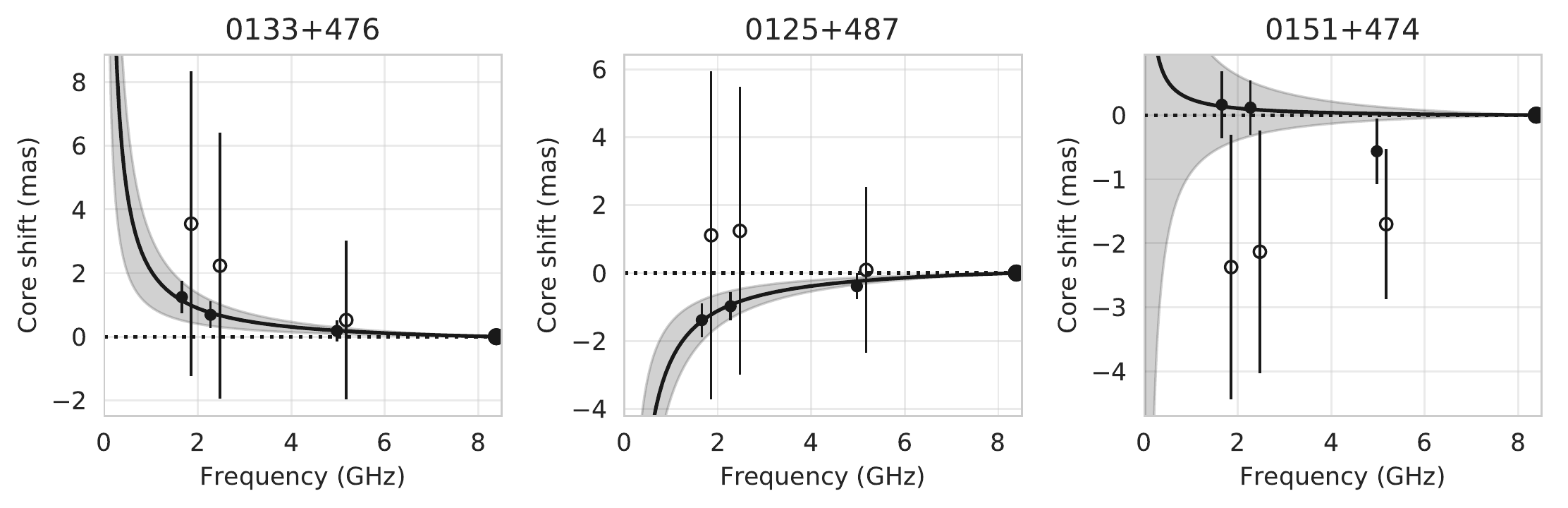}
 \includegraphics[width=0.9\textwidth]{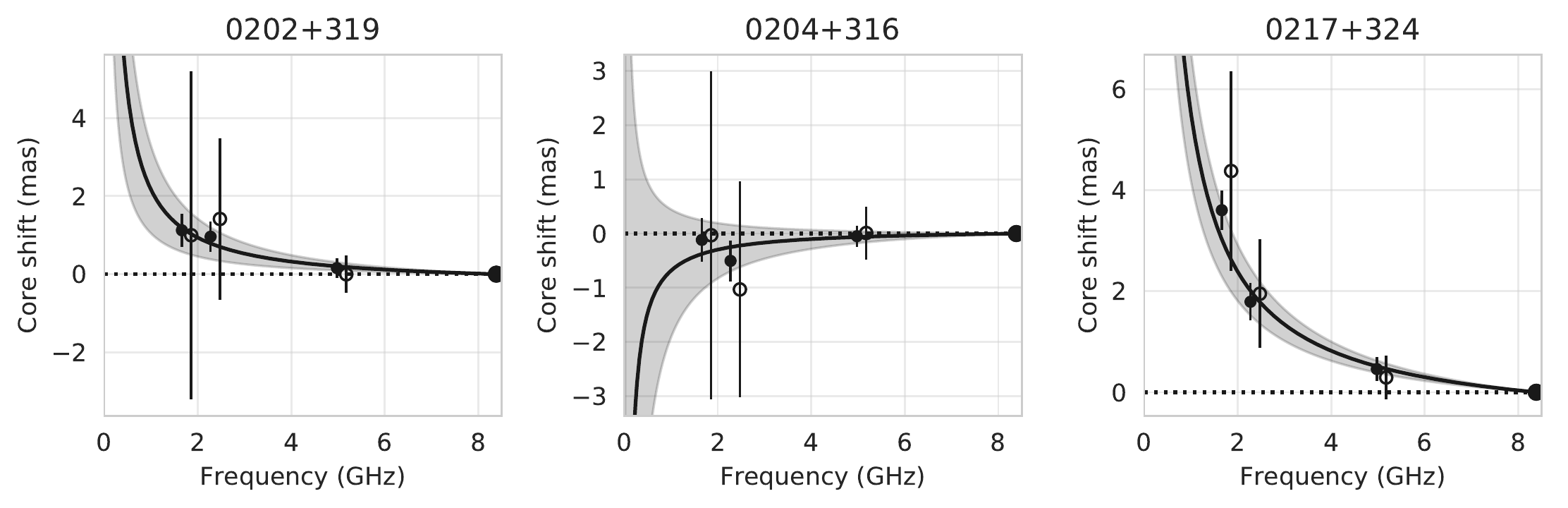}
 \includegraphics[width=0.9\textwidth]{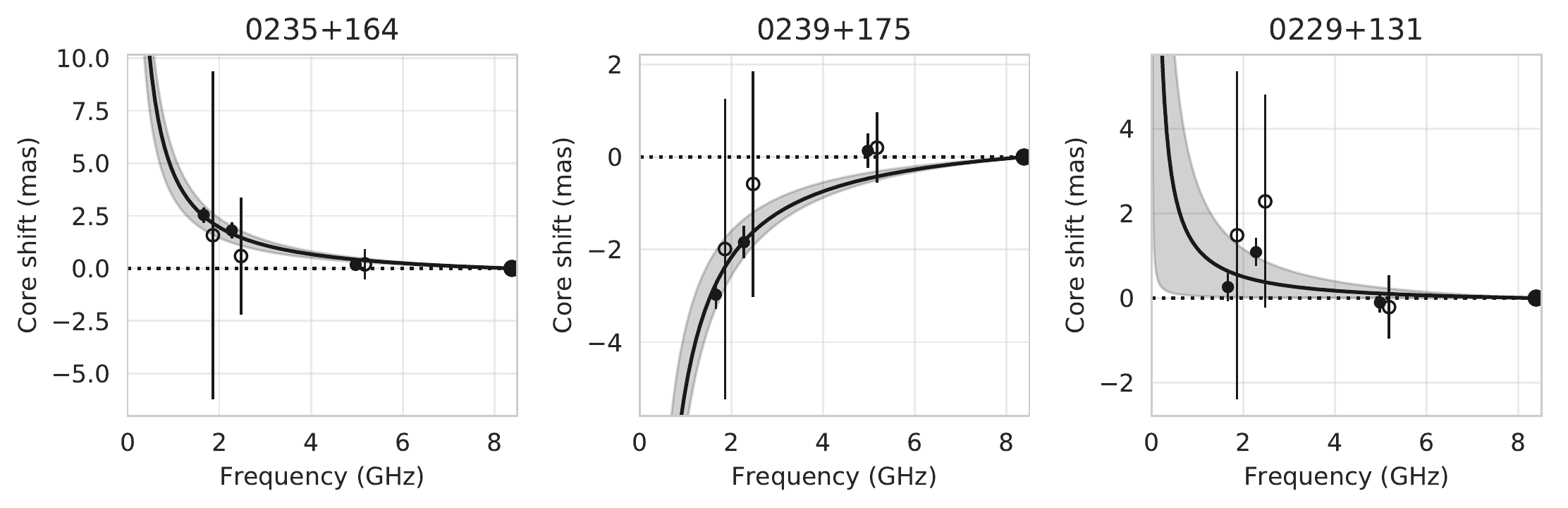}
 \includegraphics[width=0.9\textwidth]{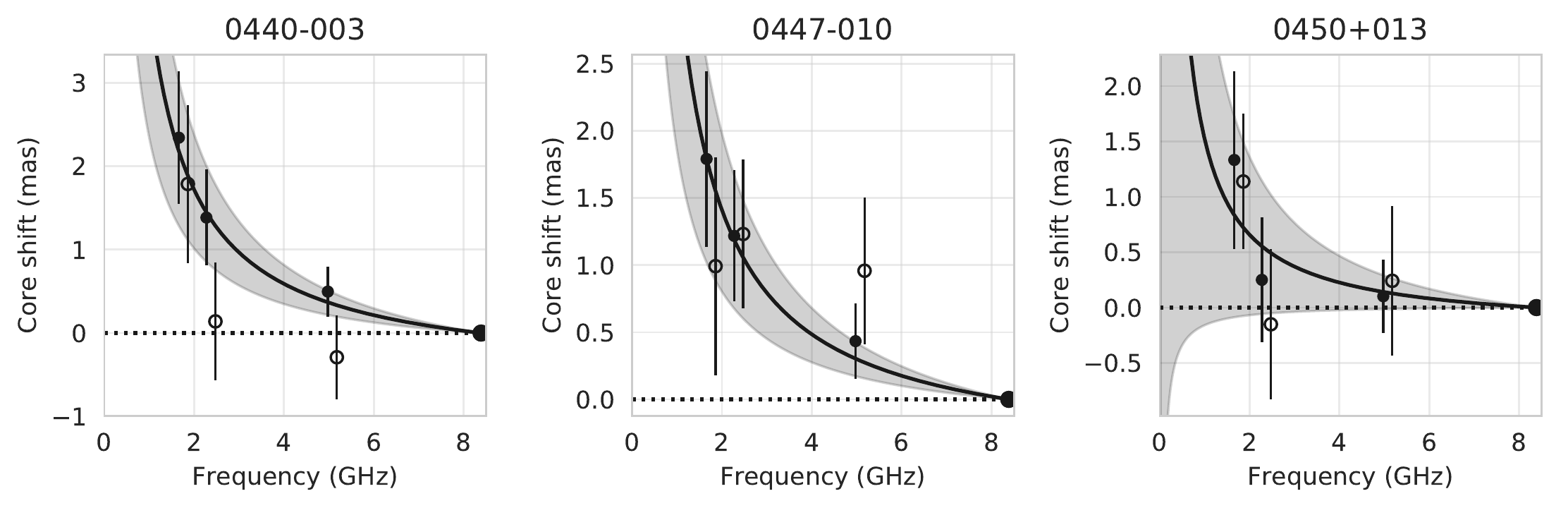}
 \caption{Frequency dependence of the core shifts relative to the X band
measured for closely spaced triplets of sources. Filled circles show the
results obtained using models with the estimated jet directions and the hollow
circles the results obtained without applying these models. The projection
of the core-shift vector onto the jet direction is shown. The curve and
shaded region correspond to the dependence $\Delta r_\textrm{core} = a +
b/\nu$, constructed using the filled circles and the 68$\%$ confidence
interval about this fit. The results were obtained via our relative
astrometry.}
 \label{fig:cs_freqdep_astrom}
\end{figure*}

\addtocounter{figure}{-1}
\begin{figure*}
 \includegraphics[width=0.9\textwidth]{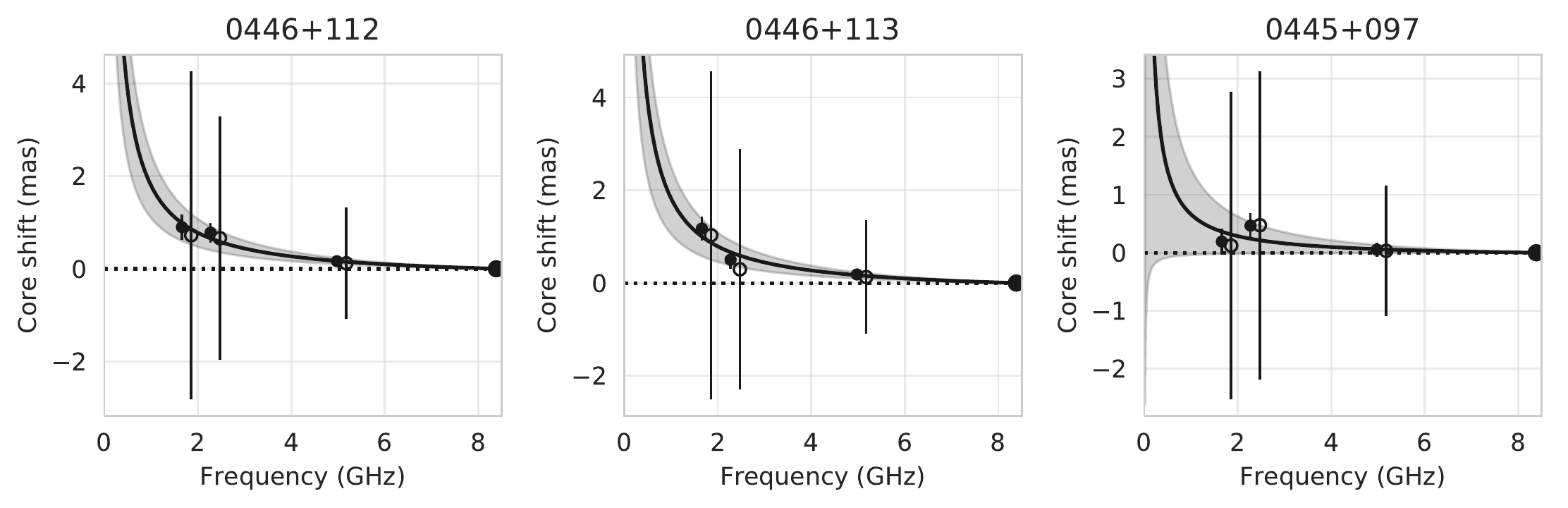}
 \includegraphics[width=0.9\textwidth]{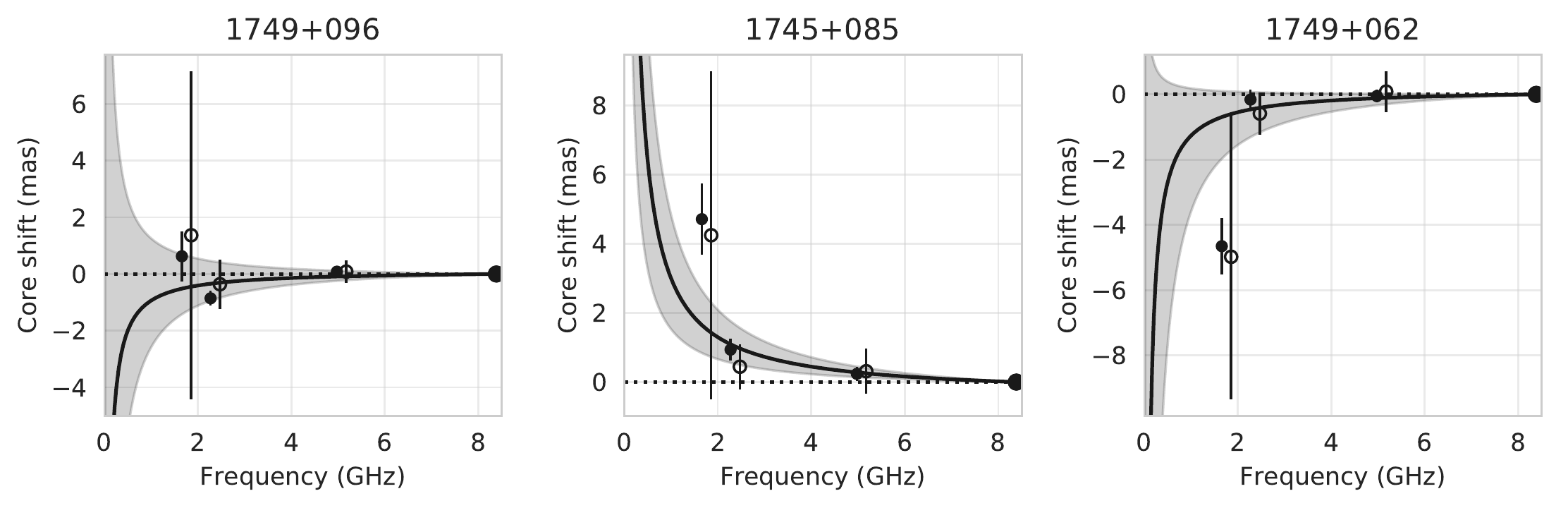}
 \includegraphics[width=0.9\textwidth]{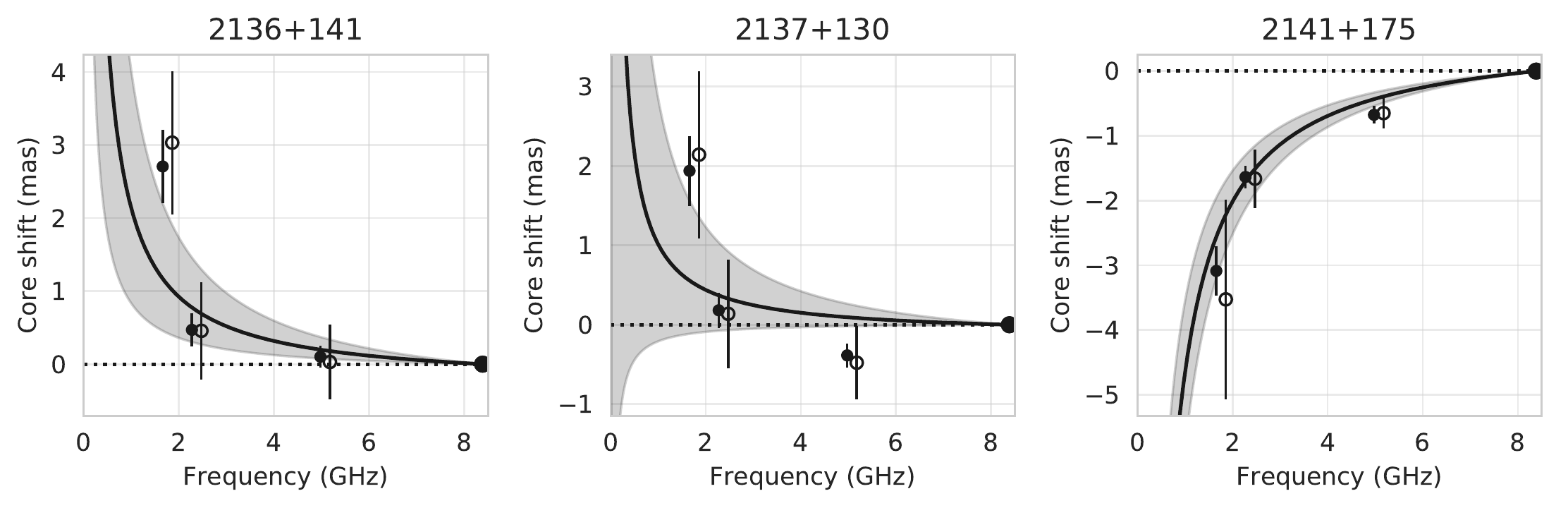}
 \includegraphics[width=0.9\textwidth]{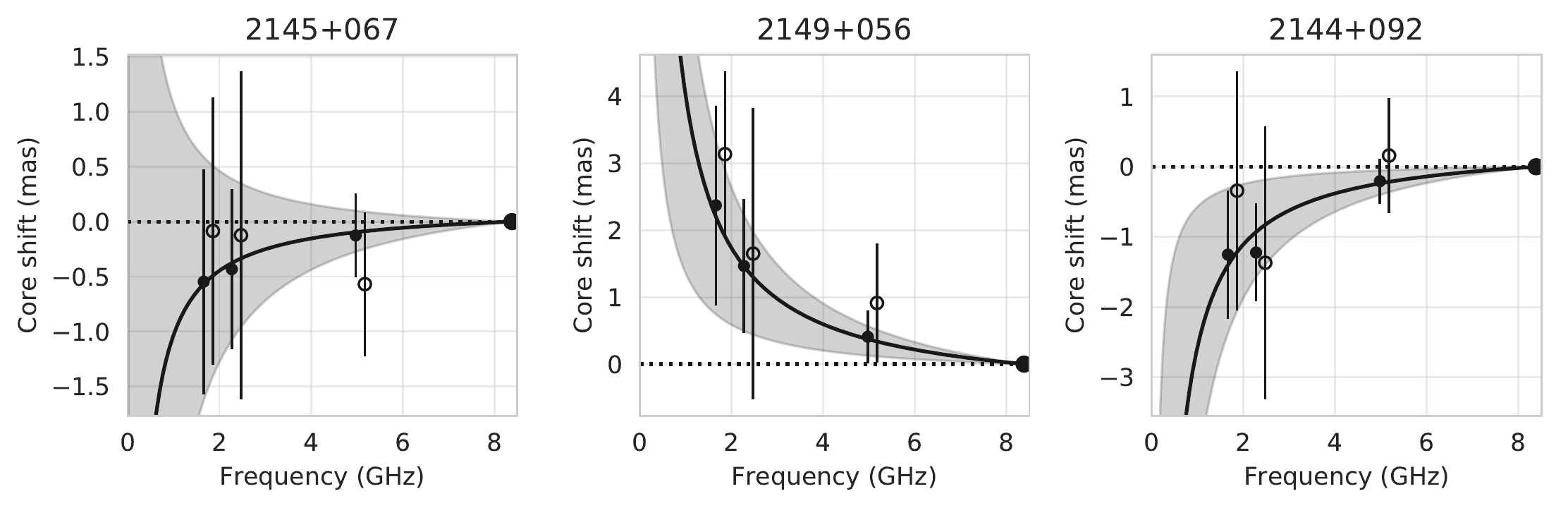}
 \caption{Continued}
\end{figure*}

\begin{figure*}
 \includegraphics[width=0.9\textwidth]{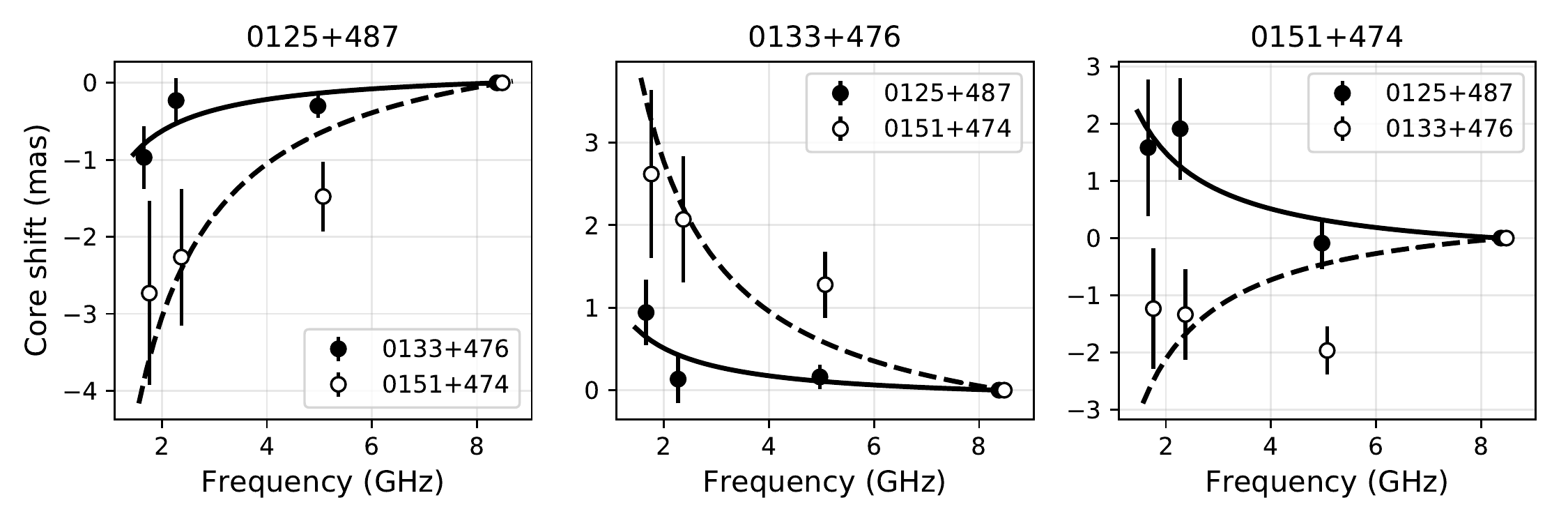}
 \includegraphics[width=0.9\textwidth]{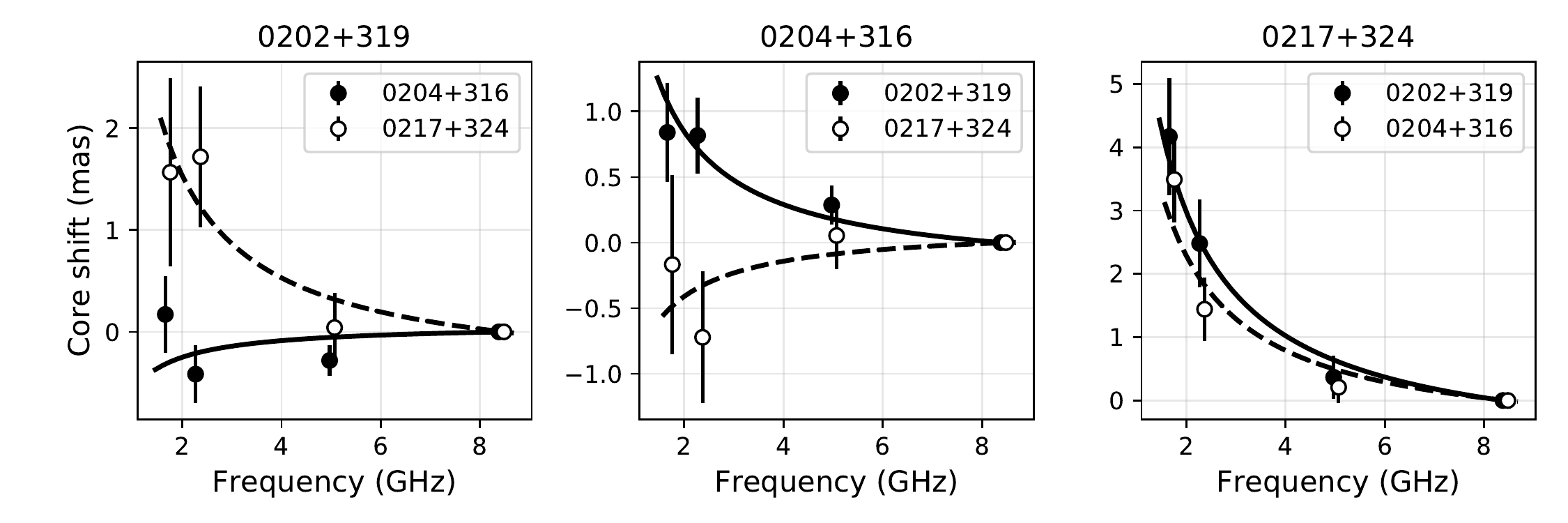}
 \includegraphics[width=0.9\textwidth]{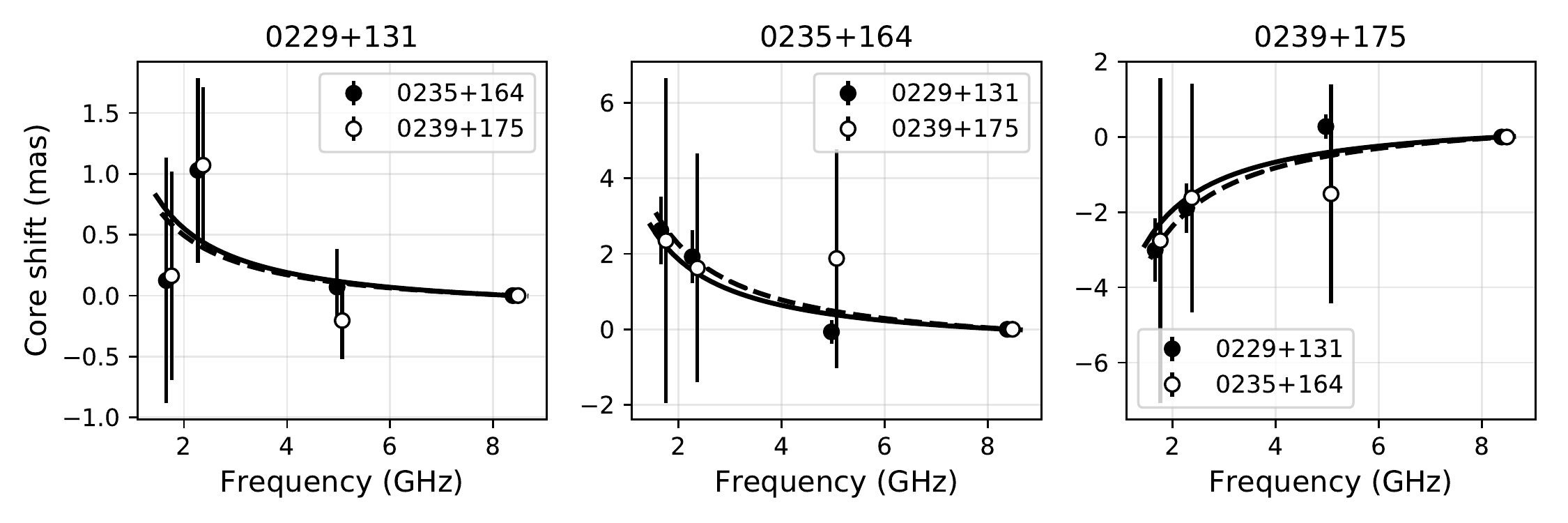}
 \includegraphics[width=0.9\textwidth]{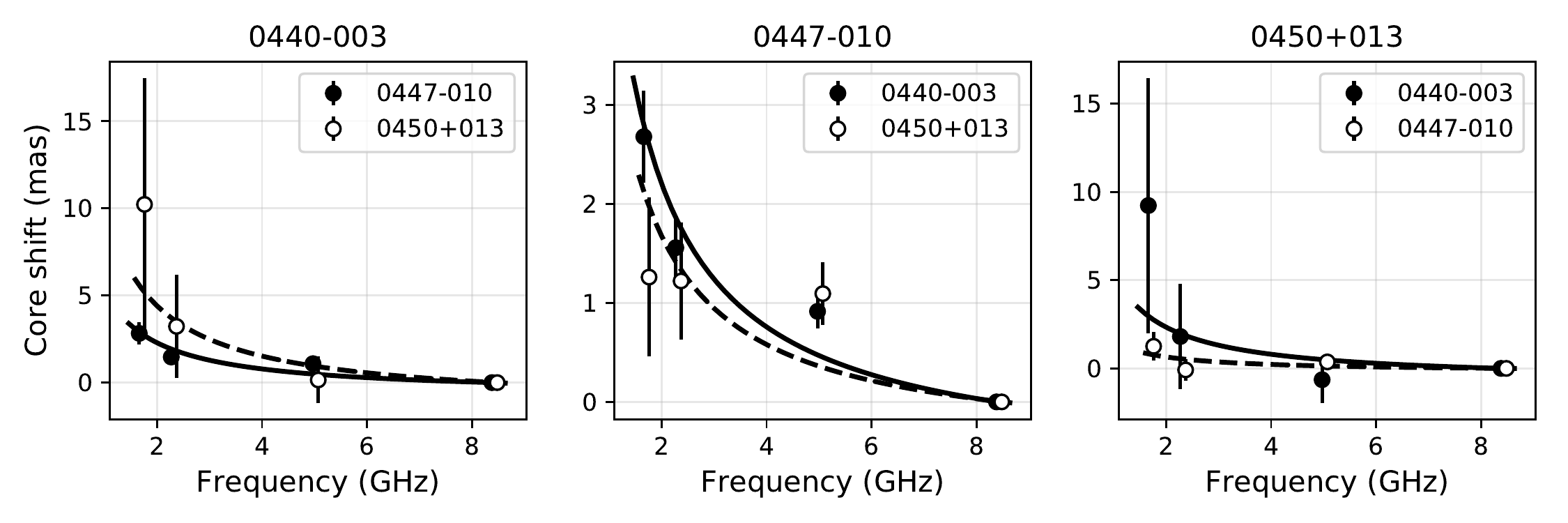}
 \caption{Frequency dependence of the core shift relative to the X band
measured by expanding the relative shift vector into components along
the jet directions of closely spaced sources considered in pairs. The
solid and dotted curves show the dependences $\Delta r_\textrm{core} =
a + b/\nu$ fitted using the filled and hollow points, respectively. The
results were obtained via our relative astrometry.}
 \label{fig:cs_vs_freq_using_jet_pa}
\end{figure*}

\addtocounter{figure}{-1}
\begin{figure*}
 \includegraphics[width=0.9\textwidth]{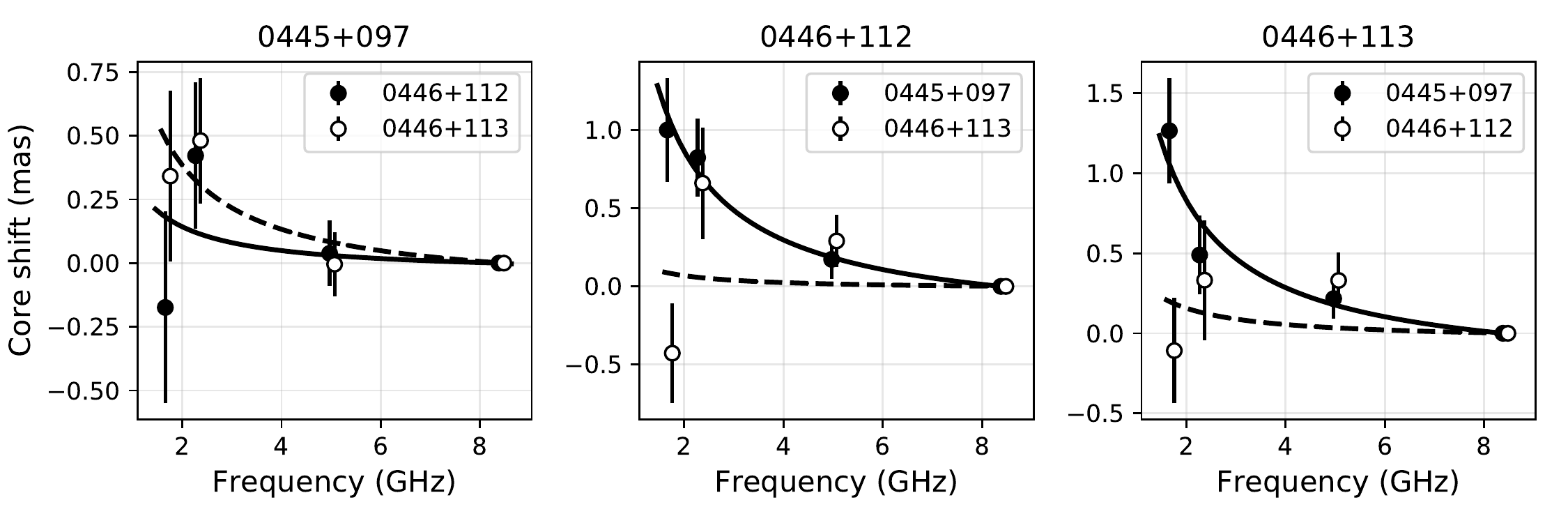}
 \includegraphics[width=0.9\textwidth]{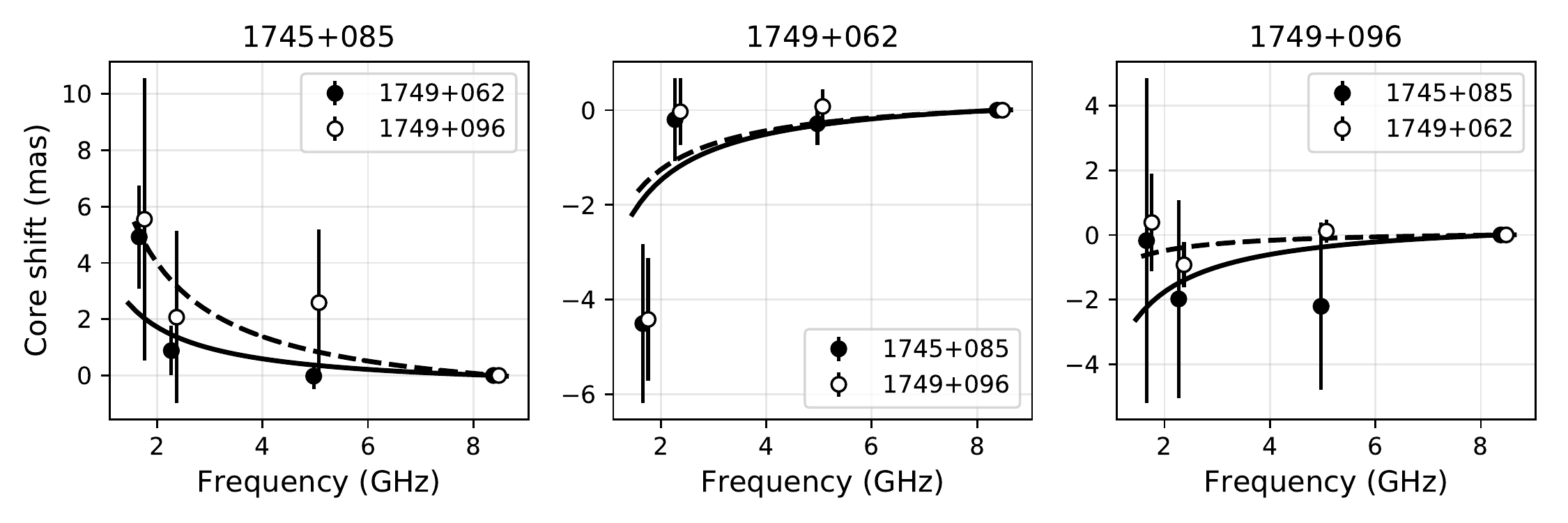}
 \includegraphics[width=0.9\textwidth]{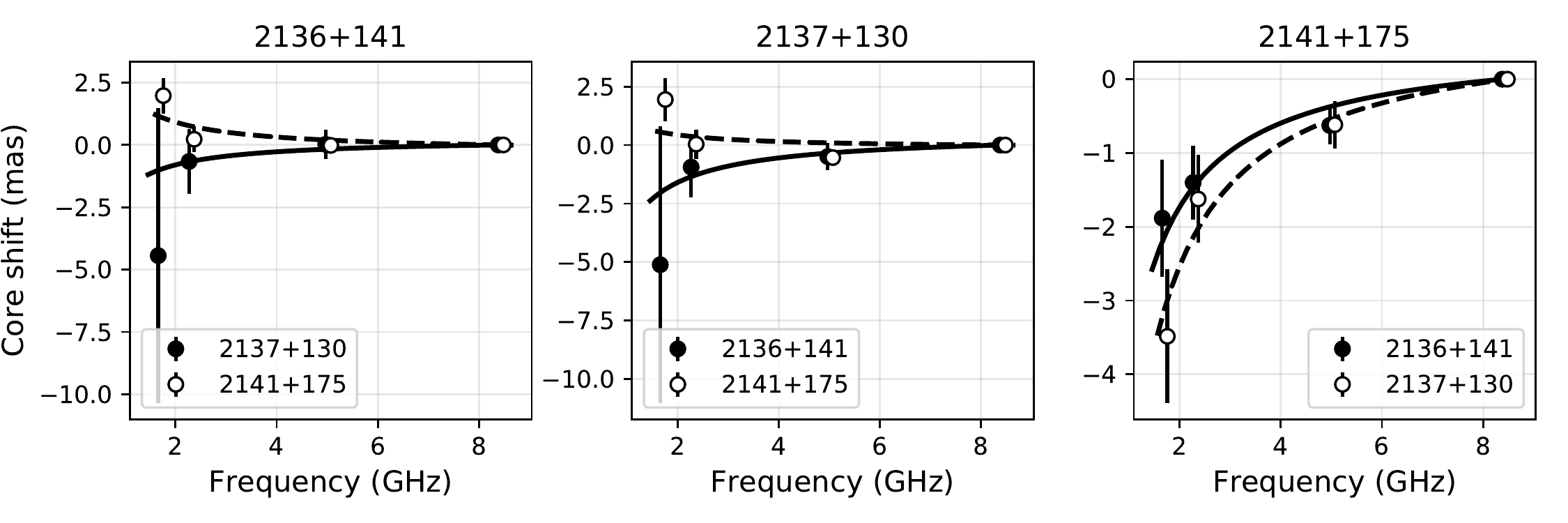}
 \includegraphics[width=0.9\textwidth]{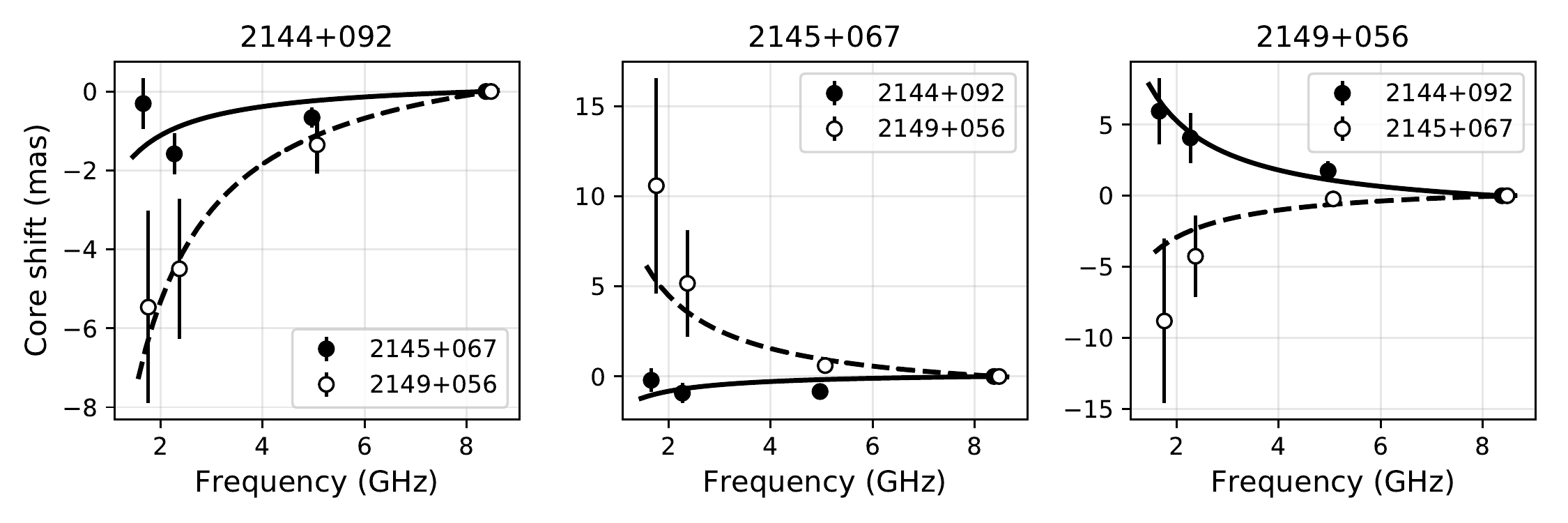}
 \caption{Continued}
\end{figure*}

\begin{table*}
\setlength{\tabcolsep}{10pt}
\caption{Frequency-dependent core shifts relative to X-band (8.4 GHz),
measured using the astrometric method (see Section~\ref{s:method_astrometry}). The values of
the coefficient $b$ in the relation $r_\mathrm{core}(\nu) = b / \nu$ are
also presented.}
\label{tab:cs_astrometry}
\begin{tabular}{c|r|r|r|r}
\hline \multicolumn{1}{c|}{\multirow{2}*{Source}} &
\multicolumn{3}{c|}{Core shift (mas)} &
\multicolumn{1}{c}{\multirow{2}*{$b$ (mas GHz)}} \\
\cline{2-4}
  & \multicolumn{1}{c|}{X $\rightarrow$ L} & \multicolumn{1}{c|}{X $\rightarrow$ S} &
  \multicolumn{1}{c|}{X $\rightarrow$ C} &  \\
\hline
0133+476 &  $1.25\pm 0.5$ &  $0.69\pm 0.4$ &  $0.18\pm 0.3$ &  $2.34 \pm 1.3$ \\
0125+487 & $-1.39\pm 0.5$ & $-0.98\pm 0.4$ & $-0.39\pm 0.4$ & $-2.92 \pm 1.3$ \\
0151+474 &  $0.16\pm 0.5$ &  $0.12\pm 0.4$ & $-0.56\pm 0.5$ &  $0.28 \pm 1.3$ \\
0202+319 &  $1.12\pm 0.4$ &  $0.96\pm 0.4$ &  $0.15\pm 0.2$ &  $2.45 \pm 1.3$ \\
0204+316 & $-0.12\pm 0.4$ & $-0.50\pm 0.4$ & $-0.05\pm 0.2$ & $-0.78 \pm 1.3$ \\
0217+324 &  $3.60\pm 0.4$ &  $1.79\pm 0.4$ &  $0.46\pm 0.2$ &  $6.26 \pm 1.5$ \\
0235+164 &  $2.54\pm 0.3$ &  $1.81\pm 0.4$ &  $0.17\pm 0.3$ &  $5.13 \pm 1.2$ \\
0239+175 & $-2.97\pm 0.3$ & $-1.84\pm 0.4$ &  $0.13\pm 0.3$ & $-5.67 \pm 1.3$ \\
0229+131 &  $0.26\pm 0.4$ &  $1.09\pm 0.3$ & $-0.11\pm 0.2$ &  $1.32 \pm 1.5$ \\
0440$-$003 &  $2.34\pm 0.8$ &  $1.38\pm 0.6$ &  $0.50\pm 0.4$ &  $4.51 \pm 1.8$ \\
0447$-$010 &  $1.79\pm 0.6$ &  $1.22\pm 0.5$ &  $0.43\pm 0.4$ &  $3.71 \pm 1.6$ \\
0450+013 &  $1.33\pm 0.8$ &  $0.25\pm 0.6$ &  $0.10\pm 0.3$ &  $1.73 \pm 1.9$ \\
0446+112 &  $0.90\pm 0.3$ &  $0.78\pm 0.2$ &  $0.16\pm 0.1$ &  $2.04 \pm 0.8$ \\
0446+113 &  $1.17\pm 0.2$ &  $0.50\pm 0.2$ &  $0.18\pm 0.1$ &  $2.07 \pm 0.8$ \\
0445+097 &  $0.19\pm 0.2$ &  $0.47\pm 0.2$ &  $0.05\pm 0.1$ &  $0.75 \pm 0.8$ \\
1749+096 &  $0.63\pm 1.0$ & $-0.86\pm 0.3$ &  $0.08\pm 0.2$ & $-1.07 \pm 2.2$ \\
1745+085 &  $4.71\pm 1.0$ &  $0.94\pm 0.3$ &  $0.24\pm 0.2$ &  $3.41 \pm 1.9$ \\
1749+062 & $-4.65\pm 0.8$ & $-0.16\pm 0.3$ & $-0.05\pm 0.2$ & $-1.44 \pm 2.1$ \\
2136+141 &  $2.71\pm 0.5$ &  $0.47\pm 0.2$ &  $0.10\pm 0.2$ &  $2.42 \pm 1.8$ \\
2137+130 &  $1.94\pm 0.4$ &  $0.18\pm 0.2$ & $-0.39\pm 0.2$ &  $1.15 \pm 1.7$ \\
2141+175 & $-3.09\pm 0.4$ & $-1.64\pm 0.2$ & $-0.68\pm 0.2$ & $-5.29 \pm 1.3$ \\
2145+067 & $-0.55\pm 1.1$ & $-0.43\pm 0.8$ & $-0.13\pm 0.4$ & $-1.17 \pm 2.3$ \\
2149+056 &  $2.37\pm 1.3$ &  $1.47\pm 0.9$ &  $0.41\pm 0.4$ &  $4.56 \pm 2.7$ \\
2144+092 & $-1.25\pm 1.0$ & $-1.22\pm 0.7$ & $-0.21\pm 0.3$ & $-2.90 \pm 2.1$ \\
\hline
\end{tabular}
\end{table*}

\section{RESULTS AND DISCUSSION}

\subsection{Relative Astrometric Measurements}

The results of our measurements of the frequency-dependent VLBI core shift
obtained via relative astrometry for simultaneous observations of triplets
of sources are presented in Table~\ref{tab:cs_astrometry}. The shifts in mas for the L (1.7~GHz),
S (2.3~GHz), and C (5.0~GHz) bands relative to the (highest-frequency) X-band
(8.4~GHz) are given for each source. Positive values correspond to shifts
of the core downward along the jet with decreasing frequency, as is predicted
theoretically. Figure~\ref{fig:cs_freqdep_astrom} shows the core shifts as a function of frequency.

We expect theoretically that $r_\textrm{core} \propto
\nu^{-1/k_{\textrm{r}}}$~\cite{Lobanov_1998}, but the measurement uncertainties hindered
our ability to estimate $k_{\textrm{r}}$. Therefore, we fitted the dependence
$\Delta r_\textrm{core} = a + b/\nu$, assuming $k_\textrm{r} = 1$, which is
a good approximation for most sources~\cite{Sokolovsky_2011}. For many sources in our sample
the estimated uncertainties exceed the measured shifts. At the same time,
the measured core shifts for 0133+476, 0202+319, 0217+324, 0235+164, 0440$-$003,
0446+112, 0446+113, 0447$-$010, and 2149+056 are in good agreement with the
dependence $\propto \nu^{-1}$. The median core shifts for these sources
were 1.79, 1.22, and 0.18~mas for L, S, and C bands, respectively, relative
to X band.

Note the reverse core shifts indicated for a number of the sources. We do
not yet have a clear explanation for this result. It is important to
understand whether this is an intrinsic astrophysical effect or the result
of factors we have not taken into account in the method used. We did not
detect this effect earlier in our core-shift measurements for large numbers
of sources using the self-referencing method~\cite{Pushkarev_2012,Plavin2018}. It is possible that
only astrometric measurements are sensitive to this effect, or that some
systematics are in the data, which we have not taken into account. Further
studies of this effect require new, better quality, sensitive core-shift
measurements. Such studies should be carried out, first and foremost, for
sources that have manifest reverse core shifts in our observations.

For comparison, we measured the core shifts using the relative astrometry
method when the relative shift vectors $\vec{CS}_{\textrm{rel}}$ for pairs
of sources were spread around the jet directions. The results are shown in
Fig.~\ref{fig:cs_vs_freq_using_jet_pa}.  Since we observed triplets of closely spaced sources
related by
a single phase solution, we obtained two measurements for each object. These
plots show that the measurements for a given source and their uncertainties
can depend strongly on which source in the pair is used to make the
measurements. This is related to the different distances between the sources
and the angles between their jet directions. Nevertheless, in most cases,
the results of these pairwise core-shift measurements agree with the results
obtained for the triplets as a whole.

\begin{table*}[t!]
\caption{Core shifts measured by alignment pairs of images at different frequencies
(Section~\ref{s:method_image})}
\label{tab:cs_image}
\newcolumntype{a}{D{.}{.}{2.2}}

\begin{tabular}{c|c|c|a|a|a|a}
\hline \multirow{3}*{Source} & \multirow{3}*{Frequency 1} &
\multirow{3}*{Frequency 2} &
\multicolumn{4}{c}{Core shift (mas)} \\
    \cline{4-7}
    & & & \multicolumn{1}{c|}{\parbox[c][1cm]{2cm}{Right Ascension}} &
    \multicolumn{1}{c|}{Declination} & \multicolumn{1}{c|}{Along jet} &
    \multicolumn{1}{c}{\parbox[c][1cm]{2cm}{Perpendicular to jet}}\\
\hline
 0133+476 &     C &     L &       -0.46 &         0.90 &      1.01 &  -0.02 \\
 0133+476 &     C &     S &       -0.22 &         0.92 &      0.92 &  -0.24 \\
 0133+476 &     X &     C &       -0.05 &         0.29 &      0.28 &  -0.09 \\
 0151+474 &     C &     L &       -0.04 &        -0.33 &      0.33 &  -0.04 \\
 0151+474 &     X &     L &       -0.10 &        -0.68 &      0.68 &  -0.09 \\
 0151+474 &     X &     S &       -0.02 &        -0.60 &      0.60 &  -0.01 \\
 0202+319 &     C &     L &        0.11 &         0.48 &      0.49 &  -0.05 \\
 0202+319 &     C &     S &        0.11 &         0.31 &      0.32 &  -0.08 \\
 0202+319 &     X &     S &       -0.02 &         0.58 &      0.58 &   0.09 \\
 0202+319 &     X &     C &       -0.03 &         0.07 &      0.07 &   0.04 \\
 0204+316 &     C &     S &        0.58 &        -0.50 &      0.74 &   0.23 \\
 0204+316 &     X &     L &        0.78 &        -0.99 &      1.25 &   0.14 \\
 0204+316 &     X &     S &        0.43 &        -1.10 &      1.16 &  -0.22 \\
 0204+316 &     X &     C &        0.19 &        -0.34 &      0.39 &  -0.02 \\
 0229+131 &     C &     L &        0.68 &         0.36 &      0.77 &  -0.00 \\
 0229+131 &     C &     S &        0.42 &         0.26 &      0.49 &   0.03 \\
 0239+175 &     C &     S &       -0.28 &         0.59 &      0.65 &   0.01 \\
 0445+097 &     C &     L &       -1.17 &        -0.21 &      1.01 &  -0.63 \\
 0445+097 &     C &     S &       -0.64 &        -0.42 &      0.76 &  -0.11 \\
 0446+113 &     C &     L &        0.72 &        -0.56 &      0.88 &   0.23 \\
 0446+113 &     C &     S &        0.14 &        -0.01 &      0.10 &   0.10 \\
 0447$-$010 &     C &     L &        0.89 &        -1.15 &      1.42 &   0.33 \\
 0447$-$010 &     C &     S &        0.65 &        -0.81 &      1.01 &   0.26 \\
 0450+013 &     C &     S &       -0.77 &        -0.53 &      0.89 &  -0.29 \\
 1745+085 &     C &     S &       -0.25 &        -0.33 &      0.40 &  -0.07 \\
 1745+085 &     X &     S &       -0.32 &        -0.51 &      0.60 &  -0.06 \\
 1745+085 &     X &     C &       -0.32 &        -0.13 &      0.26 &  -0.23 \\
 1749+062 &     X &     C &        0.20 &        -0.58 &      0.61 &   0.05 \\
 2144+092 &     S &     L &        0.21 &         0.07 &      0.22 &   0.03 \\
 2144+092 &     C &     L &        0.71 &         0.23 &      0.74 &   0.10 \\
 2144+092 &     C &     S &        0.65 &         0.06 &      0.65 &  -0.06 \\
 2144+092 &     X &     L &        1.06 &         0.25 &      1.09 &   0.06 \\
 2144+092 &     X &     S &        0.80 &         0.08 &      0.81 &  -0.06 \\
 2144+092 &     X &     C &        0.25 &        -0.03 &      0.25 &  -0.07 \\
 2145+067 &     C &     L &        0.45 &        -0.51 &      0.67 &  -0.13 \\
 2145+067 &     C &     S &        0.23 &        -0.27 &      0.35 &  -0.07 \\
\hline
\end{tabular}
\end{table*}

\begin{figure}
 \includegraphics[width=\linewidth,trim=0cm 0.3cm 0.2cm 0.2cm]{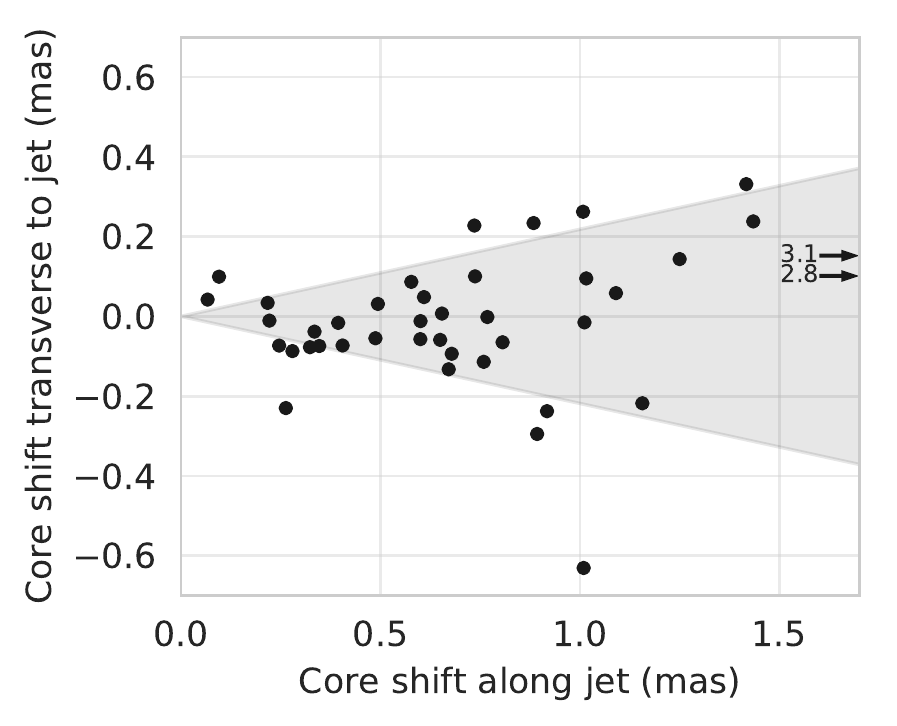}
 \caption{Measured core shifts relative to the jet directions. The arrows
show two measurements that appreciably exceed the others; the numbers next
to these arrows denote their positions along the horizontal axis.  68$\%$
of the shifts lie within $12^{\circ}$ of the horizontal direction, shaded in
the figure. The typical estimated uncertainties are 0.3~mas.}
 \label{fig:cs_alignment}
\end{figure}

\begin{figure}
 \includegraphics[width=\linewidth,trim=0.1cm 0.3cm 0.1cm 0.1cm]{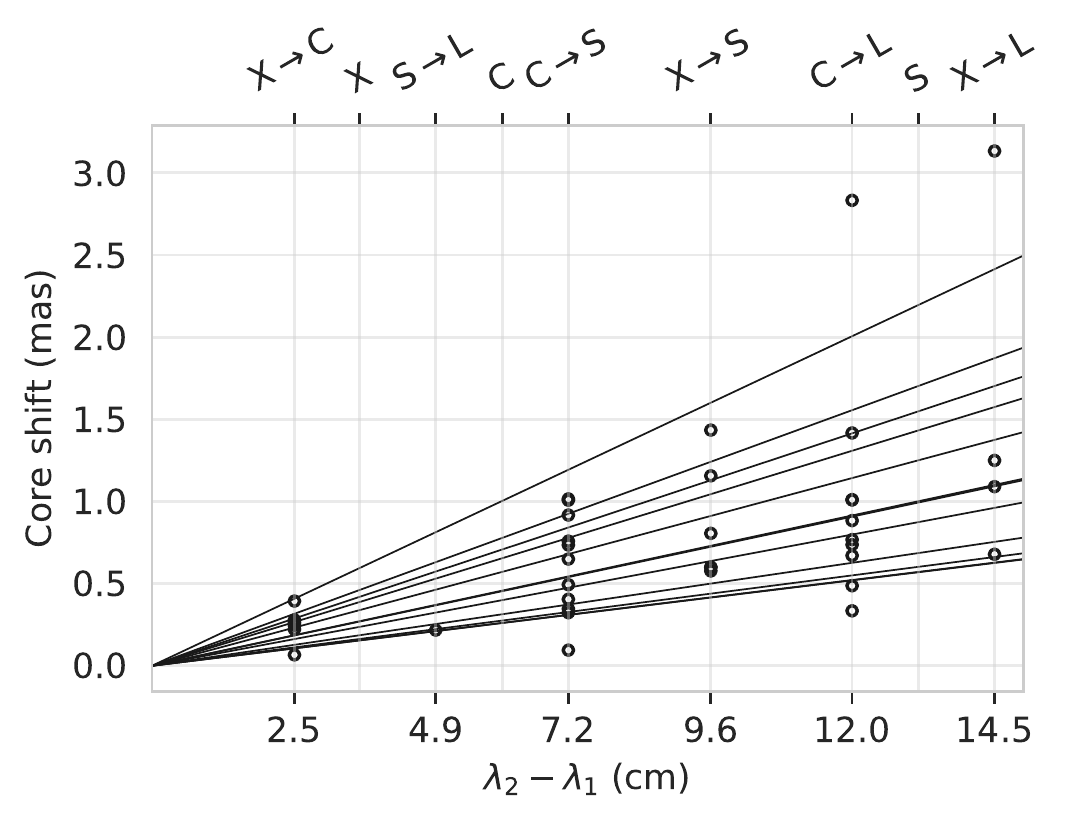}
 \caption{Dependence of the core shift measured via independent alignment
of image pairs of a given source at different bands on the difference
between corresponding wavelengths. Assuming $r_\textrm{core} \propto \nu^{-1}
\propto \lambda$, the dependence for each source is a straight line. Only
sources whose shifts were measured for more than one pair of frequencies are
shown. The individual measurements and the best-fit linear approximation for
each source are depicted. No uncertainties are shown; typical uncertainties
are 0.3~mas. The symbols X, C, and S denote the values $\lambda_2 - \lambda_1$,
corresponding to the shifts from the position at the indicated band to the
jet apex (i.e., $\lambda_1 = 0$).  The lines correspond to the sources
(from the bottom up): 0202+319, 0151+474, 0446+113, 2145+067, 0229+131,
2144+092, 1745+085, 0445+097, 0133+476, 0204+316, and 0447$-$010.}
 \label{fig:cs_freqdep}
\end{figure}

\subsection{Core-shift Measurements by Images Alignment at Different Frequencies}
\label{s:method_image}

The core shifts for sources with fairly extended structures can also be
measured by aligning images of the source obtained at pairs of frequencies,
as is described in~\cite{Plavin2018}. The reconstructed images were convolved with
identical beams corresponding to the mean beam size for all the frequencies.
The core position was determined via modeling of the source structure as
a set of circular Gaussian components (see Section~\ref{s:modeling}). It was possible
to determine the core shifts for one or more pairs of frequencies for 15
sources in this way (41 frequency pairs in all). Figure~\ref{fig:cs_alignment} shows that these
measured core shifts are in good agreement with the assumption that they
should lie along the jet.

Since the core shifts for 12 sources were measured for more than one
frequency pair, this makes it possible to study the frequency dependence of
these shifts. We fitted a model assuming a dependence of the form
$r_\textrm{core} \propto \nu^{-1/k_\textrm{r}}$ to the core-shift measurements
obtained. This fitting placed essentially no limits on the value of $k_r$,
which ranged from $0.6$ to infinity. Therefore, we assumed $k_\textrm{r} = 1$
for our subsequent analysis, in agreement with the earlier results of~\cite{Sokolovsky_2011}:
$r_\textrm{core} \propto \nu^{-1} \propto \lambda$ and $\Delta
r_\textrm{core} \propto \lambda_2 - \lambda_1$. Figure~\ref{fig:cs_freqdep} shows the dependence
of the core shift on the difference between the wavelengths used to determine it.

The typical core shifts measured between X and S bands (8 and 2~GHz) in the
current study, 0.65~mas, is in good agreement with the results
of~\cite{Kovalev_2008,Sokolovsky_2011,Plavin2018},
where the median shift $X\to S$ is estimated to be 0.44, 0.71, and 0.53~mas,
respectively.

Note that we were able to measure the shift between S and L bands (2.3 and
1.7~GHz) for only one source. This came about because these frequencies are
fairly close, and the resolution obtained at these frequencies is a factor
of three to four lower than the resolution obtained at X and C band. Two
measurements appreciably exceeding the typical core-shift values were obtained
for 0217+324: $2.8$ and $3.1$~mas for the frequency pairs $C\to L$ and
$X\to L$, respectively, shown by the arrows in Fig.~\ref{fig:cs_alignment}. A comparison of the
L-band image of this source with the other L-band images indicates that
there were no methodological errors in the measurements. This suggests that
we are seeing some more distant region of the jet at L band that is brighter
than the L-band core. This requires a separate study based on images with
higher sensitivity.

\begin{figure*}
 \includegraphics[width=0.24\textwidth]{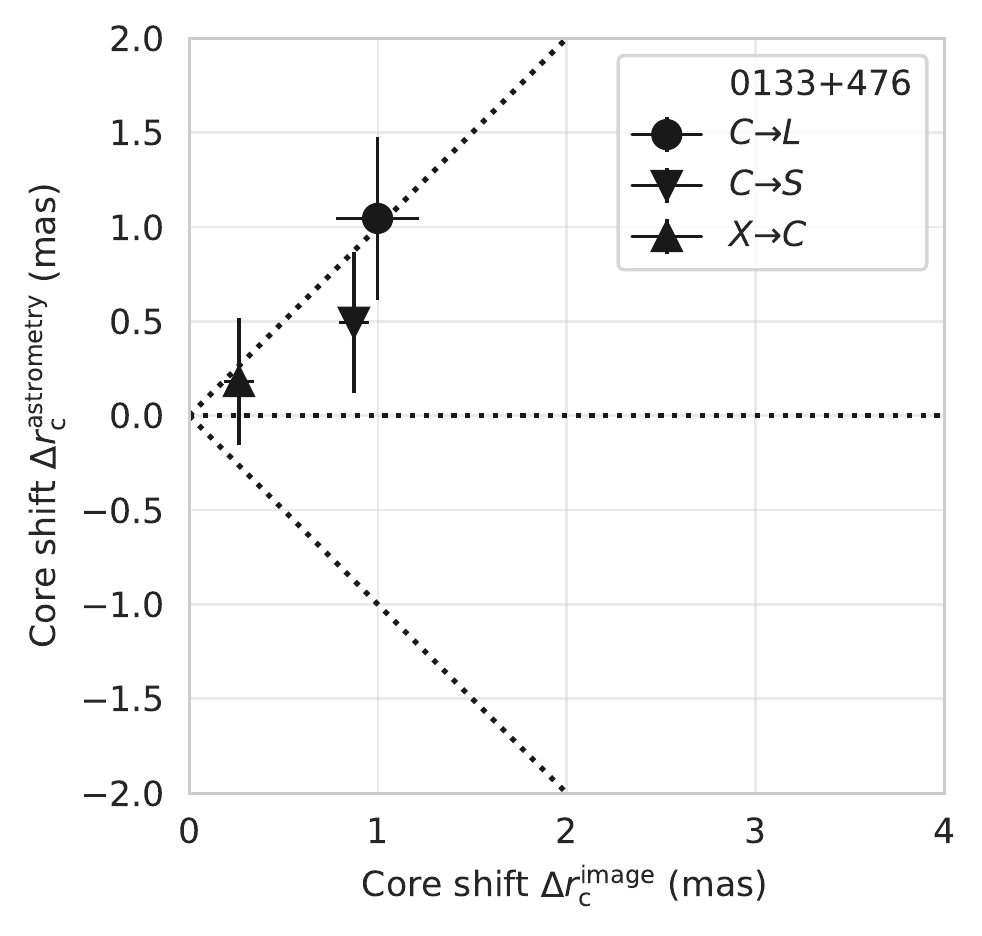}
 \includegraphics[width=0.24\textwidth]{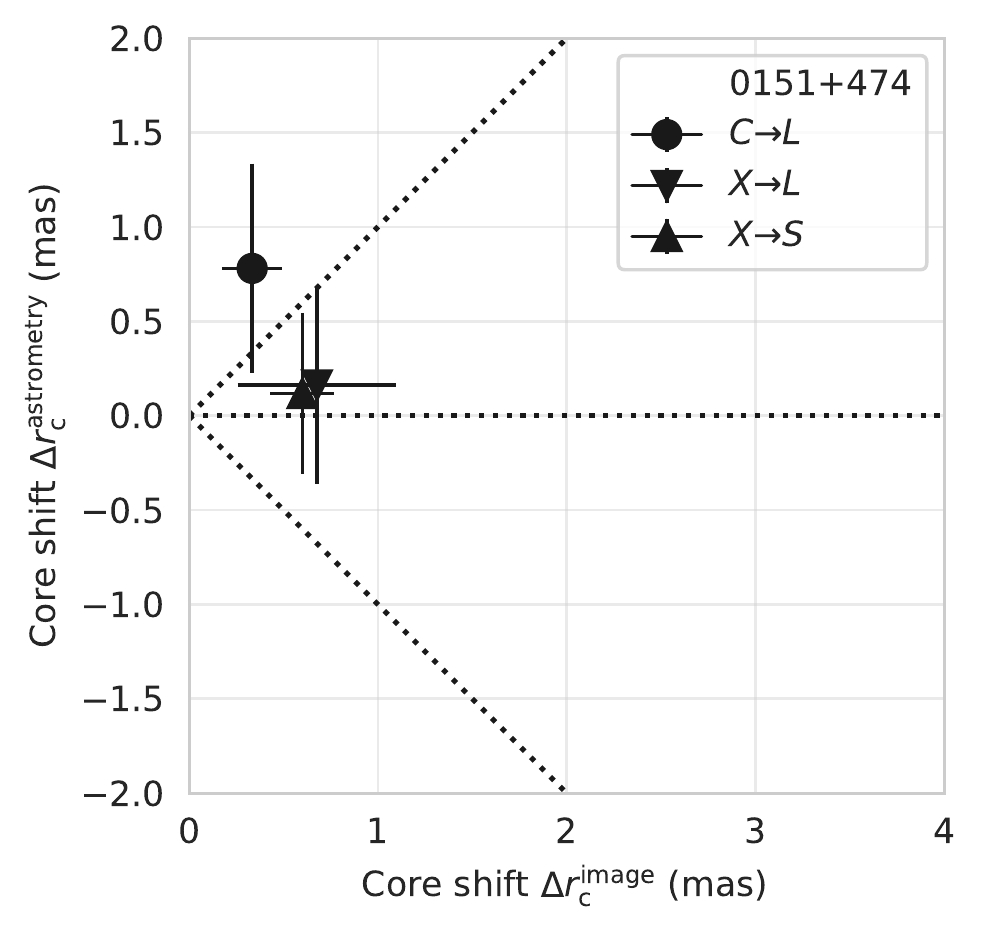}
 \includegraphics[width=0.24\textwidth]{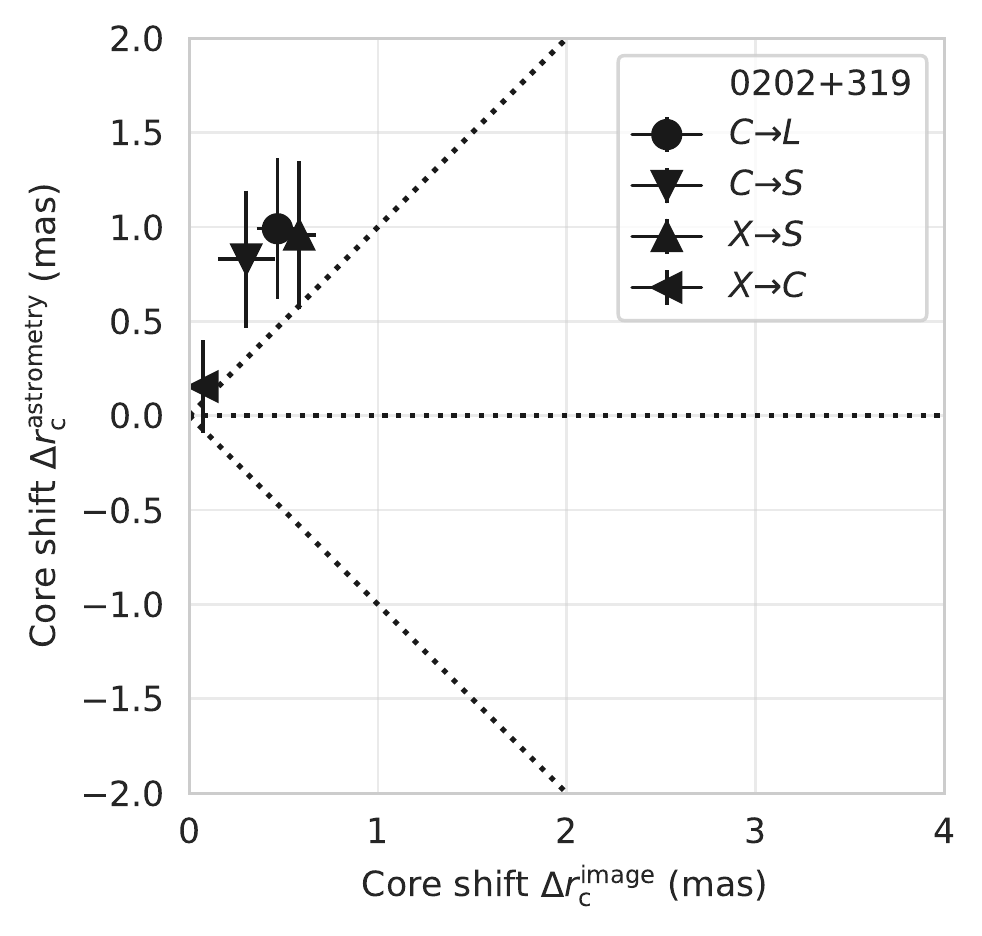}
 \includegraphics[width=0.24\textwidth]{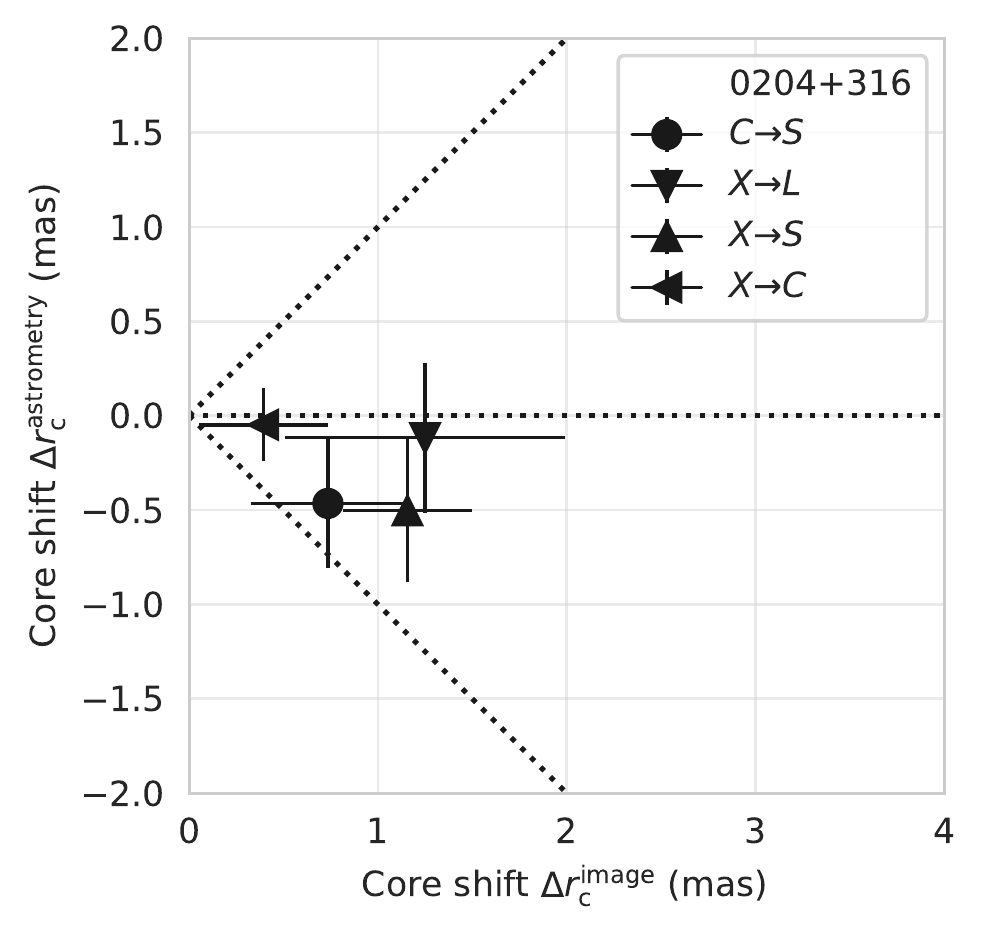}
 \includegraphics[width=0.24\textwidth]{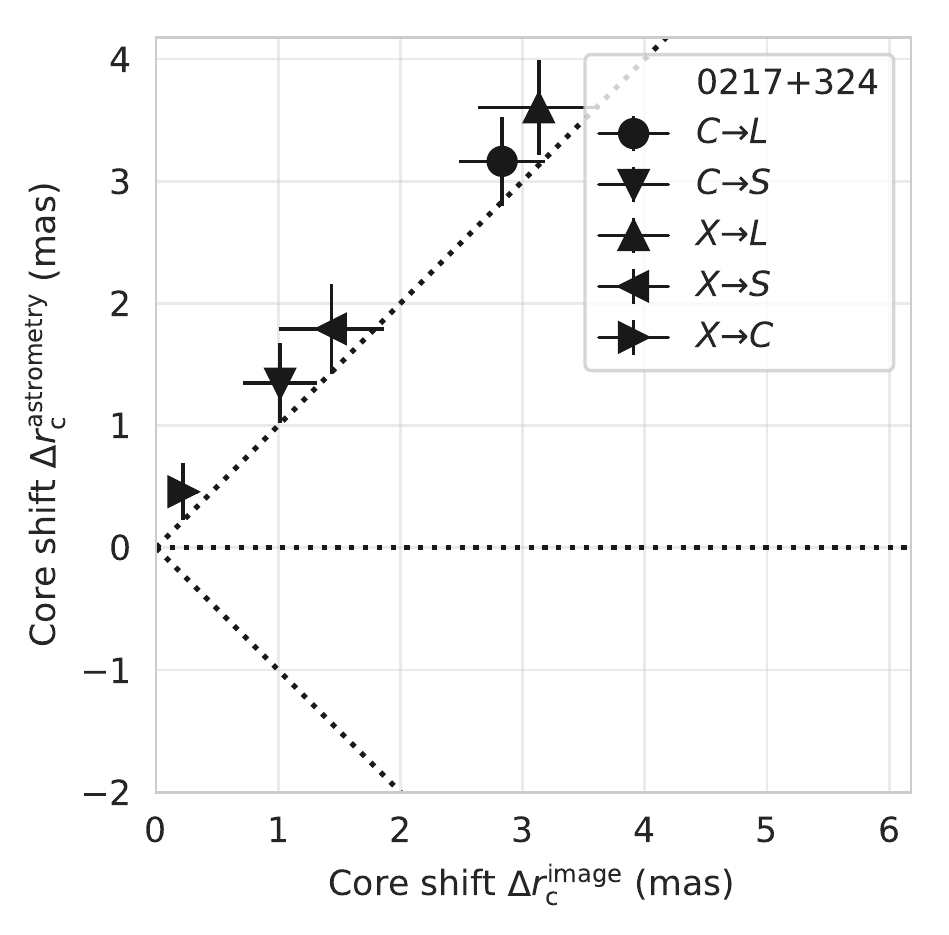}
 \includegraphics[width=0.24\textwidth]{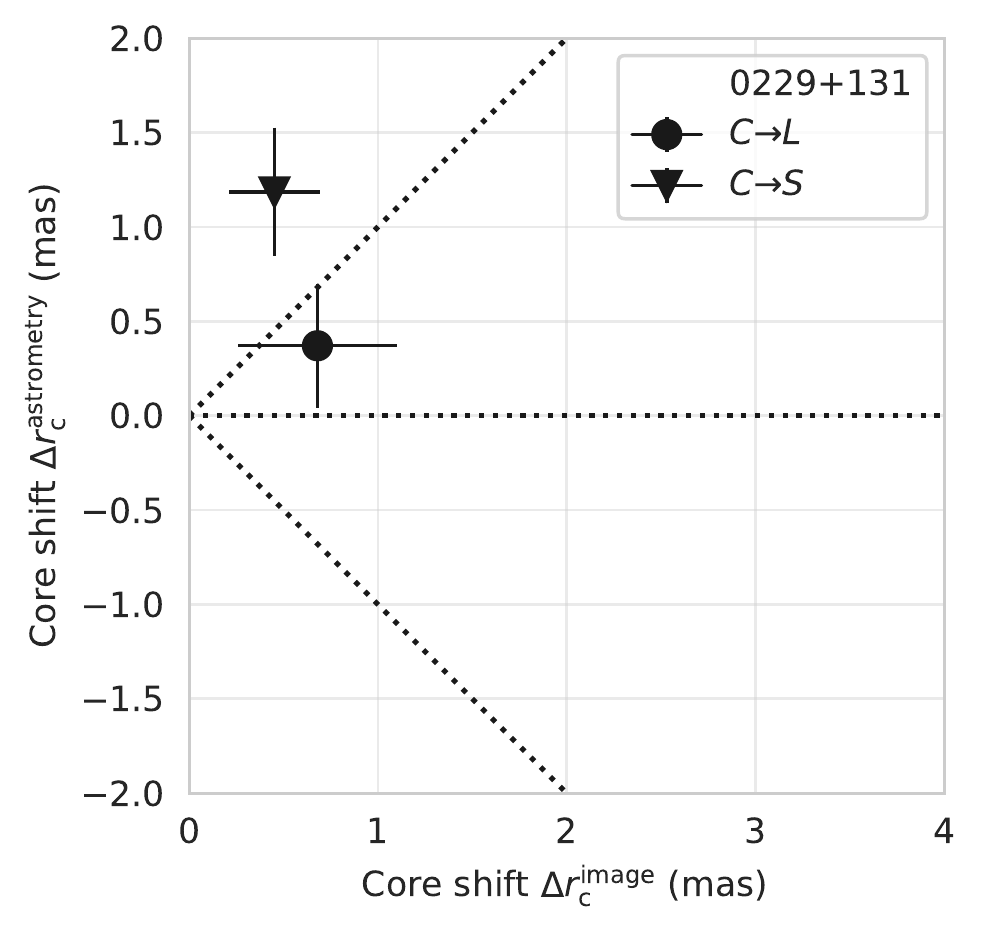}
 \includegraphics[width=0.24\textwidth]{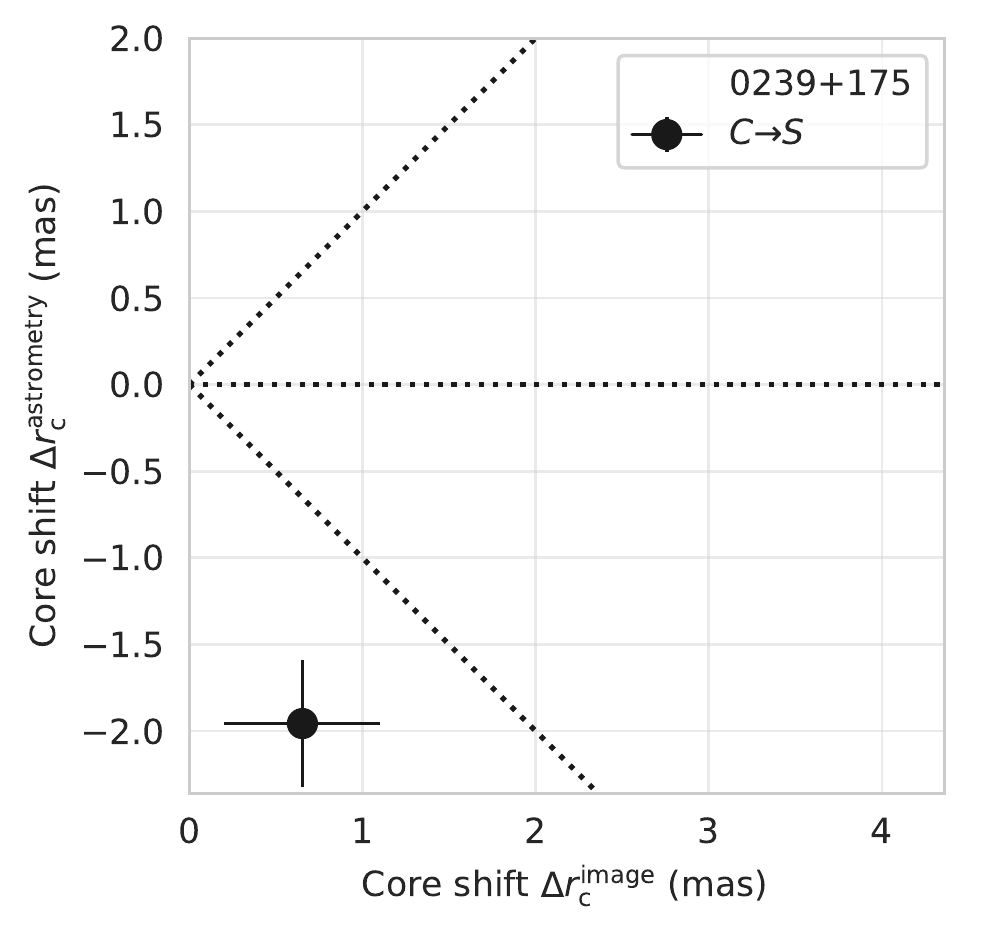}
 \includegraphics[width=0.24\textwidth]{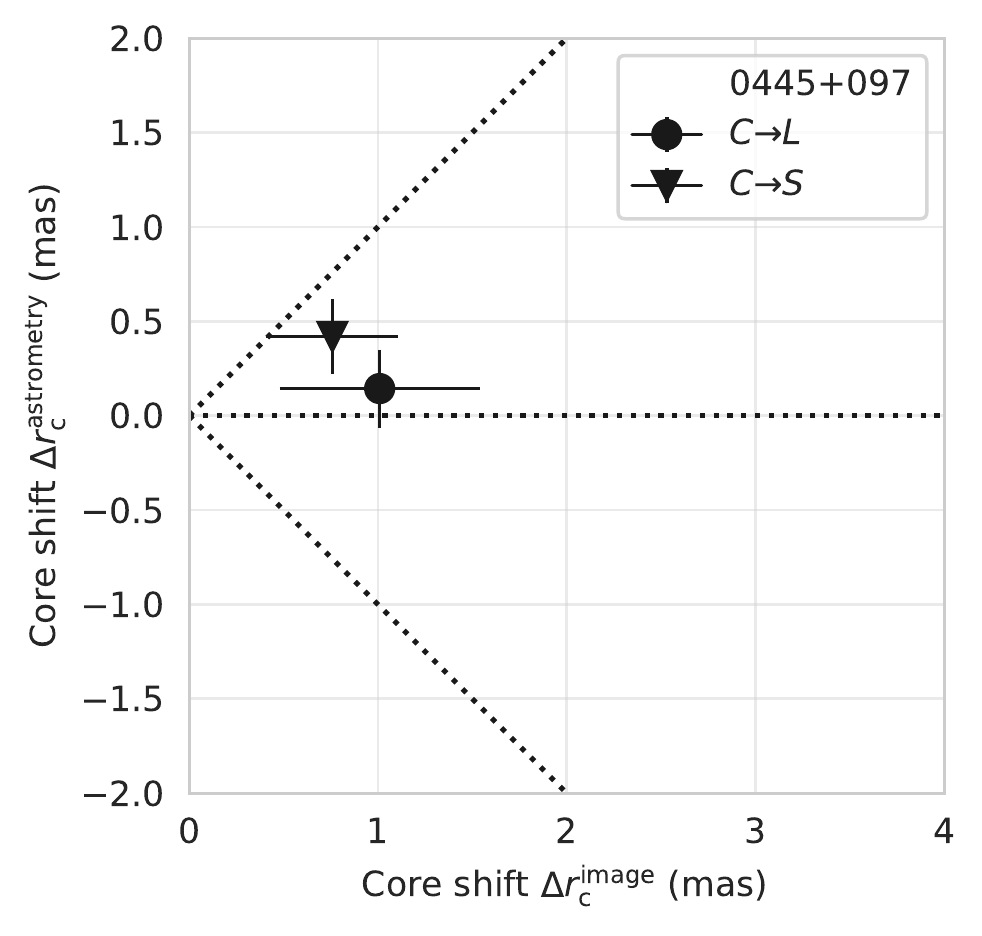}
 \includegraphics[width=0.24\textwidth]{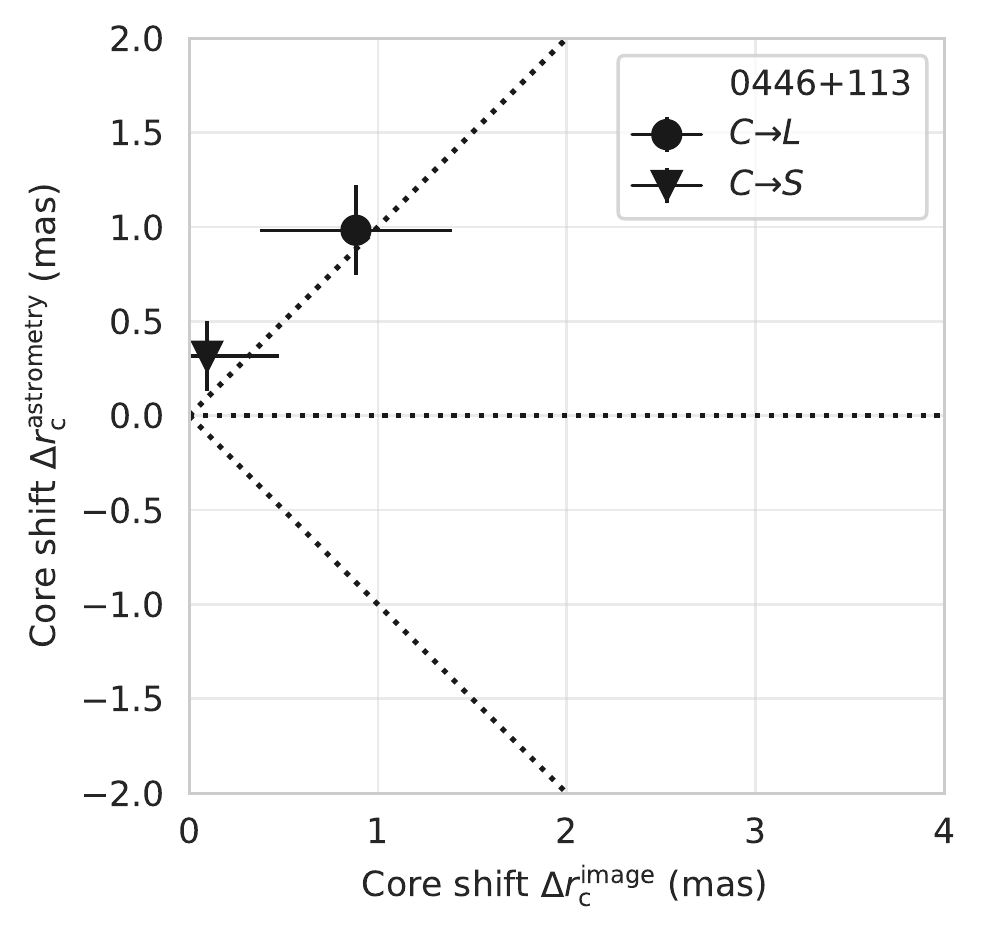}
 \includegraphics[width=0.24\textwidth]{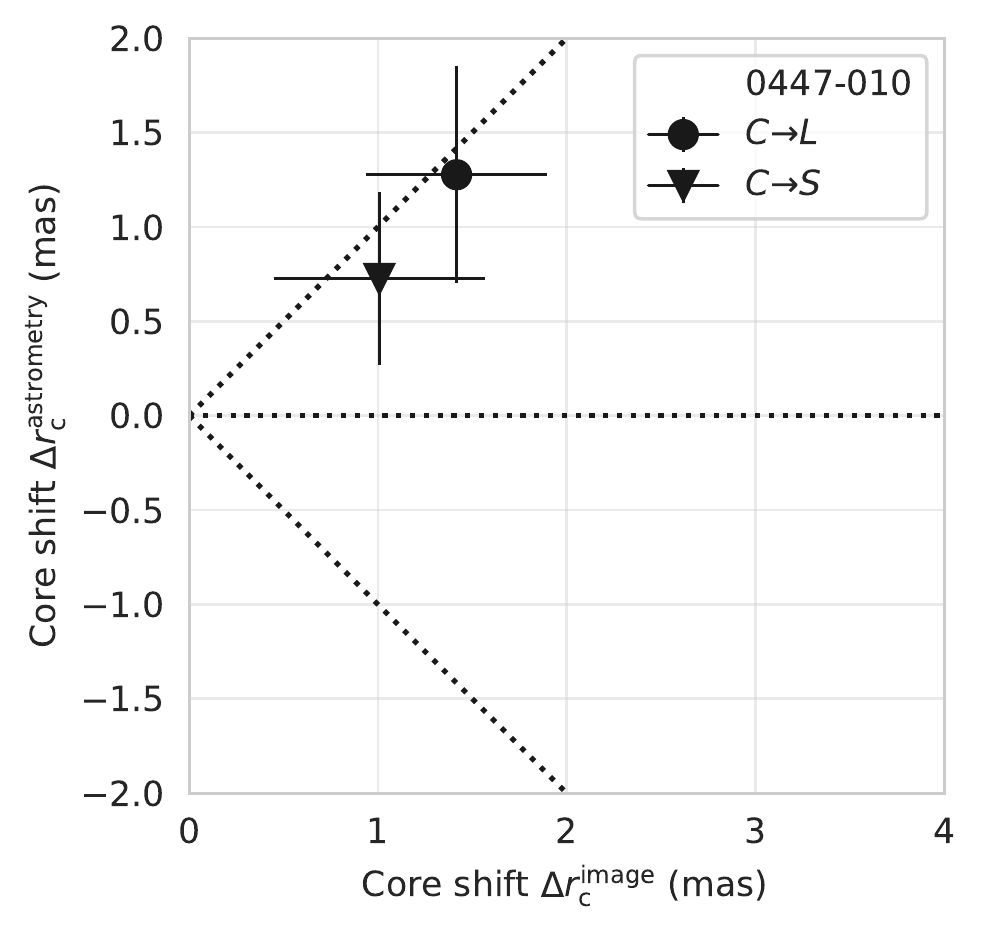}
 \includegraphics[width=0.24\textwidth]{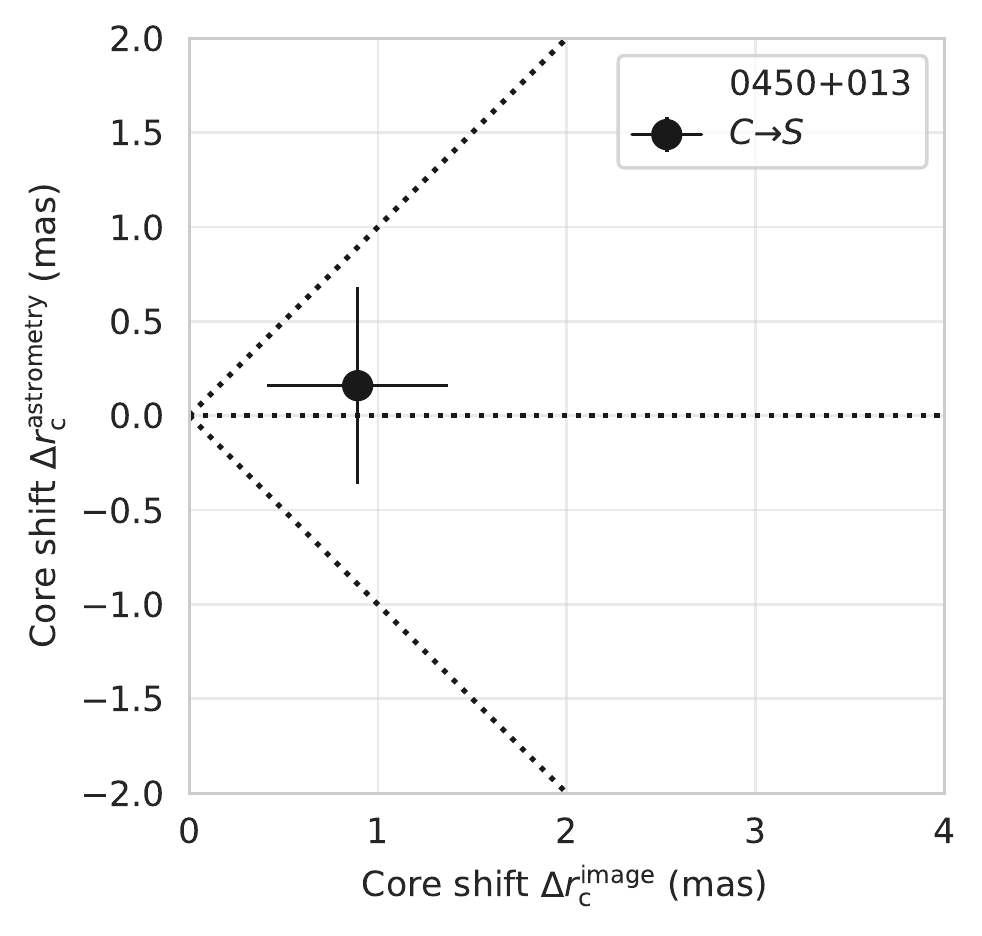}
 \includegraphics[width=0.24\textwidth]{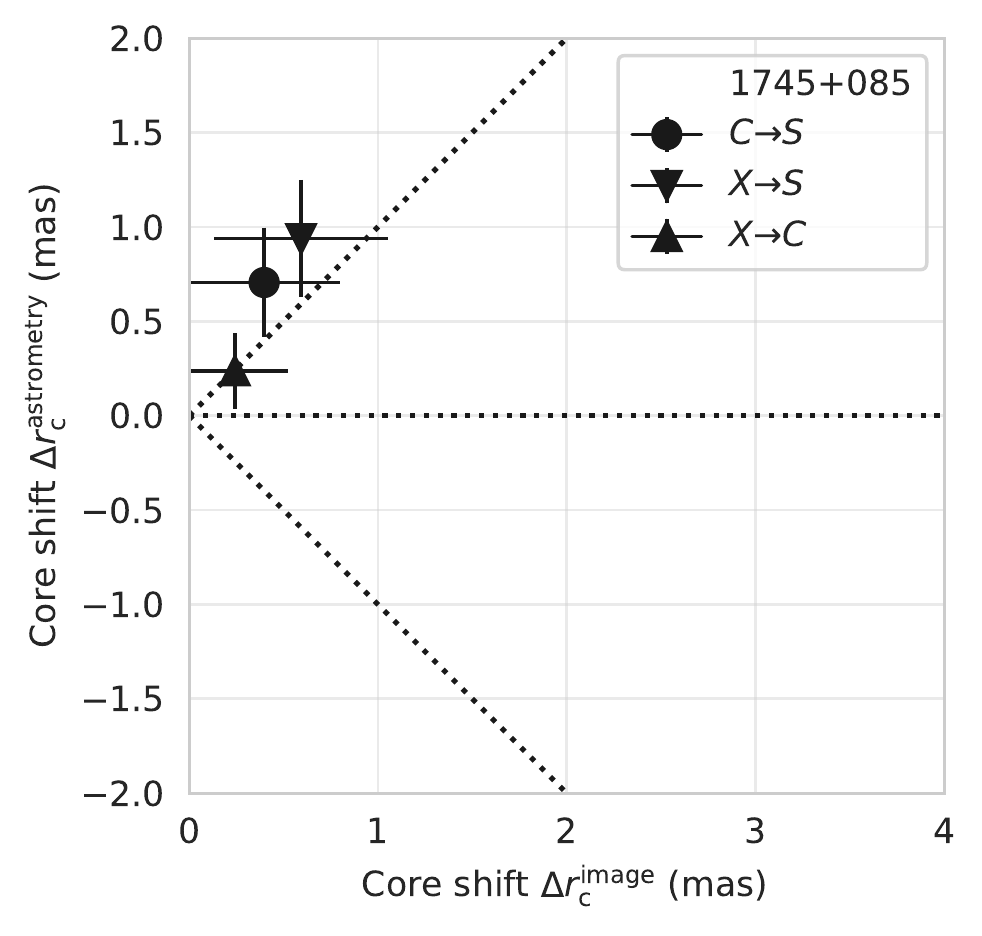}
 \includegraphics[width=0.24\textwidth]{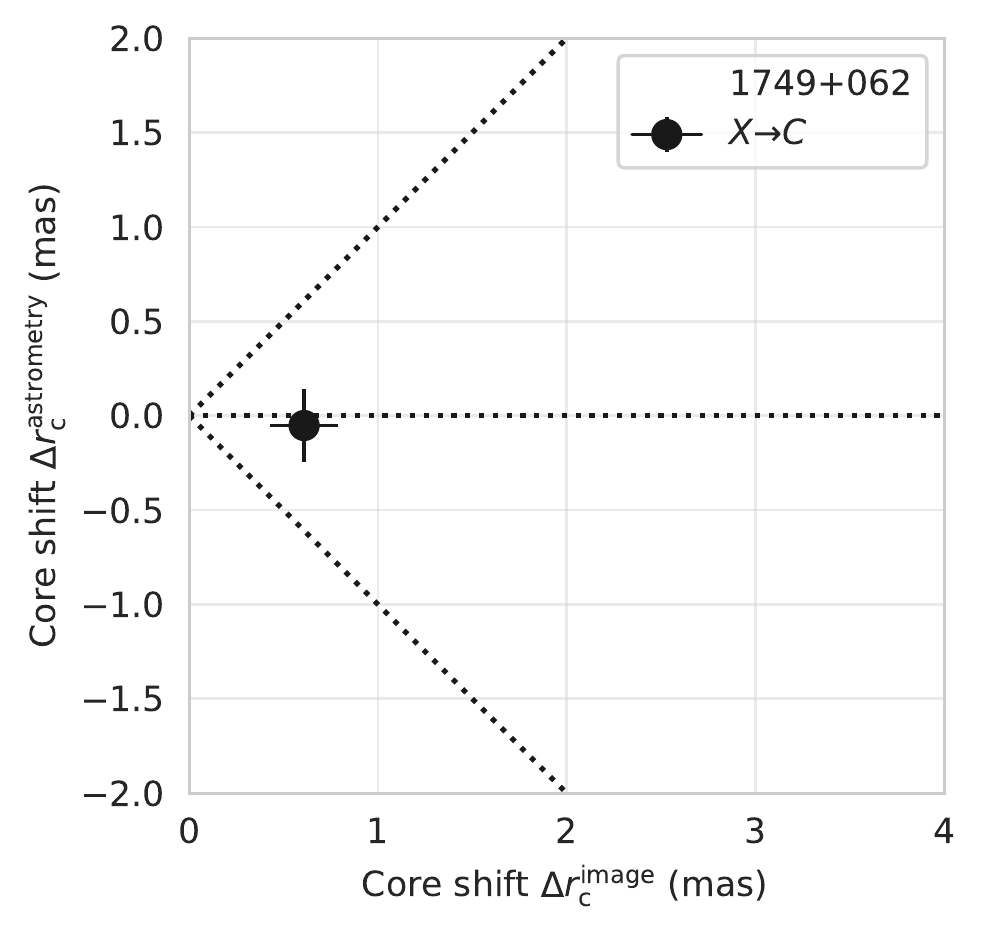}
 \includegraphics[width=0.24\textwidth]{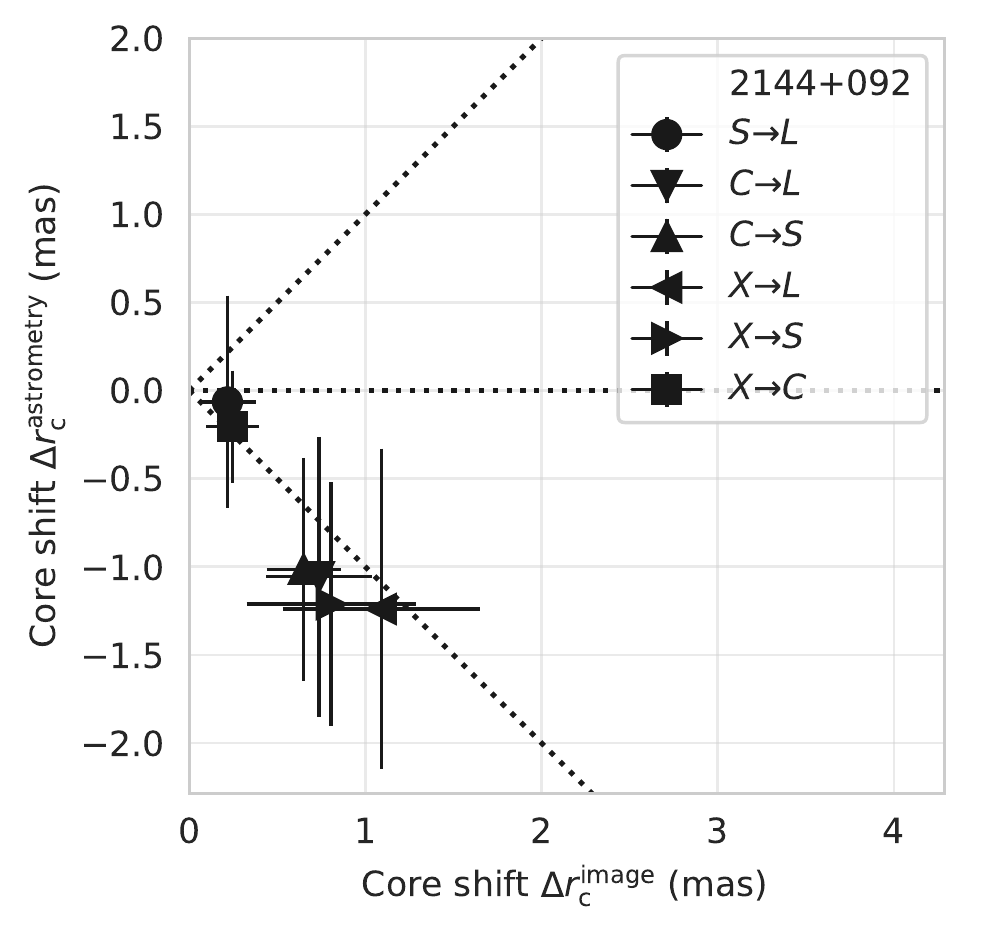}
 \includegraphics[width=0.24\textwidth]{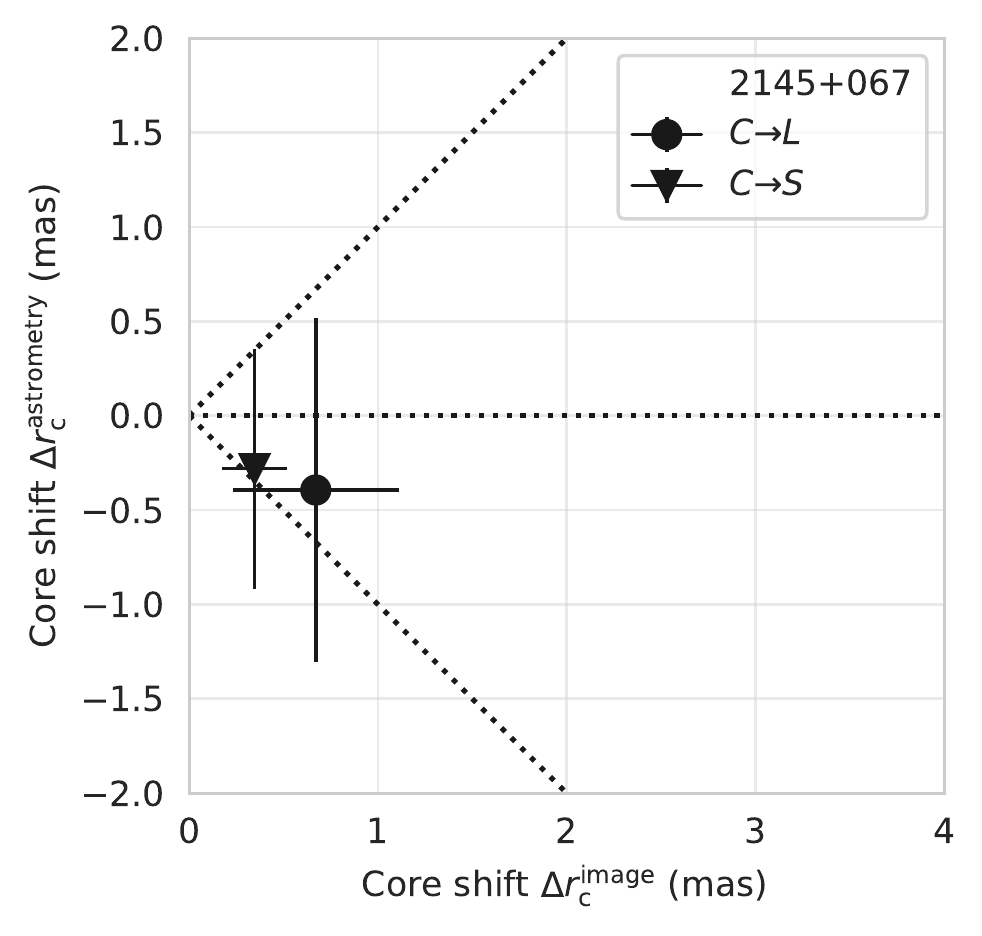}
 \caption{Comparison of the core shifts measured using two methods: images alignment at different
frequencies, $\Delta r_\textrm{c}^\textrm{image}$
(Section~\ref{s:method_image}) and applying the astrometric method $\Delta
r_\textrm{c}^\textrm{astrometry}$ (Section~\ref{s:method_astrometry},
Fig.~\ref{fig:cs_freqdep_astrom}). The dotted lines
show $\Delta r_\textrm{c}^\textrm{astrometry}=\pm \Delta
r_\textrm{c}^\textrm{image}$ and $\Delta r_\textrm{c}^\textrm{astrometry}= 0$.}
 \label{fig:rel_vs_astrom}
\end{figure*}

Figure~\ref{fig:rel_vs_astrom} presents a comparison of the results obtained using the various
methods considered. For three of five sources with good $uv$-coverage, that is,
with high declinations (the first five objects in this figure, which have
declinations ${>} 30^{\circ}$), the results for the two methods are fairly
similar and show the same frequency dependence. A variety of behavior is
seen in the remaining cases, from good agreement (e.g., 1745+085) to completely
opposite results, i.e., with the shifts directed in opposite directions
(e.g., 2144+092).

\begin{table}
\caption{Physical parameters of the sources (see Section~\ref{s:res_phys}).  $r_X$ is the
distance from the jet apex to the visible core in the X band (8~GHz) projected
onto the sky and $B_1$ is the magnetic field 1~pc from the jet apex}
\label{tab:res_phys}
\setlength{\tabcolsep}{10pt}
\begin{tabular}{c|c|c|c}
\hline
\multirow{2}*{Source} & \multicolumn{2}{c|}{$r_X$} &  \multirow{2}*{$B_1$, G} \\
\cline{2-3}
& mas &  pc &   \\
\hline
0133+476 & 0.39 & 3.03 & 1.49 \\
0151+474 & 0.16 & 1.34 & 0.72 \\
0202+319 & 0.15 & 1.33 & 0.85 \\
0204+316 & 0.42 & 3.55 & 2.07 \\
0217+324 & 0.59 &      &      \\
0229+131 & 0.24 & 2.02 & 1.53 \\
0445+097 & 0.34 & 2.88 & 2.21 \\
0446+113 & 0.15 & 1.33 & 0.82 \\
0447$-$010 & 0.46 & 2.81 & 1.04 \\
1745+085 & 0.27 &      &      \\
2144+092 & 0.27 & 2.26 & 1.26 \\
2145+067 & 0.19 & 1.51 & 0.80 \\
\hline
\end{tabular}
\end{table}

\subsection{Geometry and Physical Parameters}
\label{s:res_phys}

All of our conclusions about the physical structure of the jet presented in
this section are based on our core-shift measurements obtained by alignment of images
at different frequencies (Section~\ref{s:method_image}). The astrometric method
(Section~\ref{s:method_astrometry}) sometimes yields results that are not in agreement with our
basic assumption that core shift is due to synchrotron self-absorption at
the bases of the jets. Additional measurements and verification are
required to more fully understand those measurements.

Assuming that, on average, the position of the core as a function of the
wavelength is $r_\textrm{core} \propto \lambda$, we can estimate the
distance from the intrinsic base (apex) of the jet (corresponding to
$\lambda_1 = 0$ in the formulas above) to the core observed at a given
frequency.  The core positions at X, C, and S bands are given in Fig.~\ref{fig:cs_freqdep}.
The typical distance from the jet apex to the 8~GHz core is 0.27~mas, or
2.1~pc projected onto the sky. The intrinsic distance from the jet apex
to the 8~GHz core is ${\sim} 20$~pc for a jet with a typical angle to the
line of sight of $\theta \sim 6^{\circ}$~\cite{Hovatta2009}.

We can use the measured core shifts to estimate the magnetic-field strength
near the base of the jet. Assuming equipartition between the energies of
the magnetic field and particles, and adopting a spectral index for the
jet $\alpha={-}0.5$, which is taken to be observed at close to the critical
angle (for which the apparent component speed $\beta_\textrm{app}$ is
maximum), the magnetic field in Gauss at a distance of 1~pc from the intrinsic
jet base can be estimated as follows \cite{Pushkarev_2012}:
\begin{equation}
B_1 \approx 0.042 \Omega_\textrm{rv}^{3/4} (1+z)^{1/2}
(1+\beta_\textrm{app}^2)^{1/8} \,,
\end{equation}
where $\beta_\textrm{app}$ is the apparent speed of the jet and
$\Omega_\textrm{rv}$ a parameter corresponding to the core shift measured in
pc$\cdot$GHz. Since the dependence of $B_1$ on $\beta_\textrm{app}$
is weak ($B_1$ increases by only a factor of 1.8 as $\beta_\textrm{app}$
increases from 0 to 10), We used the fixed factor
$(1+\beta_\textrm{app}^2)^{1/8} = 1.5$, which corresponds to
$\beta_\textrm{app} = 5$. The mean magnetic field 1~pc from the intrinsic jet
apex is then $B_\text{1,mean} = 1.2$~G; values for individual
sources are presented in Table~\ref{tab:res_phys}.

\section{CONCLUSION}

We have developed two approaches to measuring the frequency-dependent core
shifts of each source in a group of closely spaced quasars, which are
related by a single phase solution derived from VLBI relative astrometry.
In one approach, the difference in the core shifts for two sources is
written in terms of contributions in the directions of the relativistic
jets of each, assuming that the shifts occur along these directions. In the
other approach, we use measurements of an arbitrary number of closely spaced
sources to estimate the core shifts taking into account \emph{a priori}
information about the jet directions, for either all or only some of the
sources.

We organized pilot observations of eight triplets of compact extragalactic
radio sources on the EVN to test these methods. Three telescopes of the
Russian Kvazar--KVO array took part in these observations. Their participation
improved the resulting measurement accuracy and the sensitivity and quality
of the reconstructed images. Unfortunately, the failure of the Hartebeeshoek
telescope in South Africa led to a loss of long baselines in the North--South
direction, which appreciably lowered the accuracy of the measurements.

We have estimated the frequency-dependent core shifts for the 24 objects
included in this study. The measured core shifts for nine of these are
significant. For these sources, the median VLBI core shifts at 1.7, 2.3, and
5.0~GHz relative to our highest frequency, 8.4~GHz, were 1.79, 1.22, and
0.18~mas, respectively. We were able to independently measure the core shifts
of a number of sources via self-referencing using extended, optically thin
regions in the source. For these, we have also estimated the distance from
the 8.4~GHz VLBI core to the intrinsic base of the relativistic jet and the
magnetic-field strength 1~pc from this base. The typical values of these
quantities for these sources were 2~pc projected onto the sky and 1.2~G.

We conclude that the relative-astrometry method can be used to measure
frequency-dependent VLBI core shifts for very compact sources for which
other methods are not suitable. This is especially important for objects
defining the highest-accuracy inertial reference frame available --- the
International Celestial Reference Frame (ICRF) --- which is based on VLBI
measurements. Achieving the required accuracy and reliability of these
measurements requires VLBI observations with good $uv$-coverage, high
sensitivity, and high angular resolution.

\section*{ACKNOWLEDGEMENTS}

We deeply thank Denise Gabuzda for translating this article into English.
This work was supported by the Russian Science Foundation (grant 16-02-10481).
The European VLBI Network is a joint facility of independent European,
African, Asian, and North American radio astronomy institutes. Scientific
results from data presented in this publication are derived from the following
EVN project code(s): EK028.

\bibliography{core_shift}

\appendix

\clearpage

\begin{center}
\newcolumntype{e}{D{.}{.}{4.0}}
\newcolumntype{a}{D{.}{.}{2.2}}
\newcolumntype{b}{D{.}{.}{2.1}}

\begin{longtable*}{c|c|e|c|e|a|a|b}
\caption{Map parameters} \label{tab:maps} \\
\hline
Source & Frequency (GHz) & \multicolumn{1}{c|}{$I_{\text{peak}}$ (mJy/beam)} & rms (mJy/beam) &
\multicolumn{1}{c|}{$S_\text{tot}$ (mJy)} &
\multicolumn{1}{c|}{$\theta_\text{maj}$ (mas)} &
\multicolumn{1}{c|}{$\theta_\text{min}$ (mas)} &
\multicolumn{1}{c}{$PA_{\text{beam}}$ (deg)} \\ \hline

\endfirsthead

\multicolumn{8}{c}%
{{\tablename\ \thetable{} -- Continued}} \\
\hline
Source & Frequency (GHz) & \multicolumn{1}{c|}{$I_{\text{peak}}$ (mJy/beam)} & rms (mJy/beam) &
\multicolumn{1}{c|}{$S_\text{tot}$ (mJy)} &
\multicolumn{1}{c|}{$\theta_\text{maj}$ (mas)} &
\multicolumn{1}{c|}{$\theta_\text{min}$ (mas)} &
\multicolumn{1}{c}{$PA_{\text{beam}}$ (deg)} \\ \hline
\endhead

\hline \hline
\endlastfoot

\multirow{4}*{0125+487} & 1.659 &   192 &  0.13 &   262 & 10.25 &  4.84 &  17.8 \\
           & 2.273 &   233 &  0.67 &   361 &  5.13 &  2.54 &  16.7 \\
           & 4.975 &   314 &  0.12 &   370 &  3.02 &  1.51 &  16.3 \\
           & 8.385 &   394 &  0.17 &   455 &  1.76 &  0.87 &  16.0 \\ \hline
\multirow{4}*{0133+476} & 1.659 &  1277 &  0.31 &  1768 &  9.11 &  4.04 &  17.8 \\
           & 2.273 &  1608 &  0.67 &  2489 &  5.34 &  2.59 &  20.1 \\
           & 4.975 &  2317 &  0.31 &  2959 &  2.70 &  1.25 &  16.4 \\
           & 8.385 &  3039 &  0.43 &  3701 &  1.55 &  0.70 &  22.6 \\ \hline
\multirow{4}*{0151+474} & 1.659 &   234 &  0.16 &   308 & 10.86 &  4.98 &  20.3 \\
           & 2.273 &   380 &  0.68 &   455 &  5.28 &  2.64 &  22.0 \\
           & 4.975 &   550 &  0.11 &   603 &  2.94 &  1.48 &  18.6 \\
           & 8.385 &   638 &  0.22 &   702 &  1.71 &  0.83 &  24.5 \\ \hline
\multirow{4}*{0202+319} & 1.659 &   729 &  0.27 &   988 & 14.21 &  4.21 &  10.8 \\
           & 2.273 &  1026 &  0.37 &  1331 &  7.33 &  2.45 &   9.5 \\
           & 4.975 &  1820 &  0.25 &  2001 &  3.75 &  1.30 &   6.7 \\
           & 8.385 &  2861 &  0.47 &  3078 &  2.29 &  0.74 &   9.6 \\ \hline
\multirow{4}*{0204+316} & 1.659 &   237 &  0.17 &   608 & 15.05 &  4.42 &  11.3 \\
           & 2.273 &   271 &  0.50 &   662 &  7.42 &  2.44 &  10.9 \\
           & 4.975 &   194 &  0.12 &   396 &  4.27 &  1.45 &   7.9 \\
           & 8.385 &   164 &  0.20 &   273 &  2.64 &  0.85 &  13.3 \\ \hline
\multirow{4}*{0217+324} & 1.659 &   457 &  0.18 &   825 & 14.47 &  4.35 &  12.1 \\
           & 2.273 &   359 &  0.59 &   765 &  7.14 &  2.47 &  11.6 \\
           & 4.975 &   160 &  0.14 &   428 &  4.24 &  1.42 &   8.8 \\
           & 8.385 &   131 &  0.14 &   320 &  2.51 &  0.82 &  14.4 \\ \hline
\multirow{4}*{0229+131} & 1.659 &  1117 &  0.50 &  1508 & 20.10 &  4.14 &   9.3 \\
           & 2.273 &  1693 &  0.85 &  1996 &  9.39 &  2.54 &   8.6 \\
           & 4.975 &  1679 &  0.29 &  2029 &  4.87 &  1.23 &   6.3 \\
           & 8.385 &  1398 &  0.39 &  1954 &  3.30 &  0.69 &  10.0 \\ \hline
\multirow{4}*{0235+164} & 1.659 &  2038 &  1.16 &  2088 & 18.05 &  3.98 &  10.4 \\
           & 2.273 &  2906 &  0.78 &  3037 &  8.97 &  2.54 &  10.6 \\
           & 4.975 &  4751 &  0.48 &  4930 &  4.27 &  1.20 &   7.0 \\
           & 8.385 &  6552 &  0.58 &  6849 &  2.75 &  0.63 &  10.6 \\ \hline
\multirow{4}*{0239+175} & 1.659 &   154 &  0.06 &   180 & 21.30 &  4.71 &  10.3 \\
           & 2.273 &   183 &  0.67 &   251 &  9.07 &  2.51 &  11.2 \\
           & 4.975 &   187 &  0.10 &   229 &  5.52 &  1.43 &   7.3 \\
           & 8.385 &   257 &  0.16 &   283 &  3.50 &  0.82 &  10.8 \\ \hline
\multirow{4}*{0440$-$003} & 1.659 &  2551 &  1.27 &  2655 & 25.53 &  4.21 &   6.1 \\
           & 2.273 &  3023 &  0.92 &  3343 & 11.80 &  2.43 &   7.3 \\
           & 4.975 &  1676 &  0.36 &  2190 &  6.24 &  1.17 &   4.4 \\
           & 8.385 &  1095 &  0.45 &  2017 &  4.09 &  0.71 &   7.2 \\ \hline
\multirow{4}*{0445+097} & 1.659 &   186 &  0.63 &   366 & 30.15 &  4.73 &   6.1 \\
           & 2.273 &   207 &  0.62 &   324 & 10.63 &  2.39 &   7.0 \\
           & 4.975 &   260 &  0.31 &   324 &  6.23 &  1.49 &   4.1 \\
           & 8.385 &   399 &  0.24 &   469 &  3.98 &  0.89 &   7.6 \\ \hline
\multirow{4}*{0446+112} & 1.659 &   726 &  0.62 &  1174 & 23.51 &  4.04 &   5.8 \\
           & 2.273 &   861 &  0.49 &  1348 & 10.33 &  2.40 &   7.3 \\
           & 4.975 &   735 &  0.17 &  1000 &  5.75 &  1.26 &   4.3 \\
           & 8.385 &   907 &  0.22 &  1232 &  3.79 &  0.78 &   7.8 \\ \hline
\multirow{4}*{0446+113} & 1.659 &   125 &  0.06 &   267 & 24.27 &  4.27 &   5.9 \\
           & 2.273 &   151 &  0.54 &   299 & 10.53 &  2.39 &   6.9 \\
           & 4.975 &   313 &  0.15 &   414 &  5.91 &  1.45 &   3.9 \\
           & 8.385 &   309 &  0.24 &   356 &  3.98 &  0.86 &   7.8 \\ \hline
\multirow{4}*{0447$-$010} & 1.659 &   146 &  0.09 &   180 & 31.45 &  4.77 &   6.2 \\
           & 2.273 &   162 &  0.60 &   218 & 11.94 &  2.38 &   7.5 \\
           & 4.975 &   115 &  0.09 &   159 &  7.20 &  1.50 &   4.5 \\
           & 8.385 &   144 &  0.17 &   178 &  4.47 &  0.89 &   6.7 \\ \hline
\multirow{4}*{0450+013} & 1.659 &    62 &  0.03 &    90 & 29.94 &  4.82 &   6.3 \\
           & 2.273 &   123 &  0.62 &   156 & 11.93 &  2.46 &   7.2 \\
           & 4.975 &   143 &  0.09 &   161 &  7.14 &  1.52 &   4.5 \\
           & 8.385 &   174 &  0.18 &   197 &  4.58 &  0.89 &   7.2 \\ \hline
\multirow{4}*{1745+085} & 1.659 &    82 &  0.06 &   246 & 28.69 &  5.80 &   2.3 \\
           & 2.273 &    75 &  0.57 &   300 & 10.59 &  2.89 &   2.2 \\
           & 4.975 &    96 &  0.12 &   211 &  6.94 &  1.97 &   2.2 \\
           & 8.385 &   127 &  0.12 &   209 &  4.18 &  1.04 &   4.6 \\ \hline
\multirow{4}*{1749+062} & 1.659 &   377 &  0.37 &   460 & 28.07 &  5.61 &   3.4 \\
           & 2.273 &   386 &  0.52 &   588 & 10.32 &  3.10 &   2.1 \\
           & 4.975 &   289 &  0.21 &   395 &  6.99 &  1.86 &   3.5 \\
           & 8.385 &   264 &  0.17 &   341 &  4.23 &  1.07 &   5.1 \\ \hline
\multirow{4}*{1749+096} & 1.659 &   996 &  0.56 &  1127 & 24.01 &  5.43 &   2.5 \\
           & 2.273 &  1928 &  0.52 &  2123 & 10.20 &  2.97 &   1.3 \\
           & 4.975 &  3346 &  0.38 &  3532 &  5.60 &  1.52 &   1.4 \\
           & 8.385 &  5736 &  1.22 &  5941 &  3.18 &  0.93 &   2.4 \\ \hline
\multirow{4}*{2136+141} & 1.659 &  1632 &  0.56 &  1767 & 19.37 &  4.34 &   8.8 \\
           & 2.273 &  2134 &  0.70 &  2536 &  9.13 &  2.60 &   7.2 \\
           & 4.975 &  2418 &  0.40 &  3095 &  4.82 &  1.19 &   5.8 \\
           & 8.385 &  2255 &  0.55 &  3580 &  3.09 &  0.67 &   9.6 \\ \hline
\multirow{4}*{2137+130} & 1.659 &   151 &  0.05 &   155 & 23.13 &  5.16 &   8.9 \\
           & 2.273 &   215 &  0.64 &   241 &  9.33 &  2.57 &   7.4 \\
           & 4.975 &   155 &  0.11 &   175 &  5.81 &  1.63 &   5.6 \\
           & 8.385 &   151 &  0.18 &   171 &  3.59 &  0.92 &   9.0 \\ \hline
\multirow{4}*{2141+175} & 1.659 &   344 &  0.16 &   449 & 20.04 &  4.85 &  10.8 \\
           & 2.273 &   357 &  0.53 &   423 &  8.75 &  2.59 &   7.7 \\
           & 4.975 &   505 &  0.15 &   564 &  4.93 &  1.53 &   6.3 \\
           & 8.385 &   608 &  0.25 &   686 &  3.18 &  0.85 &   9.2 \\ \hline
\multirow{4}*{2144+092} & 1.659 &   482 &  0.16 &   630 & 23.77 &  4.42 &   8.5 \\
           & 2.273 &   551 &  0.61 &   740 & 10.09 &  2.50 &   6.6 \\
           & 4.975 &   641 &  0.14 &   764 &  5.83 &  1.40 &   5.5 \\
           & 8.385 &   785 &  0.28 &   918 &  3.72 &  0.88 &   8.3 \\ \hline
\multirow{4}*{2145+067} & 1.659 &  2146 &  0.38 &  3705 & 21.38 &  3.70 &   8.6 \\
           & 2.273 &  3519 &  1.65 &  4962 & 10.30 &  2.41 &   7.1 \\
           & 4.975 &  4108 &  0.51 &  6175 &  5.43 &  1.01 &   6.2 \\
           & 8.385 &  4101 &  1.34 &  6847 &  3.42 &  0.62 &   8.8 \\ \hline
\multirow{4}*{2149+056} & 1.659 &   695 &  0.23 &   900 & 24.97 &  4.47 &   7.8 \\
           & 2.273 &   926 &  0.82 &  1240 & 10.42 &  2.46 &   6.5 \\
           & 4.975 &   724 &  0.19 &   907 &  6.22 &  1.35 &   5.7 \\
           & 8.385 &   552 &  0.31 &   702 &  4.03 &  0.87 &   8.8 \\
\end{longtable*}
\end{center}

\begin{figure*}[p!]

  \includegraphics[width=0.3\textwidth]{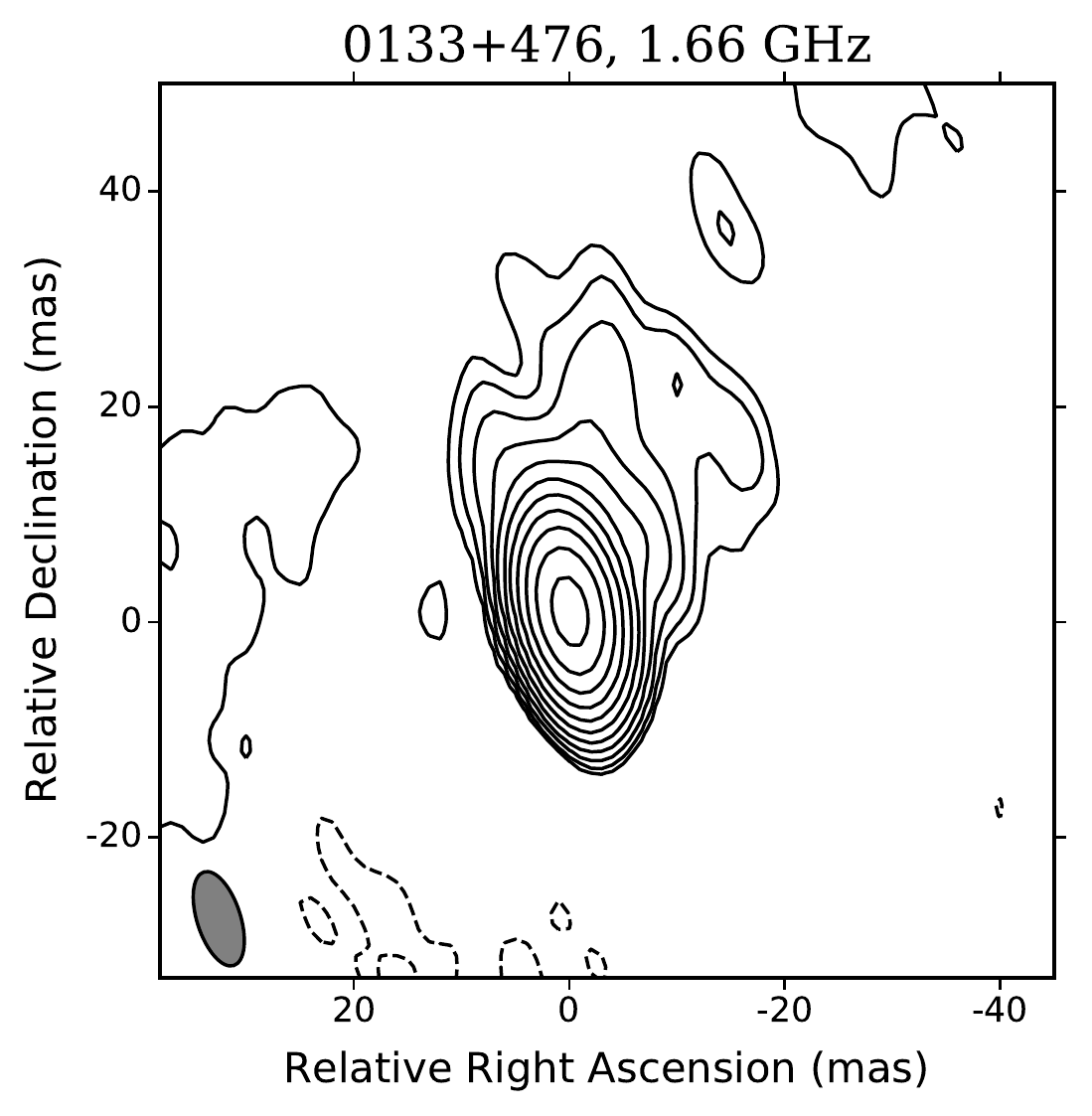}
  \includegraphics[width=0.3\textwidth]{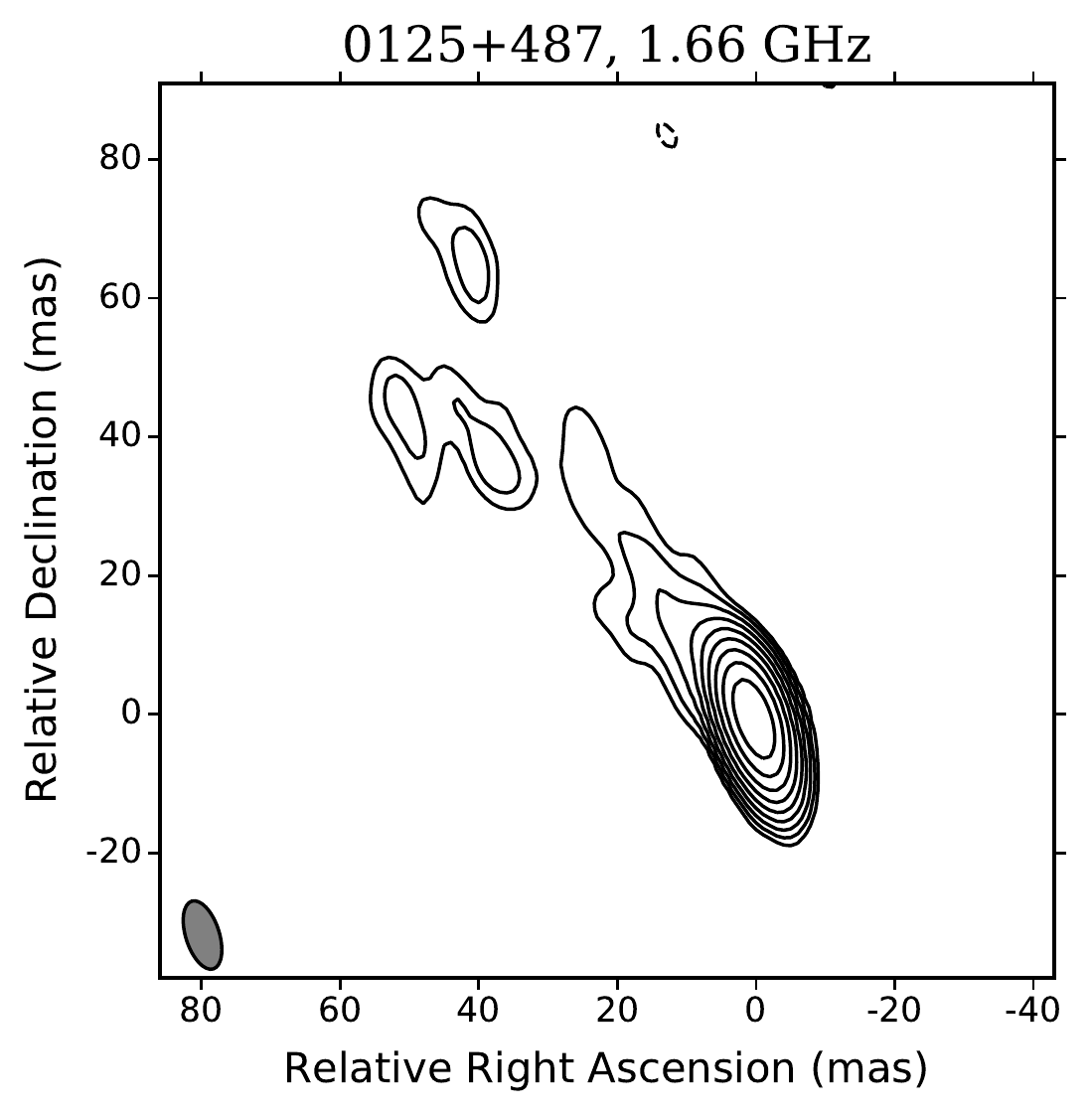}
  \includegraphics[width=0.3\textwidth]{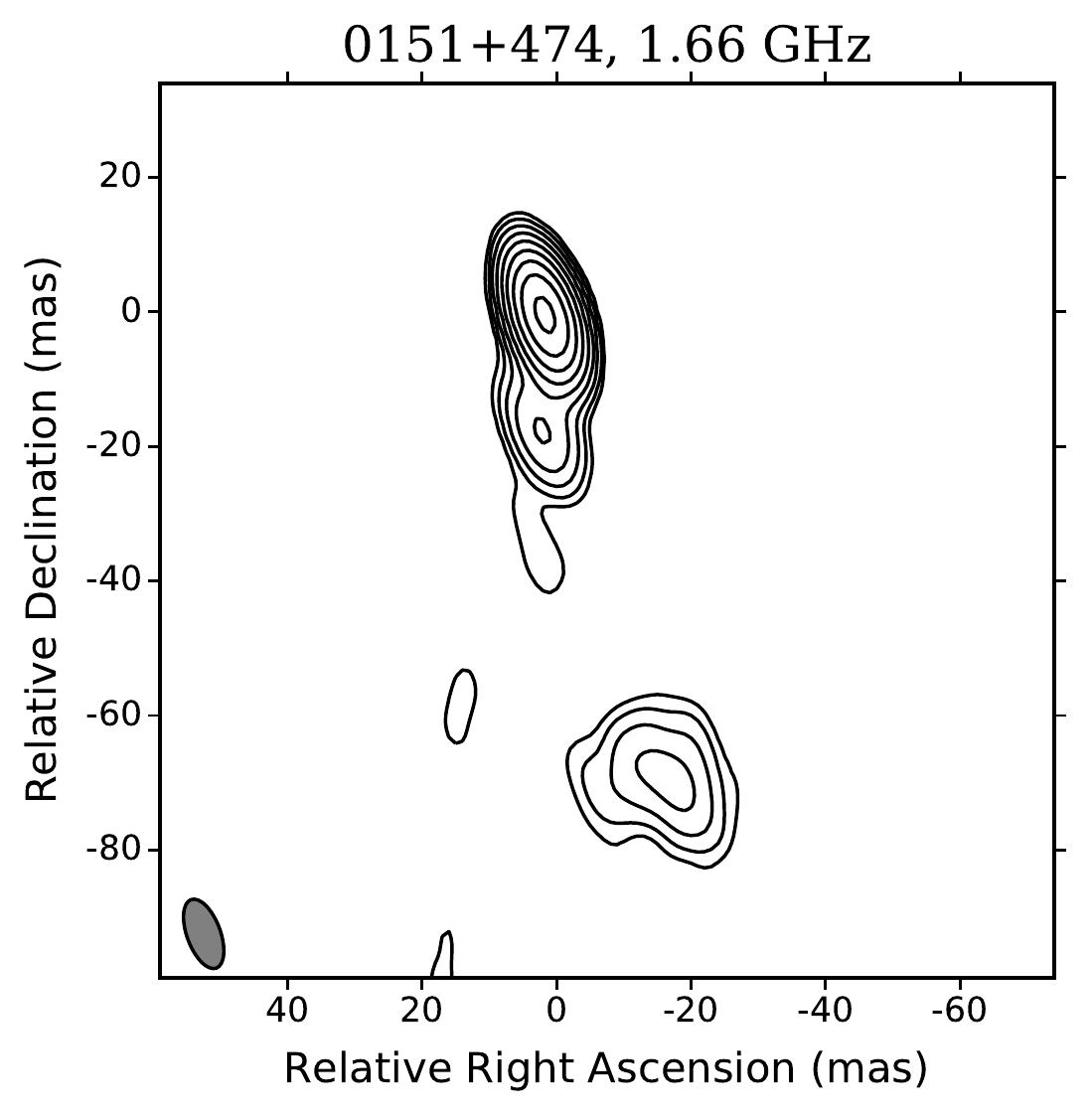}

  \includegraphics[width=0.3\textwidth]{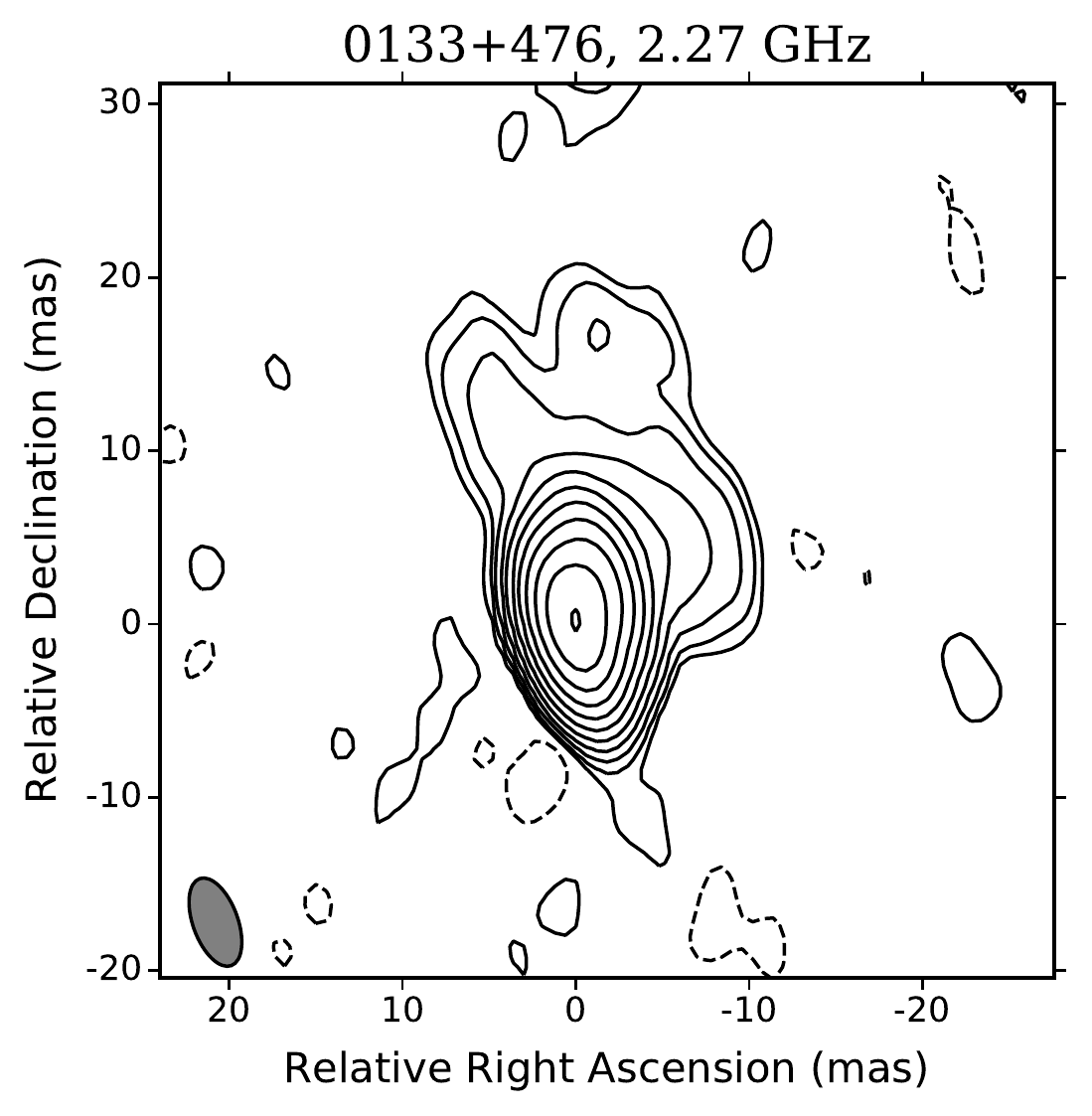}
  \includegraphics[width=0.3\textwidth]{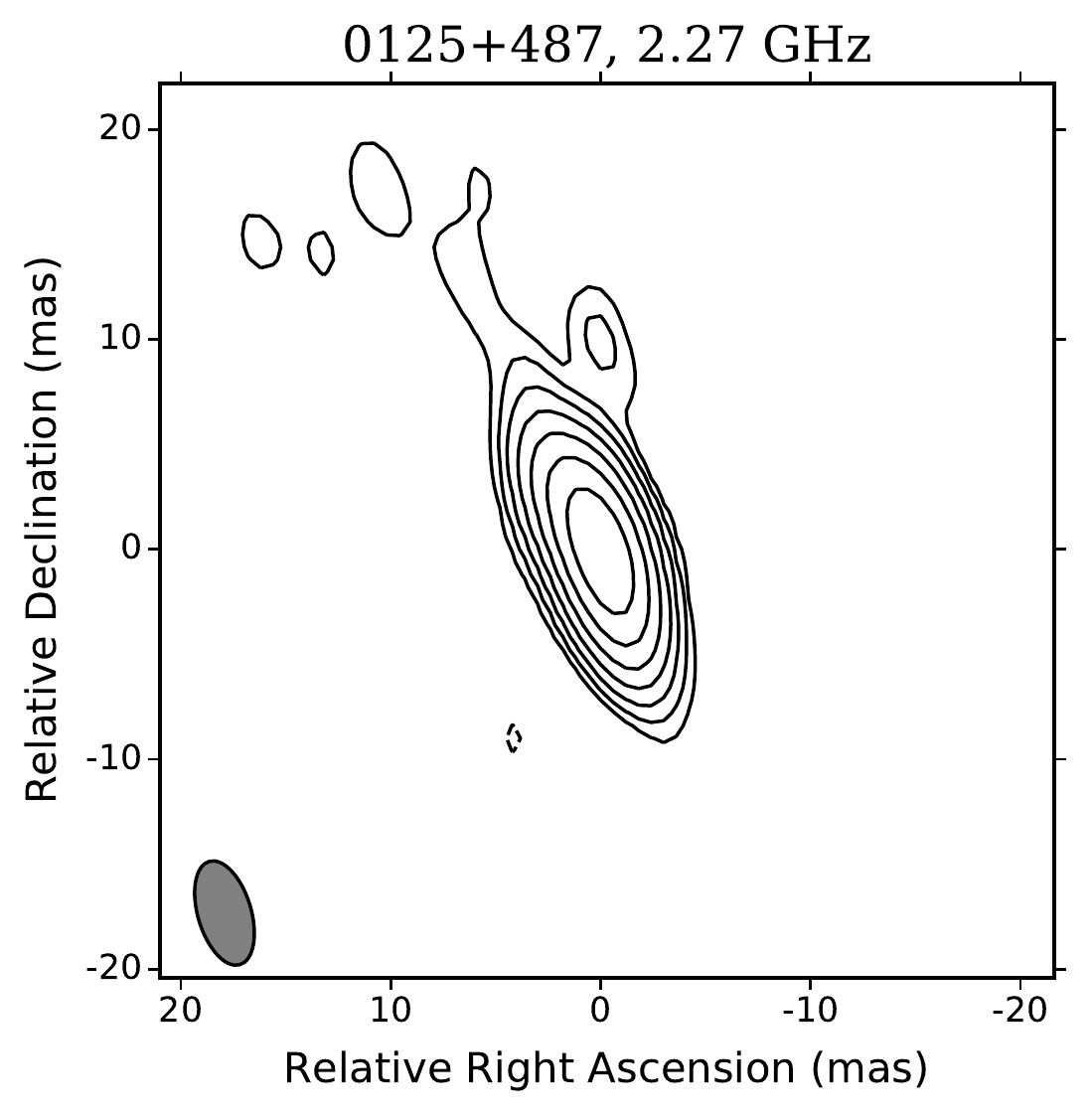}
  \includegraphics[width=0.3\textwidth]{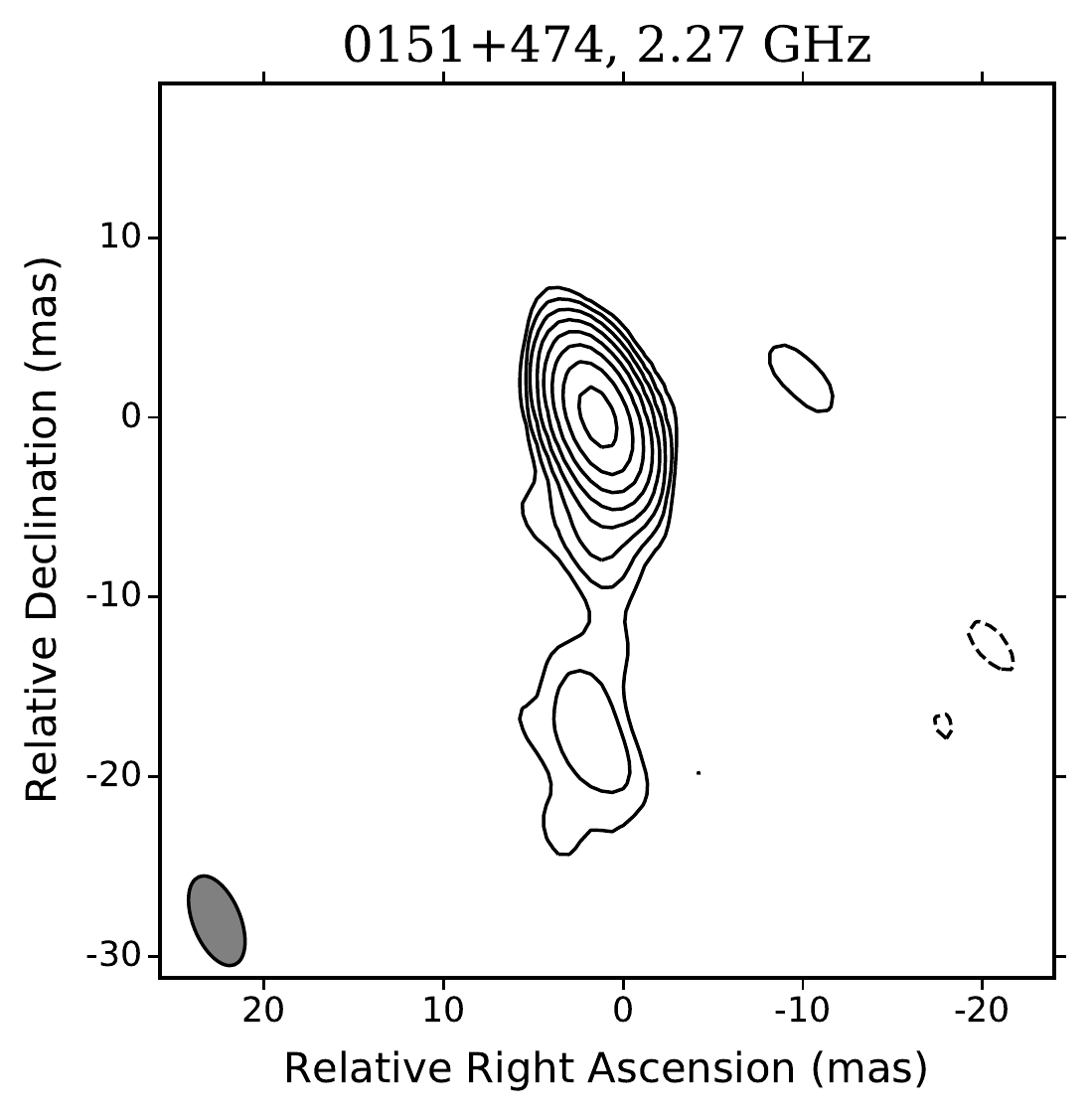}

  \includegraphics[width=0.3\textwidth]{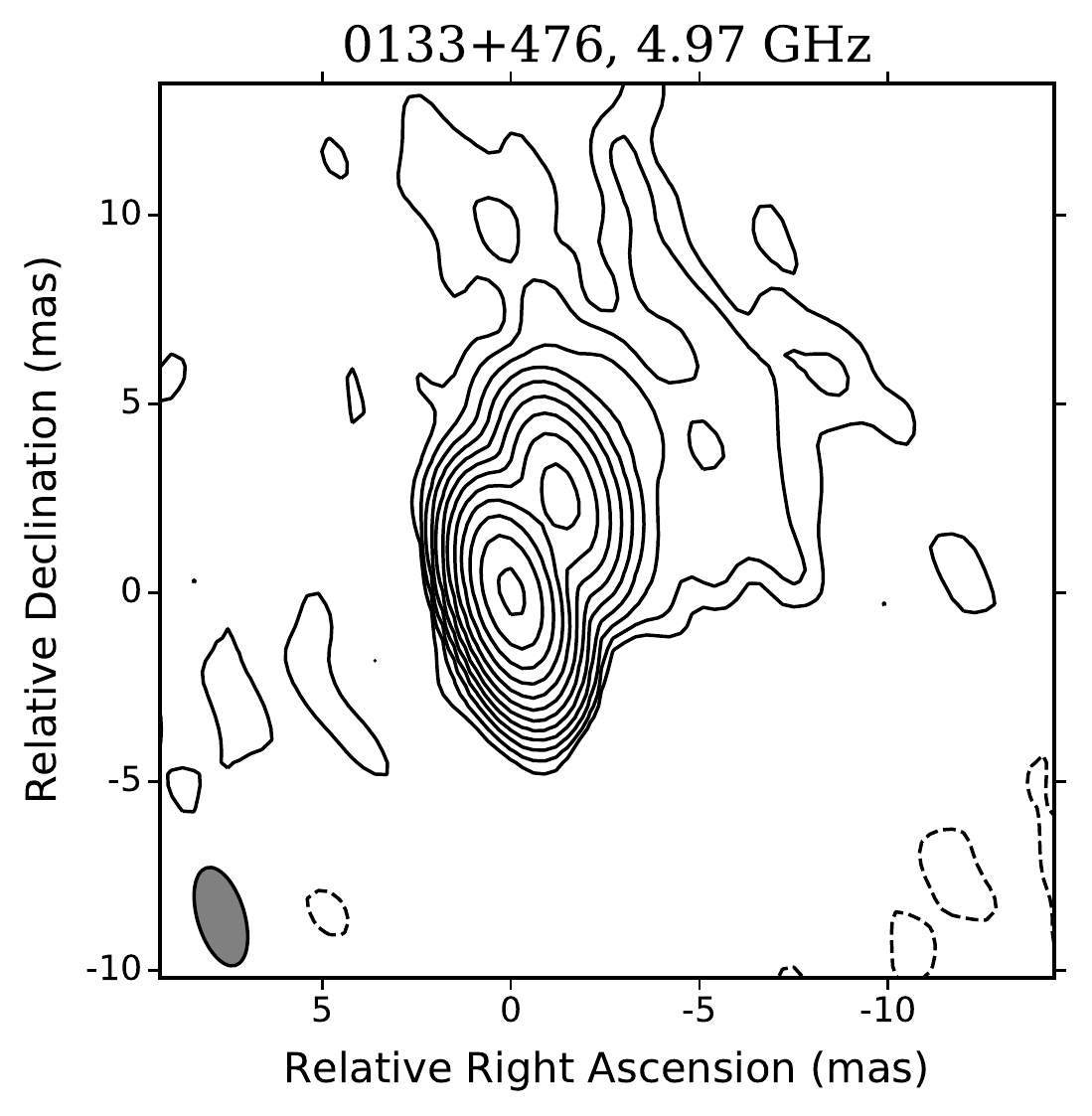}
  \includegraphics[width=0.3\textwidth]{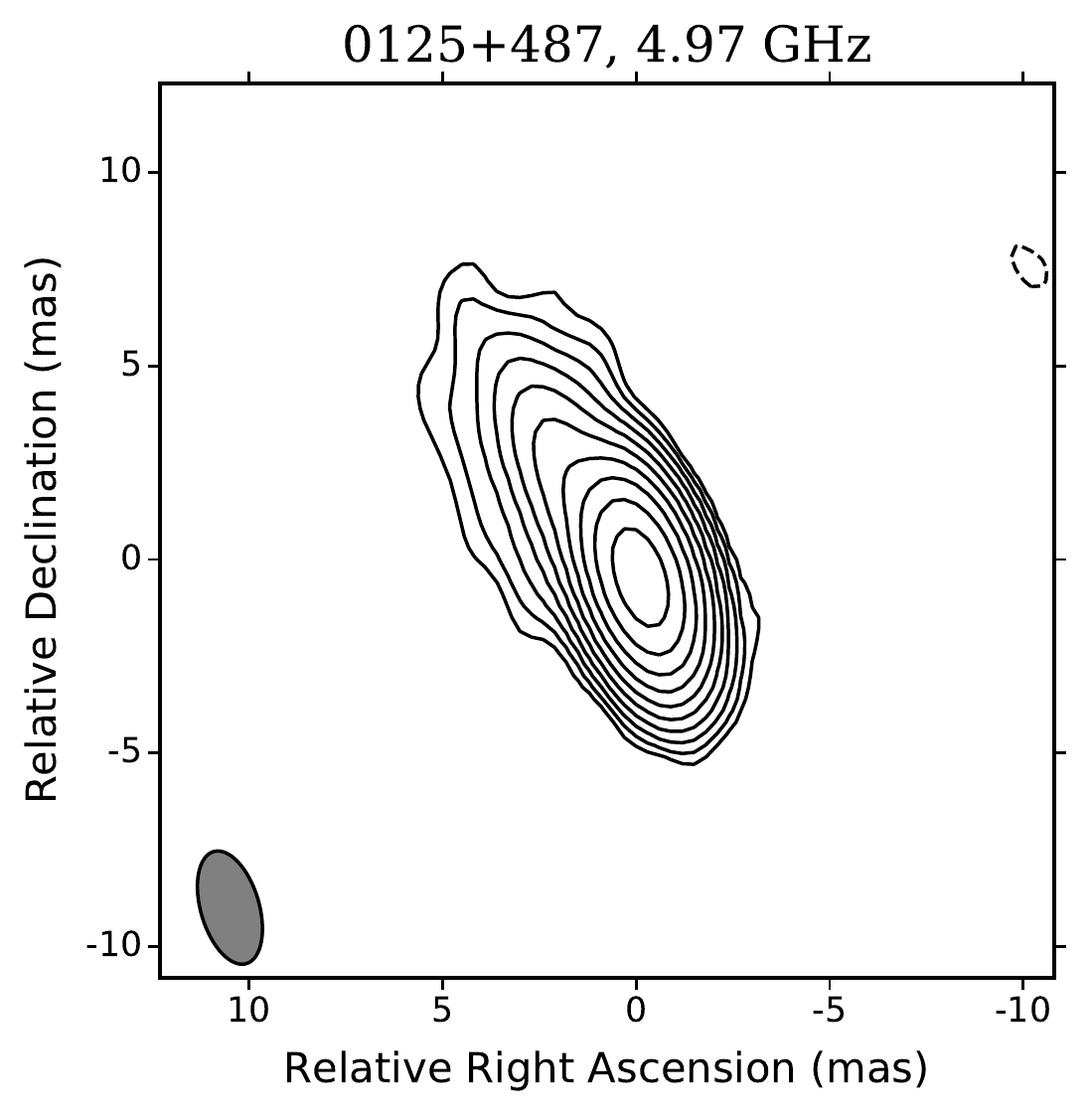}
  \includegraphics[width=0.3\textwidth]{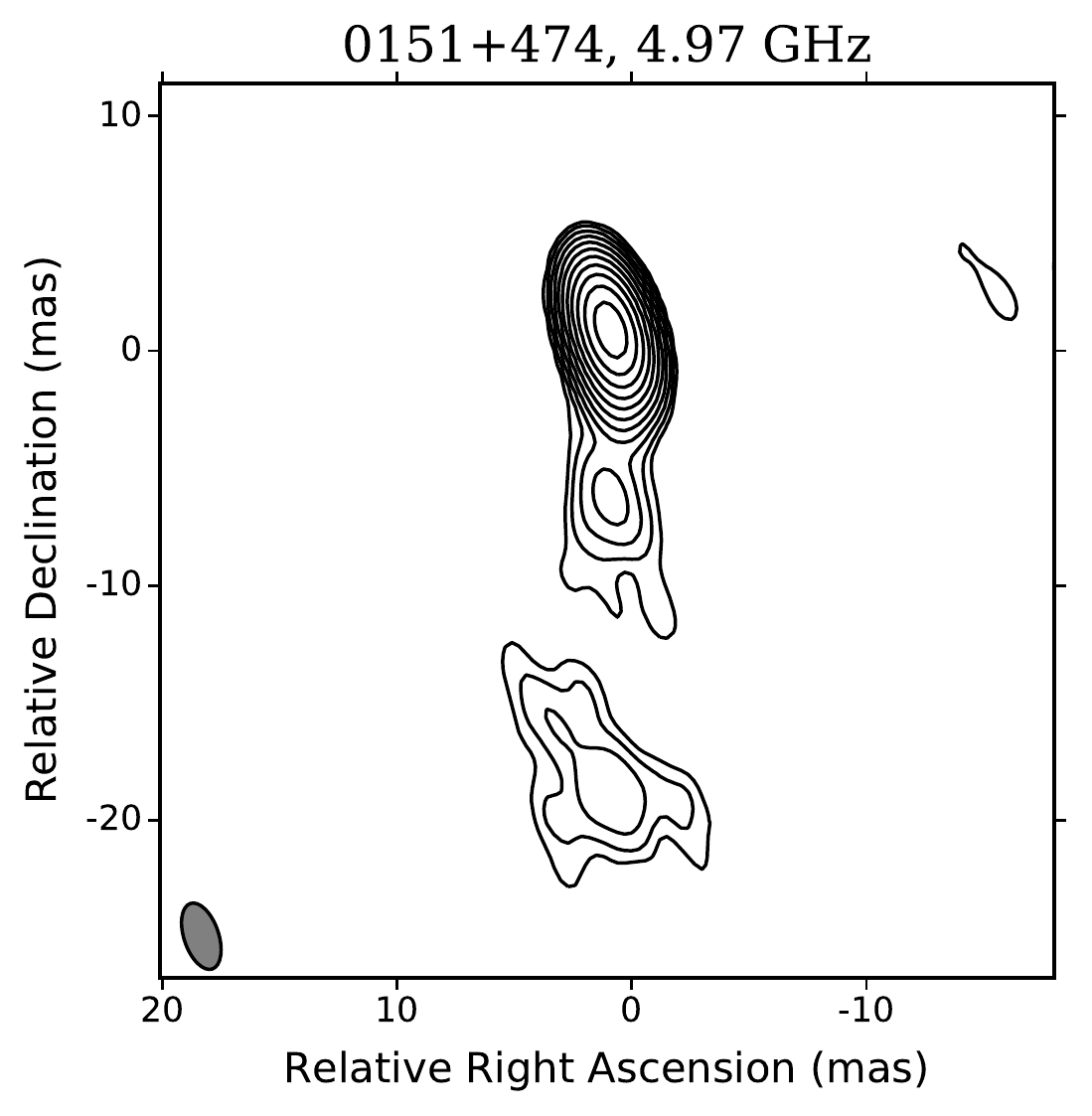}

  \includegraphics[width=0.3\textwidth]{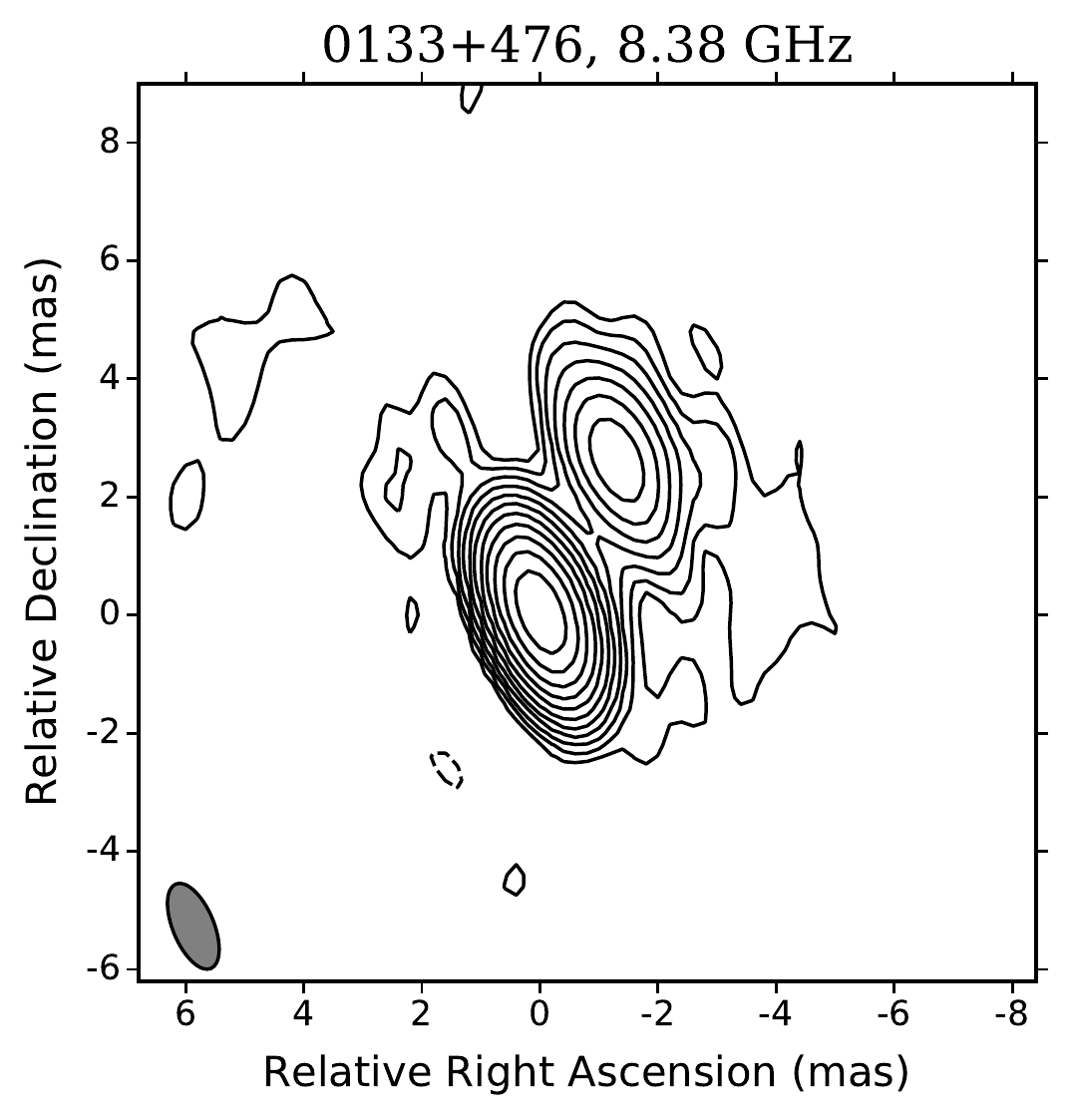}
  \includegraphics[width=0.3\textwidth]{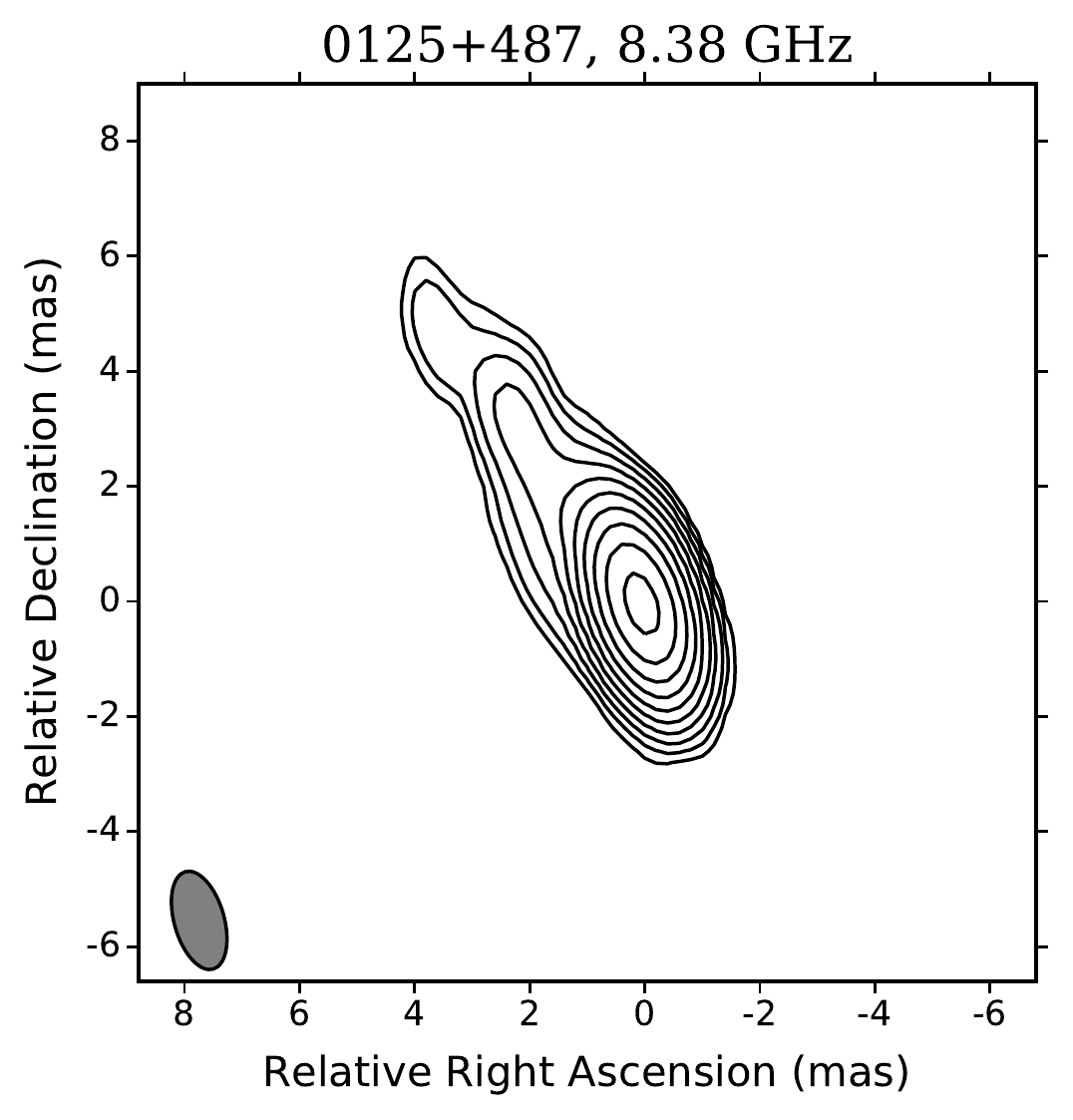}
  \includegraphics[width=0.3\textwidth]{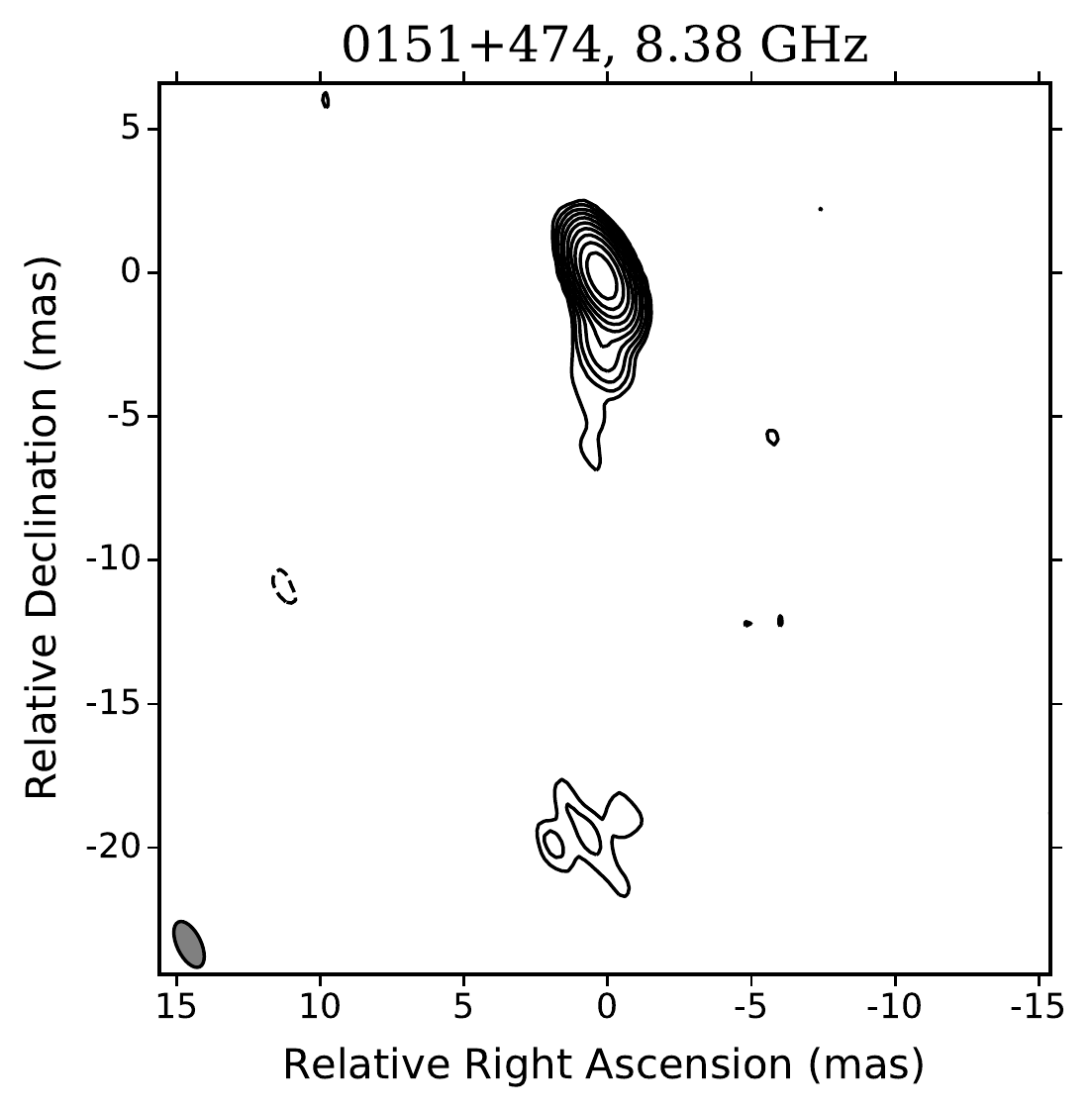}

  \caption{Total-intensity maps of all the target sources made with natural
weighting. The first contour corresponds to three times the root-mean-square
residual noise.}
  \label{fig:maps}
\end{figure*}

\addtocounter{figure}{-1}
\begin{figure*}[p!] 

  \includegraphics[width=0.3\textwidth]{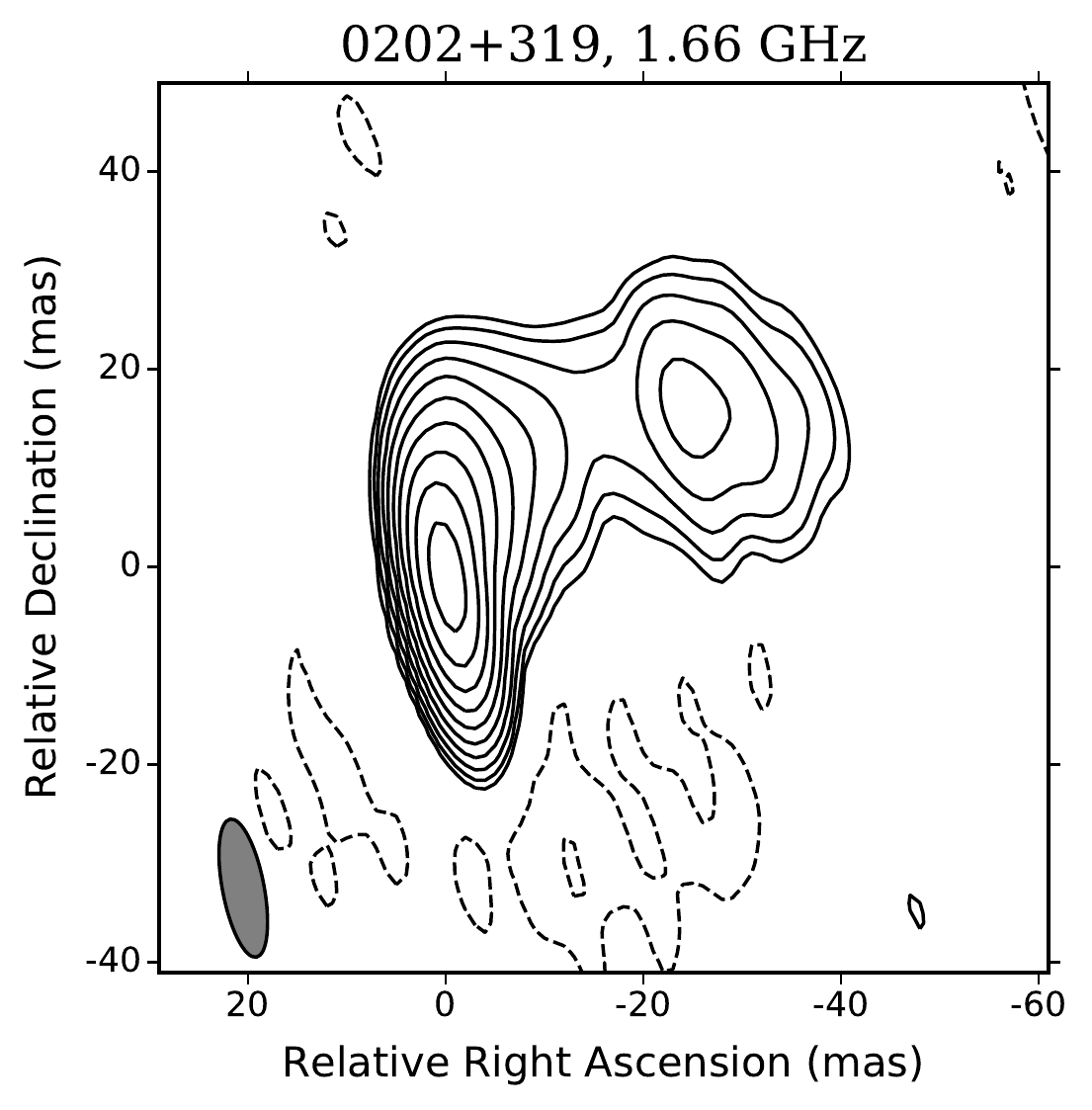}
  \includegraphics[width=0.3\textwidth]{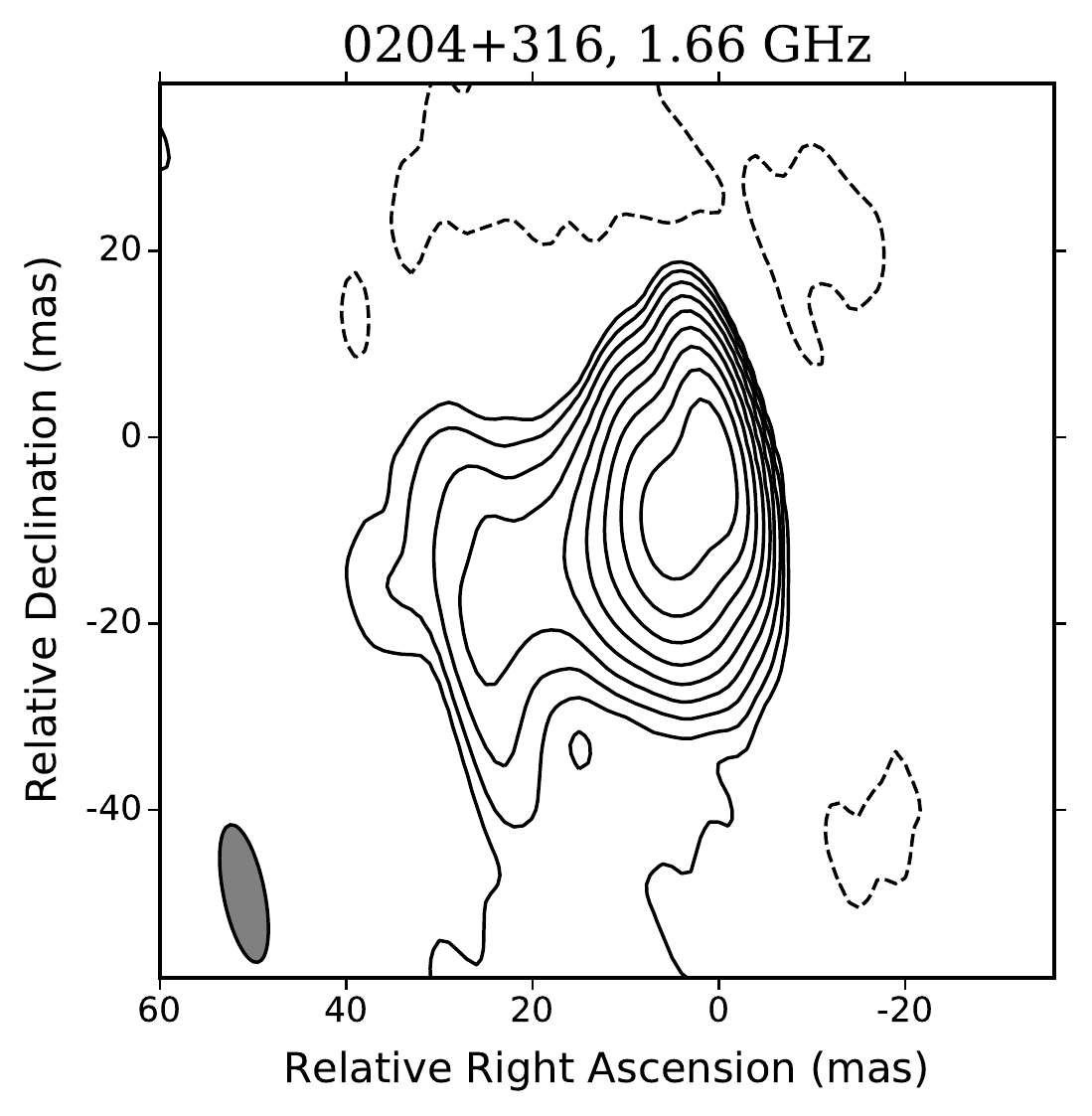}
  \includegraphics[width=0.3\textwidth]{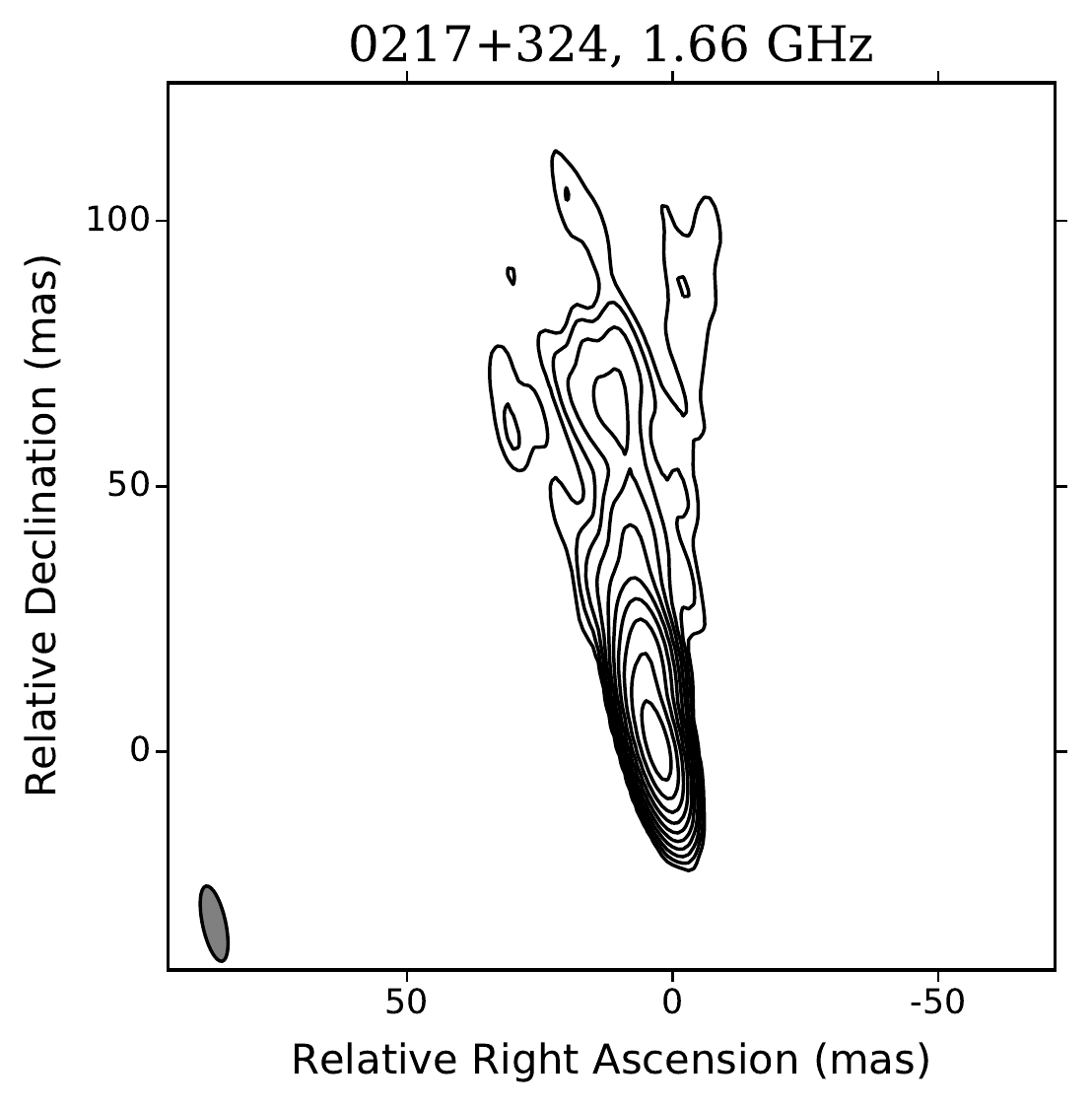}

  \includegraphics[width=0.3\textwidth]{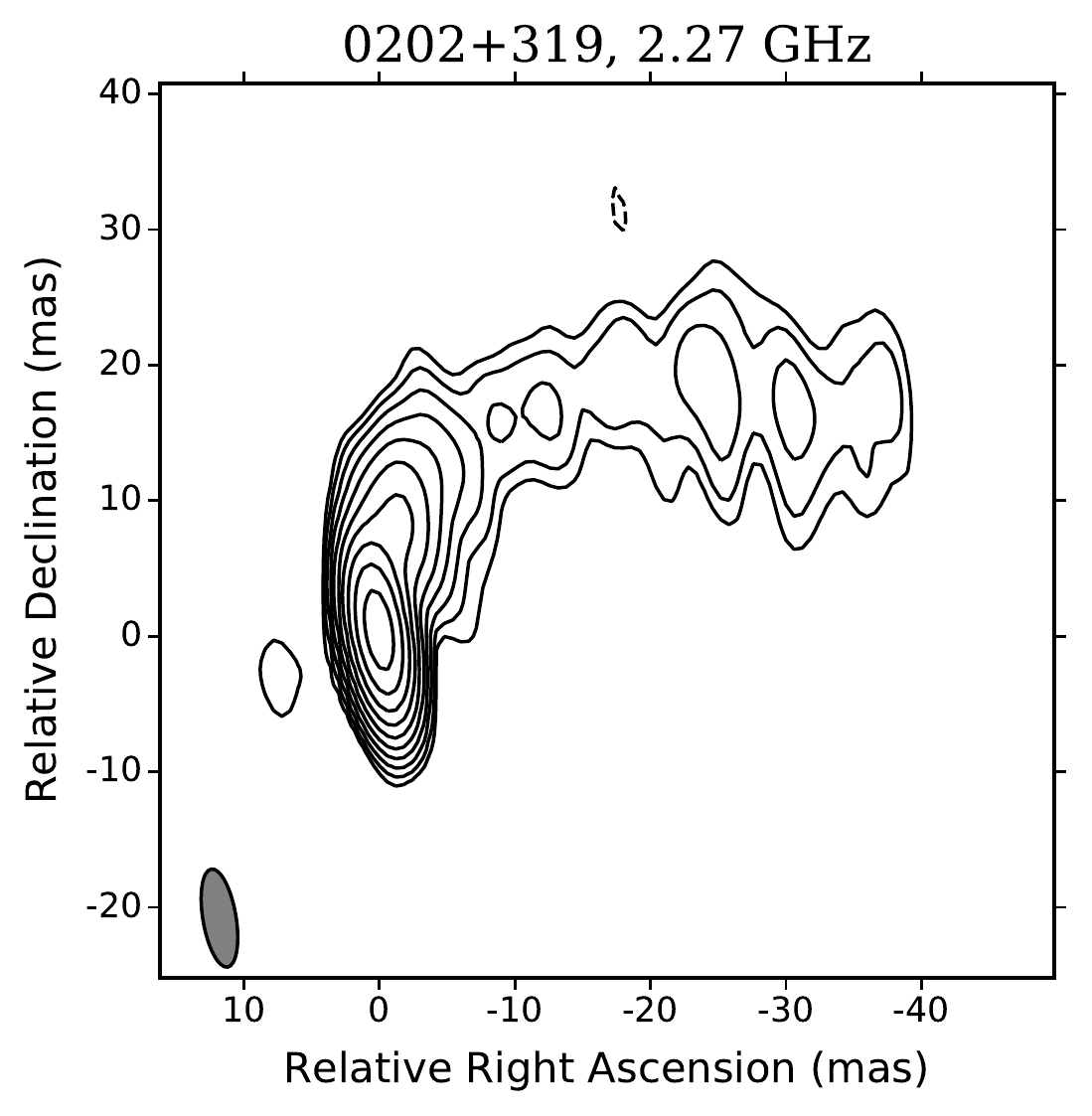}
  \includegraphics[width=0.3\textwidth]{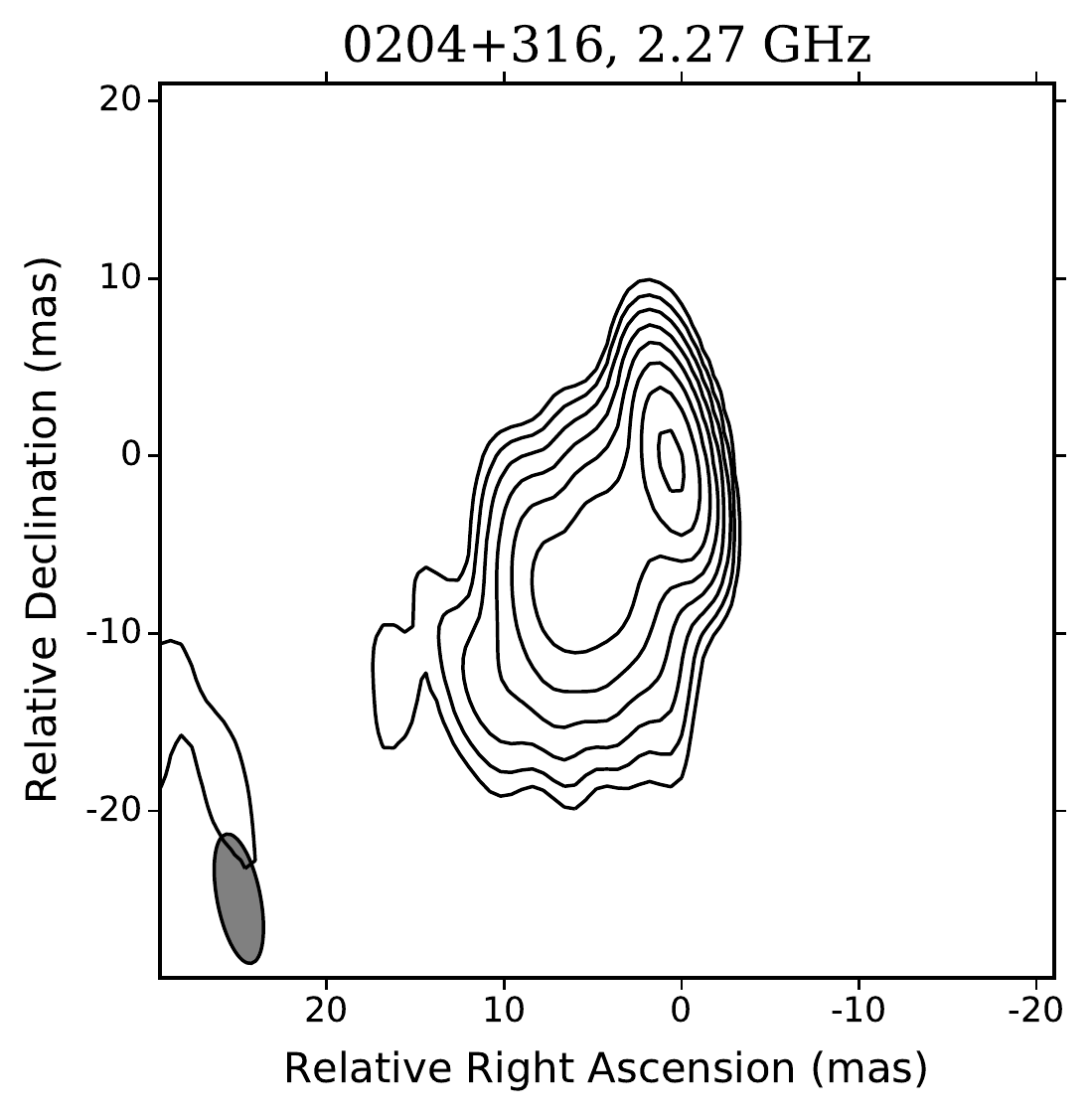}
  \includegraphics[width=0.3\textwidth]{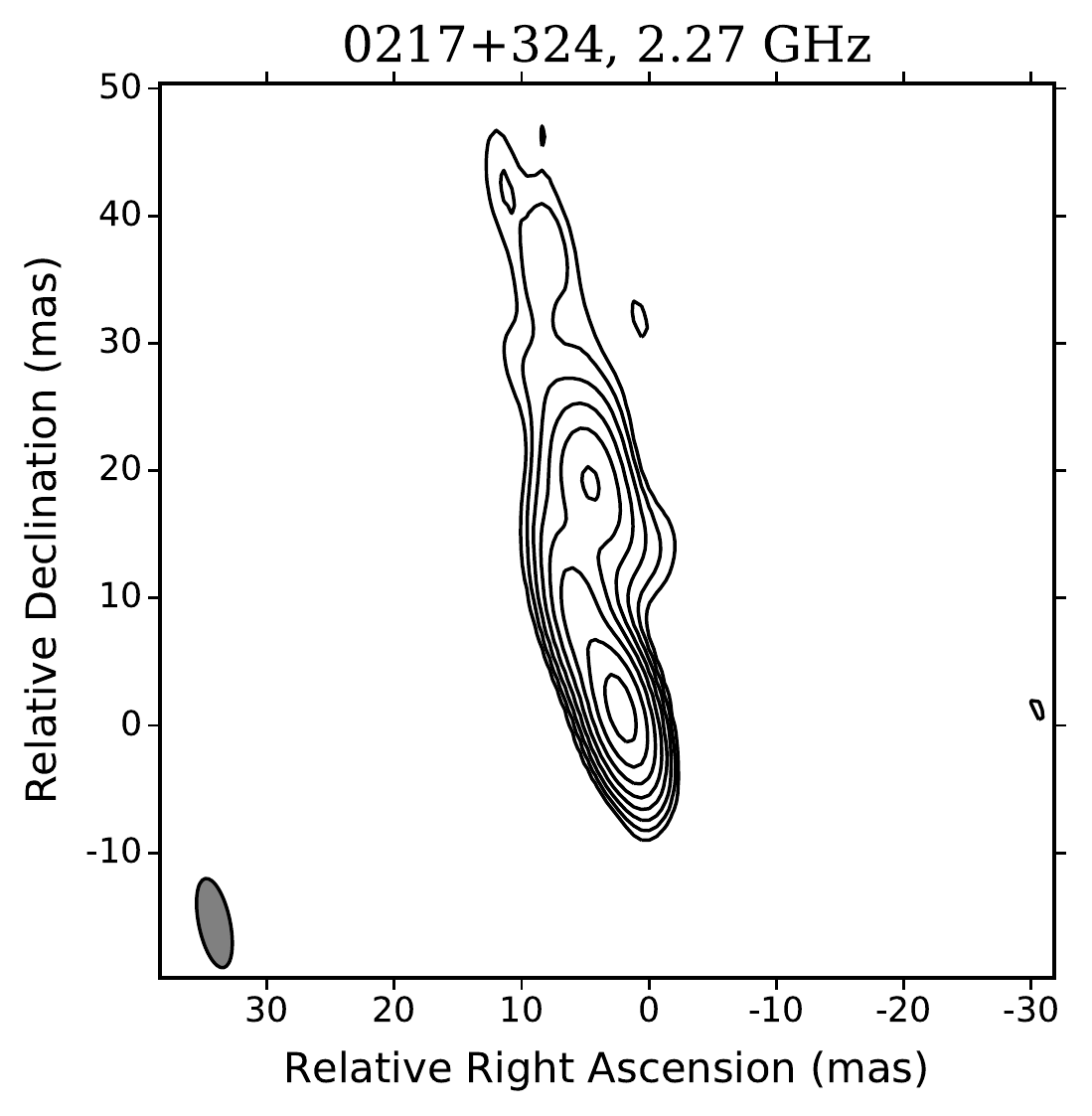}

  \includegraphics[width=0.3\textwidth]{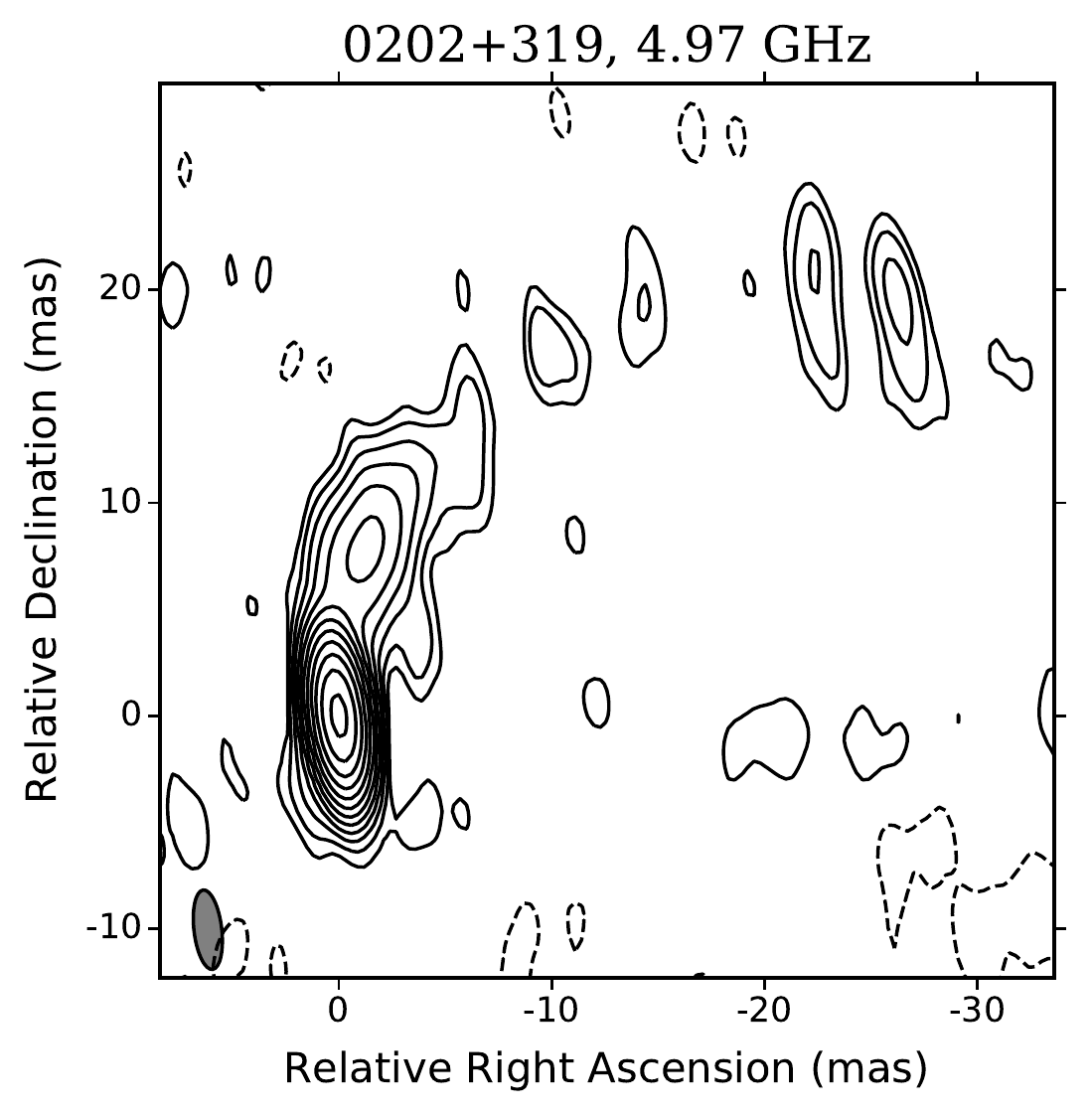}
  \includegraphics[width=0.3\textwidth]{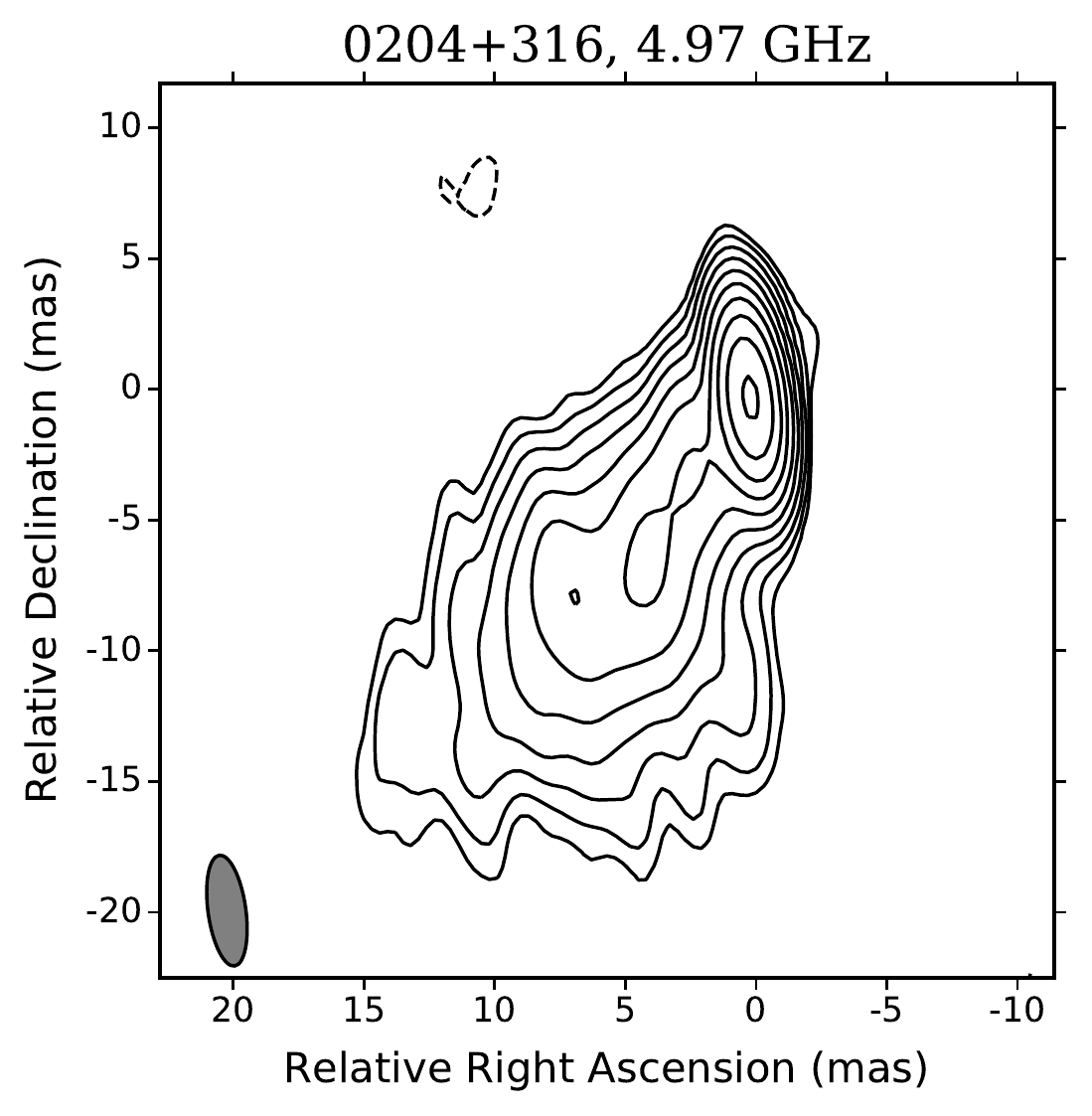}
  \includegraphics[width=0.3\textwidth]{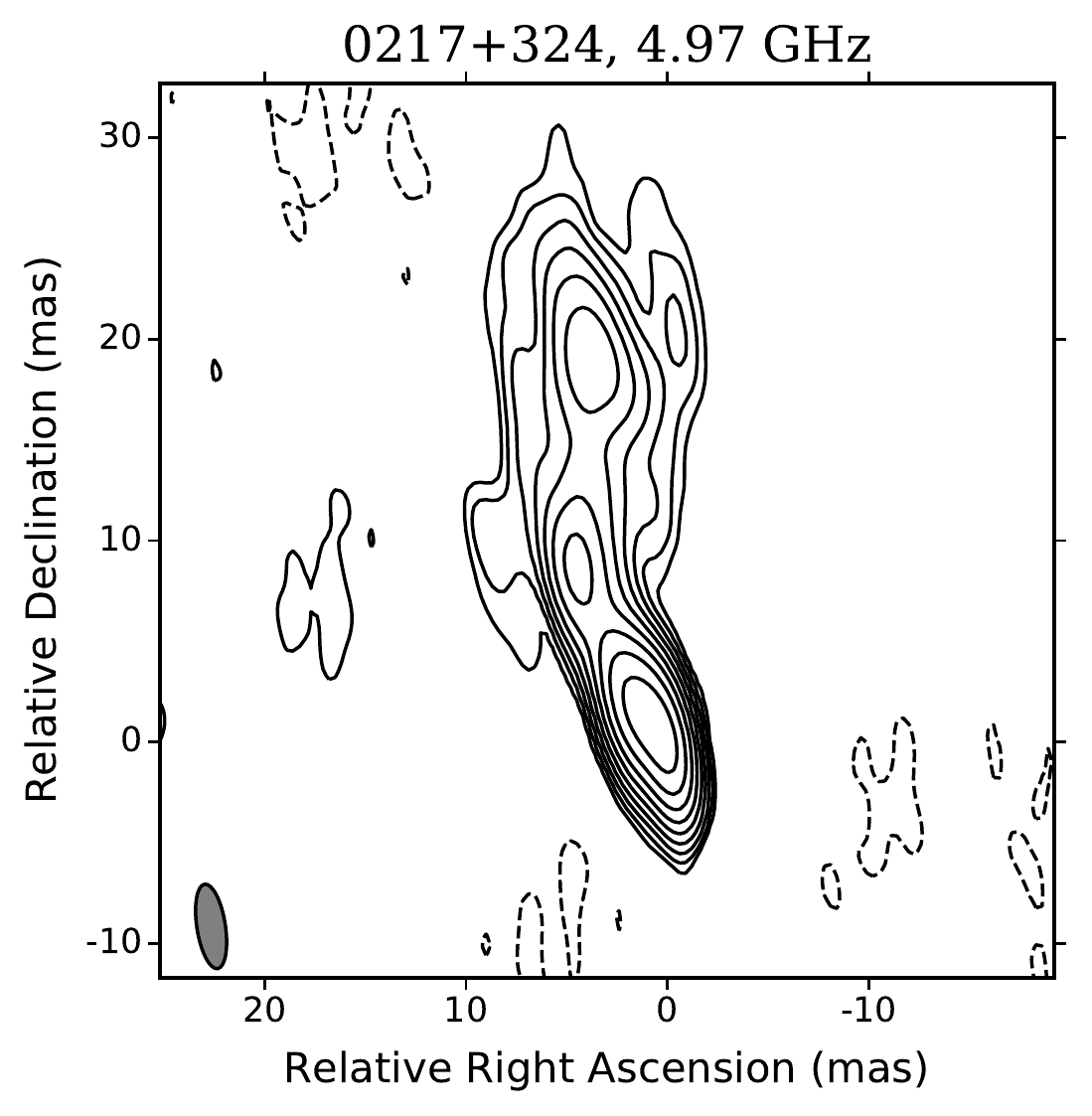}

  \includegraphics[width=0.3\textwidth]{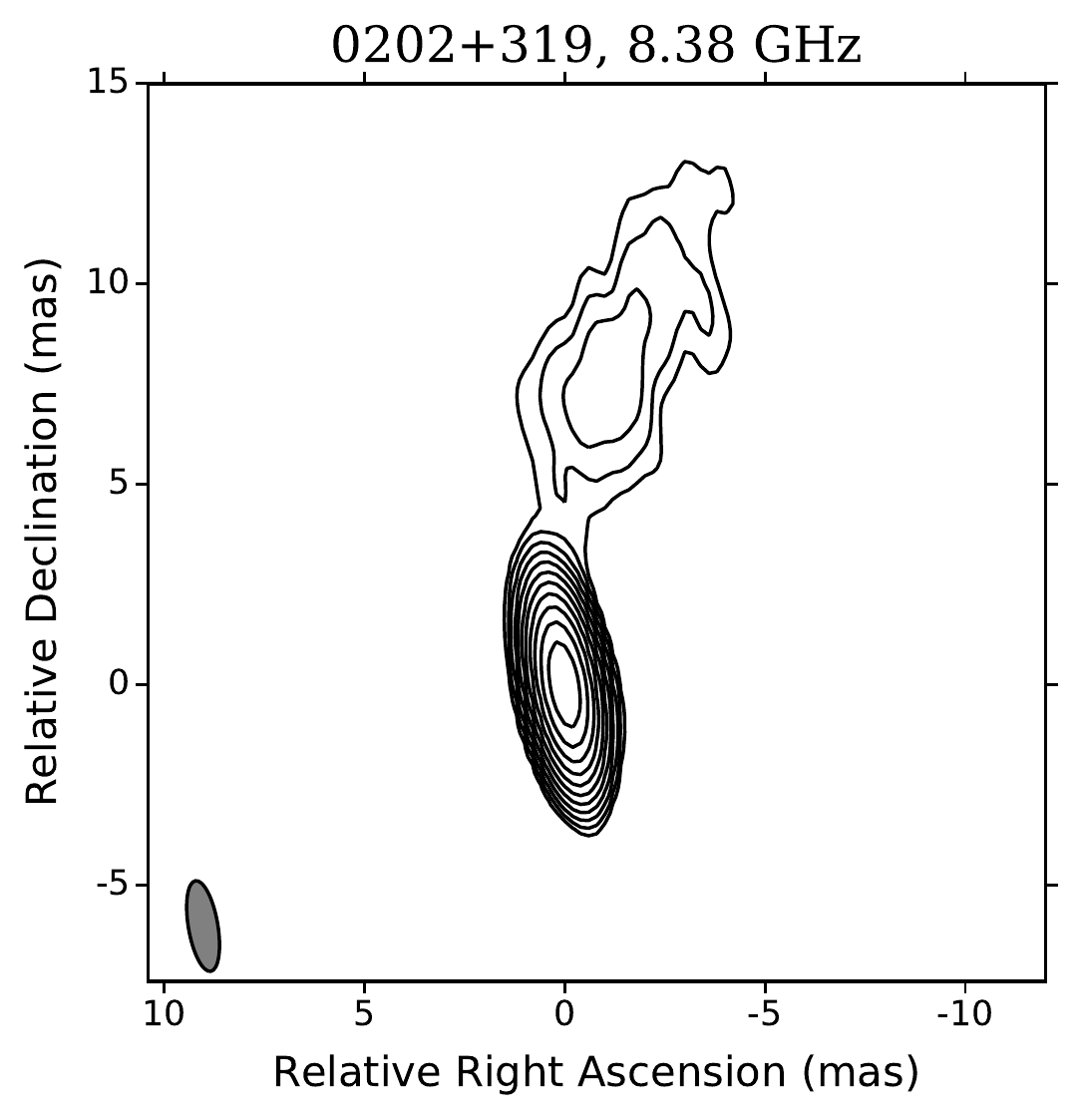}
  \includegraphics[width=0.3\textwidth]{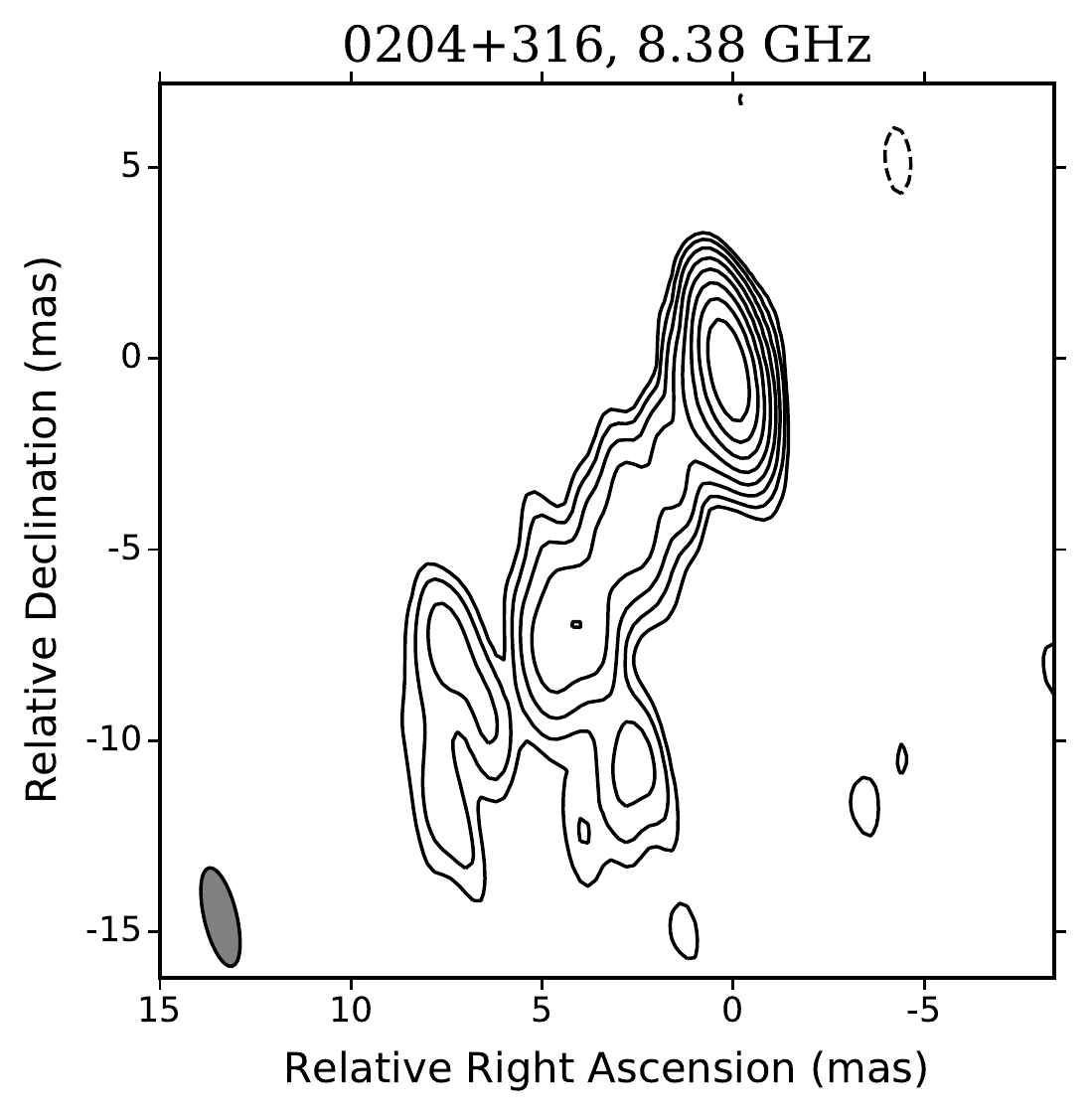}
  \includegraphics[width=0.3\textwidth]{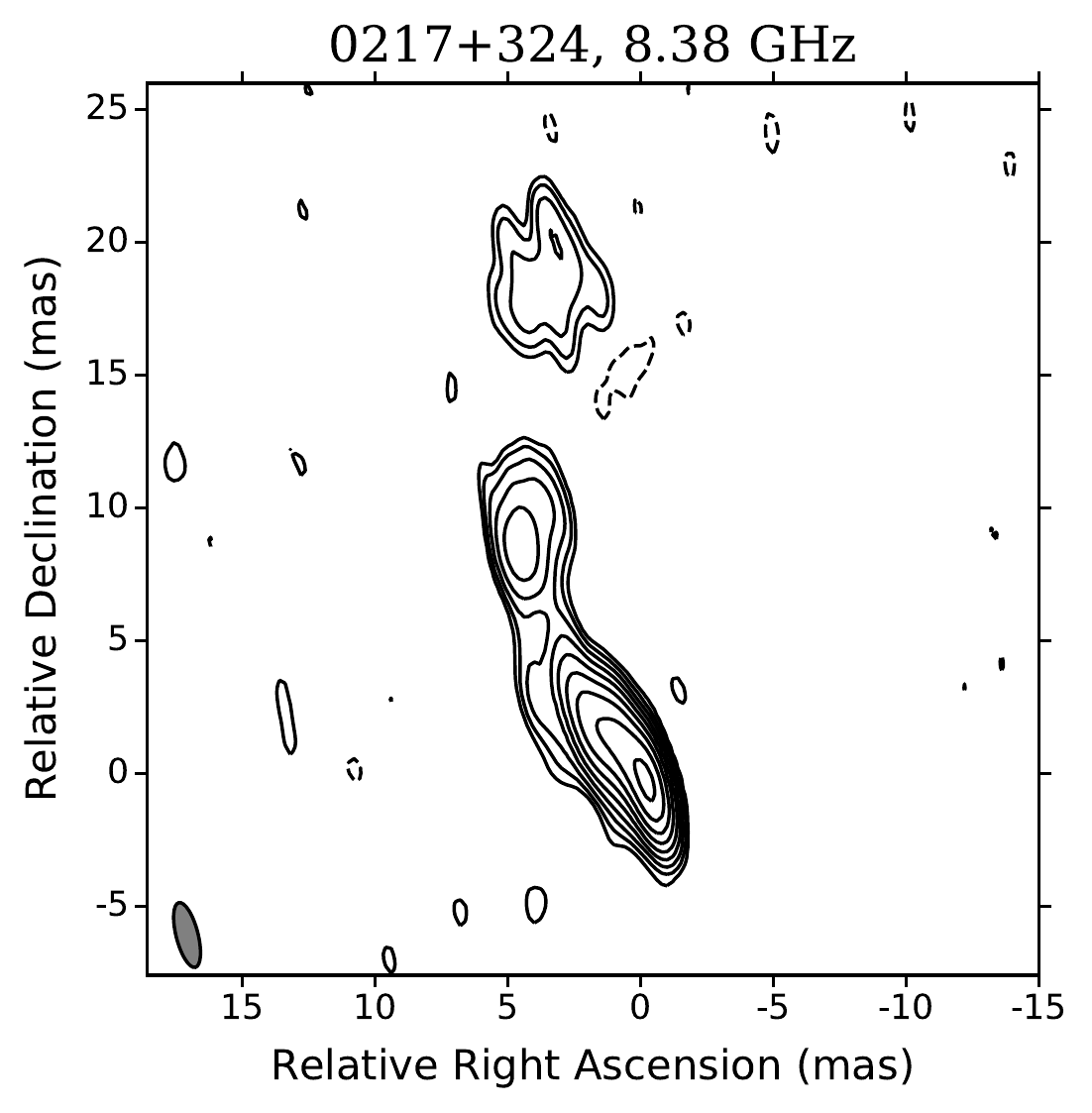}

  \caption{Continued.}
\end{figure*}

\addtocounter{figure}{-1}
\begin{figure*}[p!]

  \includegraphics[width=0.3\textwidth]{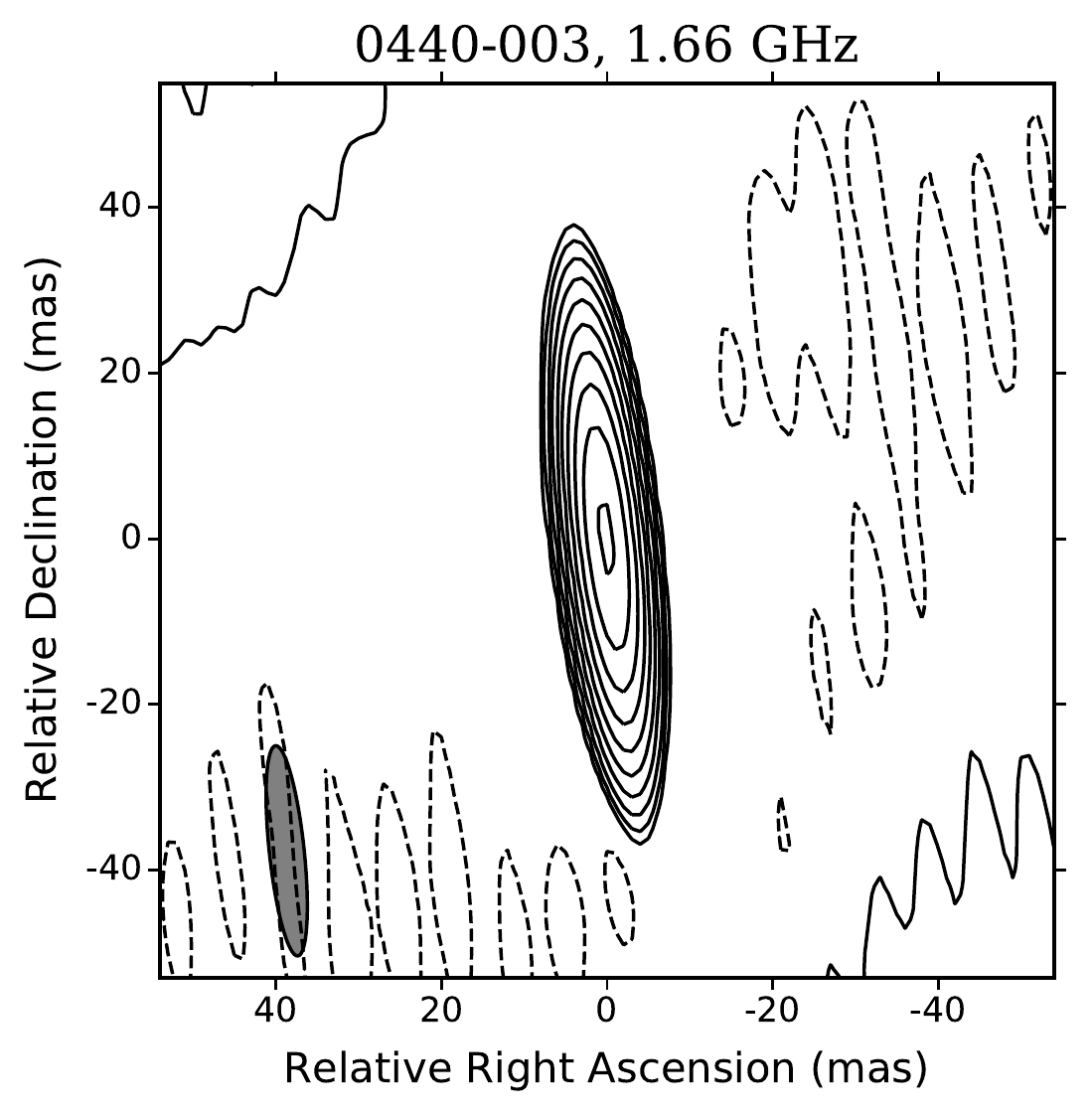}
  \includegraphics[width=0.3\textwidth]{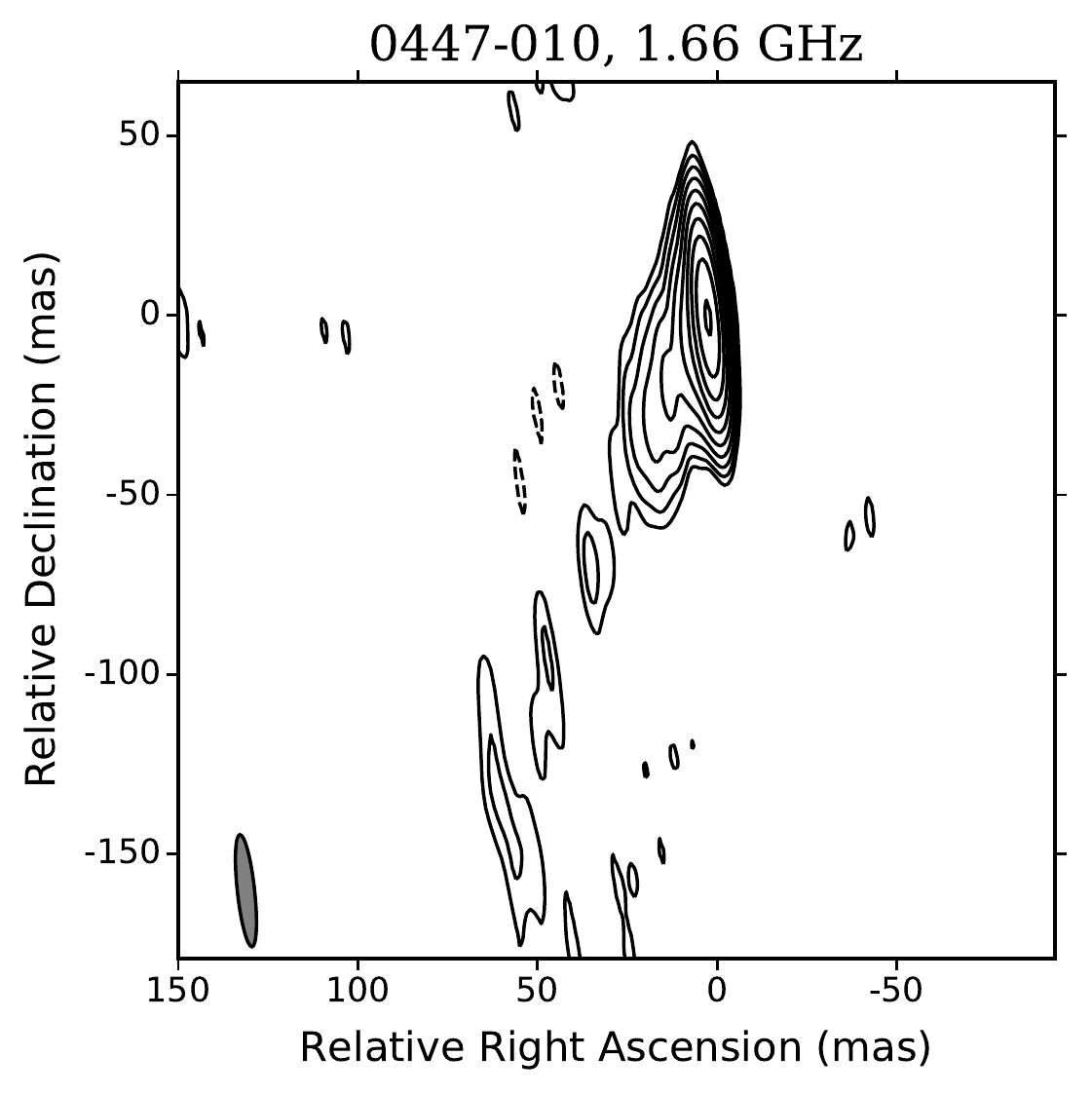}
  \includegraphics[width=0.3\textwidth]{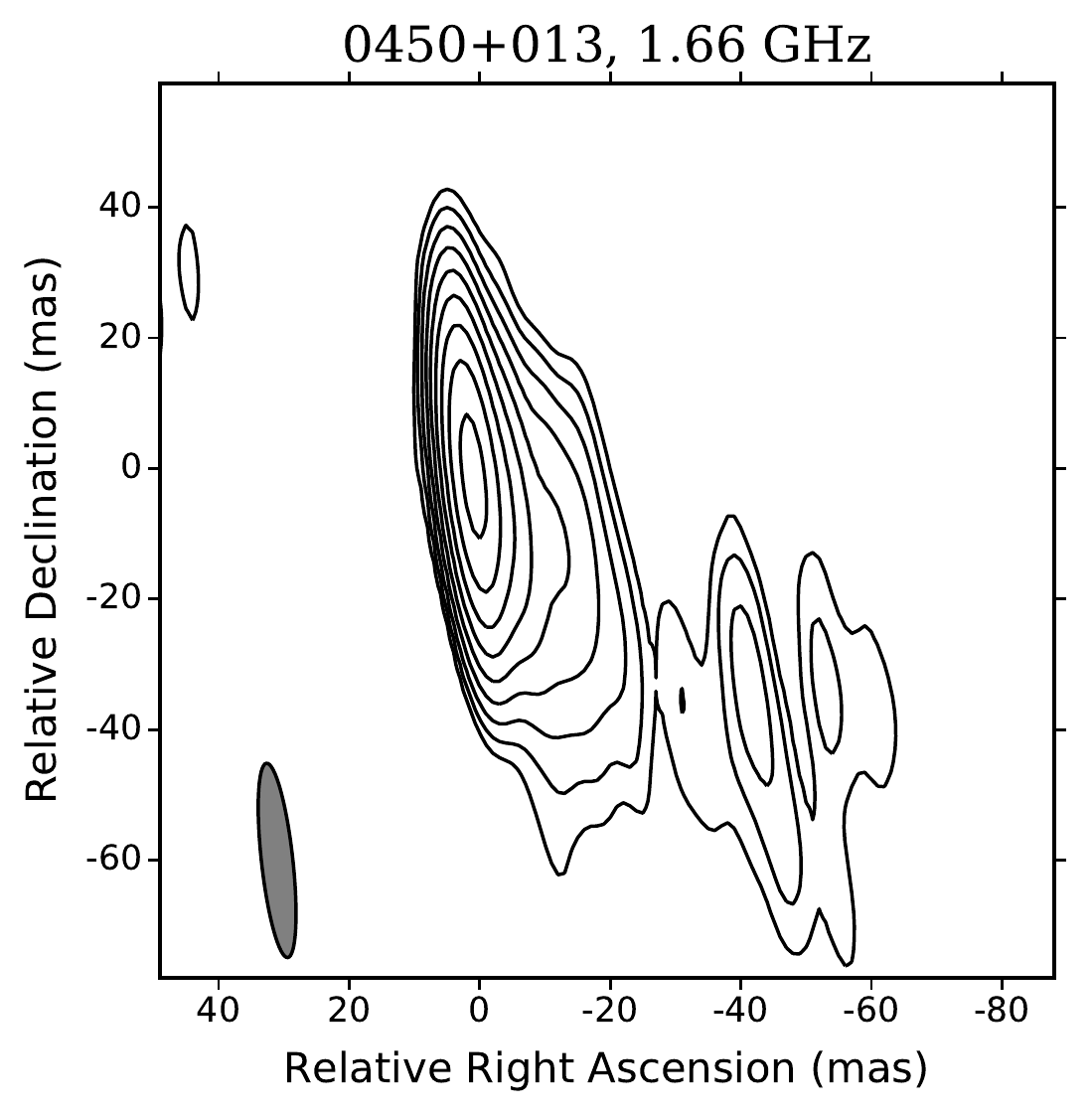}

  \includegraphics[width=0.3\textwidth]{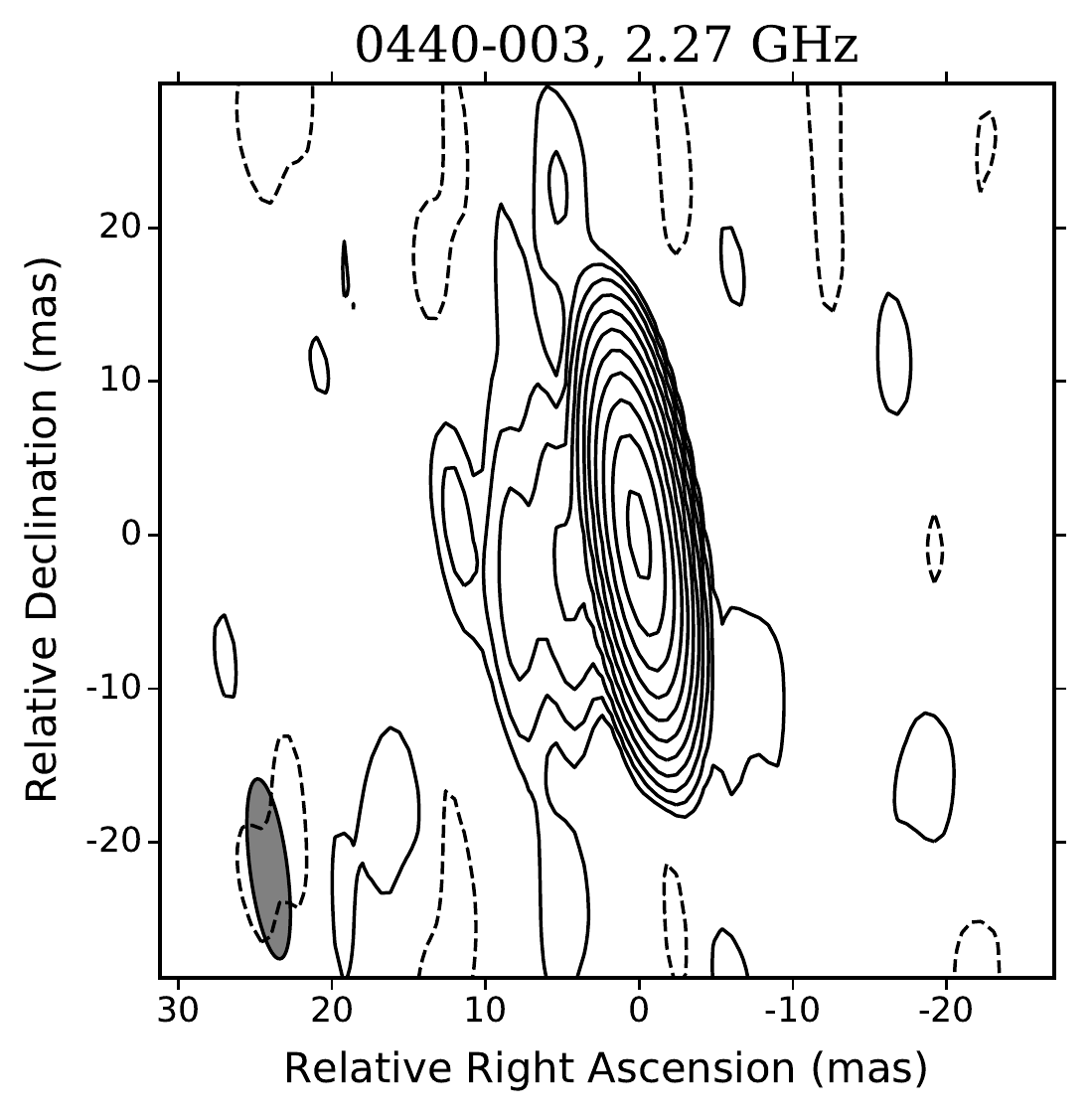}
  \includegraphics[width=0.3\textwidth]{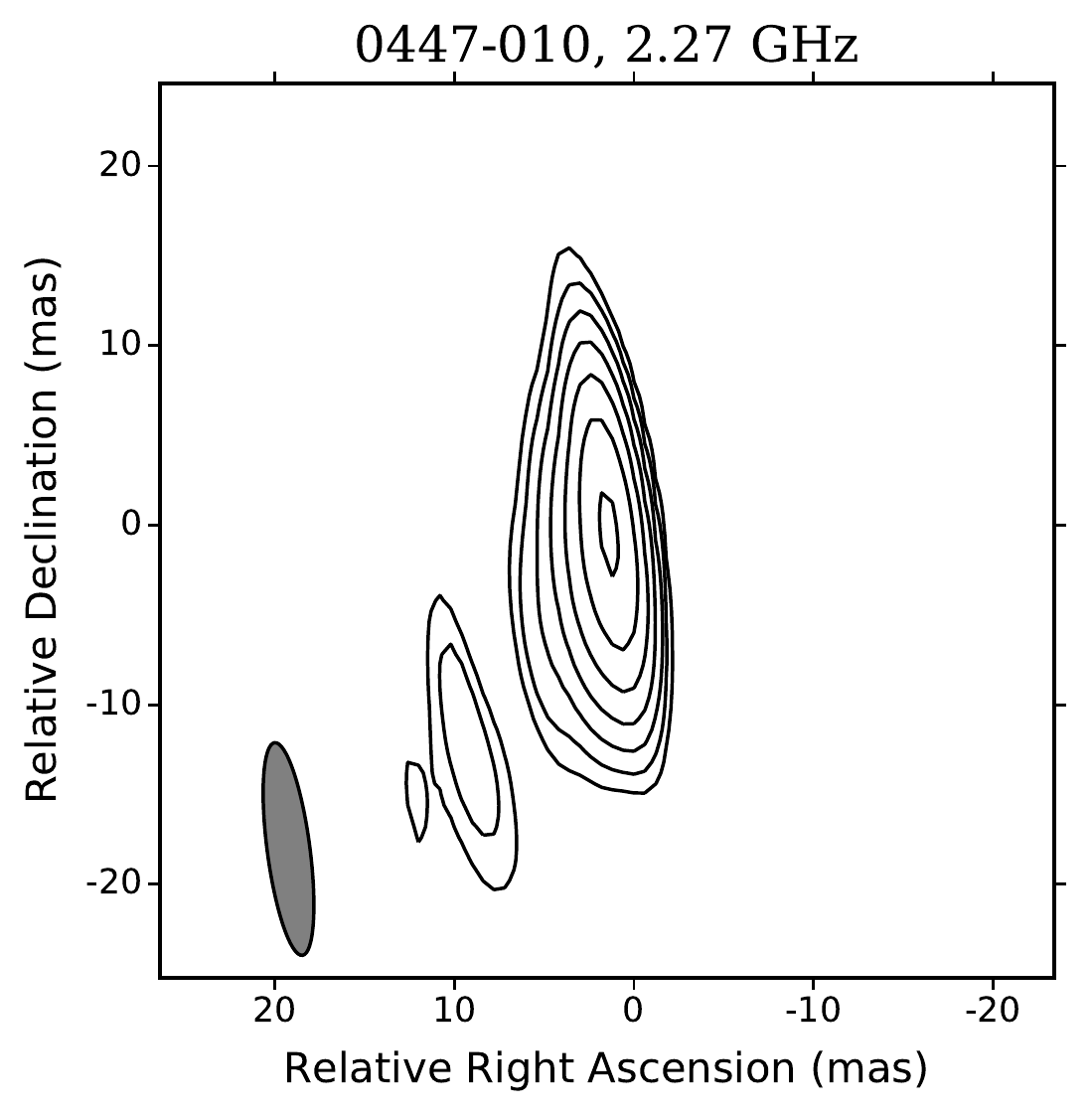}
  \includegraphics[width=0.3\textwidth]{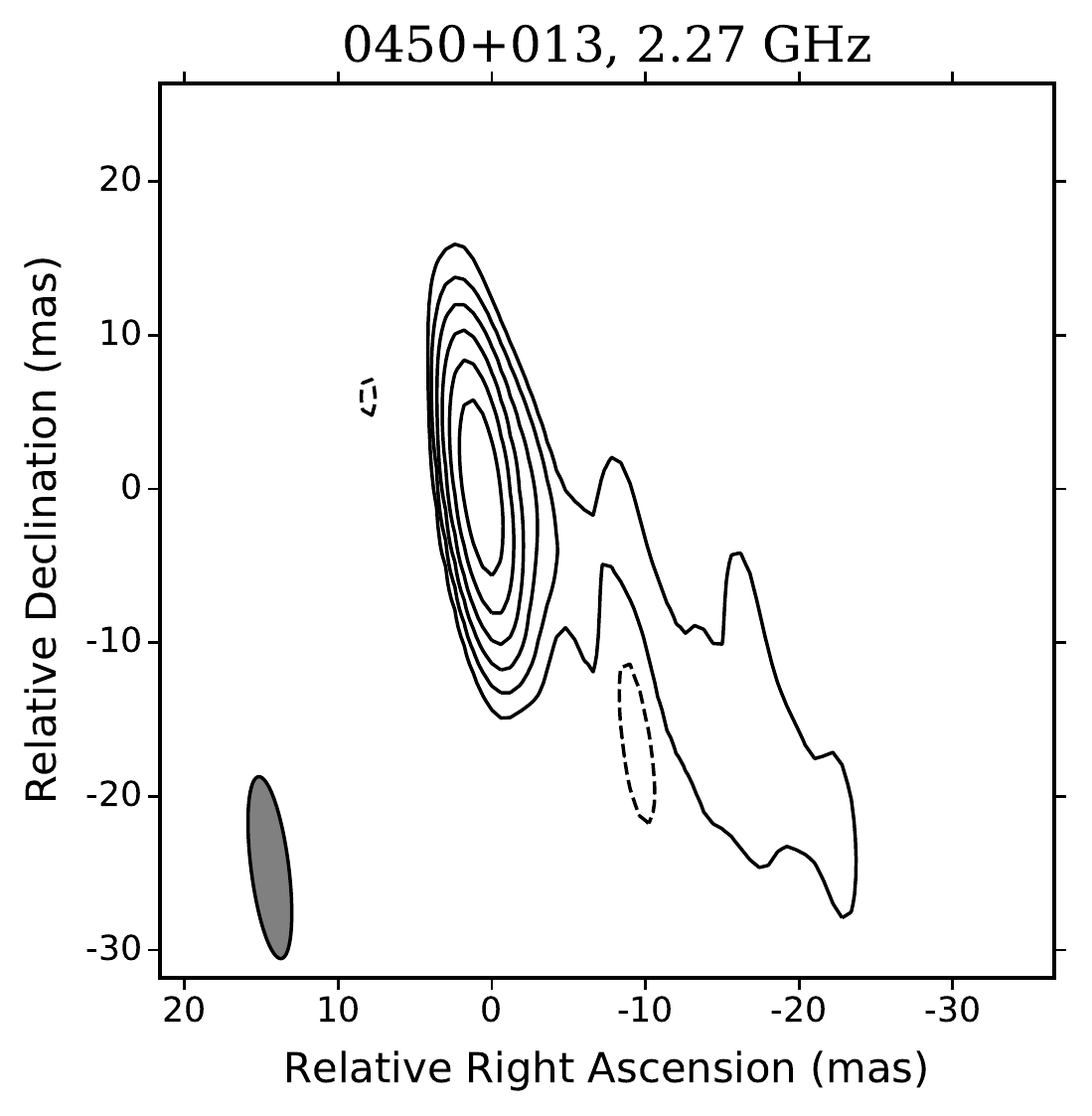}

  \includegraphics[width=0.3\textwidth]{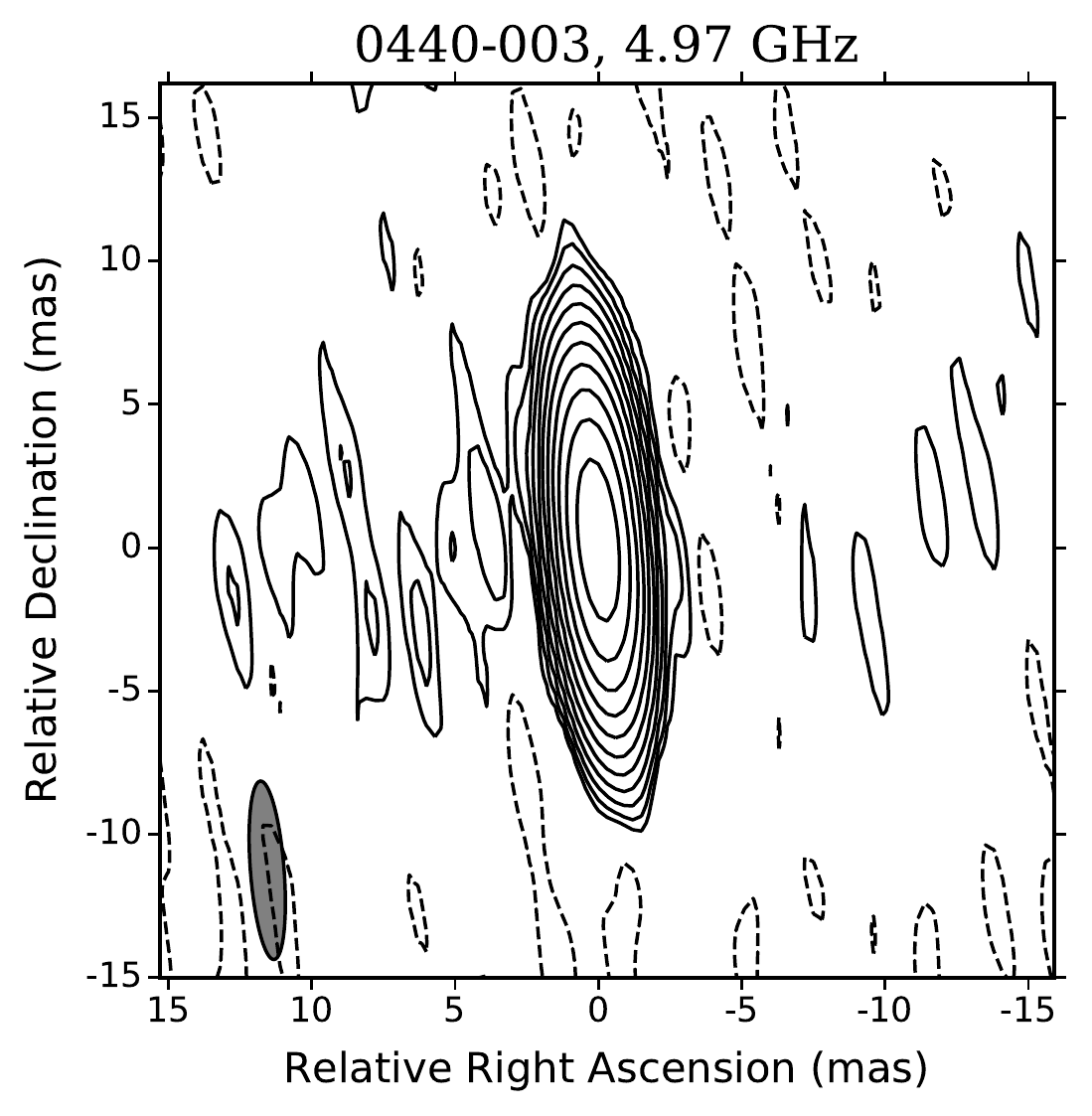}
  \includegraphics[width=0.3\textwidth]{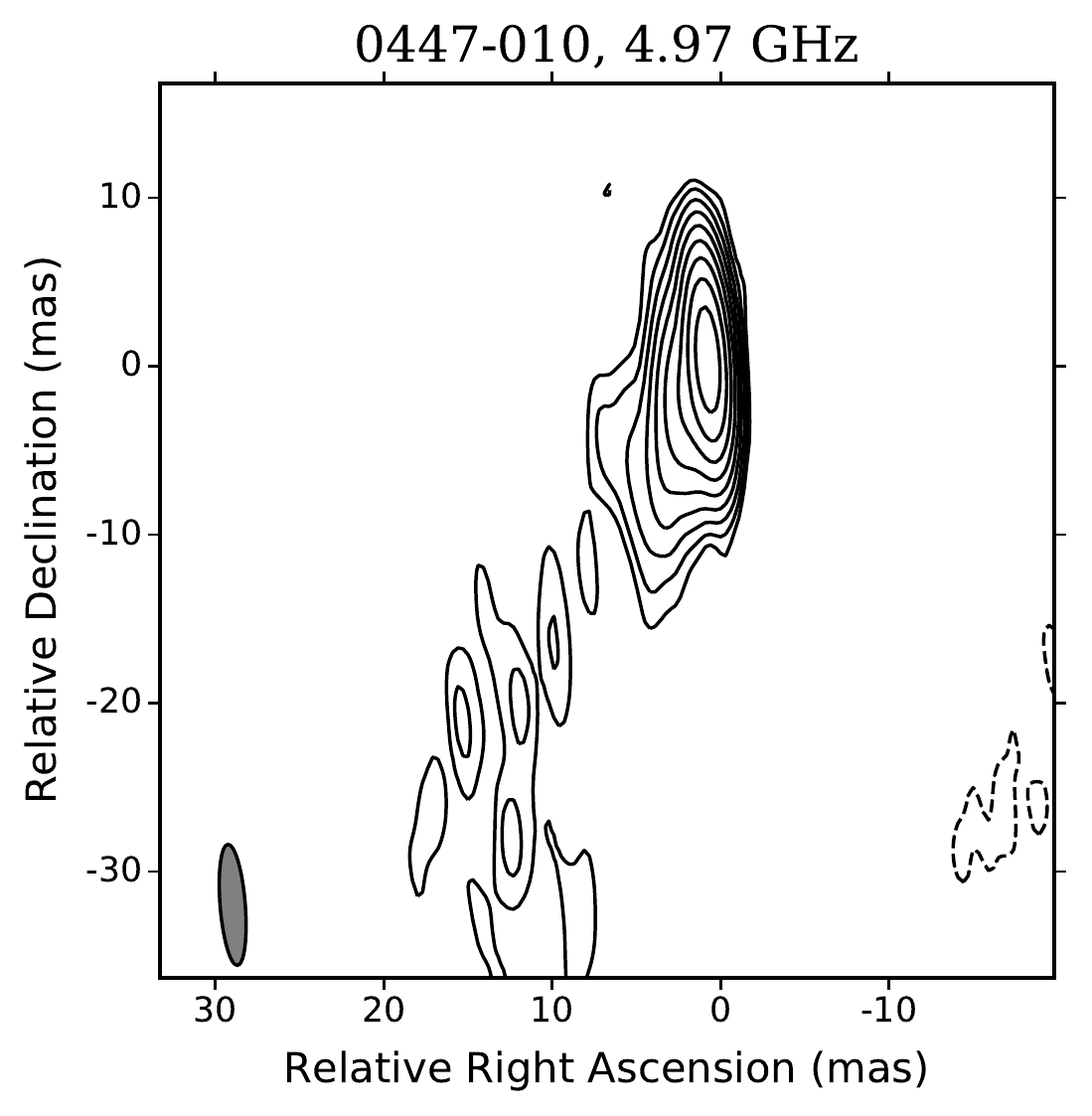}
  \includegraphics[width=0.3\textwidth]{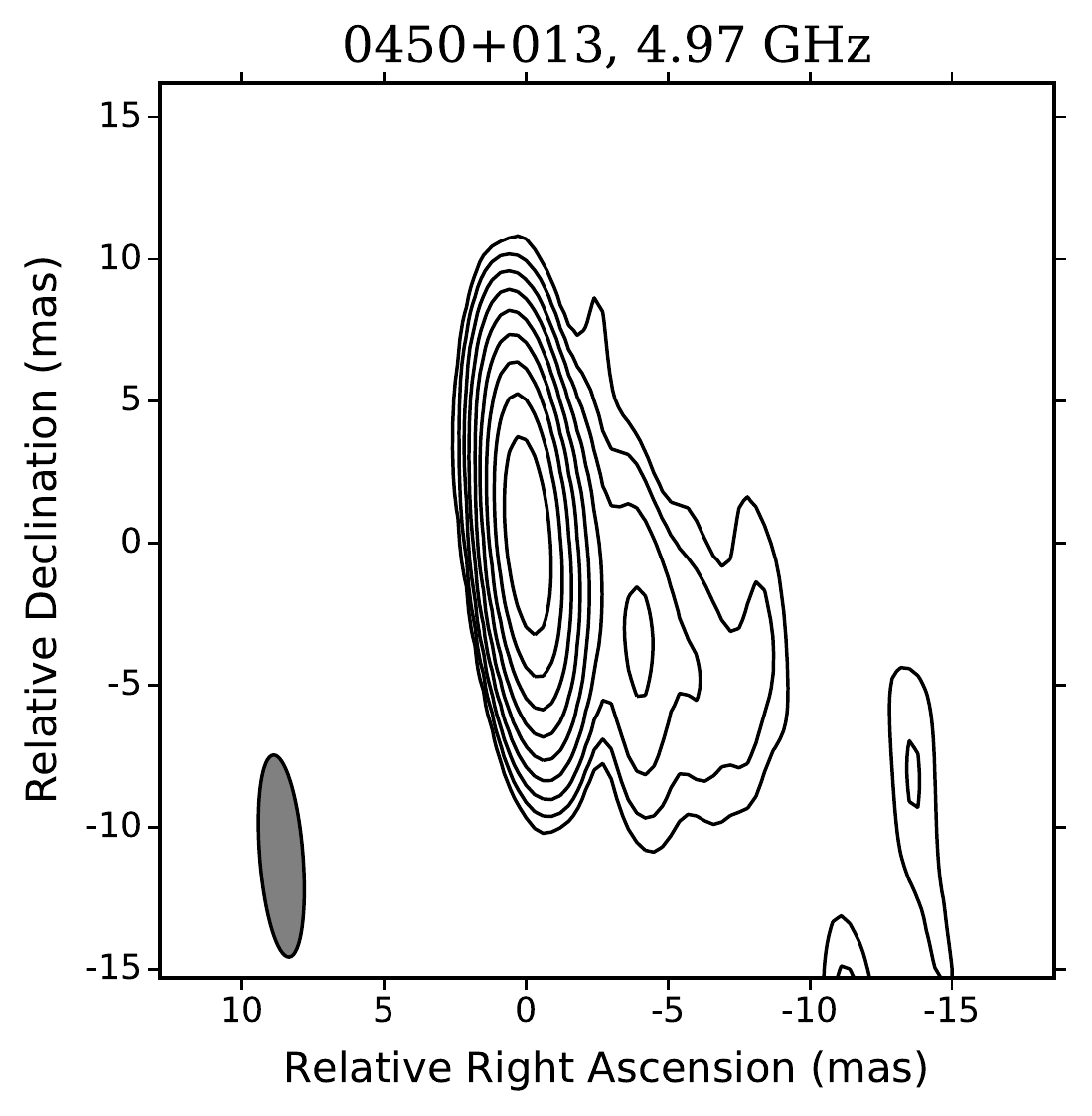}

  \includegraphics[width=0.3\textwidth]{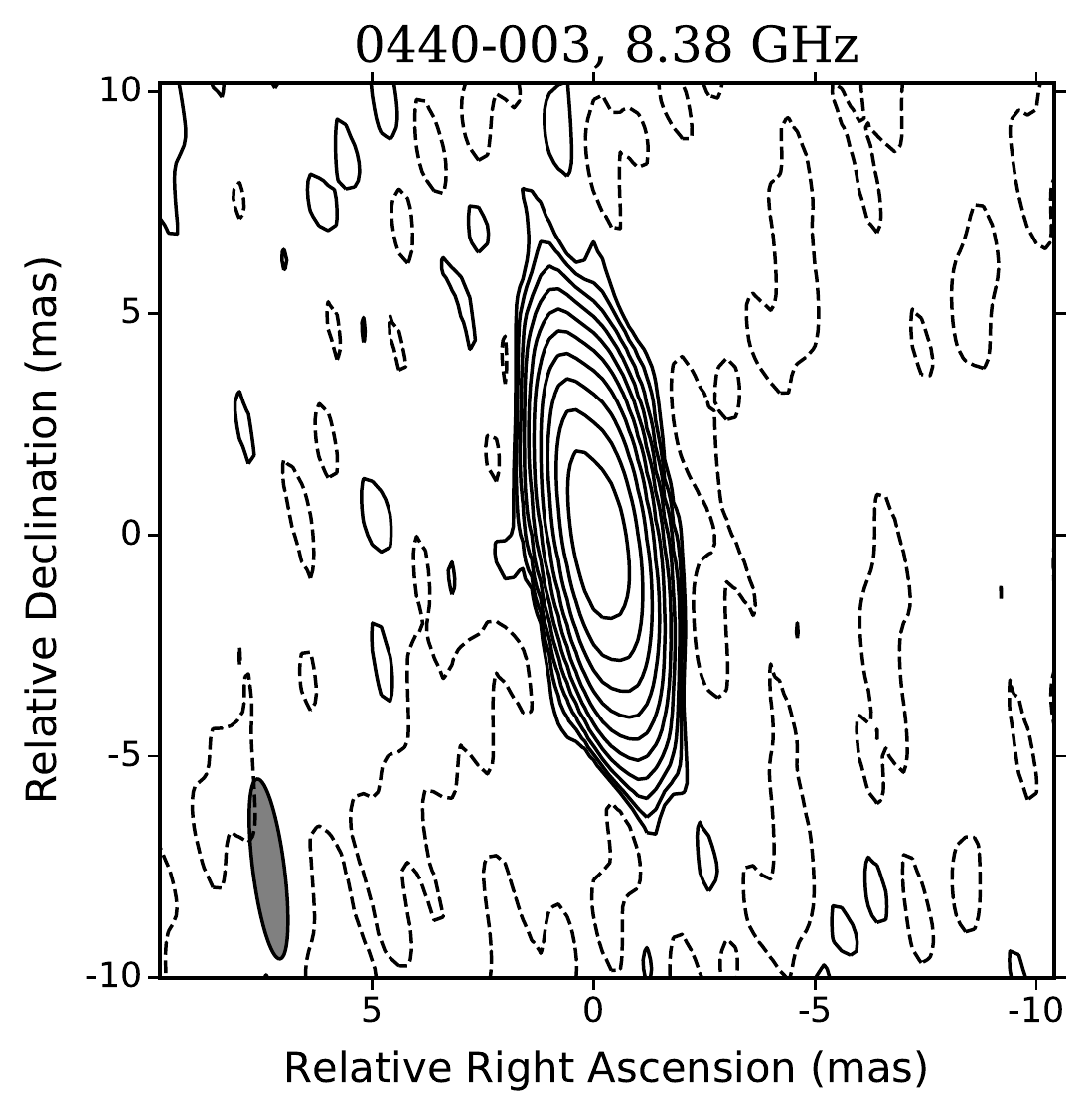}
  \includegraphics[width=0.3\textwidth]{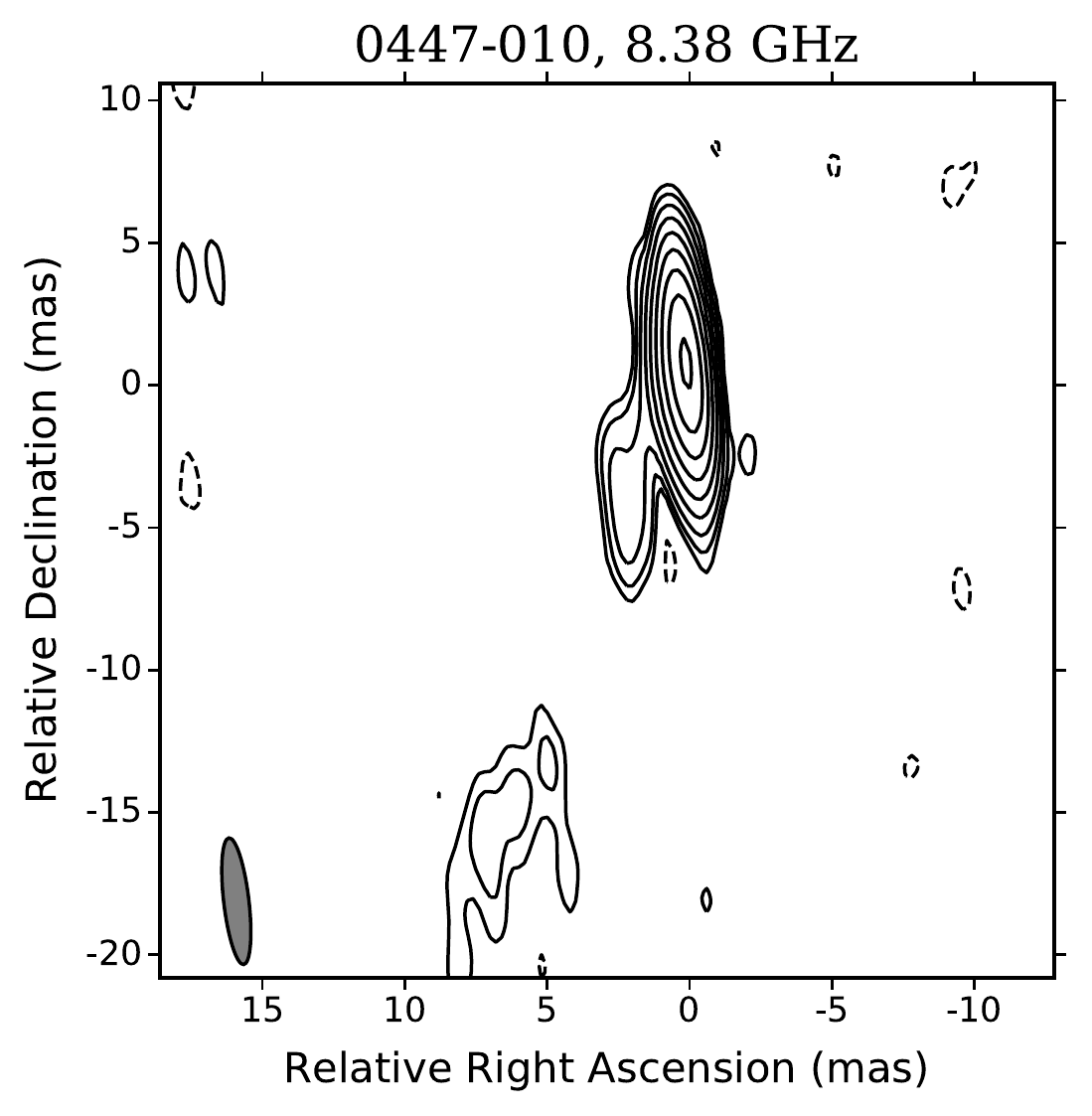}
  \includegraphics[width=0.3\textwidth]{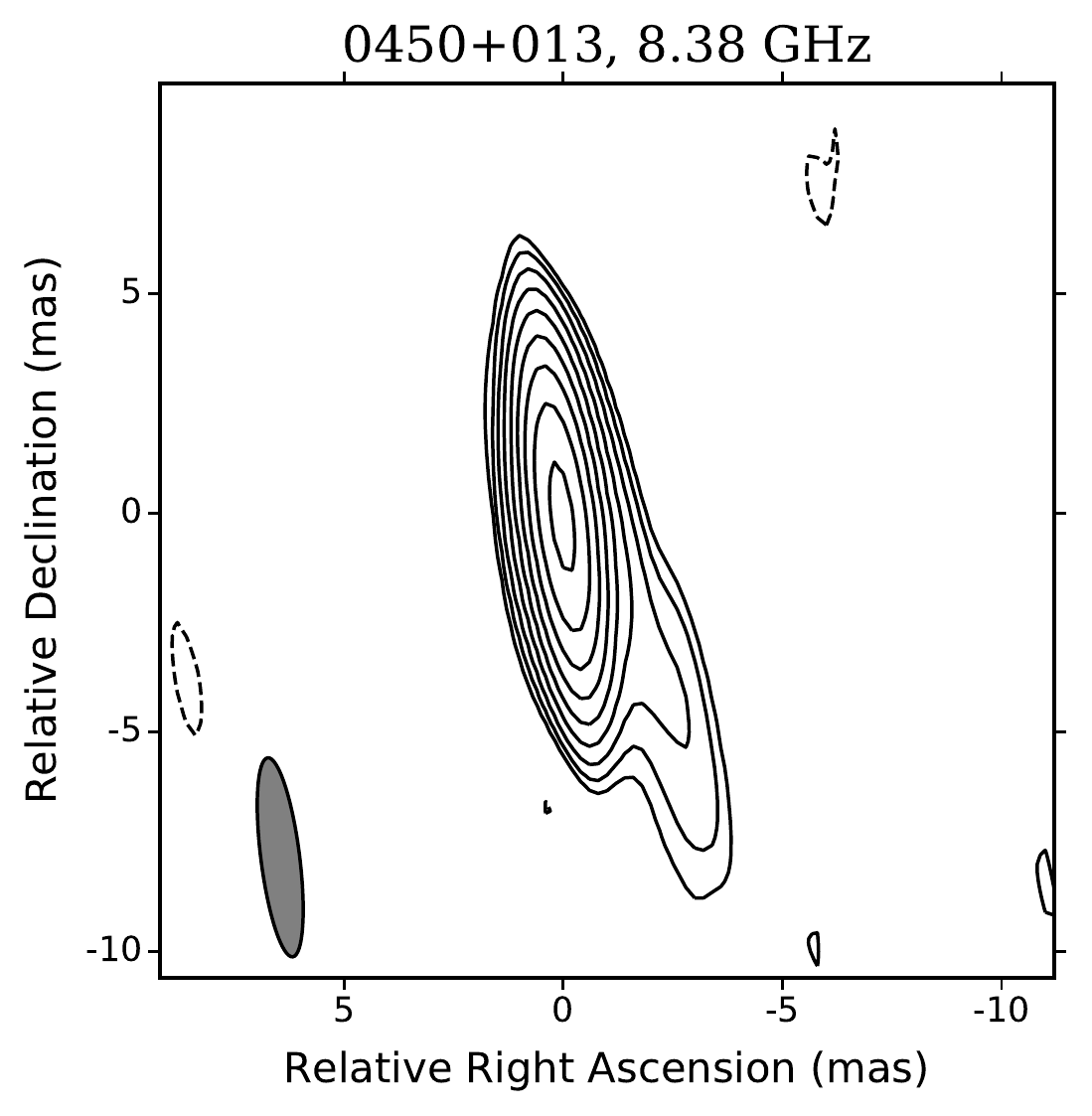}

  \caption{Continued.}
\end{figure*}

\addtocounter{figure}{-1}
\begin{figure*}[p!]

  \includegraphics[width=0.3\textwidth]{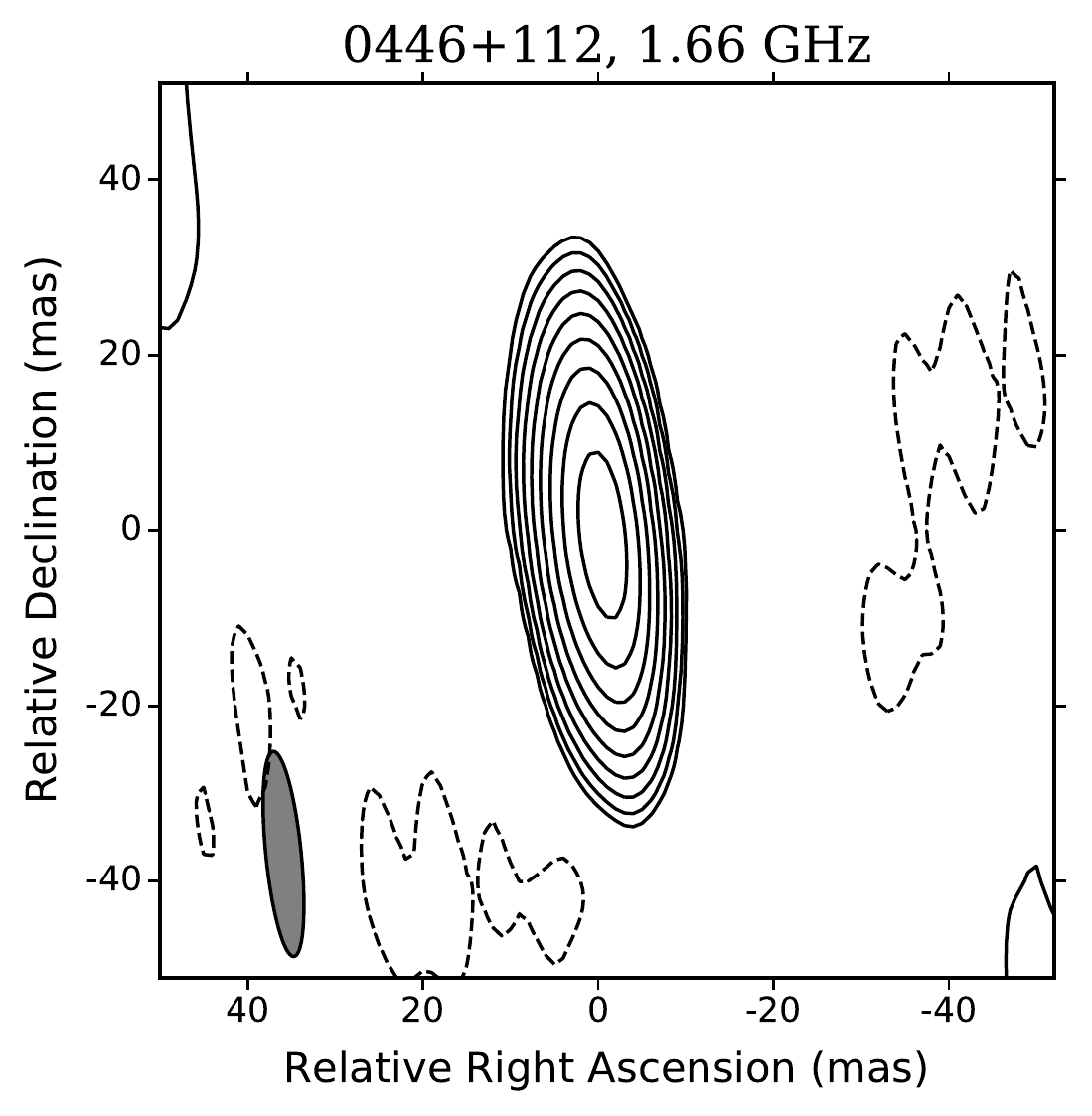}
  \includegraphics[width=0.3\textwidth]{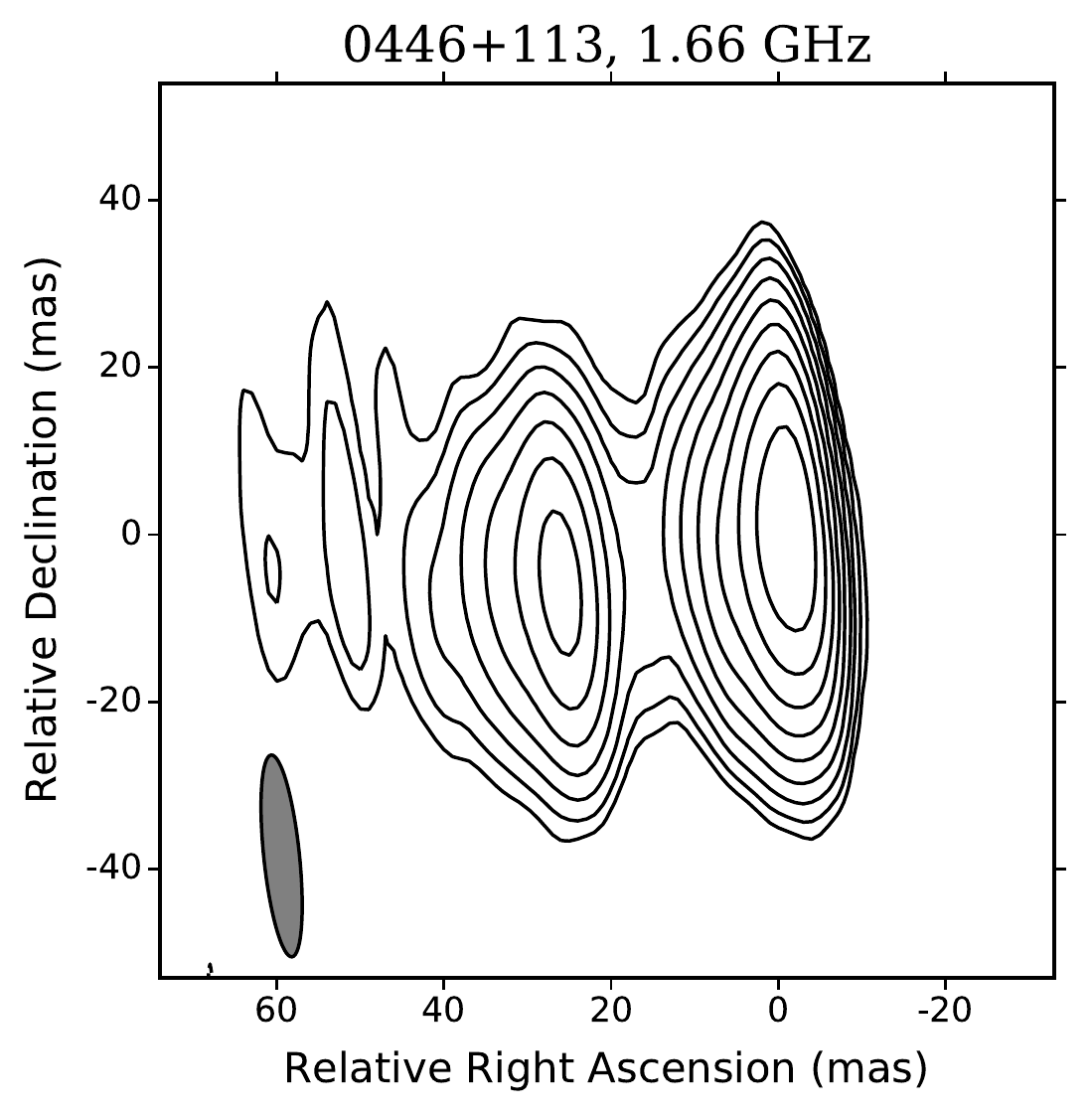}
  \includegraphics[width=0.3\textwidth]{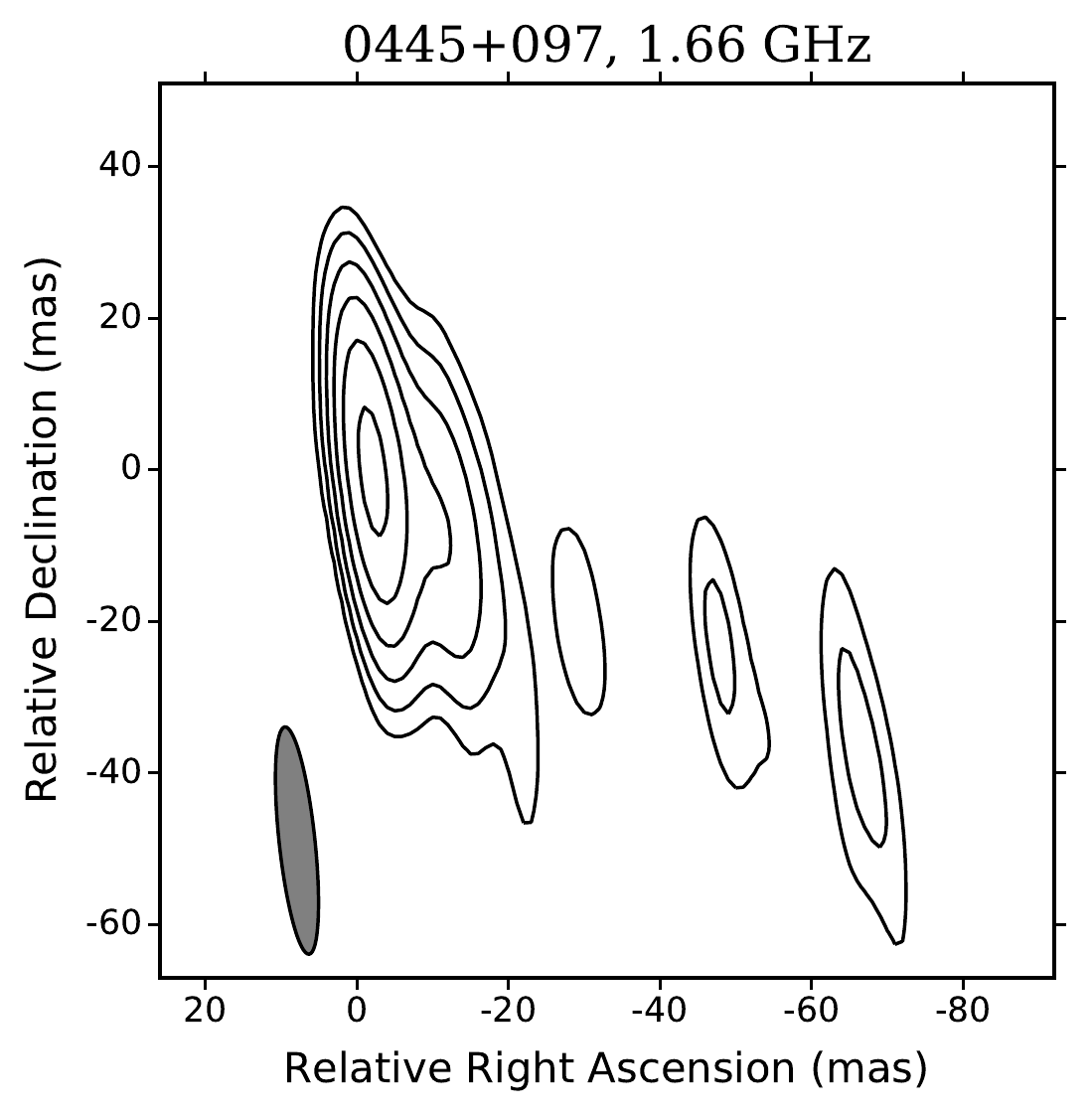}

  \includegraphics[width=0.3\textwidth]{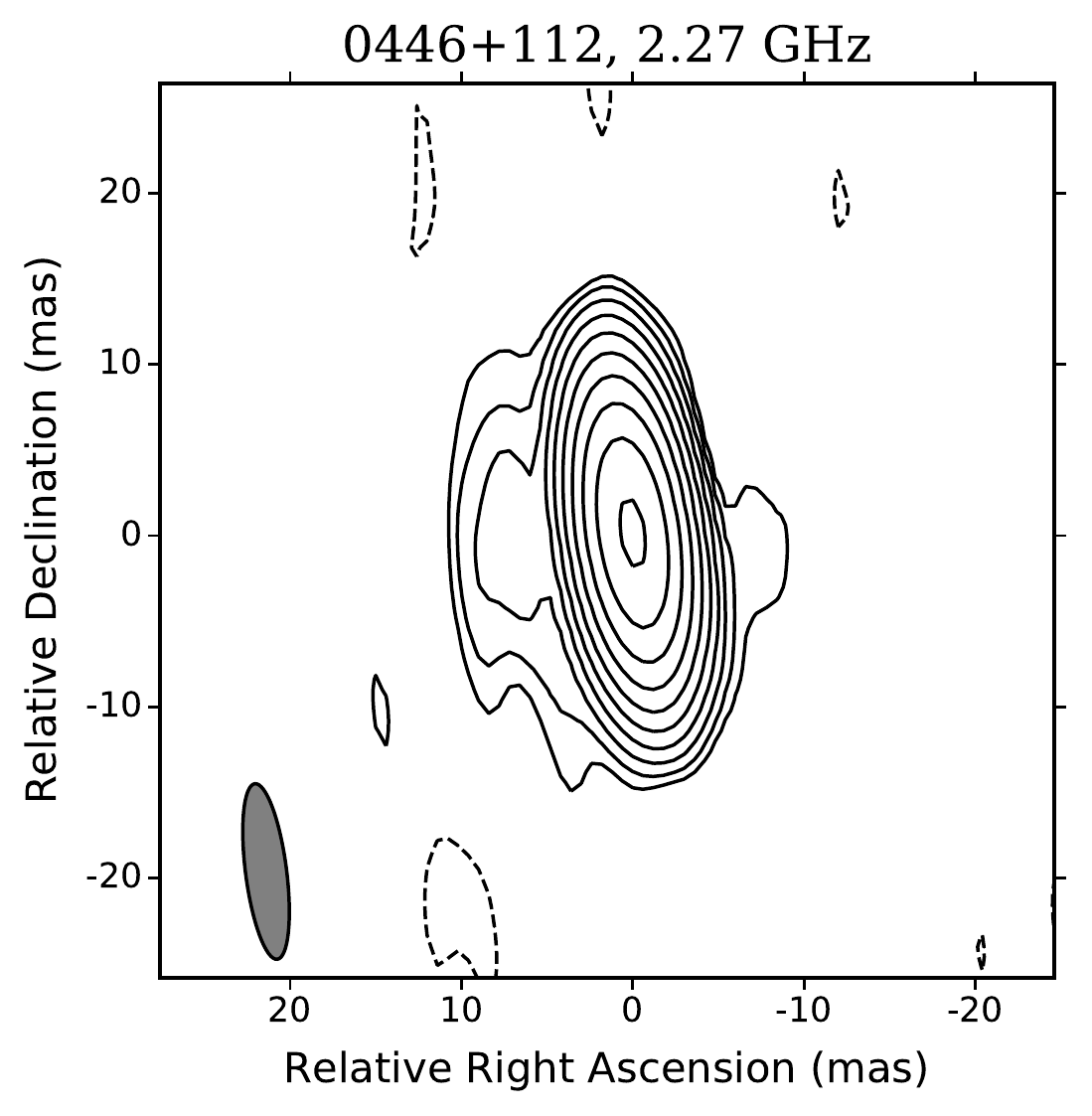}
  \includegraphics[width=0.3\textwidth]{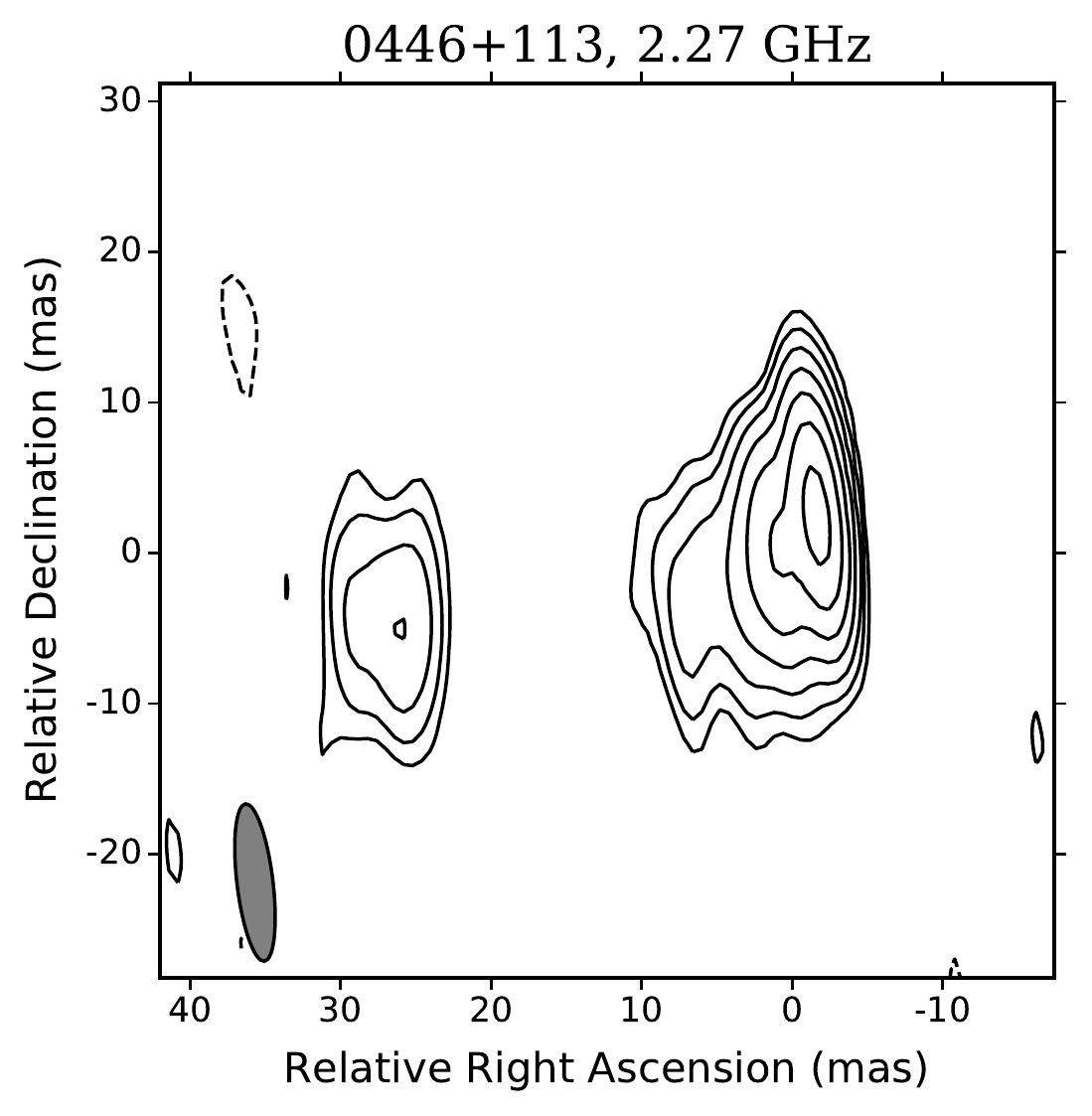}
  \includegraphics[width=0.3\textwidth]{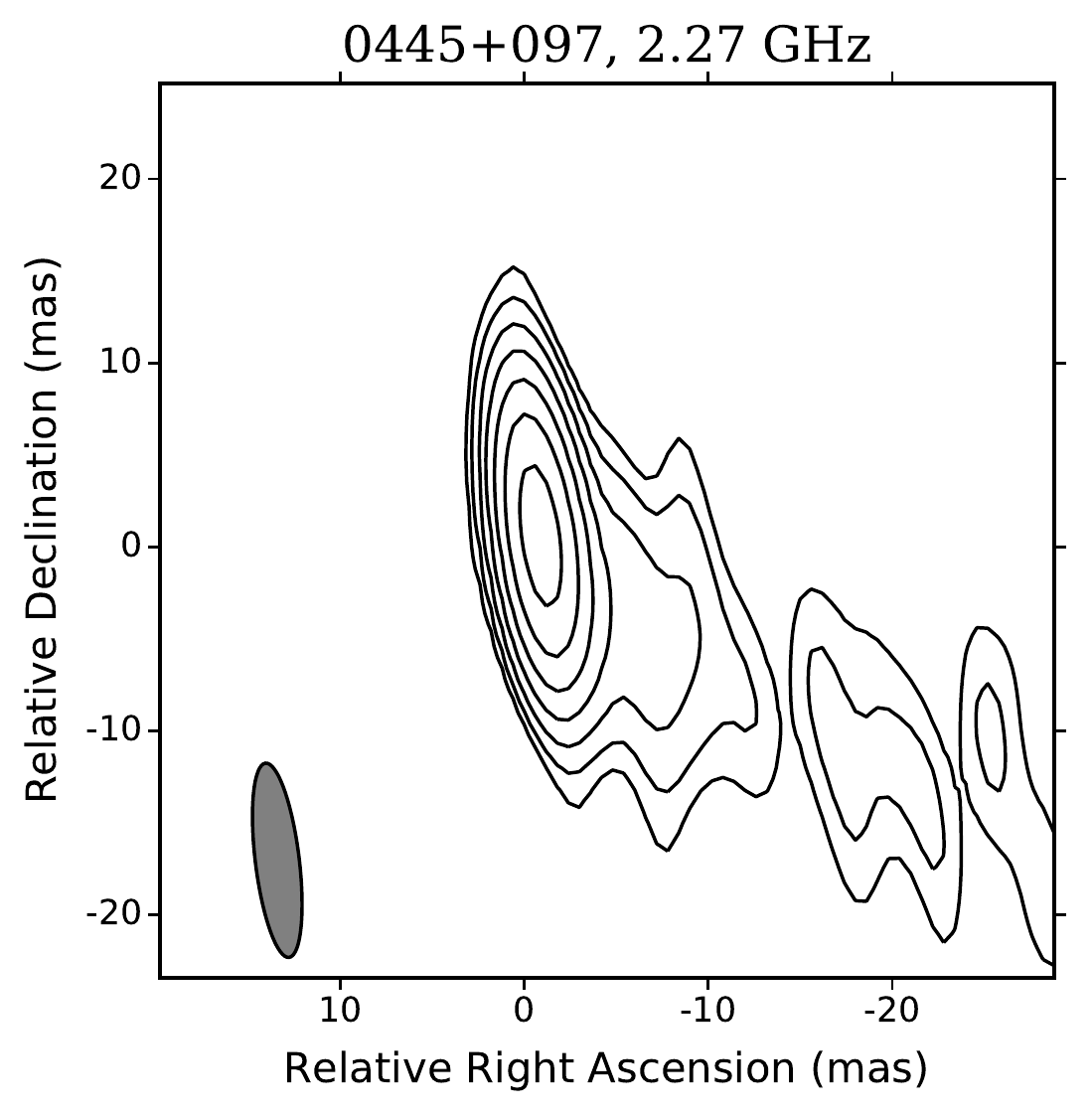}

  \includegraphics[width=0.3\textwidth]{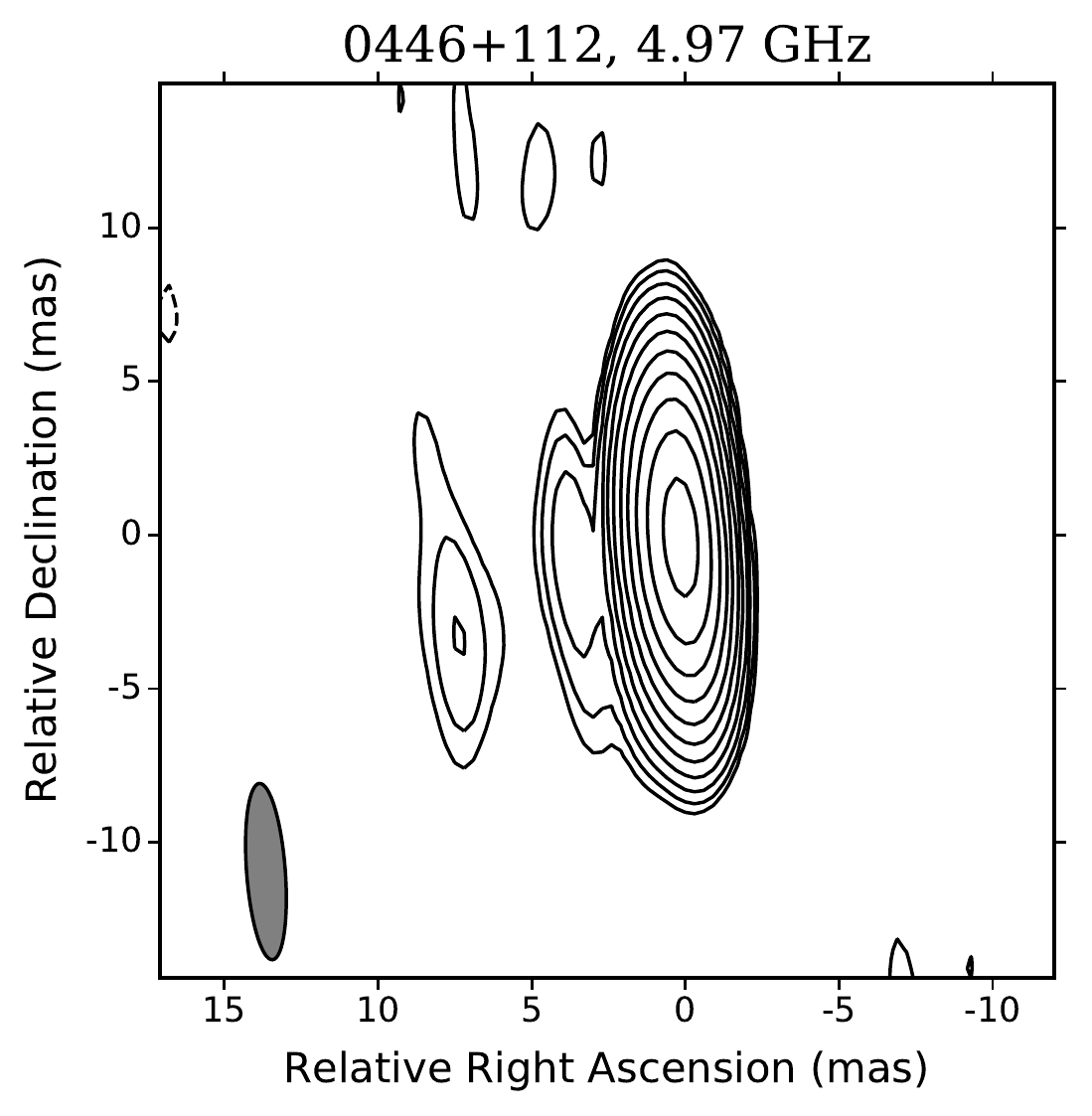}
  \includegraphics[width=0.3\textwidth]{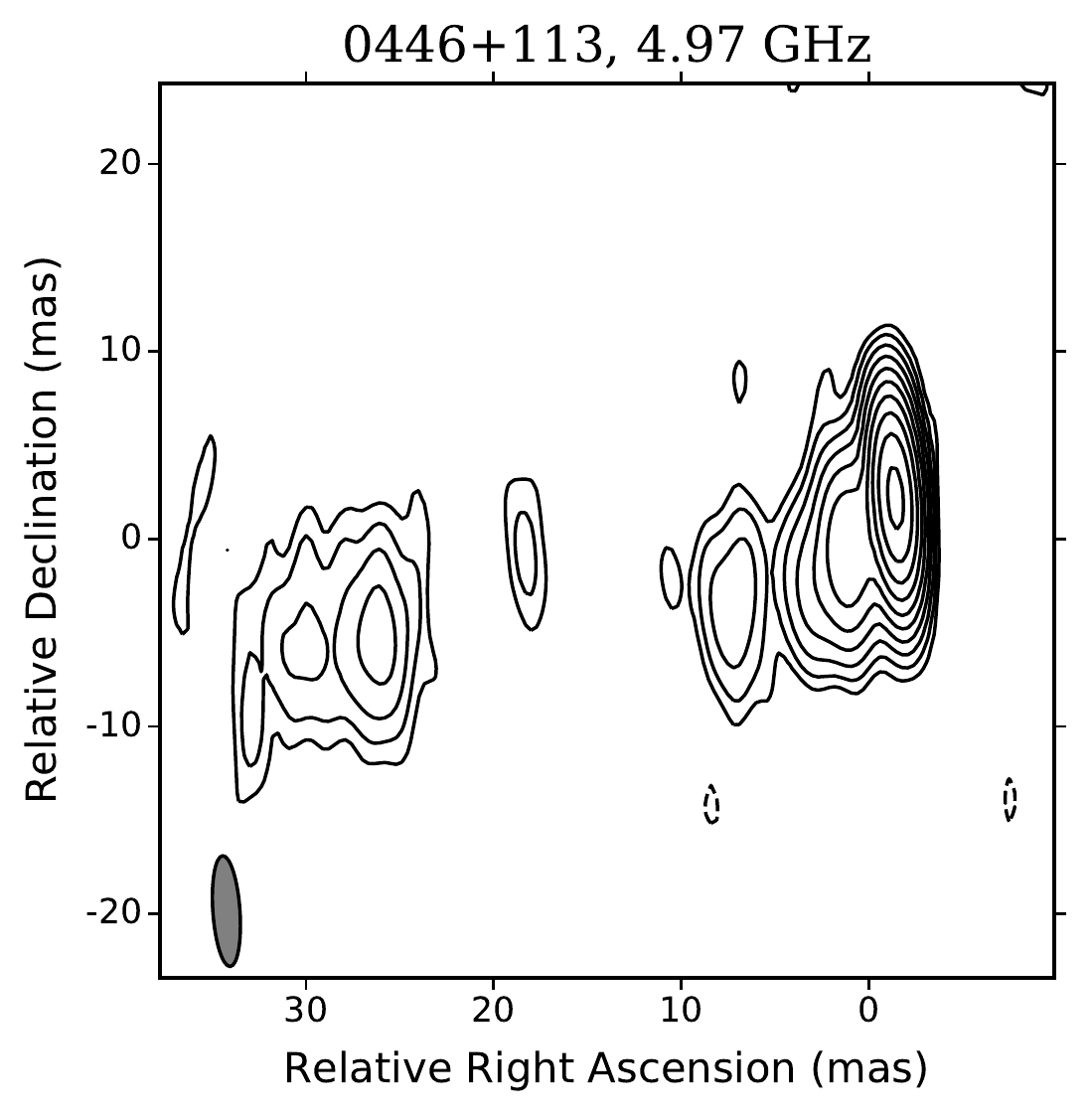}
  \includegraphics[width=0.3\textwidth]{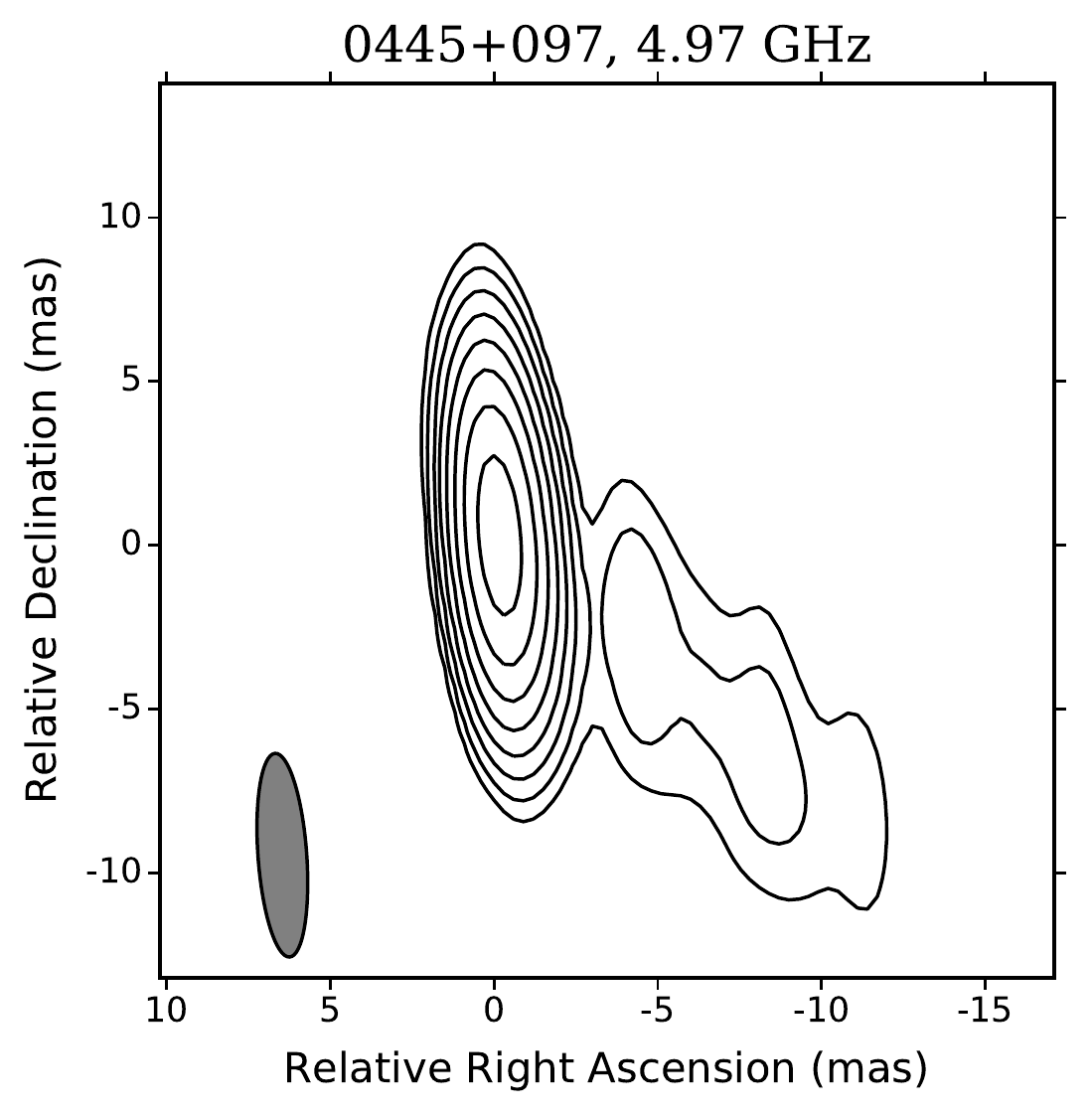}

  \includegraphics[width=0.3\textwidth]{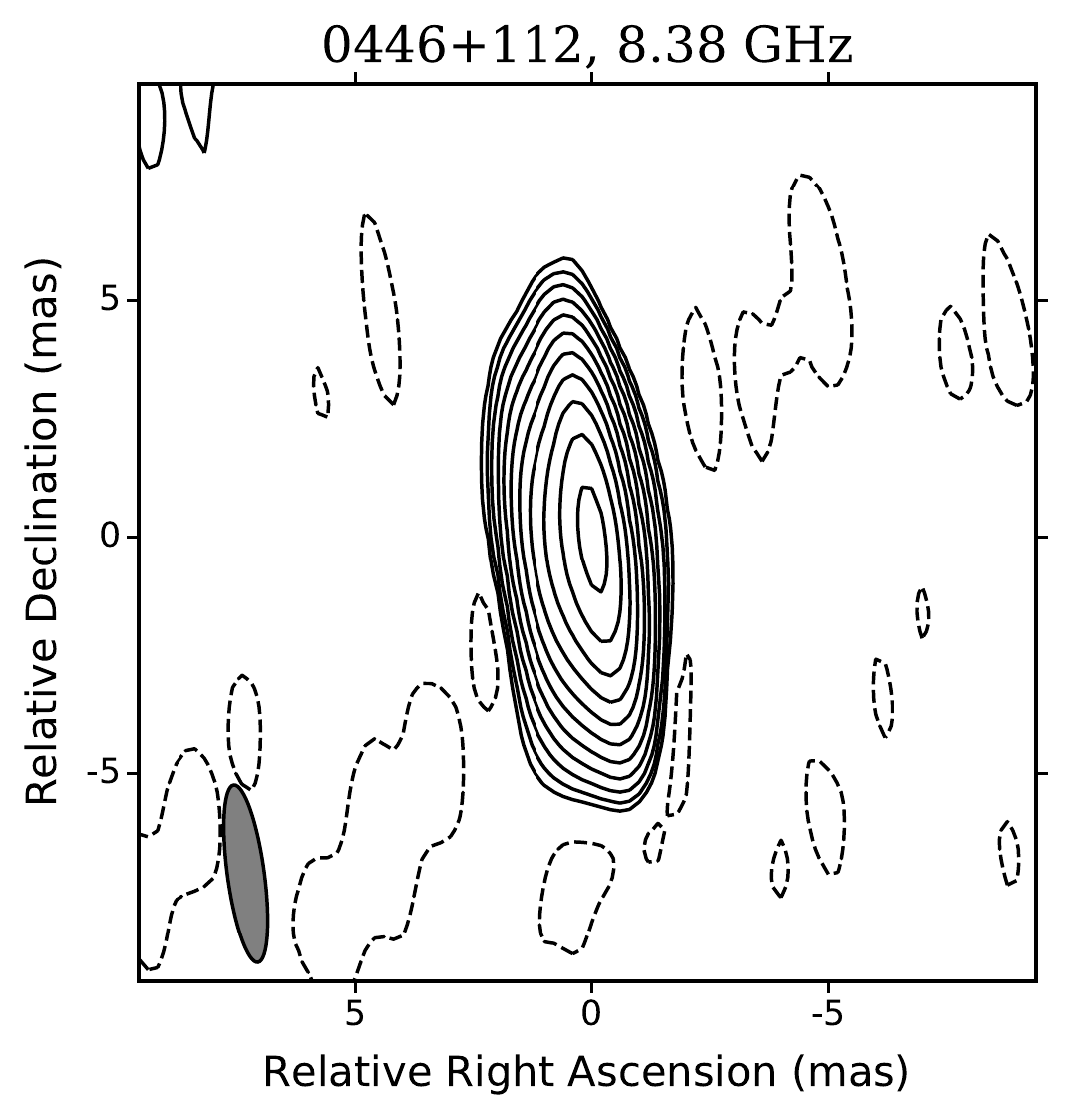}
  \includegraphics[width=0.3\textwidth]{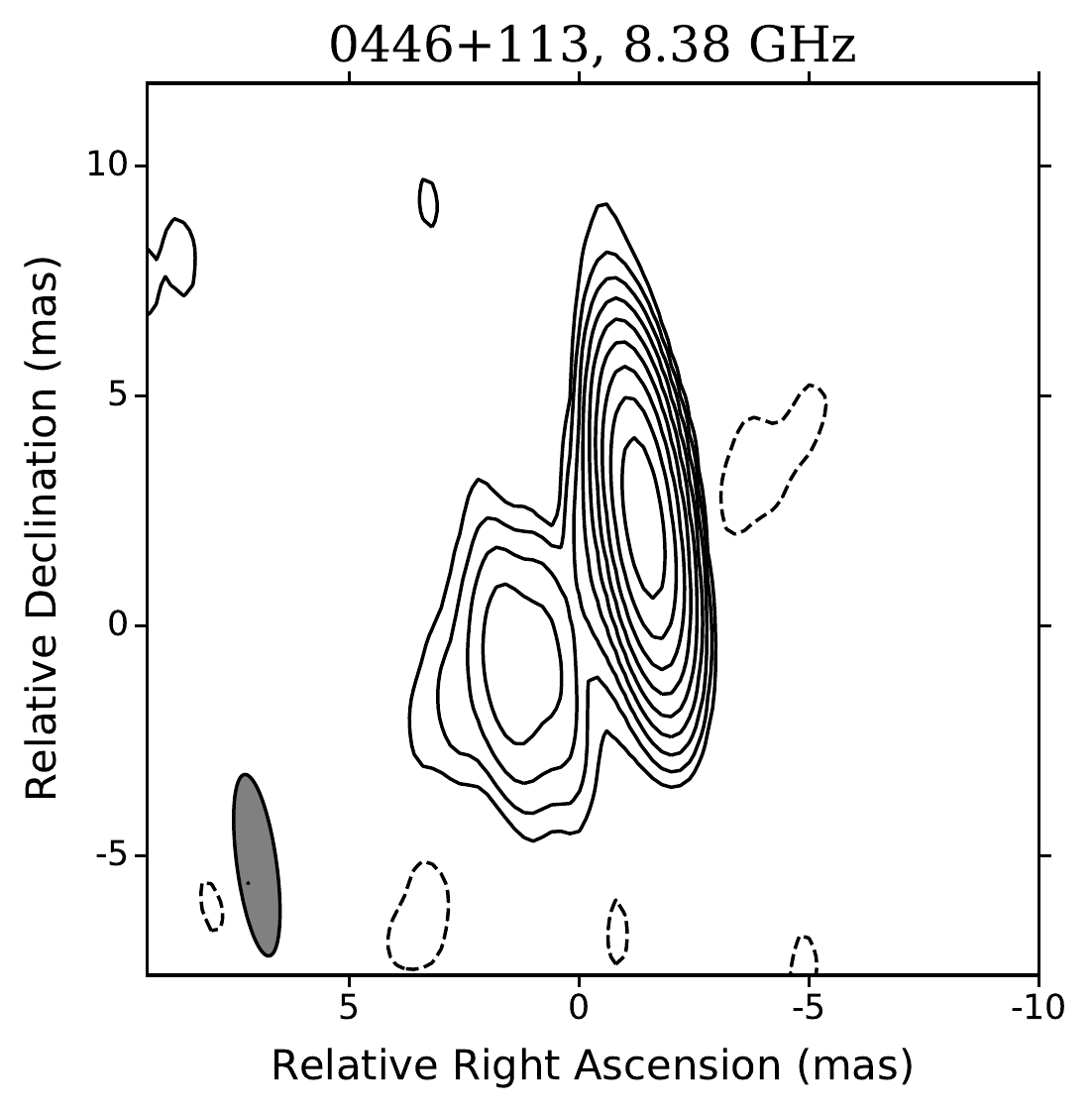}
  \includegraphics[width=0.3\textwidth]{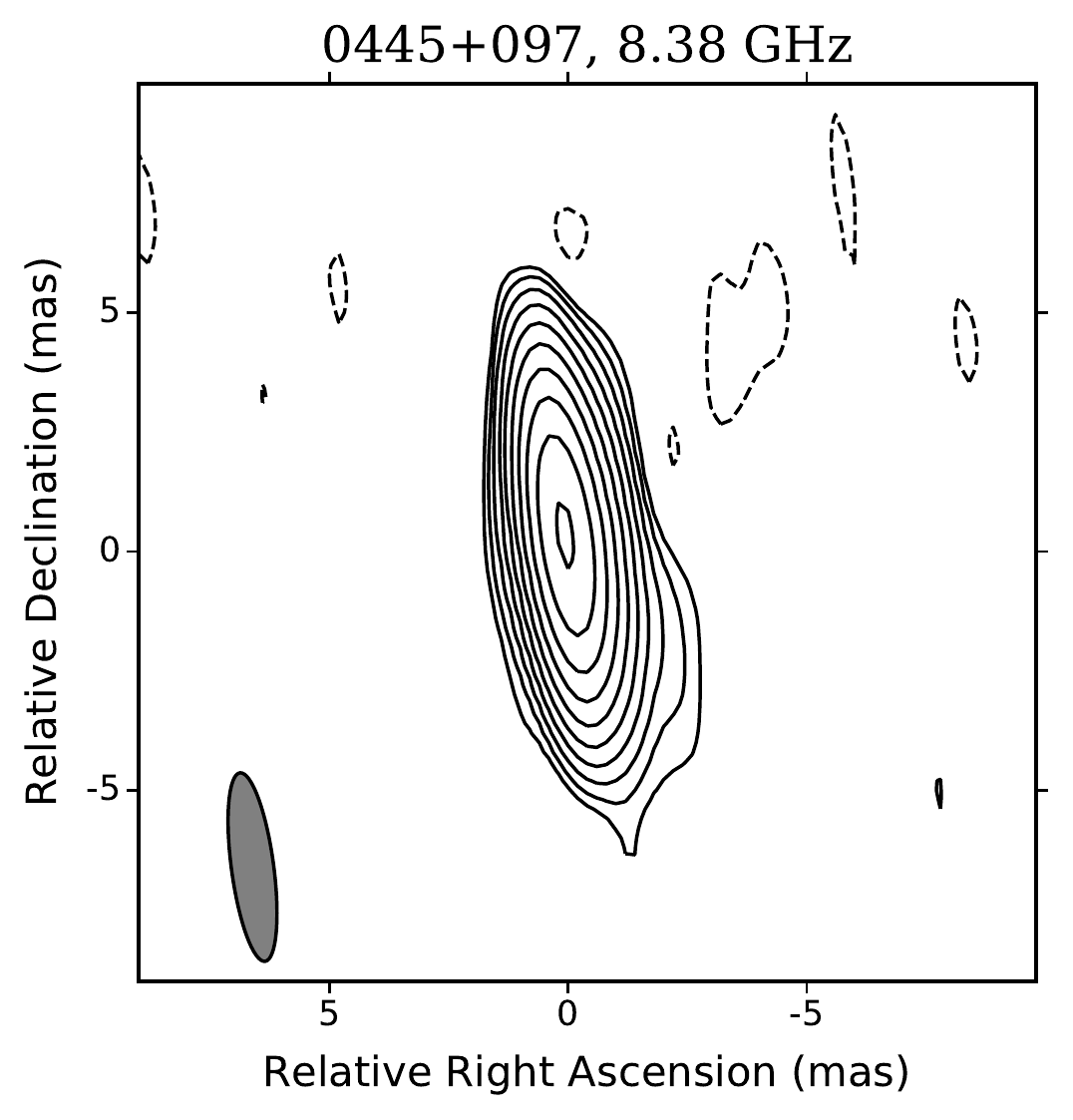}

  \caption{Continued.}
\end{figure*}

\addtocounter{figure}{-1}
\begin{figure*}[p!]

  \includegraphics[width=0.3\textwidth]{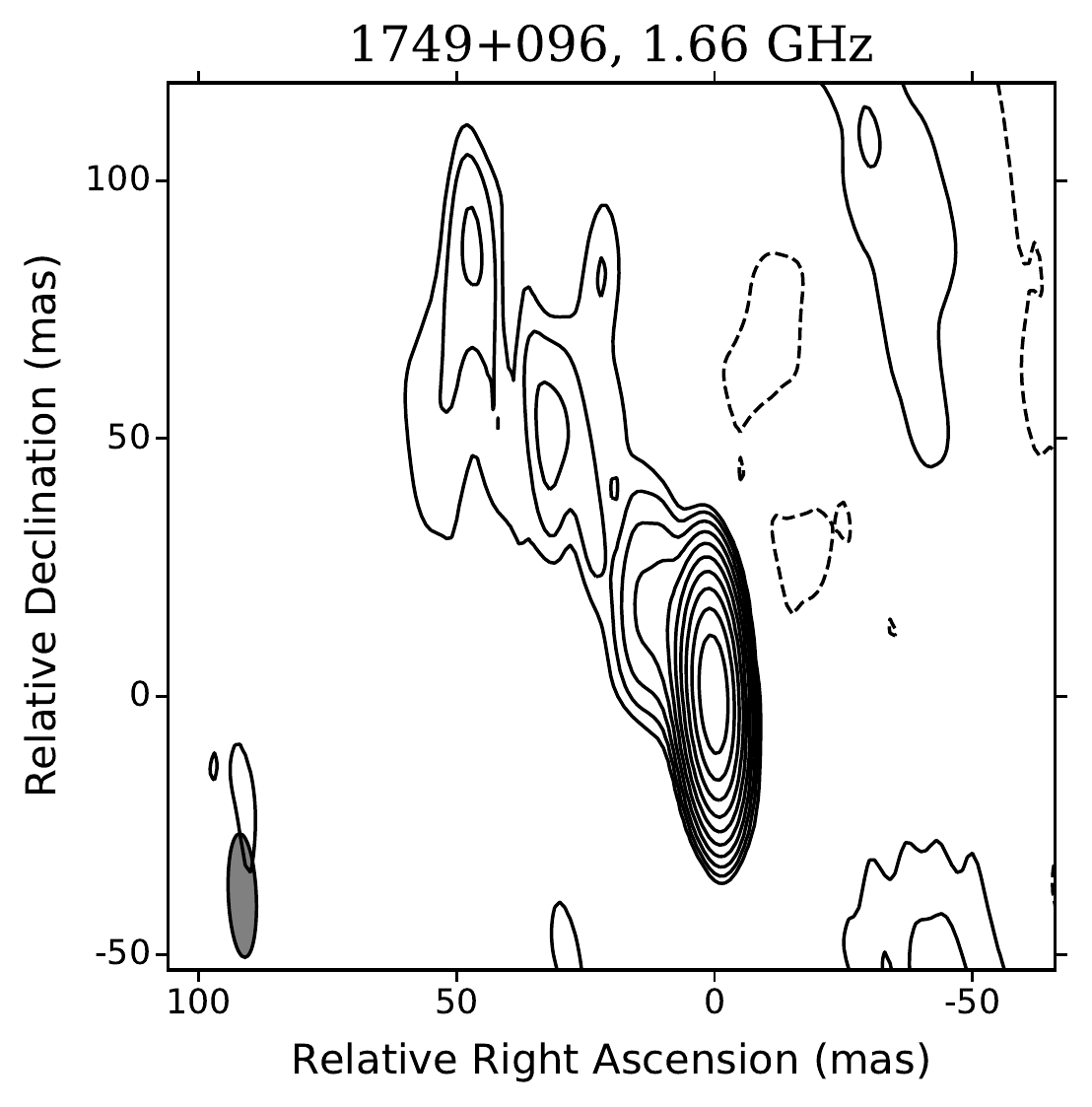}
  \includegraphics[width=0.3\textwidth]{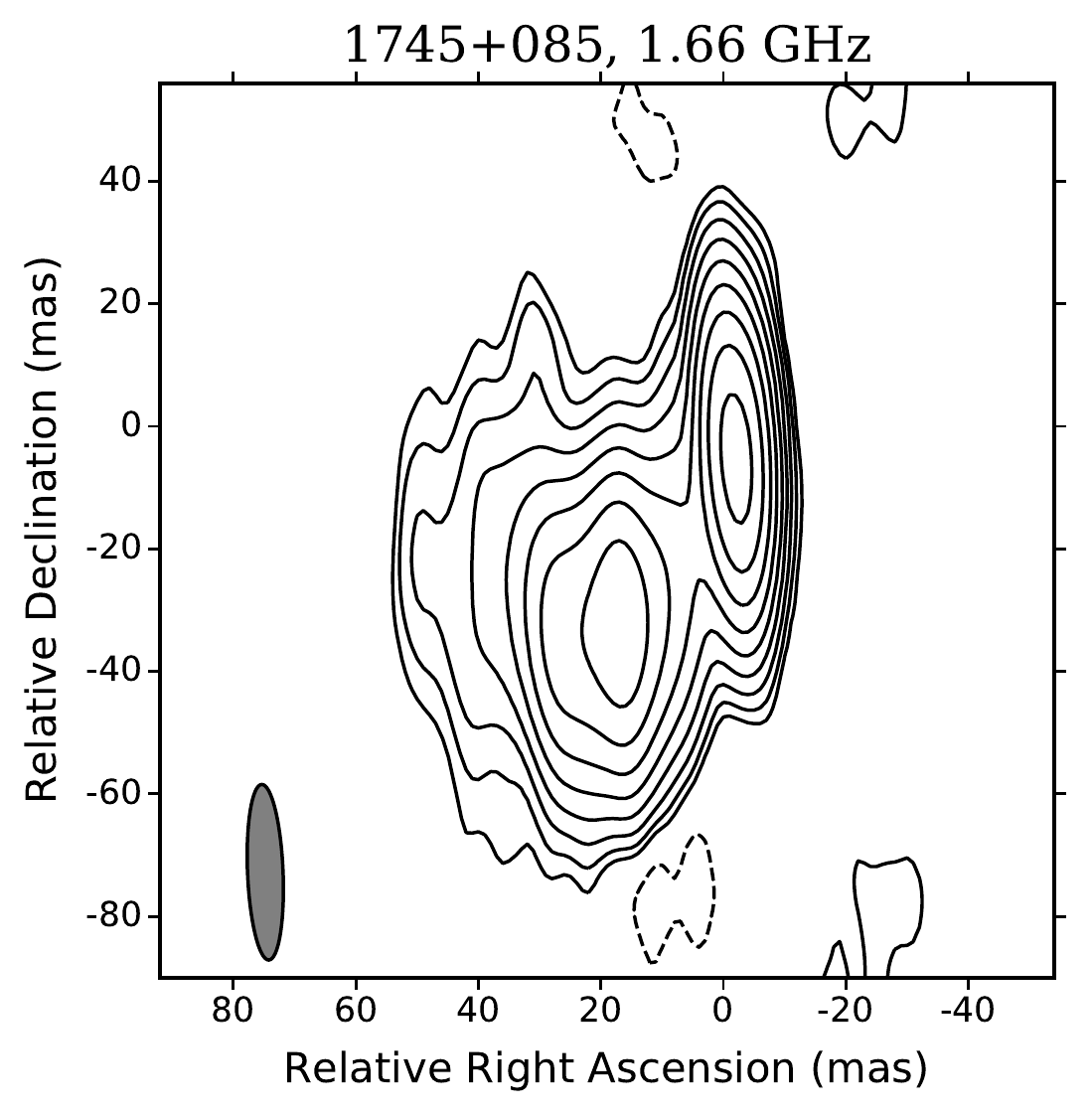}
  \includegraphics[width=0.3\textwidth]{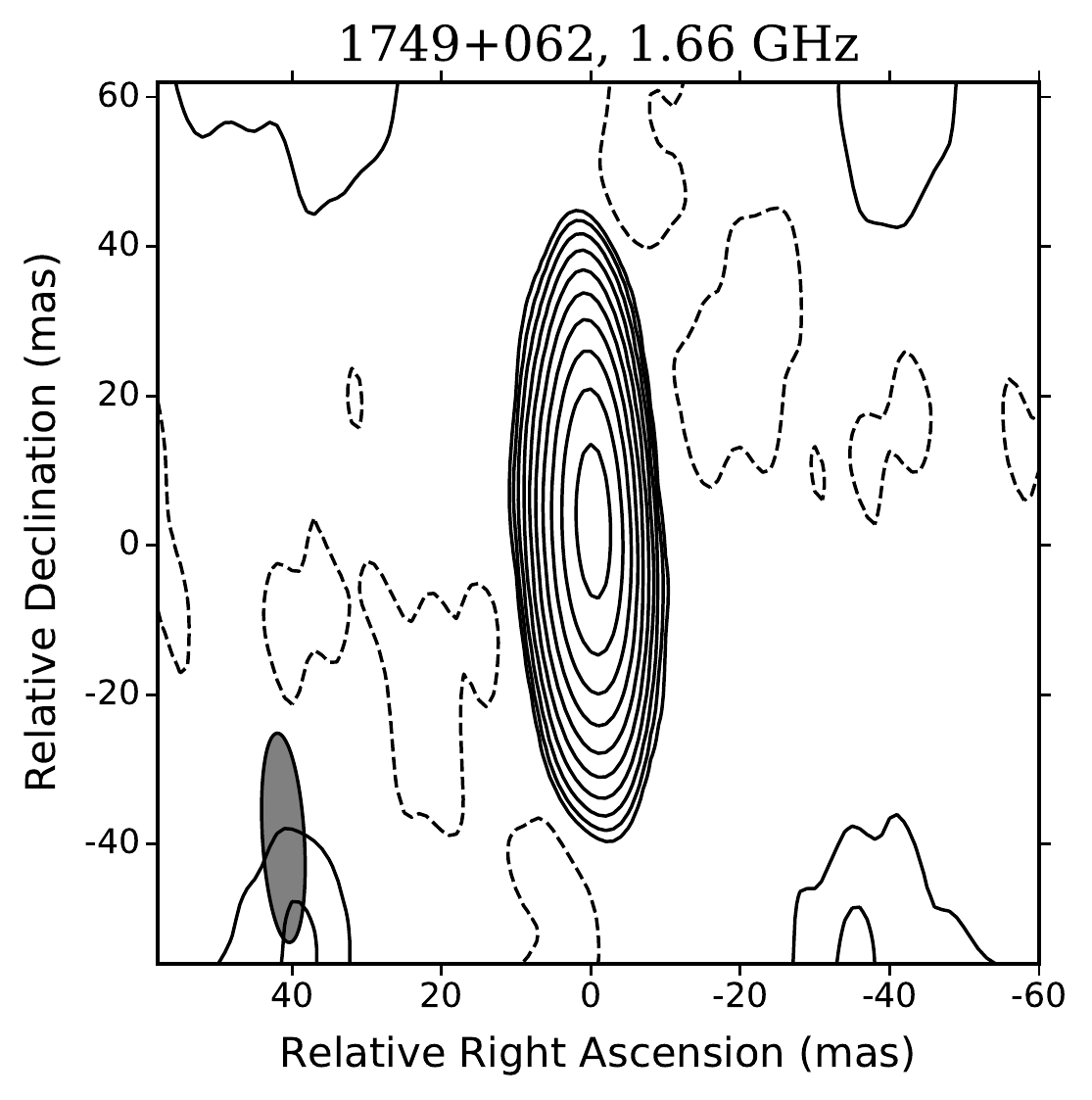}

  \includegraphics[width=0.3\textwidth]{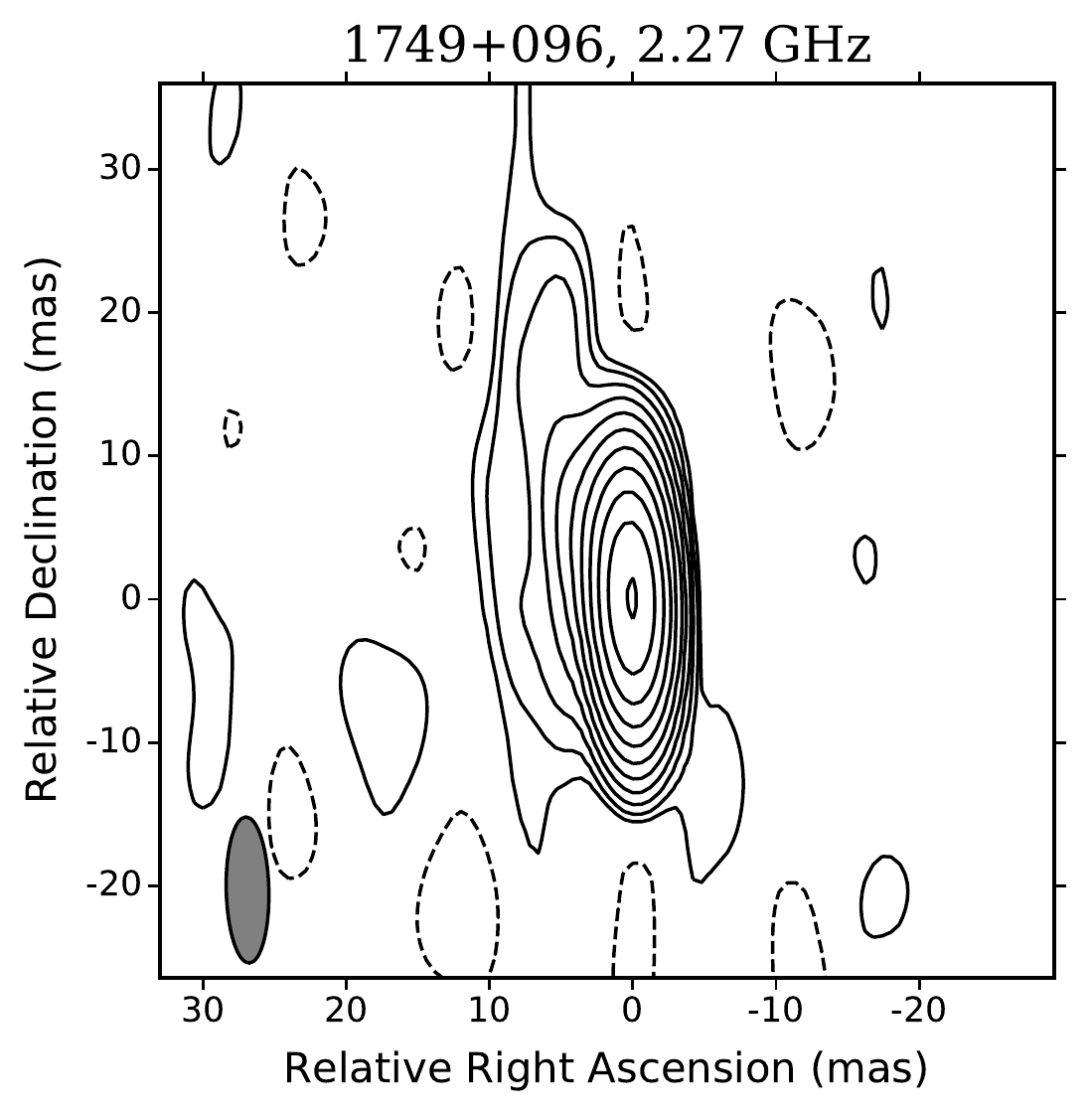}
  \includegraphics[width=0.3\textwidth]{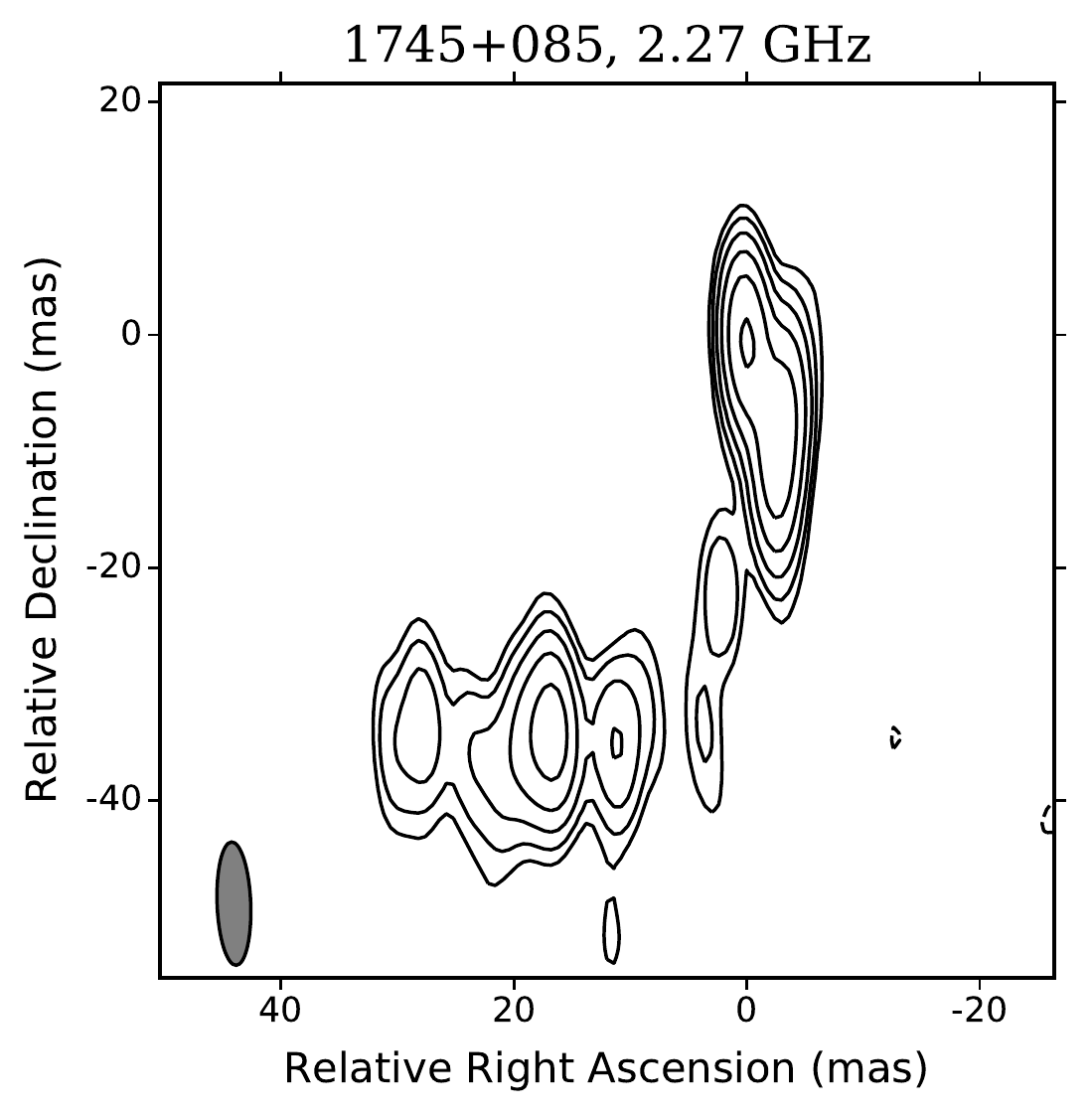}
  \includegraphics[width=0.3\textwidth]{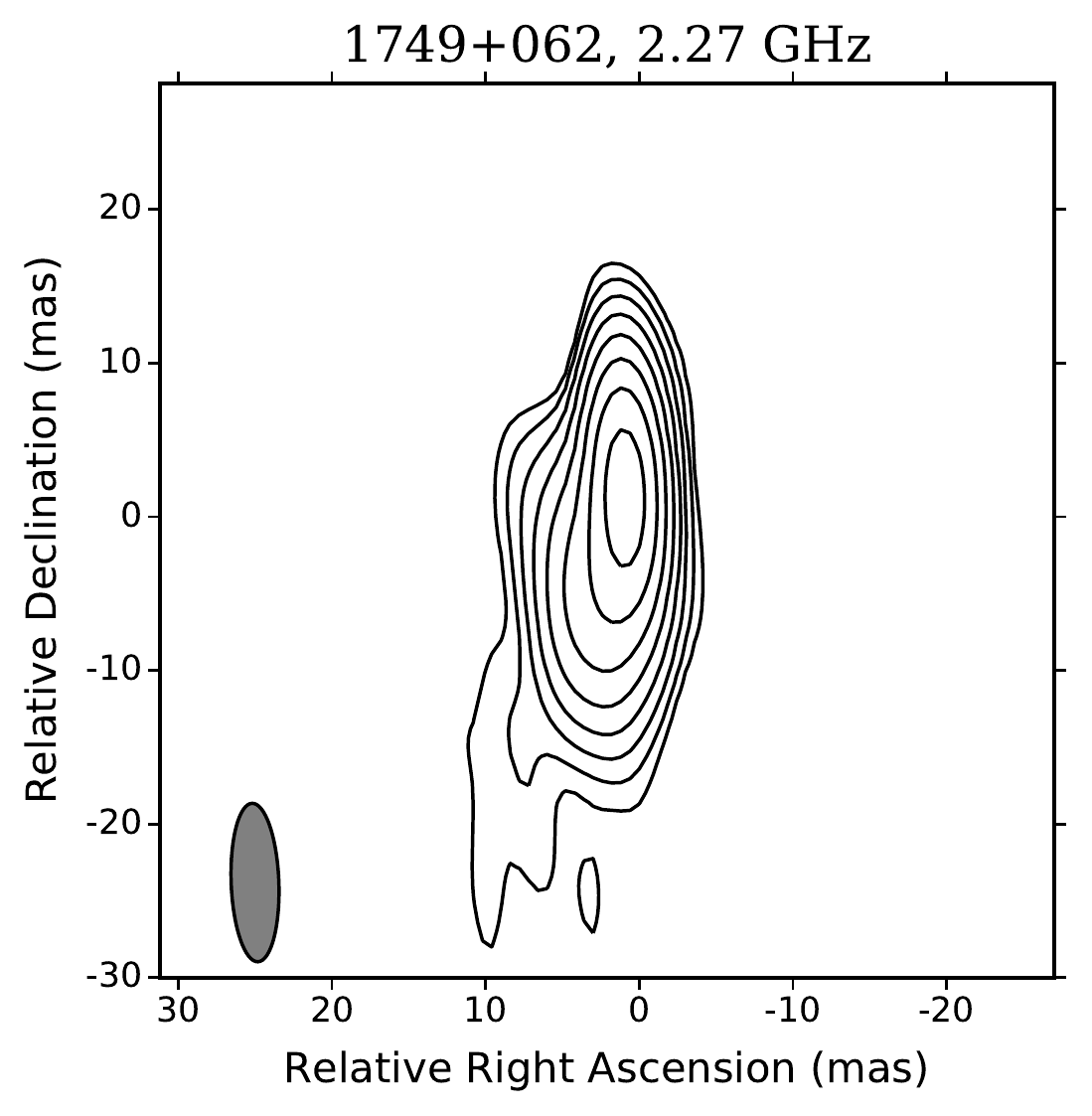}

  \includegraphics[width=0.3\textwidth]{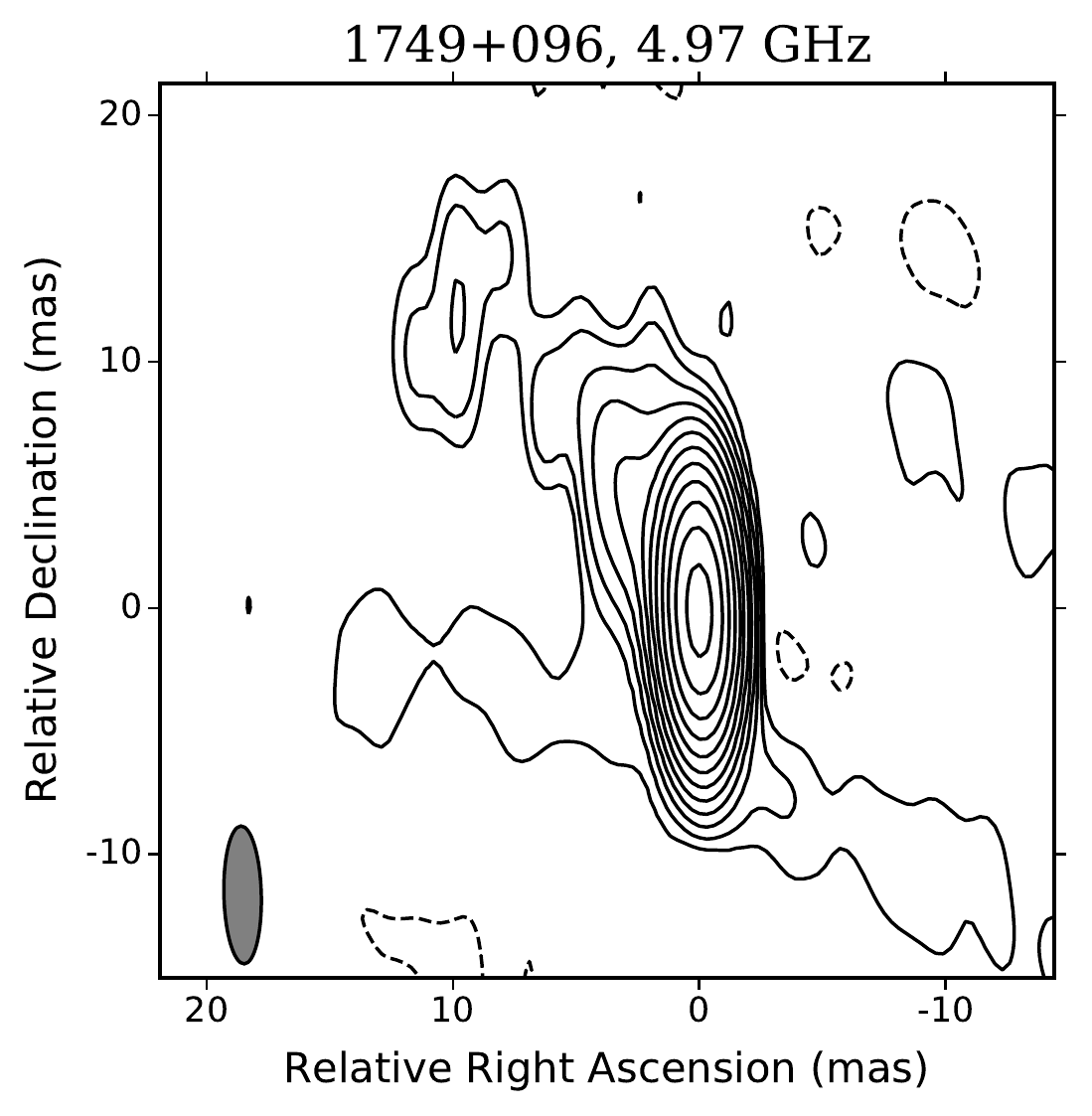}
  \includegraphics[width=0.3\textwidth]{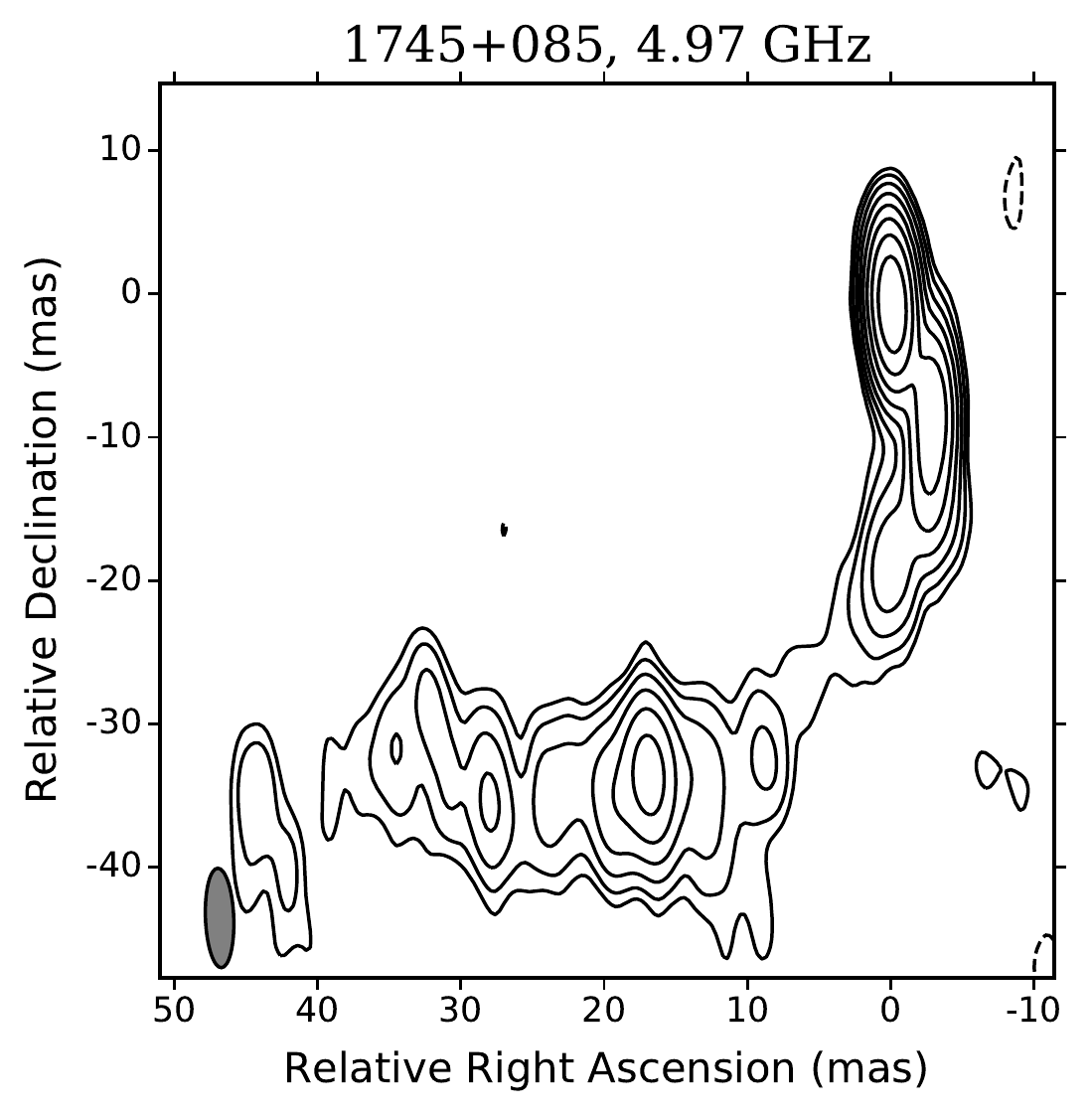}
  \includegraphics[width=0.3\textwidth]{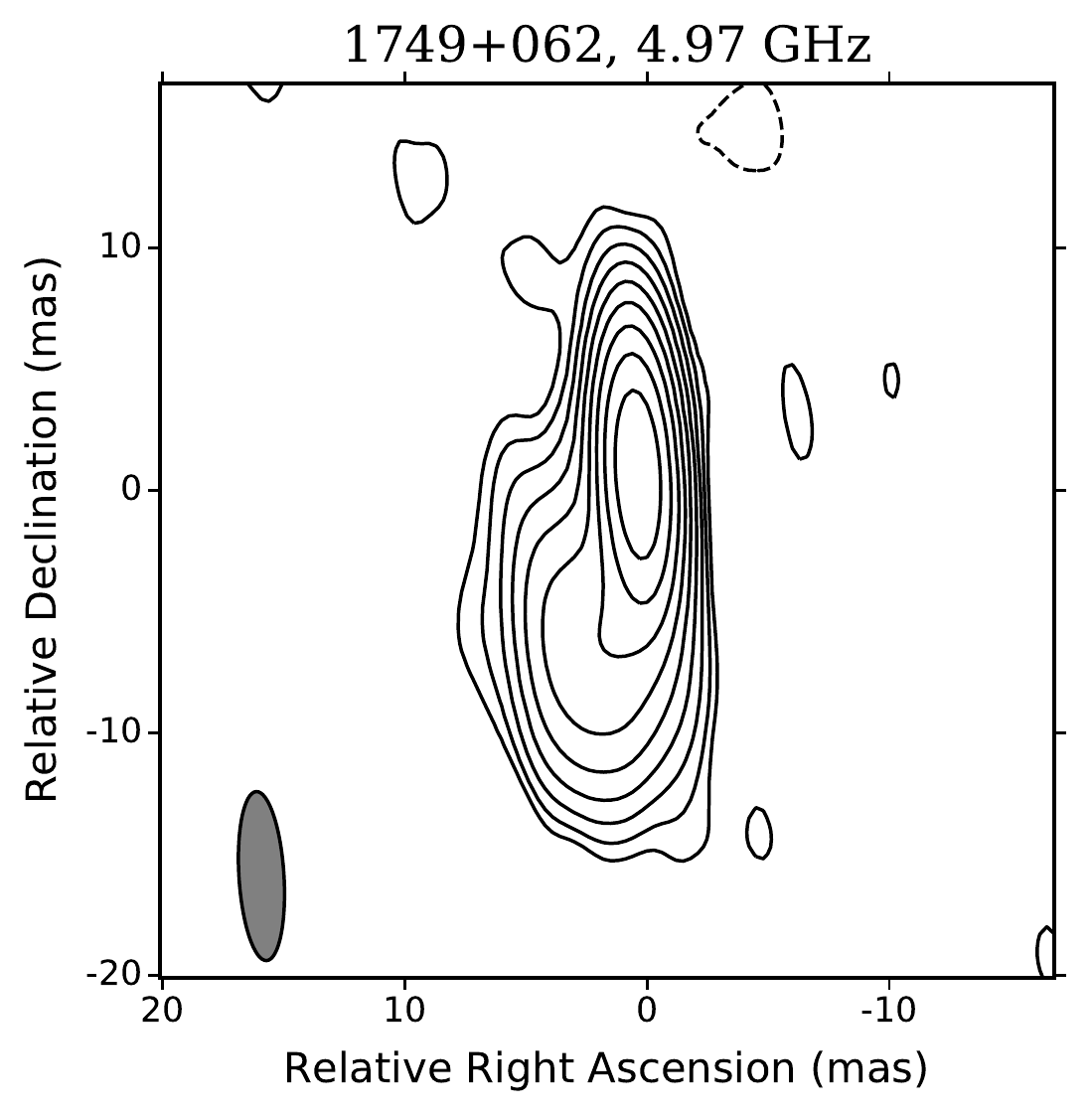}

  \includegraphics[width=0.3\textwidth]{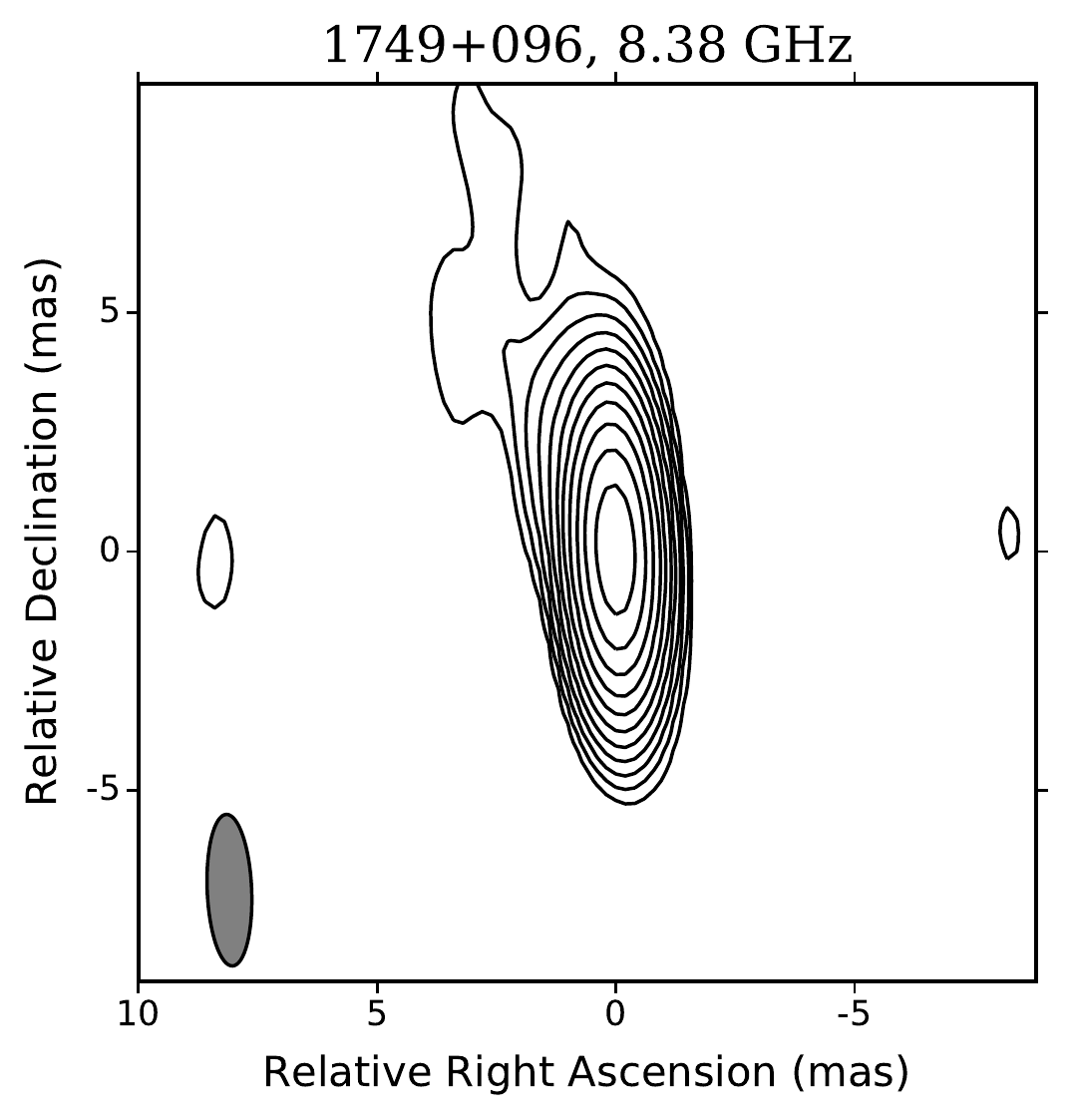}
  \includegraphics[width=0.3\textwidth]{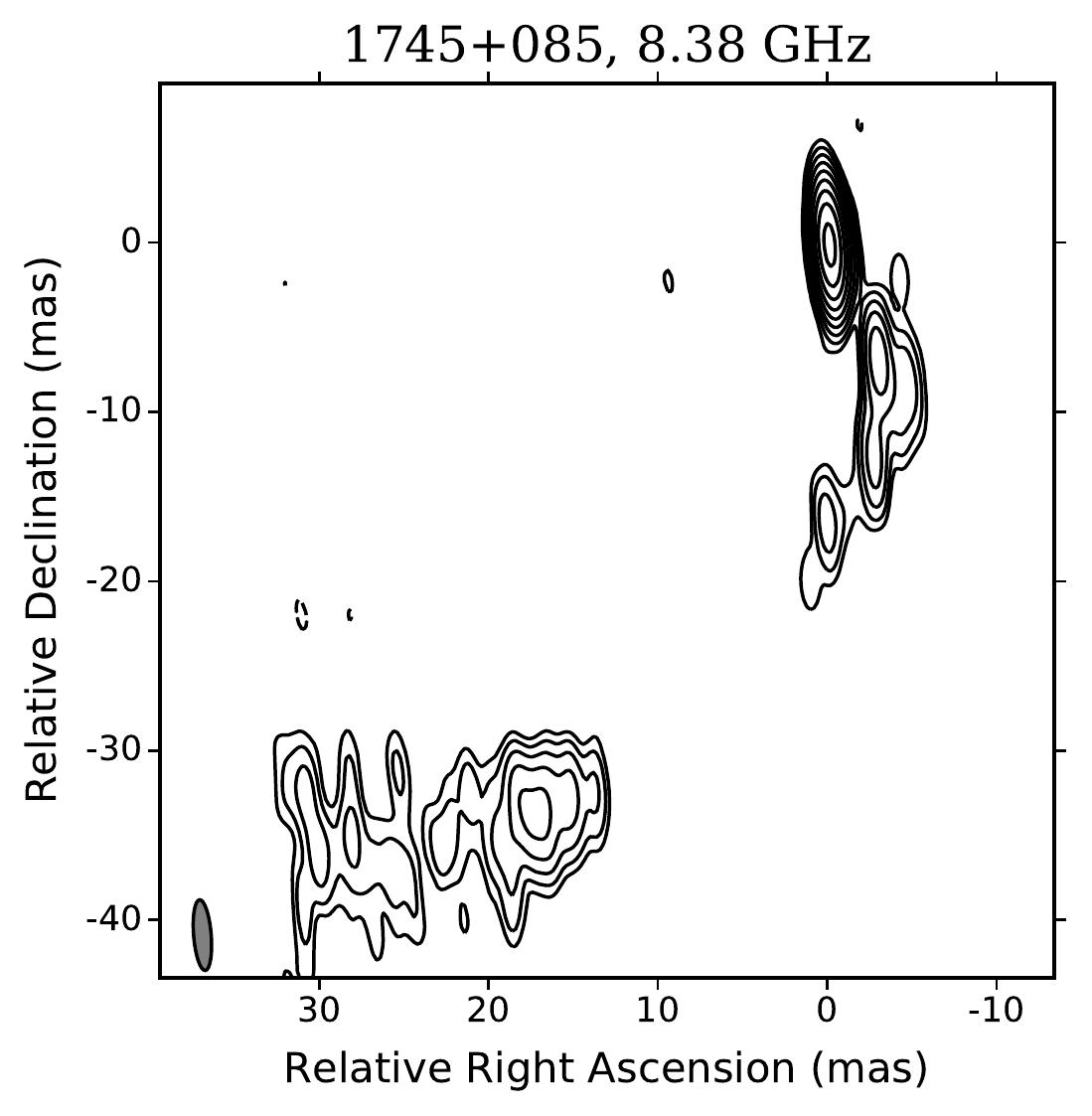}
  \includegraphics[width=0.3\textwidth]{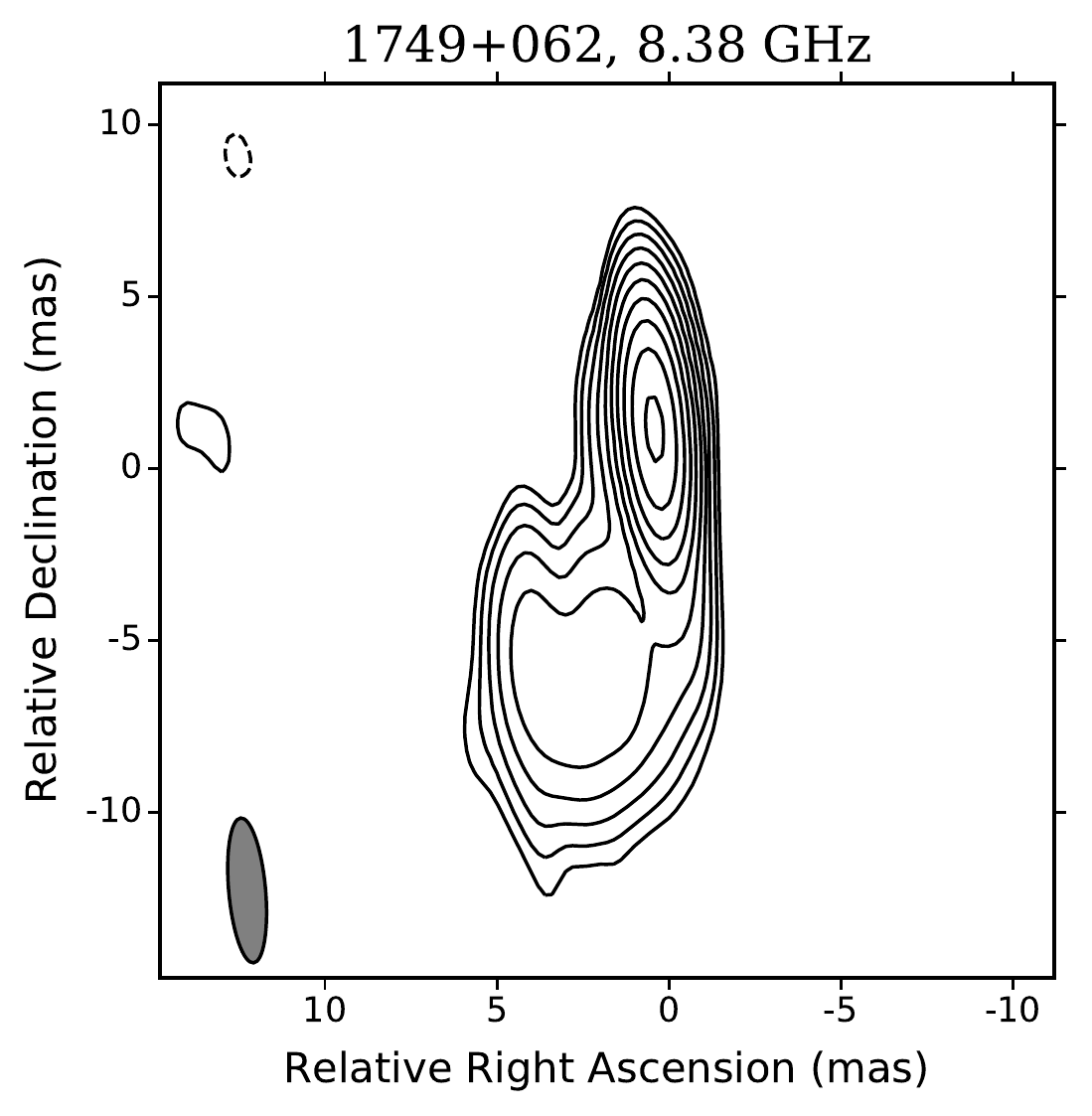}

  \caption{Continued.}
\end{figure*}

\addtocounter{figure}{-1}
\begin{figure*}[p!]

  \includegraphics[width=0.3\textwidth]{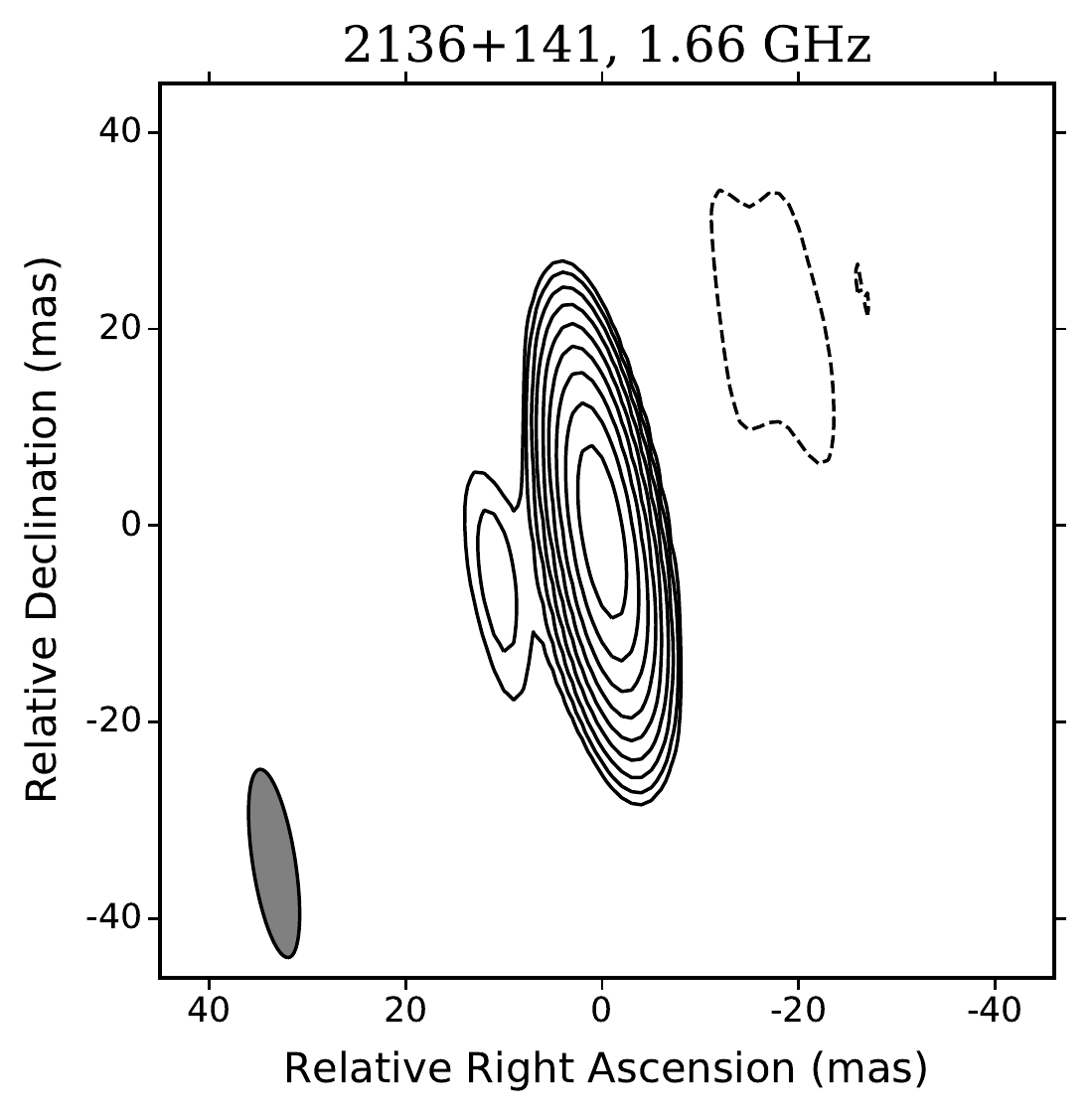}
  \includegraphics[width=0.3\textwidth]{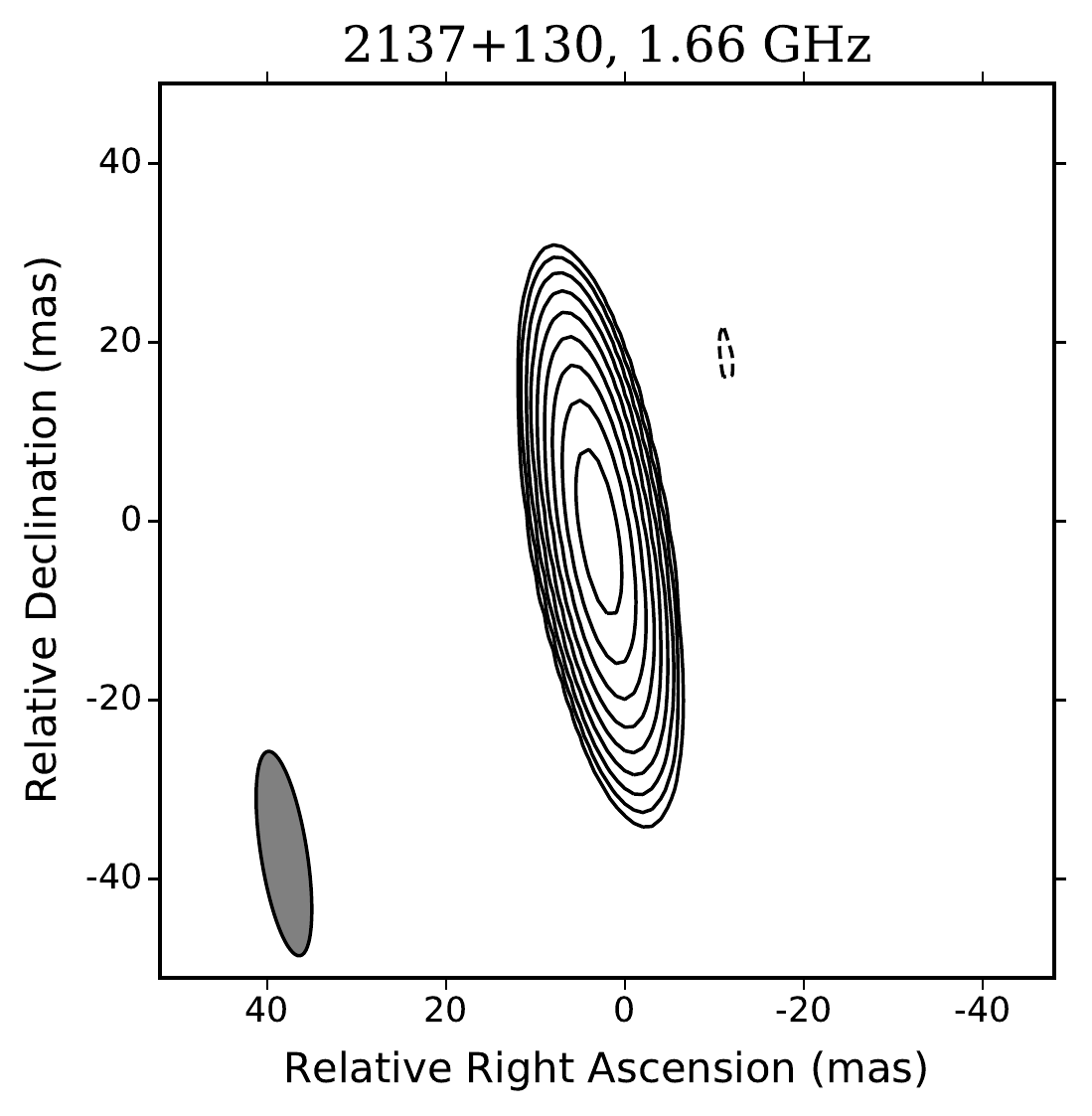}
  \includegraphics[width=0.3\textwidth]{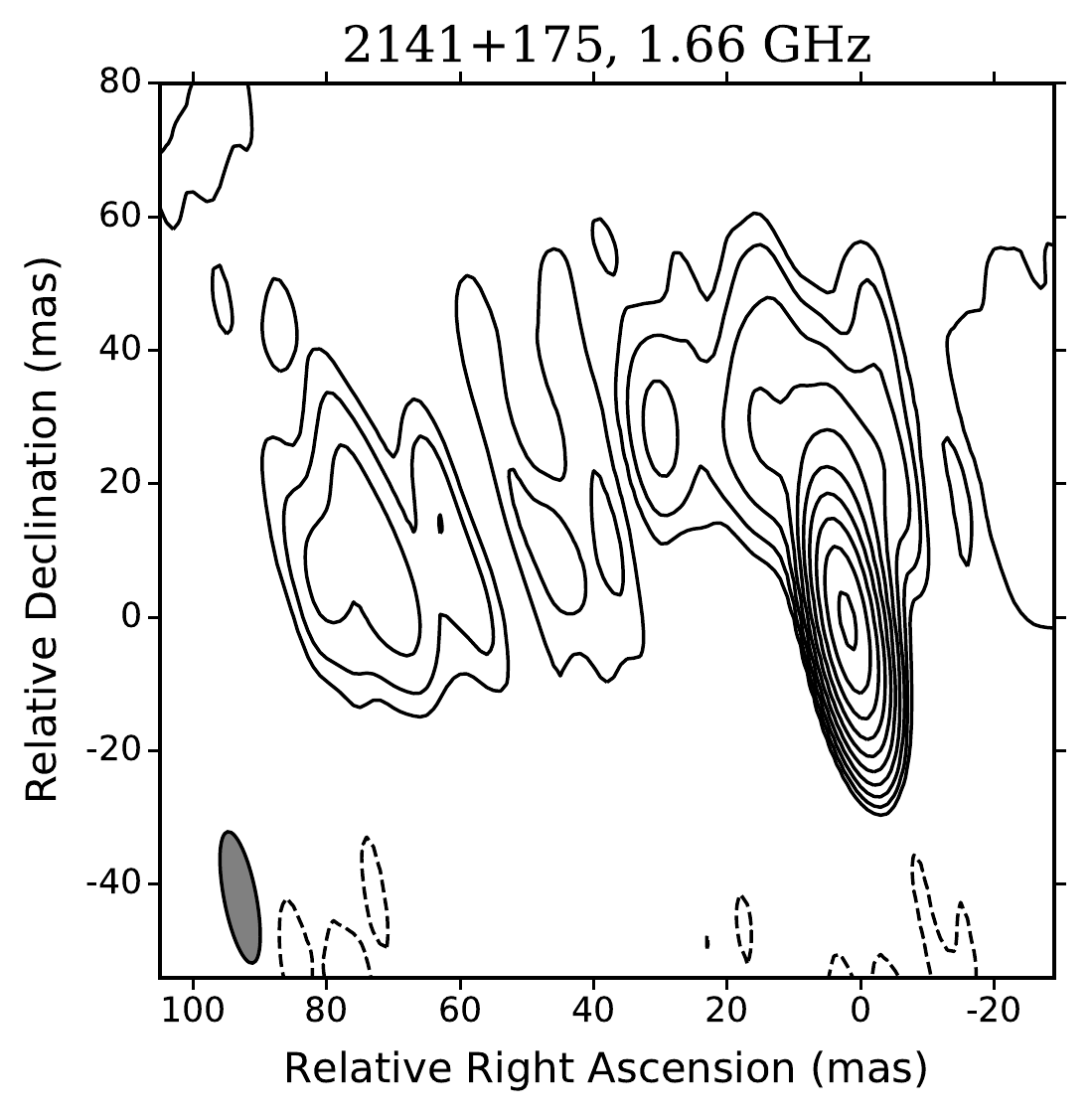}

  \includegraphics[width=0.3\textwidth]{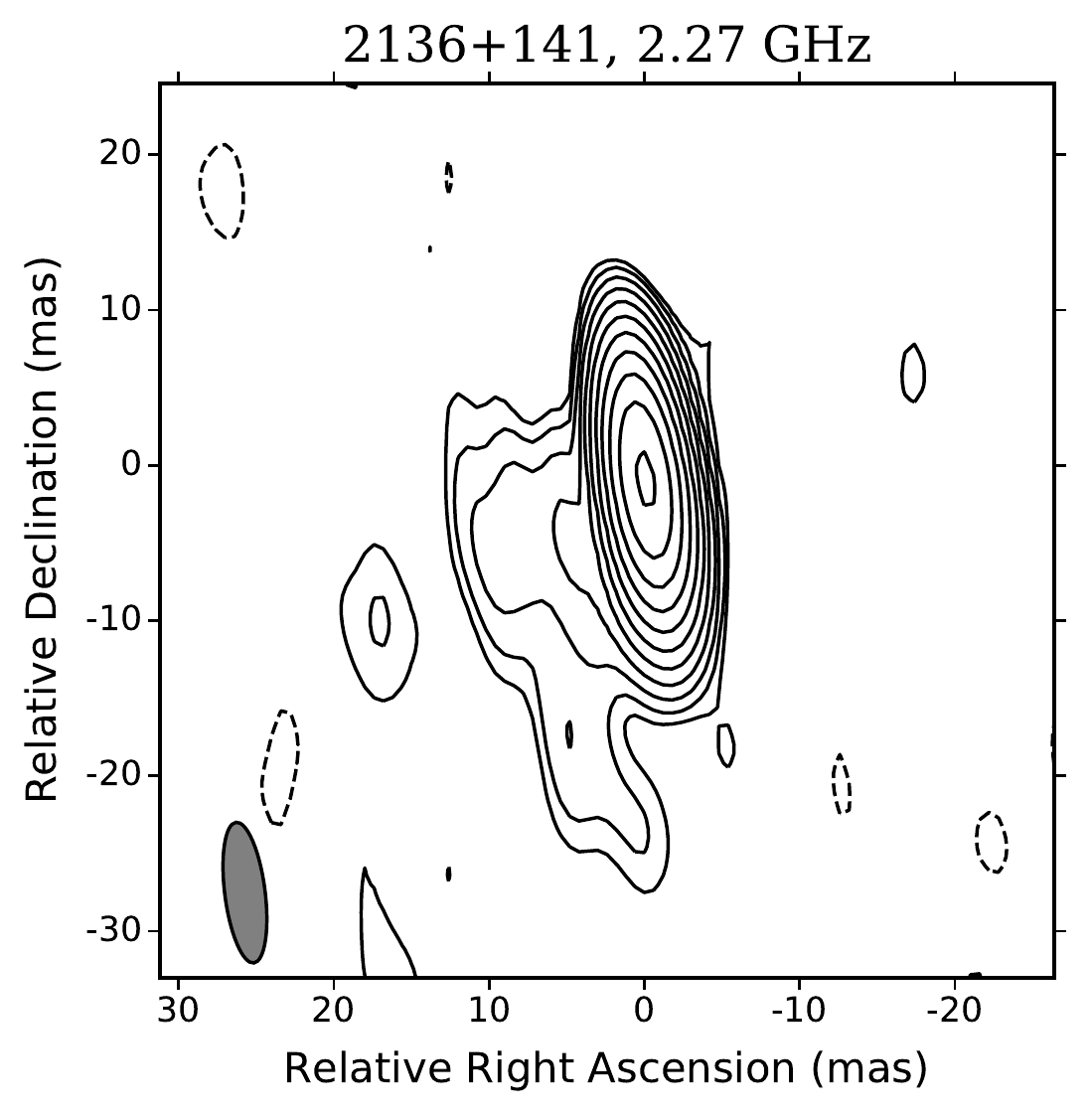}
  \includegraphics[width=0.3\textwidth]{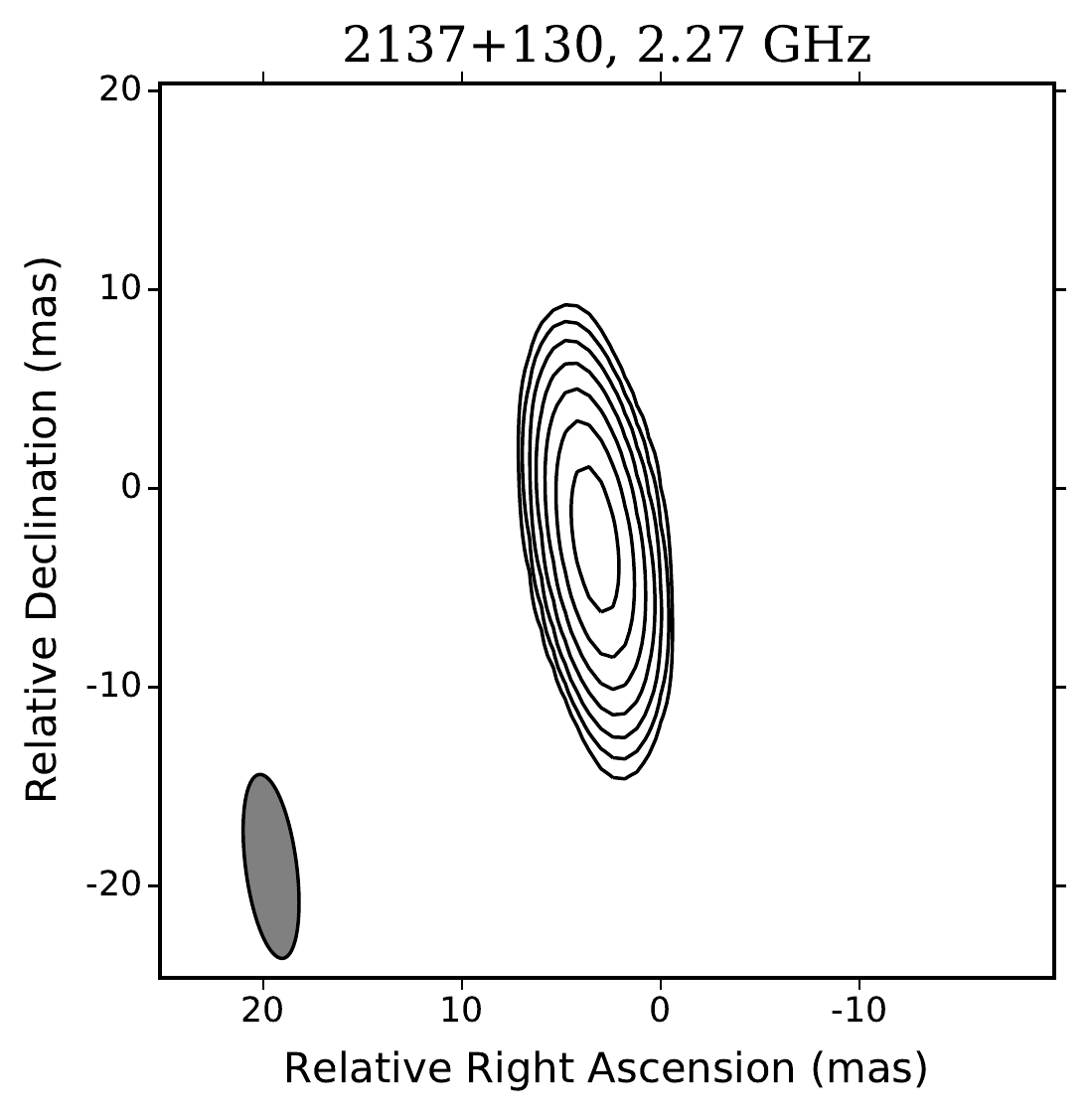}
  \includegraphics[width=0.3\textwidth]{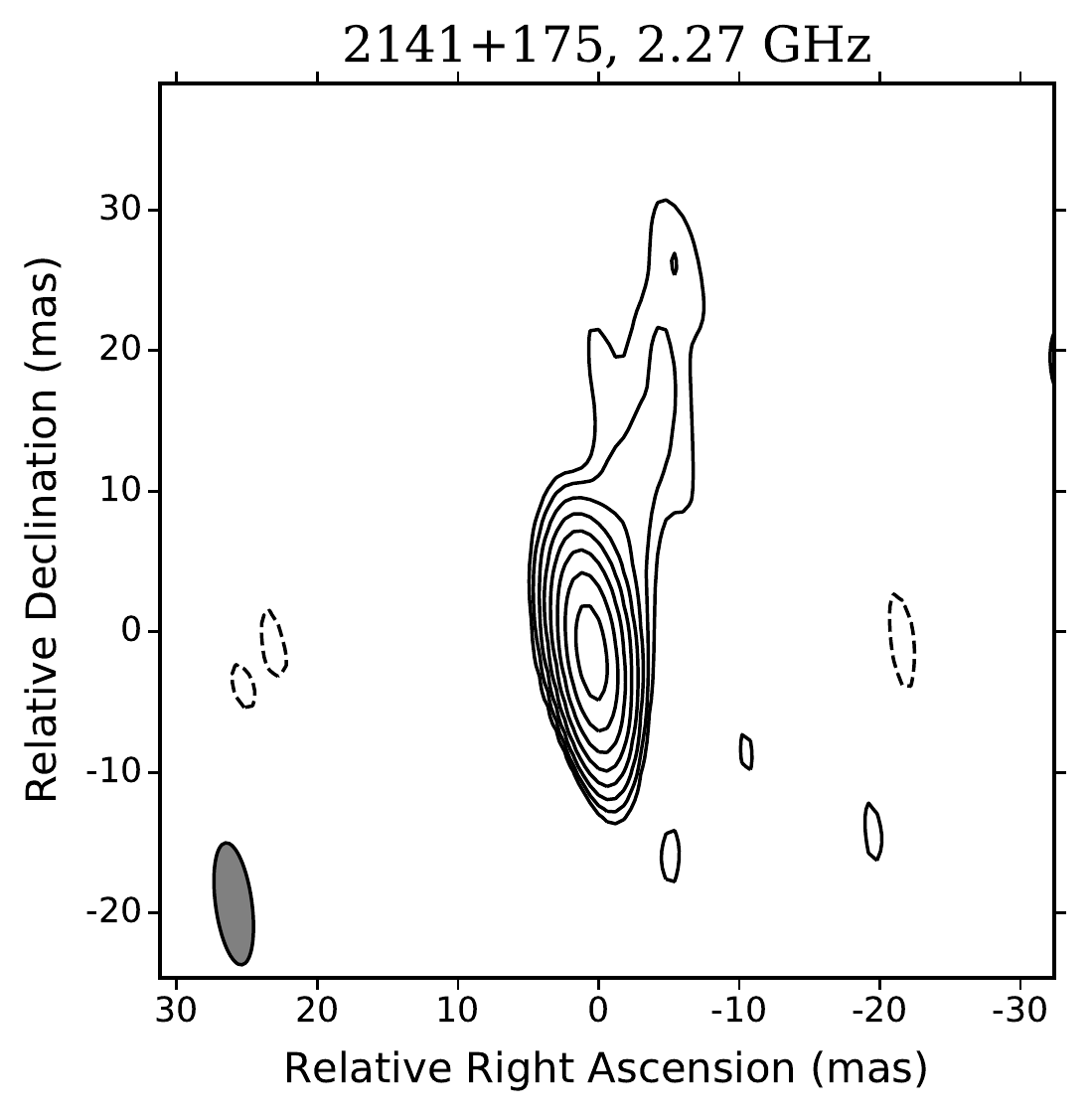}

  \includegraphics[width=0.3\textwidth]{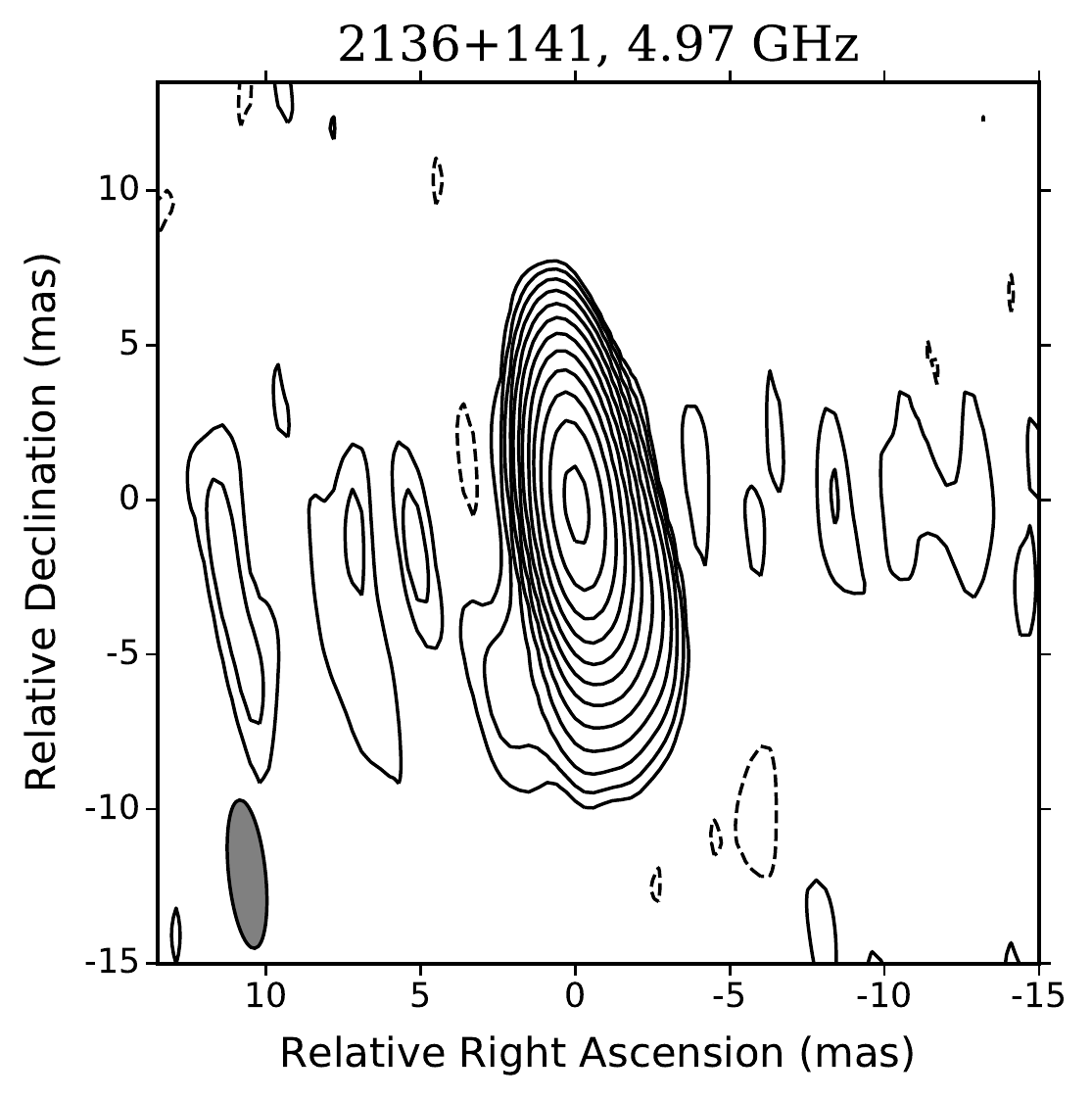}
  \includegraphics[width=0.3\textwidth]{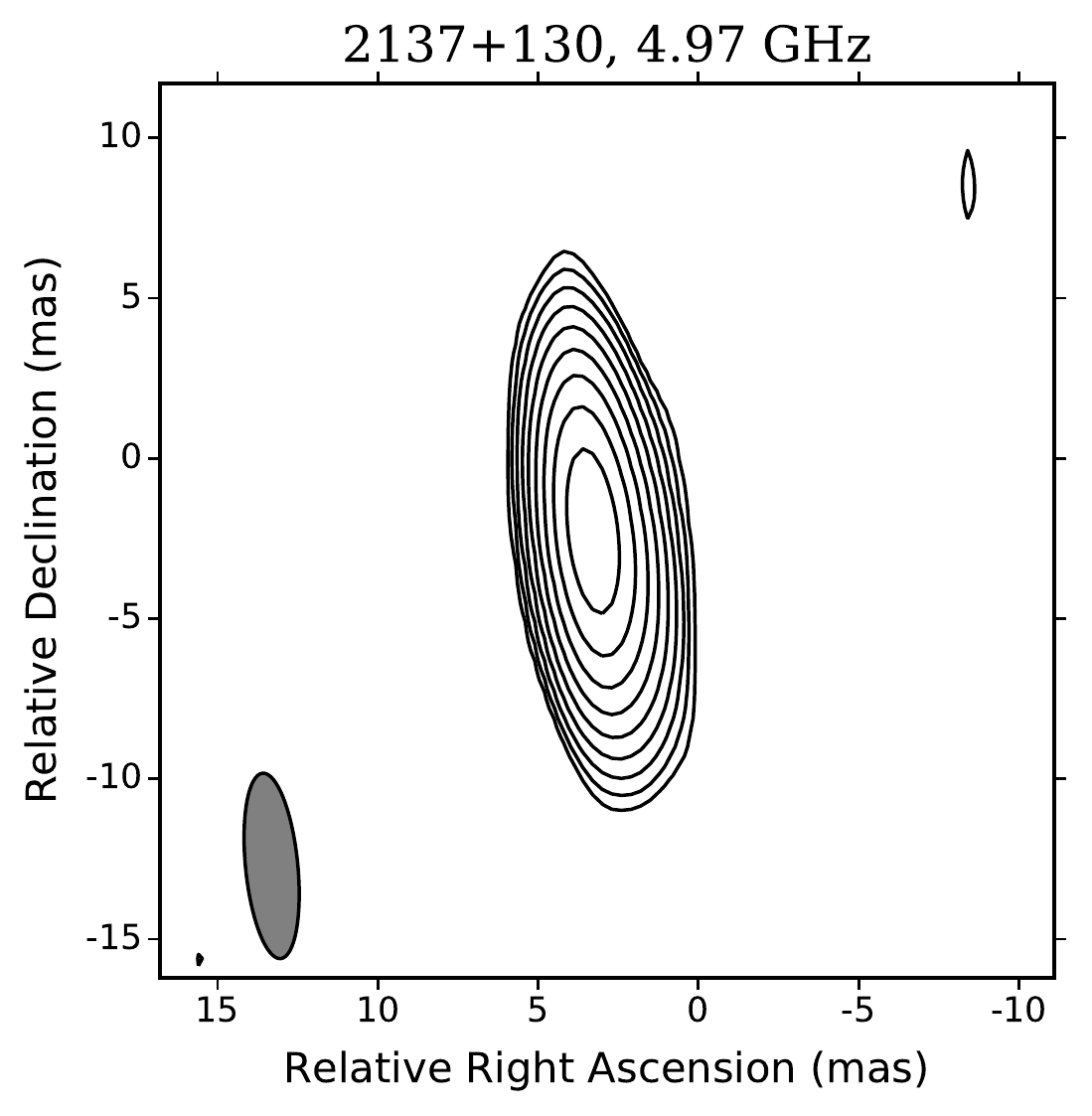}
  \includegraphics[width=0.3\textwidth]{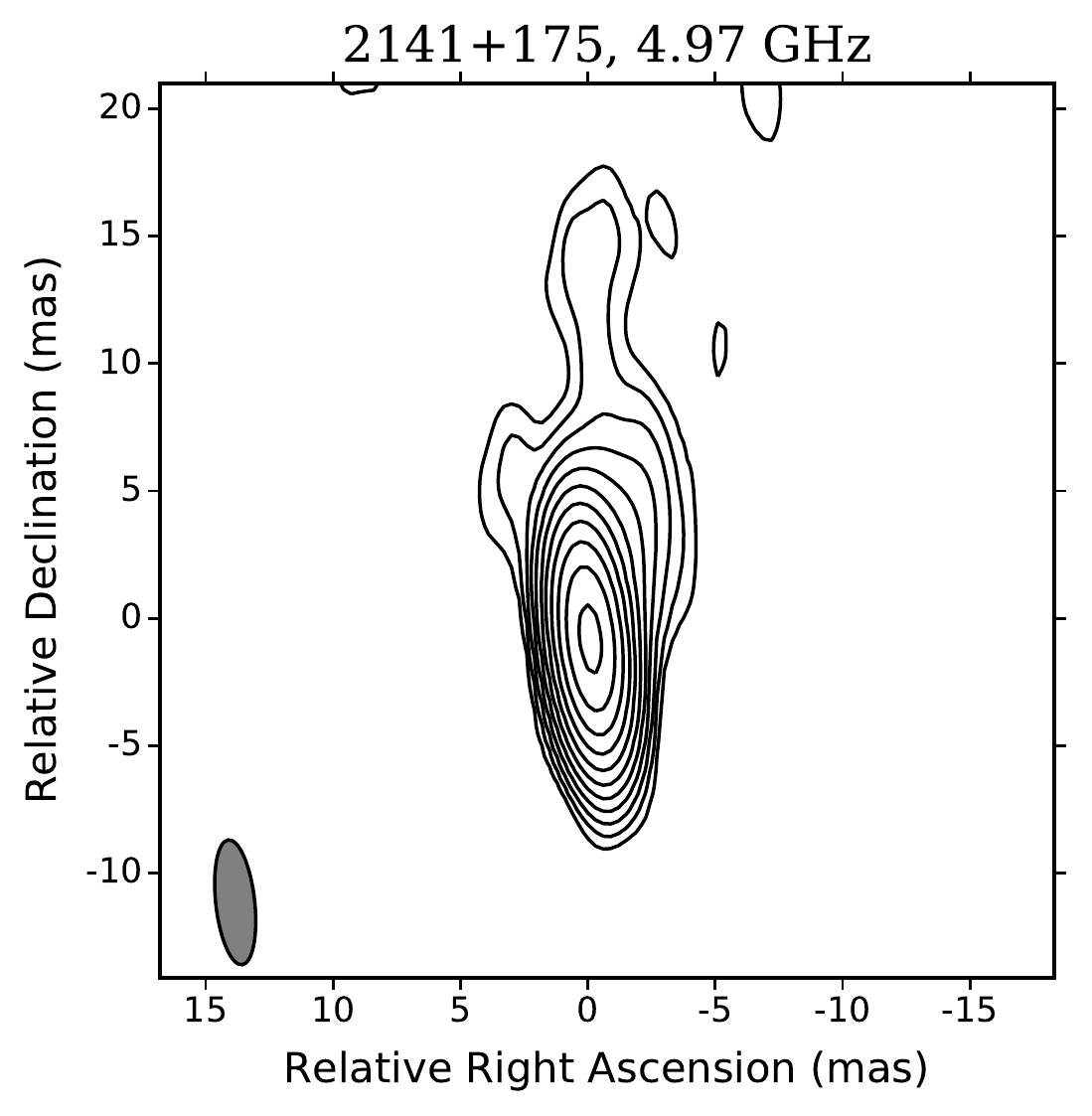}

  \includegraphics[width=0.3\textwidth]{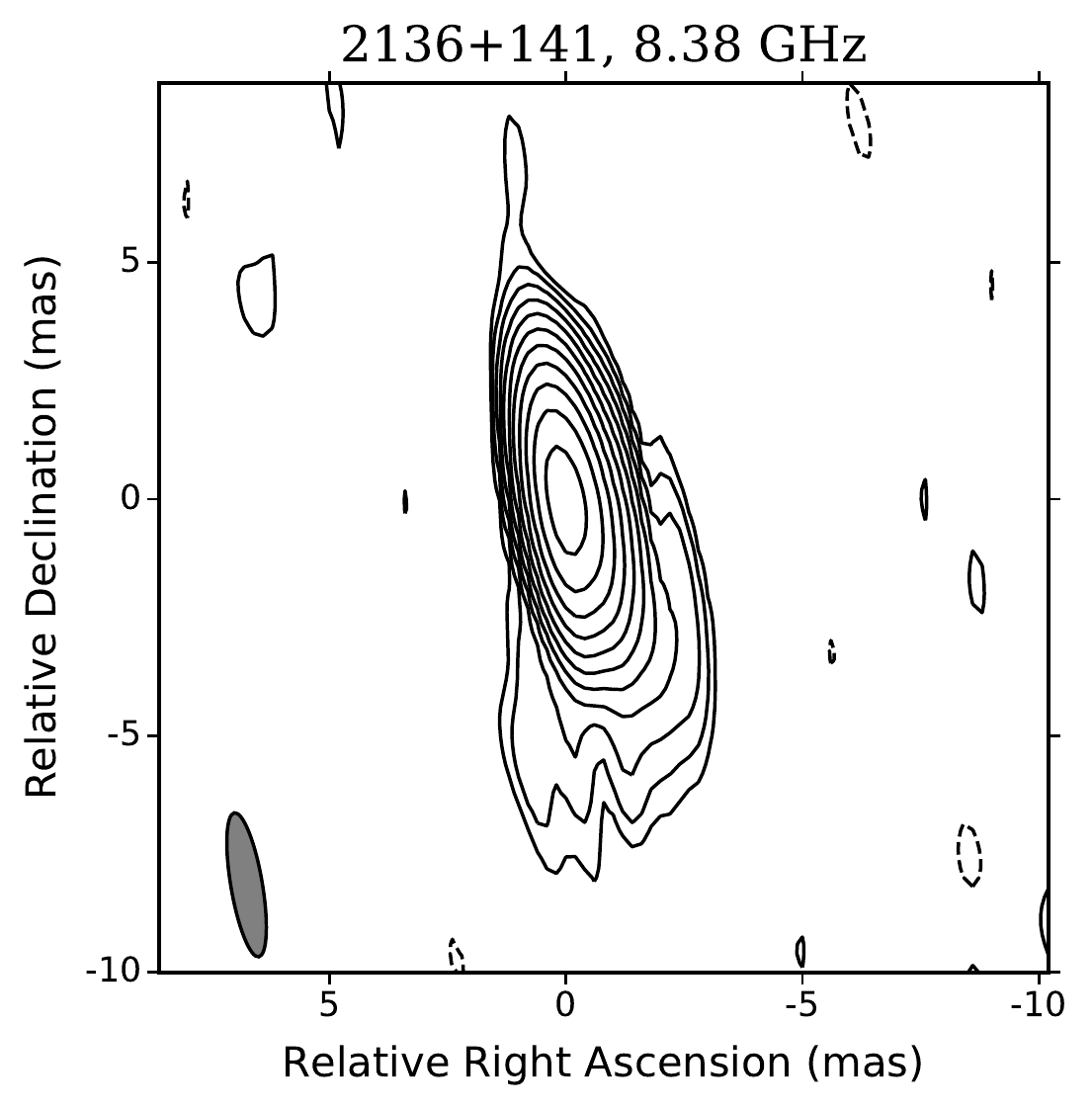}
  \includegraphics[width=0.3\textwidth]{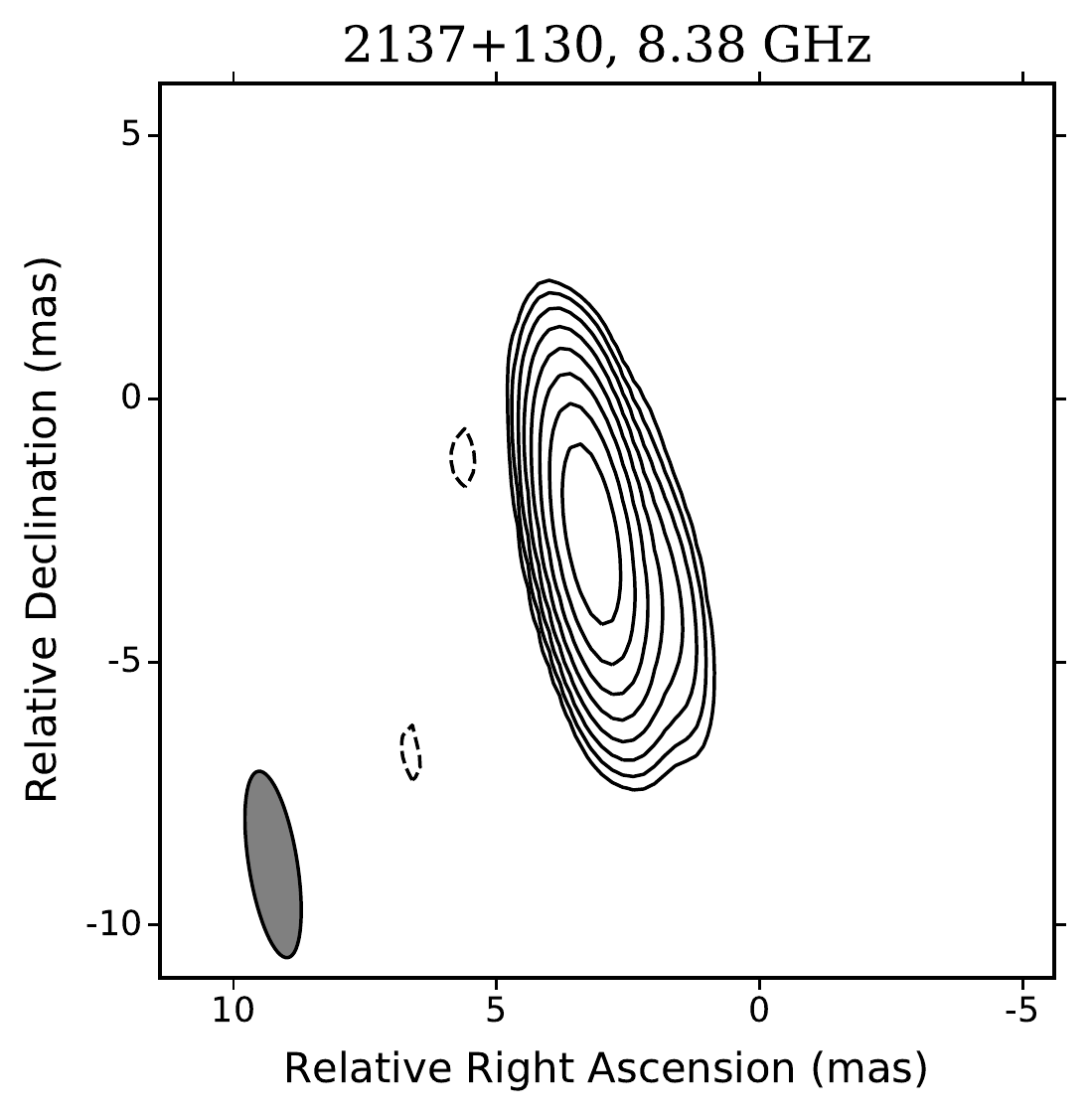}
  \includegraphics[width=0.3\textwidth]{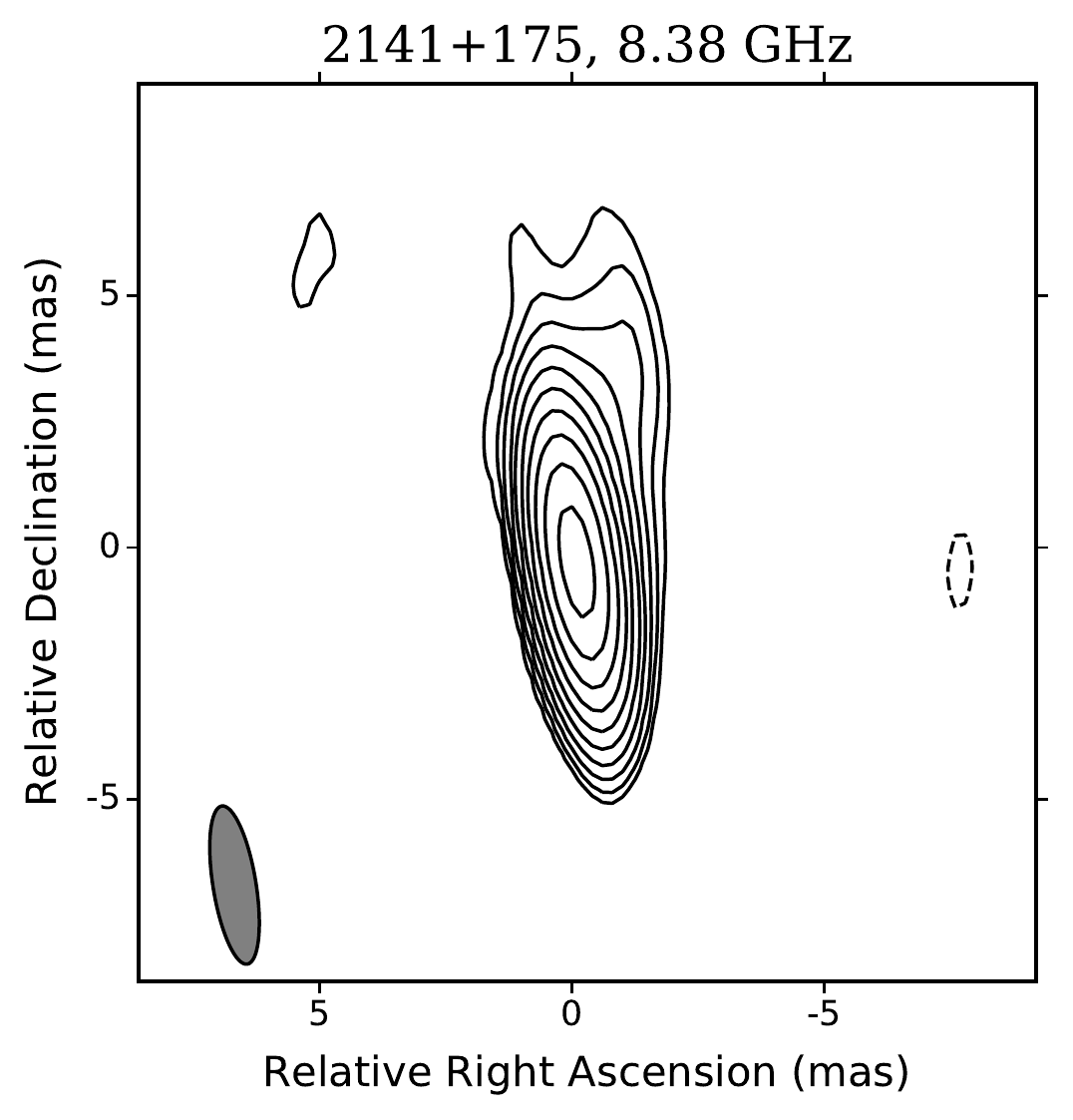}

  \caption{Continued.}
\end{figure*}

\addtocounter{figure}{-1}
\begin{figure*}[p!]

  \includegraphics[width=0.3\textwidth]{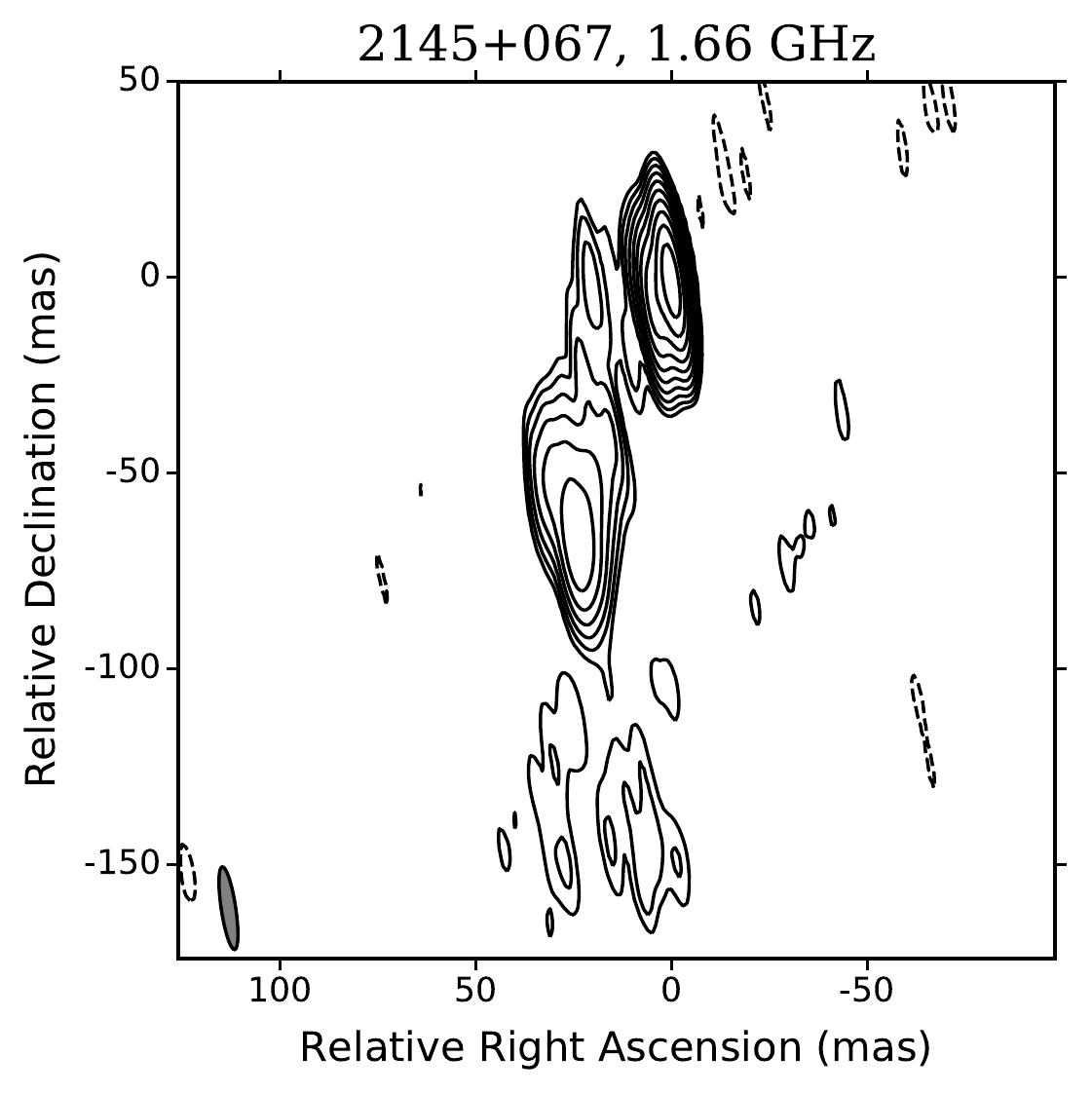}
  \includegraphics[width=0.3\textwidth]{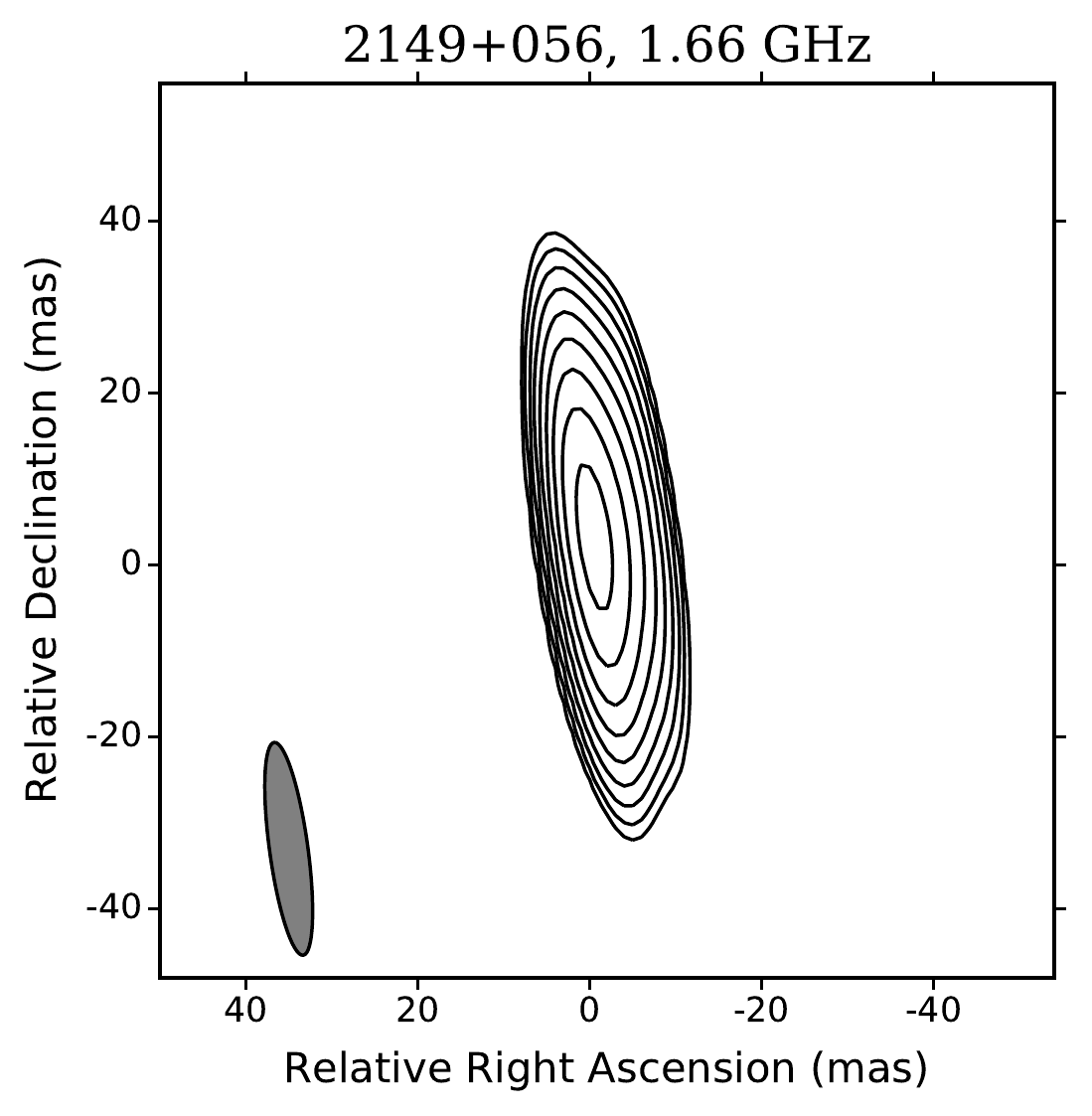}
  \includegraphics[width=0.3\textwidth]{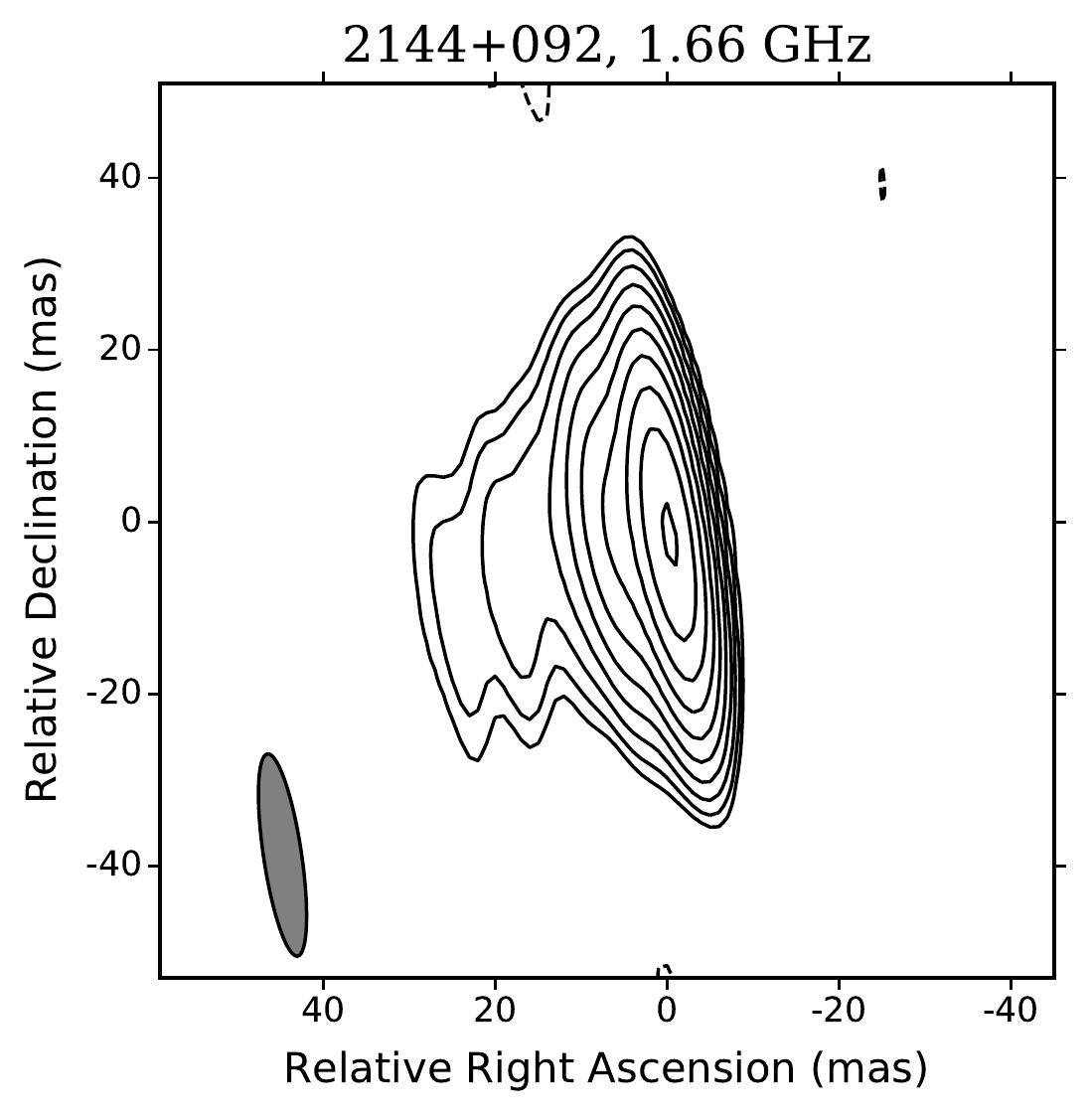}

  \includegraphics[width=0.3\textwidth]{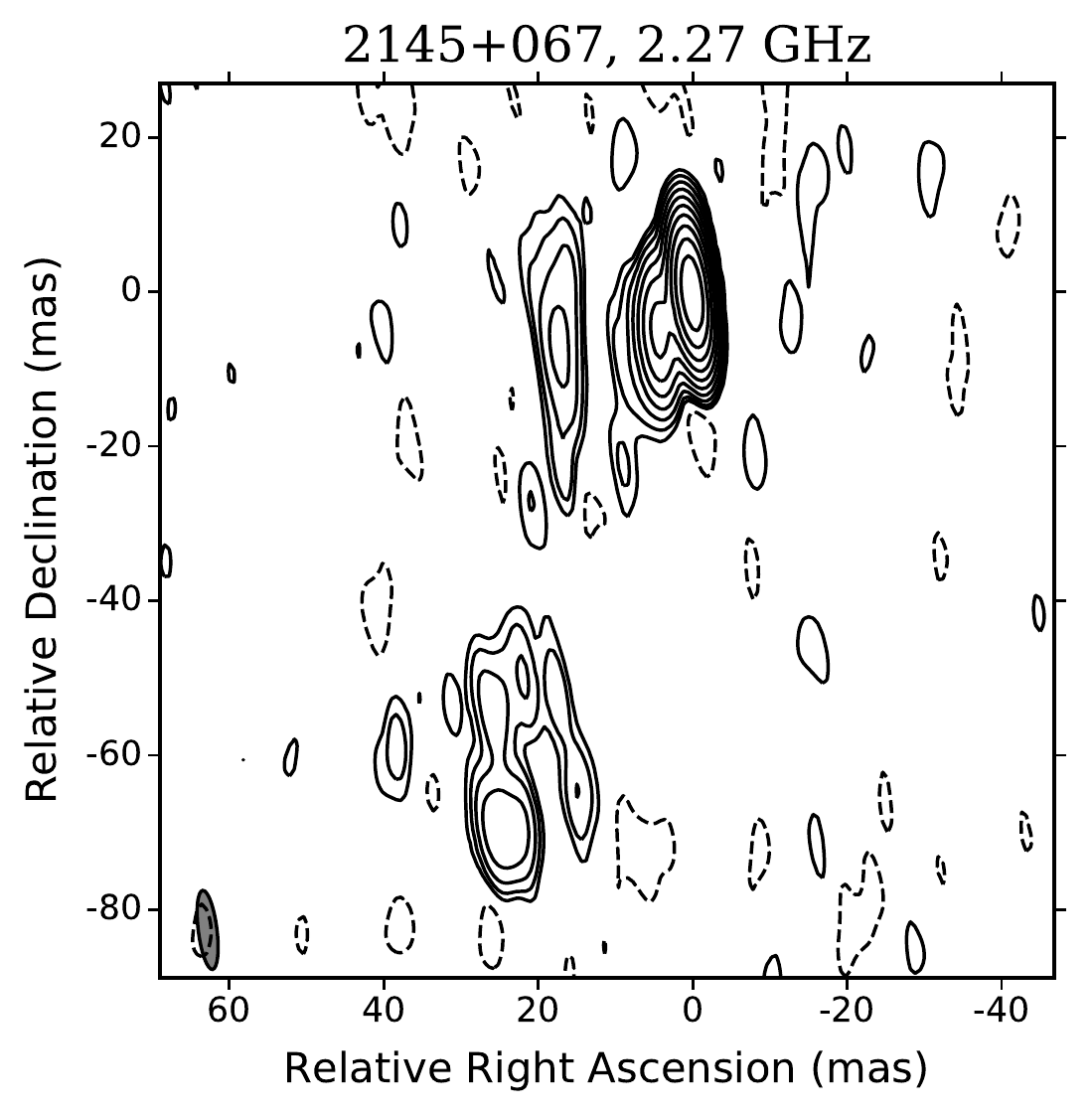}
  \includegraphics[width=0.3\textwidth]{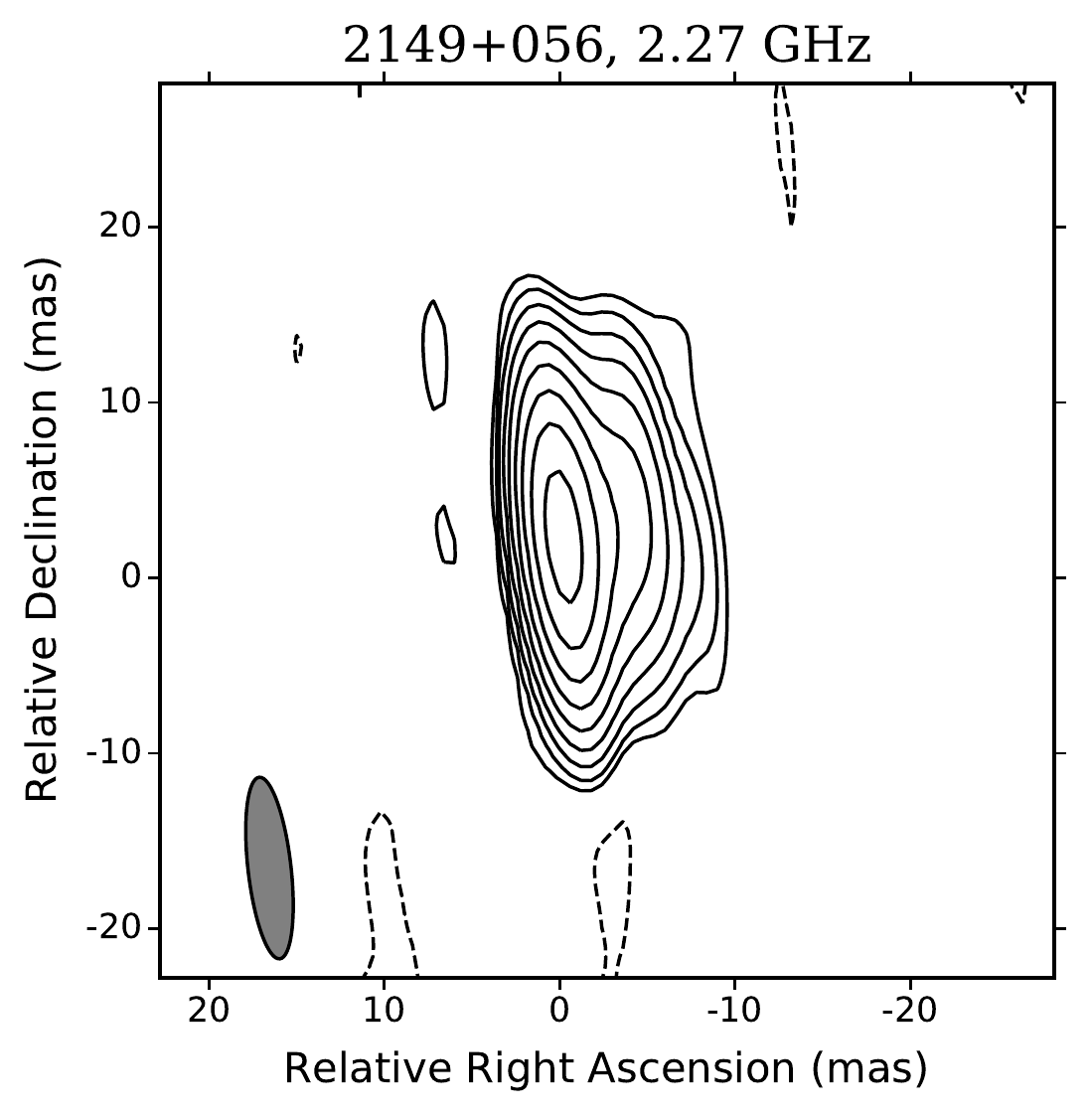}
  \includegraphics[width=0.3\textwidth]{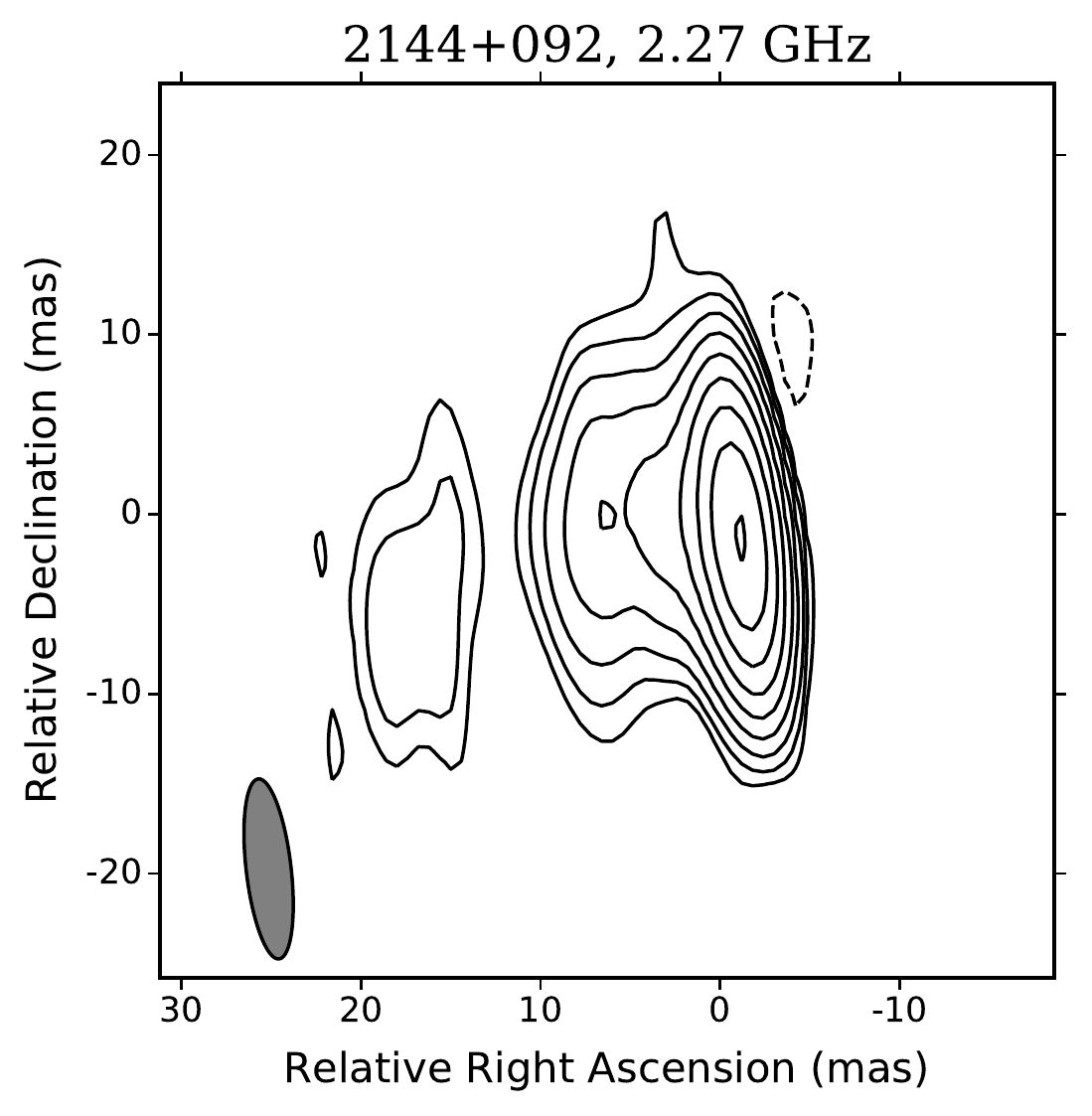}

  \includegraphics[width=0.3\textwidth]{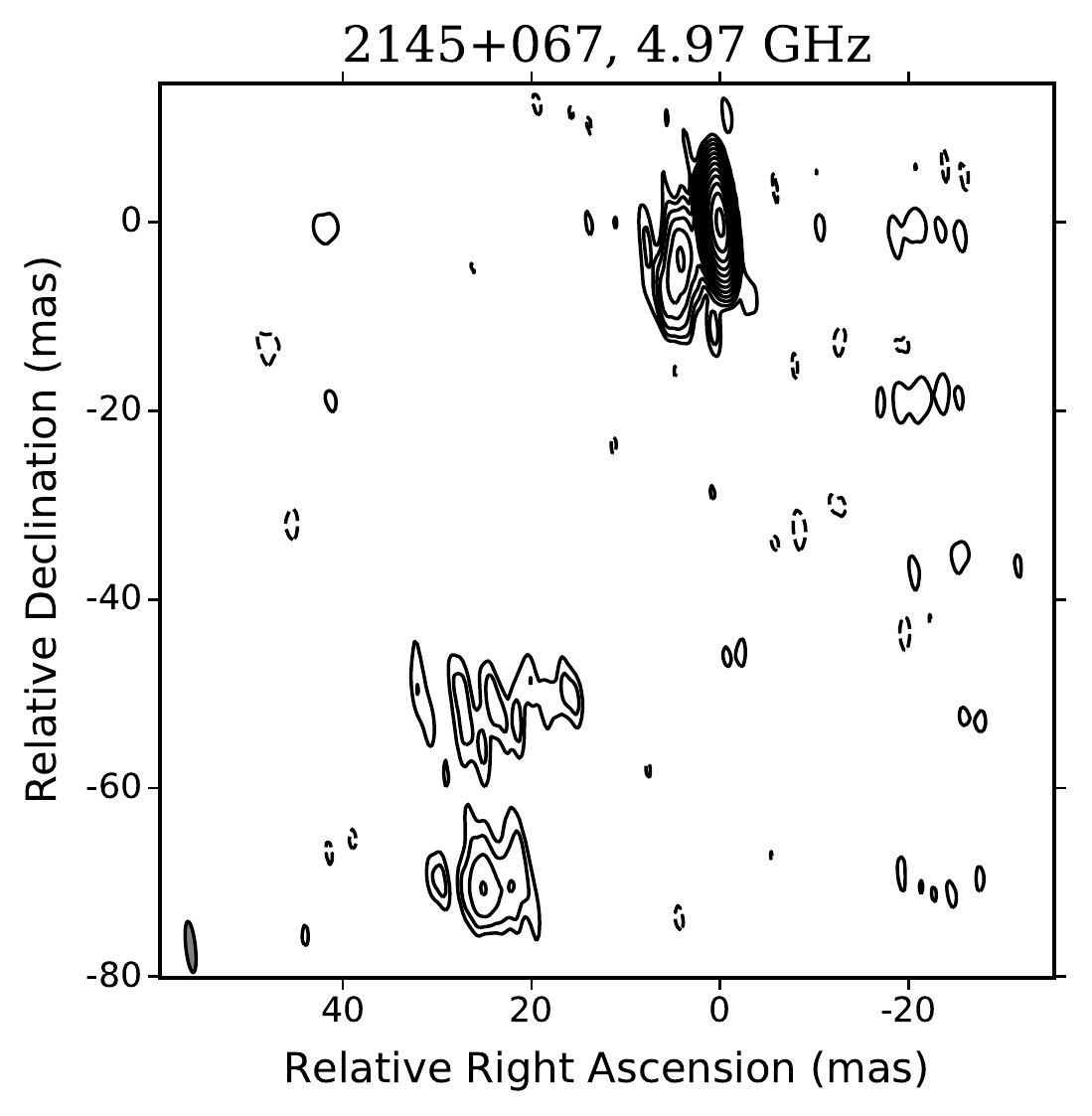}
  \includegraphics[width=0.3\textwidth]{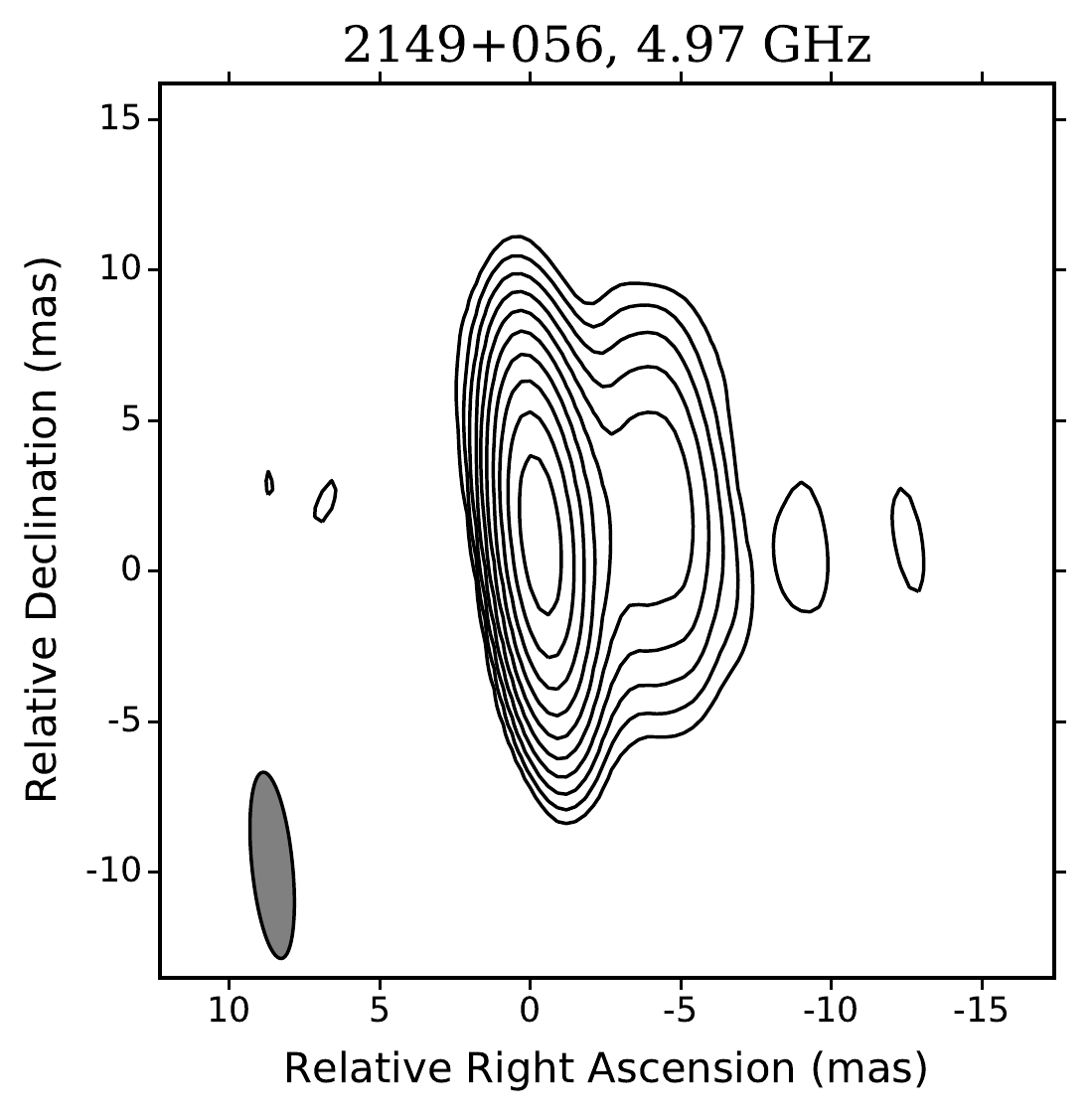}
  \includegraphics[width=0.3\textwidth]{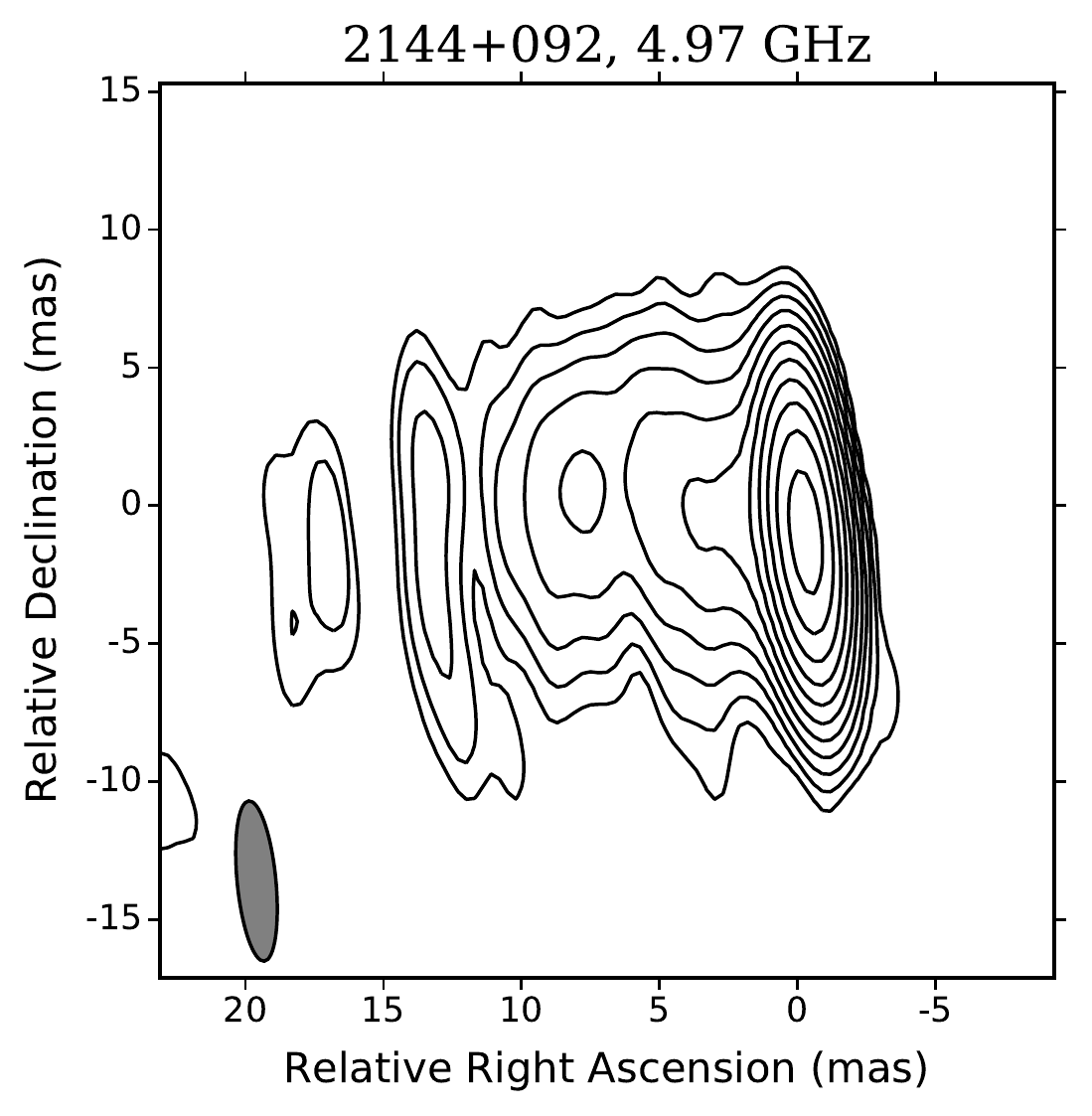}

  \includegraphics[width=0.3\textwidth]{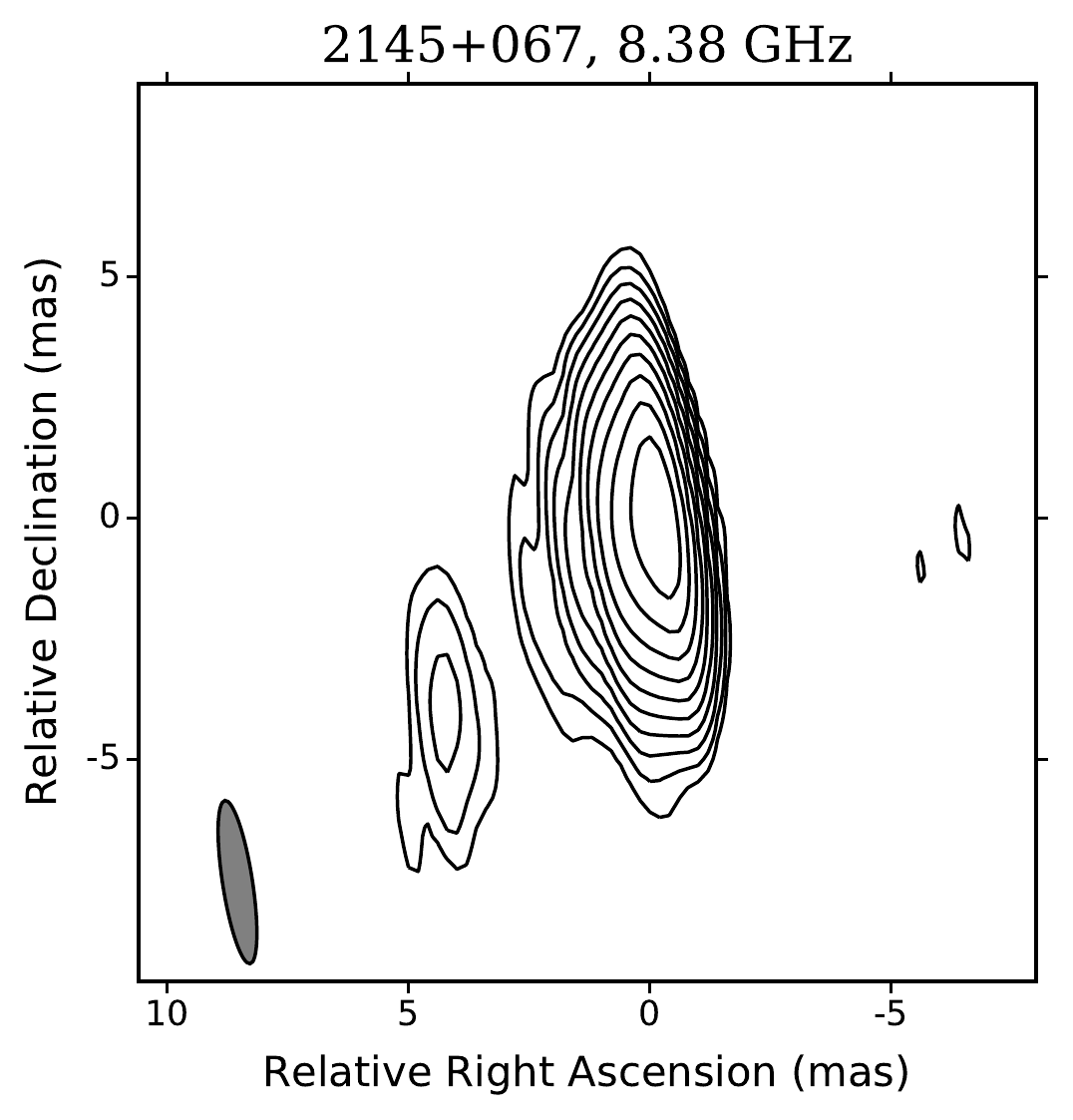}
  \includegraphics[width=0.3\textwidth]{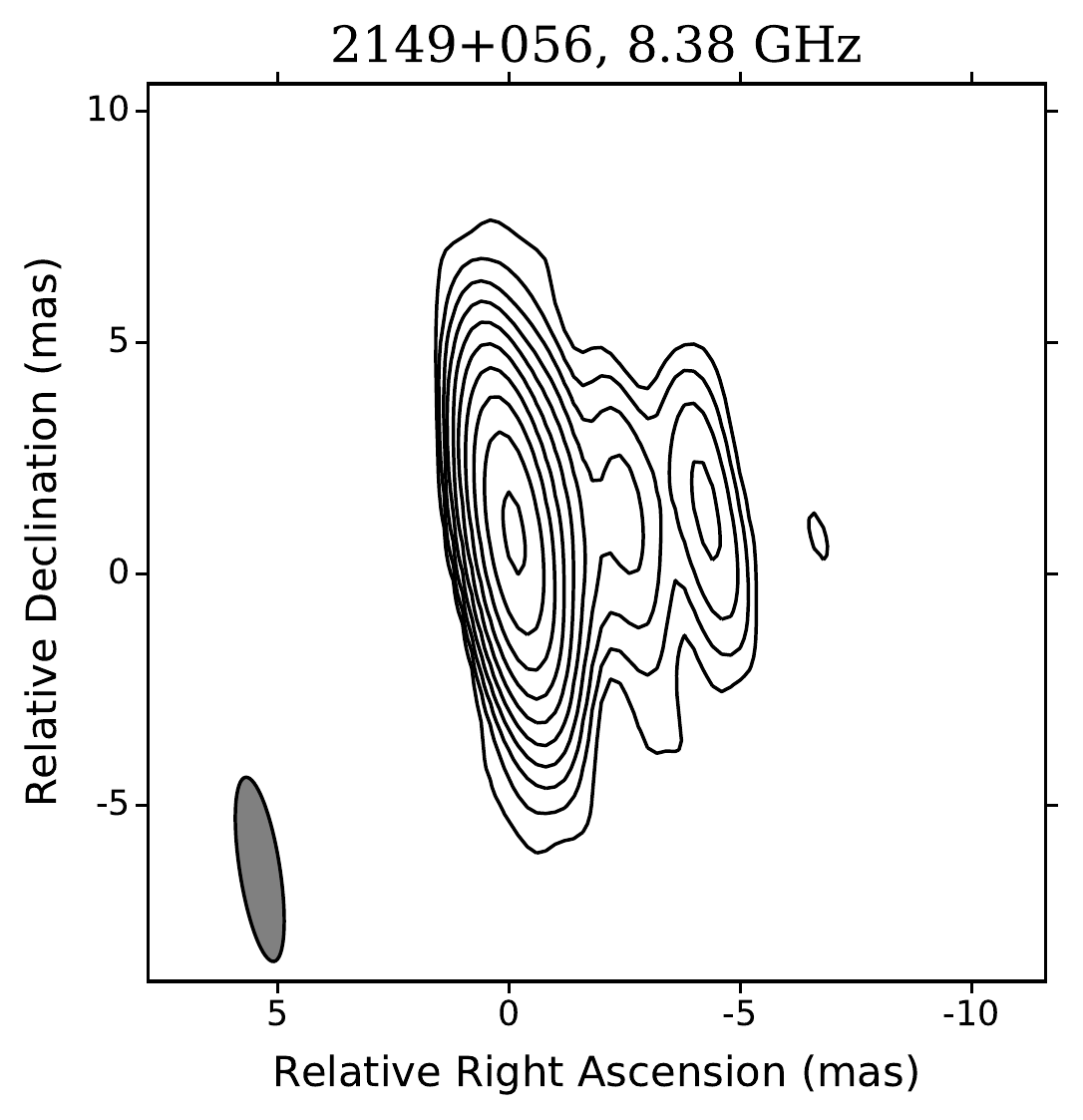}
  \includegraphics[width=0.3\textwidth]{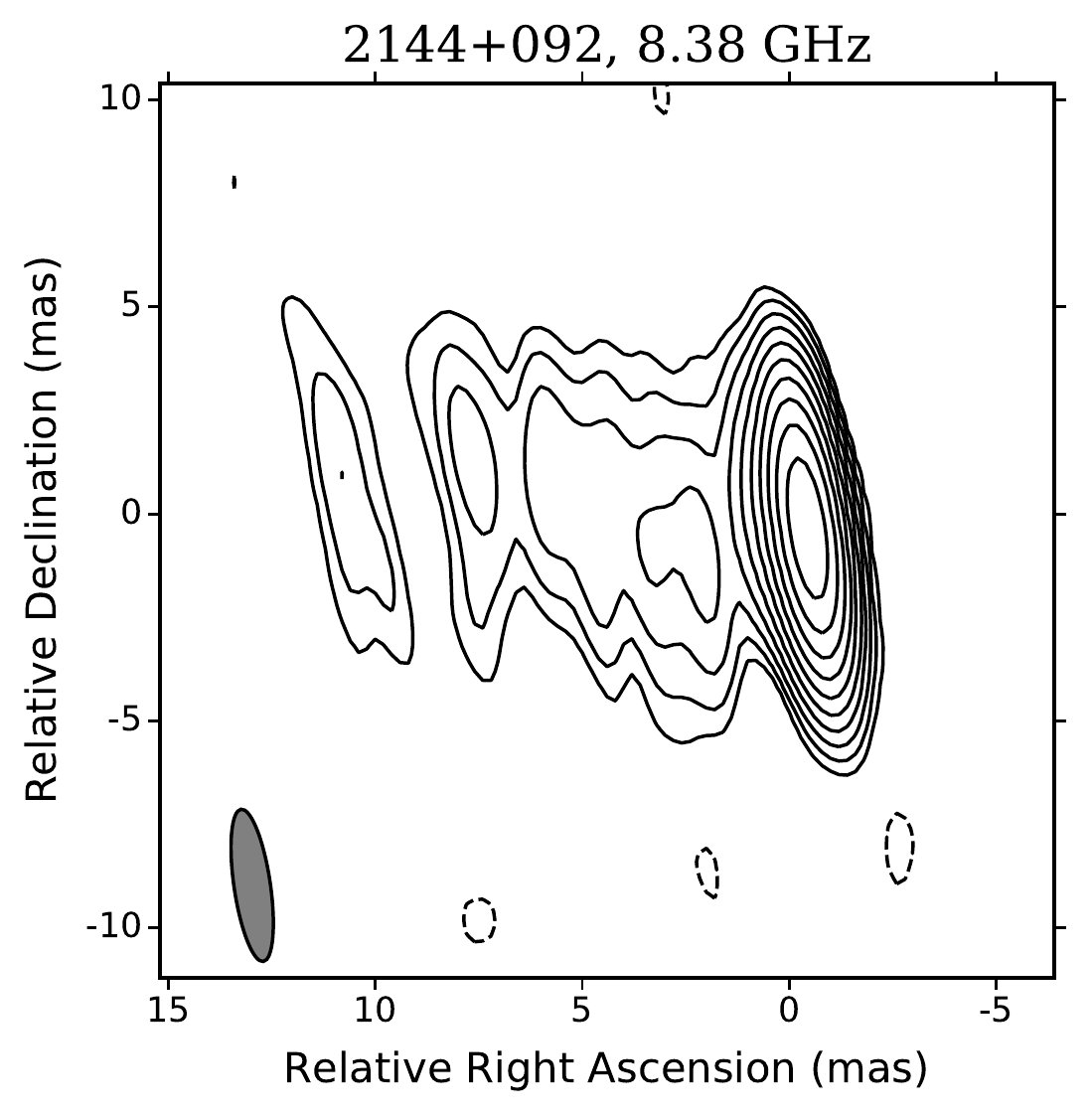}

  \caption{Continued.}
\end{figure*}

\end{document}